\documentclass[eqsecnum,
 reprint,
superscriptaddress,
 amsmath,amssymb,
 aps,
]{revtex4-2}

\usepackage{graphicx}
\usepackage{dcolumn}
\usepackage{bm}

\usepackage{url}

\DeclareMathAlphabet{\pazocal}{OMS}{zplm}{m}{n}
\usepackage{braket}
\usepackage{amsmath}
\usepackage{amssymb}
\usepackage{setspace}
\usepackage{leftidx}
\usepackage{hyperref}
\usepackage[colorinlistoftodos]{todonotes}
\usepackage{bm}
\usepackage{slashed}
\usepackage{calrsfs}
\usepackage{cancel}
\usepackage{dirtytalk}
\usepackage{calrsfs}
\usepackage{cancel}
\usepackage{dirtytalk}
\usepackage{upgreek}
\usepackage{hyperref}

\begin{document}

\title{(Non-)unitarity of strictly and partially massless fermions on de Sitter space II: an explanation based on the group-theoretic properties of the spin-3/2 and spin-5/2 eigenmodes}

\author{Vasileios A. Letsios}%
 \email{vasileios.letsios@kcl.ac.uk}
 
\affiliation{%
Department of Mathematics, University of York\\ Heslington, York, YO10 5DD, UK}  \affiliation{Present address: Department of Mathematics, King’s College London,\\  Strand, London WC2R 2LS, UK}
%


\begin{abstract}
In our previous article [Letsios 2023 J. High Energ. Phys. JHEP05(2023)015], we showed that the strictly massless spin-3/2 field, as well as the strictly and partially massless spin-5/2 fields, on $N$-dimensional ($N \geq 3 $) de Sitter spacetime ($dS_{N}$) are non-unitary unless $N=4$. The (non-)unitarity was demonstrated by simply observing that there is a (mis-)match between the representation-theoretic labels that correspond to the Unitary Irreducible Representations (UIR's) of the de Sitter (dS) algebra spin$(N,1)$ and the ones corresponding to the space of eigenmodes of the field theories. In this paper, we provide a technical representation-theoretic explanation for this fact by studying the (non-)existence of positive-definite, dS invariant scalar products for the spin-3/2 and spin-5/2 strictly/partially massless eigenmodes on $dS_{N}$ ($N \geq 3$). Our basic tool is the examination of the action of spin$(N,1)$ generators on the space of eigenmodes, leading to the following findings. For odd $N$, any dS invariant scalar product is identically zero. For even $N > 4$, any dS invariant scalar product must be indefinite. This gives rise to positive-norm and negative-norm eigenmodes that mix with each other under spin$(N,1)$ boosts. In the $N=4$ case, the positive-norm sector decouples from the negative-norm sector and each sector separately forms a UIR of spin$(4,1)$. Our analysis makes extensive use of the analytic continuation of tensor-spinor spherical harmonics on the $N$-sphere ($S^{N}$) to $dS_{N}$ and also introduces representation-theoretic techniques that are absent from the mathematical physics literature on half-odd-integer-spin fields on $dS_{N}$.

\end{abstract}

\maketitle
\begin{widetext}
\section{Introduction} \label{Introduction}
This is a technical sequel to our previous paper~\cite{Letsios_announce}, in which we constructed a `field theory-representation theory' dictionary for totally-symmetric spin-$s = 3/2, 5/2$ tensor-spinors on $N$-dimensional ($N \geq 3$) de Sitter spacetime ($dS_{N}$). Totally symmetric tensor-spinors, $\Psi_{\mu_{1}...\mu_{r}}$, of spin $s \equiv r+1/2$ on $dS_{N}$ satisfy~\cite{Deser_Waldron_null_propagation,Deser_Waldron_ArbitrarySR}
\begin{align}
   &\left( \slashed{\nabla}+M\right)\Psi_{\mu_{1}...\mu_{r}}=0  \label{Dirac_eqn_fermion_dS}\\
   & \nabla^{\alpha}\Psi_{\alpha \mu_{2}...\mu_{r}}=0, \hspace{4mm}  \gamma^{\alpha}\Psi_{\alpha \mu_{2}...\mu_{r}}=0, \label{TT_conditions_fermions_dS}
\end{align}
where $\slashed{\nabla}=\gamma^{\nu}\nabla_{\nu}$ is the Dirac operator on $dS_{N}$. {As the tensor-spinors $\Psi_{\mu_{1}...\mu_{r}}$ are totally symmetric in their tensor indices, by contracting the gamma-tracelessness condition in eq.~(\ref{TT_conditions_fermions_dS}) with $\gamma^{\mu_{2}}$ and using $ \gamma^{\mu_{2}}  \gamma^{\alpha} +  \gamma^{\alpha} \gamma^{\mu_{2}}  = 2\, g^{\mu_{2}  \alpha}$, we find that the tensor-spinors are also traceless (with respect to any two tensor indices).} (See Subsection~\ref{Subsection_spinors_and_clifford} for our convention for the gamma matrices.) From now on, we will refer to the divergence-free and gamma-tracelessness conditions in eq.~(\ref{TT_conditions_fermions_dS}) as the TT conditions.

\noindent \textbf{Main results of our previous paper.}~In our previous article~\cite{Letsios_announce}, we constructed the spin-$s=3/2,5/2$ eigenmodes of eqs.~(\ref{Dirac_eqn_fermion_dS}) and~(\ref{TT_conditions_fermions_dS}) on global $dS_{N}$ ($N \geq 3$). Then, we investigated the (mis-)match between the representation-theoretic labels that correspond to the Unitary Irreducible Representations (UIR's) of the de Sitter (dS) algebra, spin$(N,1)$, and the ones corresponding to the eigenmodes. We found that for real values of $M$ the representations are unitary. However, the main interesting result of Ref.~\cite{Letsios_announce} concerns the strictly and partially massless theories (i.e. the theories that enjoy a gauge symmetry). In particular, the strictly and partially massless theories, for which the mass parameter is tuned to the special imaginary values~\footnote{The imaginary values of $M$ in eq.~(\ref{values_mass_parameter_masslessness_fermion}) imply that the action functional for strictly/partially massless half-odd-integer-spin theories on $dS_{N}$ is not hermitian. The fact that the strictly massless spin-3/2 field theory in de Sitter spacetime has an imaginary mass parameter had been already observed in cosmological supergravity~\cite{Nieuwenhuizen}.}$M= i \tilde{M}$~\cite{Deser_Waldron_ArbitrarySR}:
\begin{align}\label{values_mass_parameter_masslessness_fermion}
    \tilde{M}^{2}=-M^{2}=\left(r-\uptau+\frac{N-2}{2}\right)^{2} \hspace{6mm}(\uptau=1,...,r),
\end{align}
were found to be non-unitary, unless $N=4$. [The analysis of the previous~\cite{Letsios_announce}, as well as of the present, papers focuses only on the cases with $r=1$ and $r=2$. However, the `field theory-representation theory' dictionary of our previous paper~\cite{Letsios_announce} suggests that our main result extends to all strictly/partially massless (totally symmetric) tensor-spinors of spin $s \geq 7/2$.] The quantity $\uptau$ is known as the depth of the strictly/partially massless field (i.e. gauge potential). The value $\uptau =1$ corresponds to strict masslessness and the values $\uptau  =2,...,r$ to partial masslessness - see Refs.~\cite{STSHS, Yale_Thesis, Deser_Waldron_null_propagation, Deser_Waldron_stability_of_massive_cosm, Deser_Waldron_phases, Deser_Waldron_partial_masslessness, Deser_Waldron_Conformal, Deser_Waldron_ArbitrarySR} for background material concerning strict and partial masslessness. 

{The imaginary mass parameter of the strictly massless spin-3/2 field (gravitino) was already known in attempts to construct cosmological Supergravity~\cite{Nieuwenhuizen}. Let us present a quick way to understand the appearance of the imaginary mass parameter for the spin-3/2 fermionic gauge potential on $dS_{N}$. The Rarita-Schwinger field $V_{\mu}$ on $dS_{N}$ satisfies~\cite{Freedman}
$$ \gamma^{\mu \rho  \sigma}  \left(   \nabla_{\rho} + \frac{\kappa}{2}  \gamma_{\rho}   \right) V_{\sigma} = 0. $$
Using $\gamma^{\mu \rho \sigma} \gamma_{\rho} = -(N-2) \gamma^{\mu \sigma} = (N-2) ( -\gamma^{\mu} \gamma^{\sigma} + g^{\mu \sigma}   )$~\cite{Freedman}\footnote{Here we denote the $dS_{N}$ metric as $g_{\mu \nu}$.}, the Rarita-Schwinger equation is expressed as
$$ \gamma^{\mu \rho \sigma}\nabla_{\rho} V_{\sigma} -\frac{\kappa \,(N-2)}{2} \gamma^{\mu} \gamma^{\alpha} V_{\alpha} +\frac{\kappa \, (N-2)}{2} V^{\mu} = 0 $$
(here we do not impose TT conditions on $V_{\mu}$). For generic real values of the parameter $\kappa$ the field $V_{\mu}$ is massive - by ``massive'' we mean that it does not enjoy a gauge symmetry. Let us now require the invariance of the Rarita-Schwinger equation under gauge transformations of the form $$ \delta V_{\mu} = \left( \nabla_{\mu}+ \alpha \, \gamma_{\mu}   \right)  \epsilon,$$ where $\alpha$ is a parameter that we have to determine, while $\epsilon$ is an arbitrary spinor gauge function. We find that the equation indeed enjoys gauge symmetry for the imaginary values $\kappa=i$, $\alpha= \tfrac{i}{2}$, as well as for the values $\kappa=-i$, $\alpha=-\tfrac{i}{2}$. In other words, for these tunings of $\kappa$ and $\alpha$, the vector-spinor field $ V_{\mu}$ becomes a (fermionic) gauge potential, known as the strictly massless spin-3/2 field~\cite{Deser_Waldron_ArbitrarySR}. In the TT gauge, the field equation becomes $$ \slashed{\nabla} V_{\mu} \pm i \frac{N-2}{2} V_{\mu} = 0,$$which corresponds to eqs.~(\ref{Dirac_eqn_fermion_dS}), (\ref{TT_conditions_fermions_dS}) with mass parameter given by~(\ref{values_mass_parameter_masslessness_fermion}) with $r= \uptau =1$ (the Rarita-Schwinger equation in the TT gauge enjoys invariance under restricted gauge transformations~(\ref{PG_TYPEI_spin3/2}) where the spinor gauge function has to satisfy a certain condition). For more details concerning the imaginary tunings~(\ref{values_mass_parameter_masslessness_fermion}) of the mass parameter for strictly/partially massless spin-$s \geq 3/2$ fermions on $dS_{N}$ see Ref.~\cite{Deser_Waldron_ArbitrarySR}.}

\subsection{Main aim and strategy of the present paper}\label{subsec_mainaim}
The main aim of this paper is to provide a technical explanation for the results of our previous paper~\cite{Letsios_announce}, and, in particular, of the main result:
\begin{itemize}
    \item \textbf{Main result:} The strictly massless spin-3/2 field (gravitino) and the strictly and partially massless spin-5/2 fields on $dS_{N}$ ($N \geq 3$) are unitary only for $N=4$.
\end{itemize}
{The methods of the present paper follow those in Higuchi's seminal paper~\cite{STSHS}. In particular, in Ref.~\cite{STSHS} Higuchi constructed the symmetric transverse-traceless tensor eigenfunctions of the Laplace-Beltrami operator on $dS_{N}$ ($N \geq 3$) by analytically continuing the symmetric tensor spherical harmonics on $S^{N}$ and studied their group-theoretic properties to determine the values of the mass parameter that correspond to UIR's of spin$(N,1)$.}

In the present paper, following Higuchi~\cite{STSHS}, our technical explanation is based on studying the (non-)existence of positive-definite, dS invariant scalar products for the spin-3/2 and spin-5/2 eigenmodes on $dS_{N}$ ($N \geq 3$) - see Subsection~\ref{subsec_mainresults} for a summary of our technical explanation. Since the strictly/partially massless theories occur for imaginary values~(\ref{values_mass_parameter_masslessness_fermion}) of the mass parameter, we will focus our representation-theoretic analysis on the case where $M$ is an arbitrary imaginary number $M=i \tilde{M}$ ($\tilde{M} \neq 0$), and we will specialise to the strictly/partially massless values~(\ref{values_mass_parameter_masslessness_fermion}) when necessary.

Our strategy is as follows:
\begin{itemize}
 \item \noindent \textbf{Construction of eigenmodes.} We obtain the TT vector-spinor eigenmodes $\Psi_{\mu_{1}}$ (spin-3/2 modes) and the TT symmetric tensor-spinor eigenmodes $\Psi_{\mu_{1} \mu_{2}}$ (spin-5/2 modes) of eq.~(\ref{Dirac_eqn_fermion_dS}) with arbitrary imaginary mass parameter $M=i \tilde{M}$ ($\tilde{M} \neq 0$) by taking advantage of the well-known fact that $S^{N}$ can be analytically continued to $dS_{N}$ (see Section \ref{section_analytic_cont}).~\footnote{In our previous work~\cite{Letsios_announce}, these eigenmodes were constructed directly on $dS_{N}$ using the method of separation of variables. In the present work, we also give details that were omitted from Ref.~\cite{Letsios_announce}.} In particular, we write down explicitly the mode solutions of the following eigenvalue equation on $S^{N}$:
\begin{align}
   &\slashed{\nabla}\psi_{\mu_{1}...\mu_{r}}=i \zeta \psi_{\mu_{1}...\mu_{r}}  \label{Dirac_eqn_fermion_SN}\\
   & \nabla^{\alpha}\psi_{\alpha \mu_{2}...\mu_{r}}=0, \hspace{4mm}  \gamma^{\alpha}\psi_{\alpha \mu_{2}...\mu_{r}}=0 \label{TT_conditions_fermions_SN},
\end{align}
where $\psi_{\mu_{1}...\mu_{r}}$ is a totally symmetric tensor-spinor of rank $r$ on $S^{N}$ which also satisfies the TT conditions~(\ref{TT_conditions_fermions_SN}) and $\slashed{\nabla}$ is the Dirac operator on $S^{N}$. The eigenvalue in eq.~(\ref{Dirac_eqn_fermion_SN}) is imaginary~\cite{Homma}, i.e. $\zeta \in \mathbb{R}$, since, as is well known, $\slashed{\nabla}^{2}$ is negative semidefinite on compact spin manifolds. We call the eigenmodes satisfying eqs.~(\ref{Dirac_eqn_fermion_SN}) and (\ref{TT_conditions_fermions_SN}) the \textbf{symmetric tensor-spinor spherical harmonics (STSSH's)}.
In the present work we study only the STSSH's with ranks $r=1$ and $r=2$ on $S^{N}$ ($N \geq 3$), where we are also going to normalise them, as well as study their transformation properties under a specific spin$(N+1)$ transformation, where spin$(N+1)$ is the Lie algebra of the isometry group of $S^{N}$. Note that the unnormalised STSSH's of rank $r=1$ - i.e. the TT vector-spinor eigenmodes of the Dirac operator $\slashed{\nabla}$ on $S^{N}$ - have been already constructed in Ref.~\cite{CHH}, but no emphasis was given on their group-theoretic properties. To our knowledge, the STSSH's of rank $r=2$ are constructed in the present paper for the first time (see Section \ref{sectn_spin5/2_solving_SN} and Appendix~\ref{Appendix_spin5/2_solving_SN}). By applying analytic continuation techniques to eqs.~(\ref{Dirac_eqn_fermion_SN}) and (\ref{TT_conditions_fermions_SN}), we will obtain eqs.~(\ref{Dirac_eqn_fermion_dS}) and (\ref{TT_conditions_fermions_dS}), respectively, on $dS_{N}$.

 \item \noindent \textbf{Transformation of eigenmodes under spin$\bm{(N,1)}$.} We study the transformation properties of the eigenmodes on $dS_{N}$ under a spin$(N,1)$ boost (i.e., we study the action of a boost generator of spin$(N,1)$ on the space of eigenmodes). The corresponding transformation formulae are obtained by analytically continuing our detailed expressions for the spin$(N+1)$ transformation formulae for STSSH's on $S^{N}$. (The spin$(N+1)$ transformation formulae of STSSH's on $S^{N}$ are derived in the present paper by explicit calculation - see~Section~\ref{section_transf_SN} and Appendix~\ref{Appendix_transfrmn_proeprties_norm_fac}.)
 \item \noindent \textbf{Investigating the existence of positive-definite, dS invariant scalar products.} By exploiting the transformation properties of the eigenmodes on $dS_{N}$ under the spin$(N,1)$ boost, we examine when their norm with respect to a (any) dS invariant scalar product can be positive-definite. {(The requirement of infinitesimal dS invariance of the scalar product coincides with the requirement of anti-hermiticity of the spin$(N,1)$ generators).}
\end{itemize}

\subsection{Technical explanation of the main result}\label{subsec_mainresults}
  The explanation of our main result is given by the following technical results/statements concerning the spin-3/2 and spin-5/2 TT eigenmodes of eq.~(\ref{Dirac_eqn_fermion_dS}) with arbitrary imaginary mass parameter $M= i \tilde{M}$ ($\tilde{M} \neq 0$):
\begin{enumerate}
    \item \textbf{For even} $\bm{N>4}$: all dS invariant scalar products for these eigenmodes must be indefinite for all imaginary $M=i \tilde{M}$ ($\tilde{M} \neq 0$). This is demonstrated by showing that both positive-norm and negative-norm mode solutions exist and they mix with each other under spin$(N,1)$ for all $\tilde{M} \neq 0$ [including the strictly and partially massless values~(\ref{values_mass_parameter_masslessness_fermion})]. \label{basic_result_Neven}
    \item \textbf{For} $\bm{N = 4}$: all dS invariant scalar products for these eigenmodes must be indefinite unless $\tilde{M}$ is tuned to the strictly/partially massless values~(\ref{values_mass_parameter_masslessness_fermion}). The solution space of the strictly/partially massless theories is divided into two spin$(4,1)$ invariant subspaces, denoted as $\mathcal{H}_{-}$ and $\mathcal{H}_{+}$, where all mode solutions in $\mathcal{H}_{-}$ have `negative helicity', while all mode solutions in $\mathcal{H}_{+}$ have `positive helicity'. Then, we introduce a specific dS invariant scalar product [eq.~(\ref{inner_prod_dSN_gamma5})] in $\mathcal{H}_{-}$ and $\mathcal{H}_{+}$. For this choice of scalar product, it happens that the norm is positive-definite in $\mathcal{H}_{-}$ and negative-definite in $\mathcal{H}_{+}$. However, group-theoretically, we are allowed to have a different scalar product for each invariant subspace (since they correspond to different irreducible representations). Thus, by a redefinition of the scalar product in $\mathcal{H}_{+}$, we can change the sign of the associated norm and make it positive-definite. This shows that $\mathcal{H}_{-}$ and $\mathcal{H}_{+}$ form a direct sum of unitary irreducible representations of spin$(4,1)$.\label{basic_result_N=4}
    
    \item \textbf{For} $\bm{N}$ \textbf{odd}: For all $M=i \tilde{M} \neq 0$ [including the strictly and partially massless values~(\ref{values_mass_parameter_masslessness_fermion})], there does not exist any dS invariant scalar product for these eigenmodes. Thus, by definition, the corresponding spin$(N,1)$ representations are not unitary. \label{basic_result_Nodd}
\end{enumerate}
These findings/statements provide a technical explanation of the results presented in our previous article~\cite{Letsios_announce}, and the main scope of the current paper is to prove them. However, our findings seem to contrast with the claims made in Ref.~\cite{Deser_Waldron_ArbitrarySR}. The non-unitarity of the strictly and partially massless spin-$s=3/2, 5/2$ fields on $dS_{N}$ for $N \neq 4$ was missed in Ref.~\cite{Deser_Waldron_ArbitrarySR}, apparently because the dS invariant norm of the corresponding eigenmodes was not examined. For recent discussions on integer-spin representations of the de Sitter group see Refs.~\cite{Mixed_Symmetry_dS, Gizem,Gizem2, Gizem3, Gizem4, charm}, while spin-1/2 representations have been discussed in Refs.~\cite{Schaub1, Schaub2}. A`field theory-representation theory' dictionary for unitary totally symmetric tensor-spinors of any half-odd-integer spin $s \geq 1/2$ on $dS_{N}$ has been proposed in our previous article~\cite{Letsios_announce}.

\noindent \textbf{Note on real values of $\bm{M}$.} In this paper, we do not discuss eigenmodes with real mass parameters on $dS_{N}$. However, all of our results (i.e. explicit expressions for the eigenmodes and spin$(N,1)$ transformation formulae) also hold for real $M$. A crucial difference between the imaginary and the real mass parameter cases on $dS_{N}$, is the dS invariant scalar product. As we will demonstrate later, in the case of imaginary mass parameter, a dS invariant scalar product is given by~(\ref{inner_prod_dSN_gamma5}) for even $N$, while there is no dS invariant scalar product for odd $N$. On the other hand, in the case of real mass parameter, the conventional Dirac-like inner product can be always defined, and is dS invariant (this inner product corresponds to the product that results by just removing $\gamma^{N+1}$ from the scalar product~(\ref{inner_prod_dSN_gamma5})). The spin-3/2 and spin-5/2 theories with real mass parameters on $dS_{N}$ are always unitary~\cite{Letsios_announce} - this is easy to check given the results and tools introduced in the present paper.
 \subsection{Reviewing the spin-0 and spin-1 cases}
{To illustrate the methods that we are going to use in the main text, we will briefly review the simpler spin-0 and spin-1 cases as studied by Higuchi~\cite{STSHS}. }

\noindent {\textbf{Spin-0 case.} We will use the spin-0 case to explain how to analytically continue the scalar eigenfunctions of the Laplace-Beltrami operator $\nabla^{\mu}\nabla_{\mu}$ from $S^{N}$ to $dS_{N}$ ($N \geq 3$). The analytic continuation techniques we are going to present here will play a central role from Section~\ref{section_analytic_cont} onwards. The reason why these analytic continuation techniques work at the eigenmode level (for any integer or half-odd-integer spin) is related to the field equation's form on $S^{N}$. This form allows for a straightforward transformation into the corresponding field equation on global $dS_{N}$ by making simple replacements~\cite{STSHS}.}

{$\bm{S^{N}~-}$ The line element of $S^{N}$~(\ref{line_element_SN}) is given by $$ds_{N}^{2} = d\theta_{N}^{2} + \sin^{2}{\theta_{N}} \,ds^{2}_{N-1},$$ where  $\pi \geq \theta_{N}   \geq 0$ and $ds^{2}_{N-1}$ is the line element of $S^{N-1}$. The latter can be similarly expressed as $$ds^{2}_{N-1} = d\theta^{2}_{N-1} + \sin^{2}{\theta_{N-1}}  \, ds^{2}_{N-2}$$ (see Section~\ref{Section_geometry_N_sphere}).}

{The scalar spherical harmonics $Y^{(L\,k_{N-1}...k_{1})}(\theta_{N}, \bm{\theta}_{N-1})$ on $S^{N}$ satisfy~\cite{STSHS}
\begin{align}  \label{scalar harmonics eqn SN}
    \nabla^{\mu}  \nabla_{\mu}Y^{(L\,k_{N-1}...k_{1})} = -L(L+N-1)\,  Y^{(L\,k_{N-1} ...k_{1})},
\end{align}
where $L=0,1,...$ is the spin$(N+1)$ angular momentum quantum number. The rest of the angular momentum quantum numbers are also integers and they satisfy $L \geq k_{N-1} \geq ... \geq k_{2} \geq |k_{1}|$ corresponding to the chain of subalgebras spin$(N+1) \supset$ spin$(N) \supset ... \supset$ spin$(3) \supset$ spin$(2)$. Equation~(\ref{scalar harmonics eqn SN}) is expressed as~\cite{STSHS}
\begin{align}\label{scalar harmonics eqn SN expand}
  \left(\frac{\partial^{2}}{\partial \theta_{N}^{2}} +(N-1)  \cot{\theta_{N}} \frac{\partial}{  \partial   \theta_{N}}   +\frac{\tilde{\Box}}{\sin^{2}{\theta_{N}}} \right)Y^{(L\,k_{N-1}...k_{1})}(\theta_{N}, \bm{\theta}_{N-1})  = -L(L+N-1)Y^{(L\,k_{N-1}...k_{1})}(\theta_{N}, \bm{\theta}_{N-1}),
\end{align}
where $\tilde{\Box}$ is the Laplace-Beltrami operator on $S^{N-1}$. The solution can be found by expressing $Y^{(L\,k_{N-1}...k_{1})}$ in the form~\cite{STSHS}
\begin{align} \label{solution scalar SN}
    Y^{(L\,k_{N-1}...k_{1})}(\theta_{N}, \bm{\theta}_{N-1}) =& \,_{N}c^{k_{N-1}}_{L} \, \left( \sin{\theta_{N}}   \right)^{-(N-2)/2}\,P^{-\left( k_{N-1} +(N-2)/2   \right)}_{L+(N-2)/2}(\cos{\theta_{N}})\nonumber\\
    &\times \tilde{Y}^{(k_{N-1}...k_{1})}(\bm{\theta}_{N-1}) ,
\end{align}
where $\tilde{Y}^{(k_{N-1}...k_{1})}(\bm{\theta}_{N-1})$ are scalar spherical harmonics on $S^{N-1}$ satisfying 
$$ \tilde{\Box}  \tilde{Y}^{(k_{N-1}...k_{1})}(\bm{\theta}_{N-1}) = -k_{N-1}  (k_{N-1} + N -2) \tilde{Y}^{(k_{N-1}...k_{1})}(\bm{\theta}_{N-1}),  $$ while $P_{\nu}^{-\mu}$ is the associated Legendre function of the first kind. The normalisation factor $_{N}c^{k_{N-1}}_{L}$ is given by~\cite{STSHS}
$$ \, _{N}c_{L}^{k_{N-1}} = \left( \frac{2L+N-1}{2}\, \frac{\Gamma(L+k_{N-1}+N-1)}{\Gamma(L - k_{N-1}+1)}   \right)^{1/2}  $$ so that the scalar spherical harmonics on $S^{N}$ are normalised as
$$ \int_{S^{N}} \, \sqrt{g}\,d\theta_{N}\,d\theta_{N-1}...d\theta_{1}\, \left(  Y^{(L\,k_{N-1}...k_{1})}(\theta_{N}, \bm{\theta}_{N-1})  \right)^{*} \,  Y^{(L'\,k_{N-1}'...k_{1}')}(\theta_{N}, \bm{\theta}_{N-1}) = \delta_{L L'}  \delta_{k_{N-1}   k_{N-1}'}...\delta_{k_{1}  \, k_{1}'},$$
while the lower-dimensional scalar spherical harmonic $\tilde{Y}^{(k_{N-1}...k_{1})}(\bm{\theta}_{N-1})$ satisfy the analogous normalisation condition on $S^{N-1}$.} 
\\
\\
{$\bm{dS_{N}~-} $ The line element for global $dS_{N}$~(\ref{dS_metric}) is $$ds^{2} = -dt^{2} + \cosh^{2}{t} \,ds^{2}_{N-1}.$$ The Klein-Gordon mode functions on $dS_{N}$ satisfy
\begin{align}  \label{KG eqn dSN}
    \nabla^{\mu}  \nabla_{\mu}\Phi^{(\mathcal{M}\,k_{N-1}...k_{1})}(t,\bm{\theta}_{N-1}) = \mathcal{M}^{2}\,  \Phi^{(\mathcal{M}\,k_{N-1} ...k_{1})}(t,  \bm{\theta}_{N-1}),
\end{align}
where the angular momentum quantum numbers satisfy $k_{N-1} \geq ... \geq k_{2} \geq |k_{1}|$. The Klein-Gordon equation~(\ref{KG eqn dSN}) is expressed as
\begin{align*}
  \left(-\frac{\partial^{2}}{\partial   t^{2}} -(N-1)  \tanh{t} \frac{\partial}{  \partial   t}   +\frac{\tilde{\Box}}{\cosh^{2}{t}} \right)\Phi^{(\mathcal{M}\,k_{N-1}...k_{1})}(t, \bm{\theta}_{N-1})  = \mathcal{M}^{2}\,\Phi^{(\mathcal{M}\,k_{N-1}...k_{1})}(t, \bm{\theta}_{N-1}),
\end{align*}
and by introducing the variable $x(t) = \pi/2 -it$, it can be equivalently expressed in the form:
\begin{align}\label{KG eqn dSN expand}
  \left(\frac{\partial^{2}}{\partial   x^{2}} +(N-1)  \cot{x} \frac{\partial}{  \partial   x}   +\frac{\tilde{\Box}}{\sin^{2}{x}} \right)\Phi^{(\mathcal{M}\,k_{N-1}...k_{1})}(t, \bm{\theta}_{N-1})  = \mathcal{M}^{2}\,\Phi^{(\mathcal{M}\,k_{N-1}...k_{1})}(t, \bm{\theta}_{N-1}).
\end{align}
It is clear that eq.~(\ref{scalar harmonics eqn SN expand}) takes the same form as the Klein-Gordon equation~(\ref{KG eqn dSN expand}) if we make the replacements $\theta_{N}  \rightarrow x(t)= \pi/2 - it$ and $-L(L+N-1)  \rightarrow \mathcal{M}^{2}$ [this stems from the fact that the $S^{N}$ line element can be analytically continued to the global $dS_{N}$ line element by making the replacement $\theta_{N} \rightarrow x(t)$]. This means that by analytically continuing the scalar spherical harmonics~(\ref{solution scalar SN}) on $S^{N}$ we can obtain the mode functions of the Klein-Gordon equation on global $dS_{N}$. In particular, the unnormalised Klein-Gordon mode functions are given by~\cite{STSHS}
\begin{align}\label{anal cont scalar spher}
  \Phi^{(\mathcal{M}\,k_{N-1}...k_{1})}(t, \bm{\theta}_{N-1}) = \left(\frac{1}{\,_{N}c^{k_{N-1}}_{L}} Y^{(L\,k_{N-1}...k_{1})}(x(t), \bm{\theta}_{N-1}) \right)  \Big|_{-L(L+N-1) \rightarrow \mathcal{M}^{2}},
\end{align}
where we have used $\Big|_{-L(L+N-1) \rightarrow \mathcal{M}^{2}}$ to denote the replacement of $-L(L+N-1)$ by $\mathcal{M}^{2}$. The normalisation factors for the analytically continued functions~(\ref{anal cont scalar spher}), as well as the values of $\mathcal{M}^{2}$ for which the corresponding spin$(N,1)$ representations are unitary can be found in Ref.~\cite{STSHS}. For various recent physical studies on the Quantum Klein-Gordon field on de Sitter spacetime see, e.g., Refs.~\cite{Maldacena2013, Choudhury2018, Kanno_2015, ChoudAgain, Baumann2020, CHOUDHURY2019114606}. }
\\
\\
\noindent {\textbf{Spin-1 case and `pure gauge' modes.} Here our main aim is to start with the spin-1 case on $S^{N}$ ($N \geq 3$), analytically continue to $dS_{N}$, and then review the standard representation-theoretic treatment~\cite{STSHS, HiguchiLinearised} of the `pure gauge' modes (i.e. zero-norm modes) that appear at the strictly massless tuning of the mass parameter. This will be the treatment we will follow in our main text for the strictly/partially massless spin-$s= 3/2, 5/2$ fields. (The spin-$s=3/2, 5/2$ `pure gauge' modes are discussed in detail in Subsection~\ref{subsection_PG}, as well as in Section~\ref{section_(non)unitarity} and Appendix~\ref{appendix_pure_gauge_modes}. However, as we will see, for odd $N \geq 3$ there is no dS invariant notion of norm for tensor-spinors with an imaginary mass parameter. Nevertheless, the `pure gauge' modes can be identified as discussed in Subsection~\ref{subsection_PG}.)}

{In the case of a free field that enjoys a gauge symmetry - such as the strictly massless spin-1 field (photon) or the strictly/partially massless spin-$s = 3/2, 5/2$ fields - on $dS_{N}$, the theory has physical mode solutions and `pure gauge' mode solutions. The `pure gauge' modes are non-normalisable on $dS_{N}$ (since they have zero dS invariant norm), but they form an invariant subpace under dS transformations (i.e. by applying dS transformations on `pure gauge' modes we obtain only `pure gauge' modes)~\cite{STSHS}. However, by applying spin$(N,1)$ boosts on physical modes we find linear combinations of other physical modes plus contributions from `pure gauge' modes. If, on the other hand, we apply spin$(N)$ transformations on physical modes we obtain only physical modes without `pure gauge' contributions~\cite{STSHS}. From the Quantum Field-Theoretic viewpoint, this means that the physical Wightman 2-point function of a strictly massless field (this has a mode sum expression in terms of physical modes), such as the graviton 2-point function discussed in Ref.~\cite{AtsushiHiguchi_2003}, is only spin$(N)$ invariant and not dS invariant. However, in the present article, we focus on the spin$(N,1)$ representations formed by mode functions on $dS_{N}$. If both physical and `pure gauge' modes are present in the solution space, then, given a dS invariant scalar product, all `pure gauge' modes can be identified with zero~\cite{STSHS, AtsushiHiguchi_2003}. We are allowed to do this because, using any dS invariant scalar product, the `pure gauge' modes are found to be orthogonal to themselves as well as to all physical modes~\footnote{See Refs.~\cite{STSHS, HiguchiLinearised}, Subsection~\ref{subsection_PG} and Section~\ref{section_analytic_cont}.}. This means that the dS invariant scalar product is also invariant under the gauge transformations that preserve the field equation~\cite{HiguchiLinearised}. Thus, unlike the case of the physical 2-point function, at the level of the mode functions dS symmetry is present: the action of spin$(N,1)$ is defined on equivalence classes of mode solutions with the equivalence relation: any two mode solutions belong to the same equivalence class if their difference is `pure gauge'.}

{Below we discuss the transverse vector spherical harmonics on $S^{N}$ and we present the two different types of eigenmodes: the type-$I$ and the type-$I \! I$ modes. Then, by performing analytic continuation, we will obtain the transverse spin-1 modes with arbitrary mass parameter on $dS_{N}$. For the strictly massless value of the mass parameter, the analytically continued type-$I$ modes will be identified with `pure gauge' modes, while the analytically continued type-$ I \! I$ modes will be identified with physical modes that form the corresponding spin$(N,1)$ UIR.}
\\
\\
\noindent {$\bm{S^{N}~ -}$ The transverse vector spherical harmonics on $S^{N}$ satisfy~\cite{STSHS}
\begin{align} \label{LB op on vector-harm SN}
    \nabla^{\alpha}  \nabla_{\alpha} h_{\mu}^{(A; L\, k_{N-1}\, \tilde{\rho})}  = \Big(-L (L+N-1)+1  \Big)\, h_{\mu}^{(A; L \, k_{N-1}\, \tilde{\rho})}, ~~~\nabla^{\mu }h_{\mu}^{(A; L\, k_{N-1}\, \tilde{\rho})}=0,
\end{align}
where the angular momentum quantum numbers take the values $L=1,2,...$ and $k_{N-1} = 1,2,...,L$, while $\tilde{\rho}$ represents other representation-theoretic labels. Here the vector index $\mu$ takes the values $\mu = \theta_{N}$ and $\mu = \theta_{j}$, where $\theta_{j}$ is a vector index on $S^{N-1}$ with $j = 1,2,...,N-1$. As in the spin-$3/2$ case discussed in Section~\ref{sectn_spin3/2_solving_SN}, we use the label $A$ to denote the two types of transverse eigenmodes on $S^{N}$~\cite{STSHS}}: 
\begin{itemize}
    \item  {The type-$I$ modes are denoted as $h^{(I; L\, k_{N-1} \tilde{\rho})}_{\mu}(\theta_{N}, \bm{\theta}_{N-1})$. These are constructed in terms of scalar spherical harmonics $\tilde{Y}^{(k_{N-1}\,\tilde{\rho})}(\bm{\theta}_{N-1})$ on $S^{N-1}$. We have~\cite{STSHS}
\begin{align} \label{type-I vec sphere harmonic}
    h^{(I; L\, k_{N-1} \tilde{\rho})}_{\theta_{N}}(\theta_{N}, \bm{\theta}_{N-1}) =& \,_{N}C^{(I)}_{L\,k_{N-1}} ~\,_{N}c^{k_{N-1}}_{L} \, \left( \sin{\theta_{N}}   \right)^{-1-(N-2)/2}\,P^{-\left( k_{N-1} +(N-2)/2   \right)}_{L+(N-2)/2}(\cos{\theta_{N}})\,\,~ \tilde{Y}^{(k_{N-1}\,\tilde{\rho})}(\bm{\theta}_{N-1}) \\
     =& \,_{N}C^{(I)}_{L\,k_{N-1}} ~ \left( \sin{\theta_{N}}   \right)^{-1}\,{Y}^{(L\,k_{N-1}\,\tilde{\rho})}(\theta_{N}, \bm{\theta}_{N-1}),
\end{align}
where in the second line we have used eq.~(\ref{solution scalar SN}), while $_{N}C^{(I)}_{L\,k_{N-1}}$ is a normalisation factor. The rest of the components $h^{(I; L\, k_{N-1} \tilde{\rho})}_{\theta_{j}}$ can be found using the diverge-freedom of $h^{(I; L\, k_{N-1} \tilde{\rho})}_{\mu}$.}

    \item  {The type-$I\!I$ modes are denoted as $h^{(I\!I\text{-}\tilde{A}; L\, k_{N-1} \tilde{\rho})}_{\mu}(\theta_{N}, \bm{\theta}_{N-1})$, where the label $\tilde{A}$ on $S^{N-1}$ corresponds to $A$ on $S^{N}$. In particular, the type-$I\!I$ modes $h^{(I\!I\text{-}\tilde{A}; L\, k_{N-1} \tilde{\rho})}_{\mu}$ are constructed in terms of type-$\tilde{A}$ transverse vector spherical harmonics on $S^{N-1}$, $\tilde{h}^{(\tilde{A}; k_{N-1} \tilde{\rho})}_{\theta_{j}} (\bm{\theta}_{N-1})$. The latter satisfy~\cite{STSHS}
\begin{align} \label{LB op on vector-harm SN-1}
    \tilde{\Box} \tilde{h}_{\theta_{j}}^{(\tilde{A};  k_{N-1}\, \tilde{\rho})}  = \Big(-k_{N-1} (k_{N-1}+N-2)+1  \Big)\, \tilde{h}_{\theta_{j}}^{(\tilde{A};   k_{N-1}\, \tilde{\rho})}, ~~~\tilde{\nabla}^{\theta_{j} } 
 \tilde{h}_{\theta_{j}}^{(\tilde{A}; k_{N-1}\, \tilde{\rho})}=0,
\end{align}
    where $\tilde{\nabla}_{\theta_{j}}$ is the covariant derivative on $S^{N-1}$ and $\tilde{\Box} = \tilde{\nabla}_{\theta_{j}}    \tilde{\nabla}^{\theta_{j}}$. We have~\cite{STSHS}
\begin{align}\label{type-II vec sphere harmonic}
   {h}^{(I\!I \text{-} \tilde{A}; L\, k_{N-1} \tilde{\rho})}_{\theta_{N}}(\theta_{N}, \bm{\theta}_{N-1}) =& 0 \nonumber \\
    {h}^{(I\!I \text{-} \tilde{A}; L\, k_{N-1} \tilde{\rho})}_{\theta_{j}}(\theta_{N}, \bm{\theta}_{N-1}) =&_{N}C^{(I\!I)}_{L\,k_{N-1}}\, ~ _{N}c^{k_{N-1}}_{L} \, \left( \sin{\theta_{N}}   \right)^{1-(N-2)/2}\,P^{-\left( k_{N-1} +(N-2)/2   \right)}_{L+(N-2)/2}(\cos{\theta_{N}})\,\,~ \tilde{h}_{\theta_{j}}^{(\tilde{A};k_{N-1}\,\tilde{\rho})}(\bm{\theta}_{N-1}),
\end{align}
where $_{N}C^{(I\!I)}_{L\,k_{N-1}}$ is a normalisation factor.}
\end{itemize}
{On $S^{N}$ both type-$I$ and type-$I \! I$ transverse modes have positive norm and they together form the representation space for a spin$(N+1)$ UIR. The spin$(N+1)$ invariant positive-definite inner product on $S^{N}$ is given by
\begin{align}
    \int_{S^{N}} \, \sqrt{g}\,d\theta_{N}\,...d\theta_{1}\, \left(  {h}^{(A; L\, k_{N-1} \tilde{\rho})}_{\mu}(\theta_{N}, \bm{\theta}_{N-1}) \right)^{*} \,  {h}^{({A}'; L'\, k_{N-1}' \tilde{\rho}')\mu}(\theta_{N}, \bm{\theta}_{N-1}) = \delta_{L L'}  \,\delta_{k_{N-1}   k_{N-1}'} \,\delta_{\tilde{\rho} \tilde{\rho}'} \,\delta_{A A'},
\end{align}
while the lower-dimensional vector spherical harmonics~(\ref{LB op on vector-harm SN-1}) satisfy the analogous normalisation condition on $S^{N-1}$. The normalisation factors in eqs.~(\ref{type-I vec sphere harmonic}) and (\ref{type-II vec sphere harmonic}) are given by~\cite{STSHS}
\begin{align}
    _{N}C^{(I)}_{L\,k_{N-1}} =\left( \frac{k_{N-1}(k_{N-1}+N-2)}{(L+1)(L+N-2)}   \right)^{1/2} ,\hspace{10mm}_{N}C^{(I\!I)}_{L\,k_{N-1}}=1.
\end{align}}

\noindent {$\bm{dS_{N}~-}$ The mode functions for the transverse spin-1 field on $dS_{N}$ satisfy~\cite{STSHS}
    \begin{align} \label{spin-1 eqn dSN}
    \nabla^{\alpha}  \nabla_{\alpha} \mathcal{A}_{\mu}^{(A; \mathfrak{L} \, k_{N-1}\, \tilde{\rho})} (t, \bm{\theta}_{N-1}) = \mathcal{M}^{2}_{(\mathfrak{L})}\,~ \mathcal{A}_{\mu}^{(A; \mathfrak{L} \, k_{N-1}\, \tilde{\rho})}(t, \bm{\theta}_{N-1})  , ~~~\nabla^{\mu }\mathcal{A}_{\mu}^{(A; \mathfrak{L} \, k_{N-1}\, \tilde{\rho})} (t, \bm{\theta}_{N-1})=0,
\end{align}
where $k_{N-1} = 1,2,...$, while we have parametrised $\mathcal{M}_{(\mathfrak{L})}$ using the complex number $\mathfrak{L}$ as 
\begin{align}
    \mathcal{M}_{(\mathfrak{L})}^{2} = -\mathfrak{L}   (\mathfrak{L}   + N-1)   +1 .
\end{align} 
(The unitarily allowed values of $\mathfrak{L}$ can be found in Ref.~\cite{STSHS}). The labels $A, k_{N-1}, \tilde{\rho}$ have the same meaning as on $S^{N}$. As explained in the spin-0 case, the spin-1 mode functions satisfying eqs.~(\ref{spin-1 eqn dSN}) on $dS_{N}$ can be obtained by analytically continuing the transverse vector spherical harmonics on $S^{N}$. In particular, by making the replacements $\theta_{N} \rightarrow x(t)  = \pi/2 -i t$ and $L \rightarrow \mathfrak{L}$ in the type-$I$ (\ref{type-I vec sphere harmonic}) and type-$I \! I$ (\ref{type-II vec sphere harmonic}) modes we obtain the solutions of eq.~(\ref{spin-1 eqn dSN}) as
\begin{align}\label{spin-1 modes on dSN}
 \mathcal{A}_{\mu}^{(A; \mathfrak{L} \, k_{N-1}\, \tilde{\rho})}(t, \bm{\theta}_{N-1}) = \left(\,_{N}c_{\mathfrak{L}}^{k_{N-1}}~ _{N}C^{(A)}_{\mathfrak{L}\,k_{N-1}}\right)^{-1}  \,h^{(A; \mathfrak{L}\, k_{N-1} \tilde{\rho})}_{\mu}(x(t), \bm{\theta}_{N-1}) ,
\end{align}
where $A  \in \{I , I\!I \text{-}\tilde{A}   \}$.
In this expression $\mu$ takes the values $\mu = x(t)=\pi/2-it$ and $\mu={\theta_{j}}$, where $\mathcal{A}_{x}^{(A; \mathfrak{L} \, k_{N-1}\, \tilde{\rho})}= i \, \mathcal{A}_{t}^{(A; \mathfrak{L} \, k_{N-1}\, \tilde{\rho})}$.}

{Let us now discuss the normalisation of the mode functions~(\ref{spin-1 modes on dSN}). Let $\mathcal{A}^{(1)}_{\mu}$ and $\mathcal{A}^{(2)}_{\nu}$ be any two of the mode functions in eq.~(\ref{spin-1 modes on dSN}) and define the scalar product~\cite{STSHS}
\begin{align}\label{KG scalar product spin-1}
    \left( \mathcal{A}^{(1)}, \mathcal{A}^{(2)}   \right)_{KG} =-i\, \cosh^{N-1}{t}\, \int_{S^{N-1}} \Bigg( \left(   \mathcal{A}^{(1)\mu} \right)^{*}\, \nabla^{t}  \mathcal{A}^{(2)}_{\mu} - \left( \nabla^{t}\,\mathcal{A}^{(1)\mu} \right)^{*}\,   \mathcal{A}^{(2)}_{\mu} \Bigg),
\end{align}
where the integration is over the Cauchy surface $S^{N-1}$. This scalar product is both time-independent and dS invariant~\cite{STSHS}.
The norm of the mode functions with respect to the scalar product~(\ref{KG scalar product spin-1}) has been calculated in Ref.~\cite{STSHS} as
\begin{align}\label{norm spin-1 type-I}
     \left( \mathcal{A}^{(I; \mathfrak{L} \, k_{N-1}\, \tilde{\rho})},  \mathcal{A}^{(I; \mathfrak{L} \, k_{N-1}'\, \tilde{\rho}')}\right)_{KG} = \frac{2}{\Gamma(k_{N-1}-\mathfrak{L})  \,  \Gamma(k_{N-1}+\mathfrak{L}+N-1)} \, \frac{-N+2-\mathfrak{L} (\mathfrak{L} +N-1)  }{k_{N-1} (k_{N-1} +N-2)} \delta_{k_{N-1}    k_{N-1}'}   \,   \delta_{\tilde{\rho}     \tilde{\rho}'}
\end{align}
and
\begin{align}\label{norm spin-1 type-II}
     \left( \mathcal{A}^{(I\! I \text{-} \tilde{A}; \mathfrak{L} \, k_{N-1}\, \tilde{\rho})},  \mathcal{A}^{(I\! I \text{-} \tilde{A}'; \mathfrak{L} \, k_{N-1}'\, \tilde{\rho}')}\right)_{KG} = \frac{2}{\Gamma(k_{N-1}-\mathfrak{L})  \,  \Gamma(k_{N-1}+\mathfrak{L}+N-1)}  \delta_{k_{N-1}    k_{N-1}'}   \,   \delta_{\tilde{\rho}     \tilde{\rho}'} \,   \delta_{\tilde{A}   \tilde{A}'}.
\end{align}
The spin$(N,1)$ representation formed by the mode functions is unitary if both norms~(\ref{norm spin-1 type-I}) and (\ref{norm spin-1 type-II}) are positive. From the norm~(\ref{norm spin-1 type-I}) of the type-$I$ modes we find that unitarity requires 
$$  -N+2-\mathfrak{L} (\mathfrak{L} +N-1)>0 .$$
(The values of $\mathfrak{L}$ that satisfy this condition have been classified in eqs. (9.20a), (9.20b) of Ref.~\cite{STSHS}.)}

\noindent  {However, there is also the interesting strictly massless case with $\mathfrak{L} = -1$ for which the type-$I$ modes have zero norm. This means that for $\mathfrak{L} = -1$ (i.e. $\mathcal{M}^{2}_{(\mathfrak{L})}= N-1$) the type-$I$ modes are `pure gauge' modes. The existence of `pure gauge' modes reflects the fact that the field equation~(\ref{spin-1 eqn dSN}) enjoys a gauge symmetry for $\mathfrak{L}=-1$: if $\mathcal{A}_{\mu}$ satisfies eq.~(\ref{spin-1 eqn dSN}) with $\mathfrak{L}=-1$, then the gauge transformation $\delta \mathcal{A}_{\mu} = \nabla_{\mu}a$ with $\nabla^{\nu}\nabla_{\nu} a =0$ is a symmetry of the equation. In fact the type-$I$ modes in eq.~(\ref{spin-1 modes on dSN}) with $\mathfrak{L}=-1$ can be expressed in the form $\nabla_{\mu} a^{(k_{N-1}\, \tilde{\rho})}$~\cite{STSHS}, where the scalar gauge functions satisfy $ \nabla^{\nu}  \nabla_{\nu}  a^{( k_{N-1}\, \tilde{\rho})} =0 $. On the other hand, the norm~(\ref{norm spin-1 type-II}) of the type-$I \!I$ modes is positive, and these modes are identified with the physical modes of the theory. According to the discussion in the passages before eq.~(\ref{LB op on vector-harm SN}), for $\mathfrak{L} = -1$ the type-$I$ modes are identified with zero, while the type-$ I\!I$ modes form a strictly massless UIR of spin$(N,1)$.\footnote{For recent discussions on the quantum strictly massless spin-1 field on $dS_{4}$ see Ref.~\cite{Fukelman2024}.}}

\subsection{Outline of the paper, notation and conventions}
The rest of the paper is organised as follows. In Section~\ref{Section_geometry_N_sphere}, we begin by presenting the Christoffel symbols, vielbein fields and spin connection components on $S^{N}$ in geodesic polar coordinates. Then, we present the basics about gamma-matrices and tensor-spinor fields on $S^{N}$. We also review the eigenspinors of the Dirac operator on $S^{N-1}$. In Section~\ref{section_functions describing the dependence on theta_N}, we present the functions that describe the dependence of the STSSH's on the geodesic distance ($\theta_{N}$) from the North Pole of $S^{N}$. In Section~\ref{sectn_spin3/2_solving_SN}, we write down explicitly the unnormalised STSSH's of rank 1 on $S^{N}$ (which have been constructed in Ref.~\cite{CHH}). In Section~\ref{sectn_spin5/2_solving_SN}, we write down explicitly the unnormalised STSSH's of rank 2 on $S^{N}$ (which we construct in Appendix~\ref{Appendix_spin5/2_solving_SN}). In Section~\ref{section_transf_SN}, we use the Lie-Lorentz derivative~\cite{Ortin} in order to study the transformation properties of the STSSH's of rank $r$ ($r\in \set{1,2}$) on $S^{N}$ under a spin$(N+1)$ transformation and we give their normalisation factors. In Section~\ref{section_analytic_cont}, we begin by obtaining the vector-spinor and rank-2 symmetric tensor-spinor TT eigenmodes of the Dirac operator with arbitrary imaginary mass parameter on $dS_{N}$ by analytically continuing the STSSH's of rank 1 and rank 2, respectively, on $S^{N}$. Then, we identify the `pure gauge' modes of the strictly/partially massless spin-3/2 and spin-5/2 theories on $dS_{N}$. In Section~\ref{section_(non)unitarity}, we derive the main result of this paper (i.e. we prove statements \ref{basic_result_Neven}, \ref{basic_result_N=4} and \ref{basic_result_Nodd} listed in Subsection~\ref{subsec_mainresults}), by studying the transformation properties of the TT eigenmodes of eq.~(\ref{Dirac_eqn_fermion_dS}) with arbitrary imaginary mass parameter under a spin$(N,1)$ boost. More specifically, in Subsection~\ref{Subsection_reps_Neven_part1}, we show that all dS invariant scalar products must be indefinite
for even $N > 4$ (i.e. we prove statement~\ref{basic_result_Neven}). Also, for even $N \geq 4$, we show that the `pure gauge' modes in the strictly/partially massless theories with spin $s \in \set{3/2, 5/2}$ have zero norm with respect to any dS invariant scalar product. Then, for $N=4$, we show that the requirement for dS invariance of the scalar product does not imply the indefiniteness of the norm if and only if the imaginary mass parameter $M=i\tilde{M}$ (with $\tilde{M} \neq 0$) takes the strictly/partially massless values~(\ref{values_mass_parameter_masslessness_fermion}). We also find that for the strictly/partially massless theories with spin $s \in \set{3/2, 5/2}$ on $dS_{4}$, the eigenmodes with negative helicity and the ones with positive helicity separately form irreducible representations of spin$(4,1)$ (the unitarity of these irreducible representations is proved in Subsection~\ref{Subsection_explicit_calculationnorms}).
In Subsection~\ref{Subsection_explicit_calculationnorms}, we calculate the norms of the eigenmodes on $dS_{N}$ (for even $N \geq 4$) with respect to a specific dS invariant scalar product and we verify statement~\ref{basic_result_Neven} (which was proved in the previous Subsection) and we also prove statement~\ref{basic_result_N=4}.  Subsection~\ref{Subsection_reps_Nodd} concerns the case with $N$ odd and we prove statement~\ref{basic_result_Nodd}. Finally, in Section~\ref{summary and discussions}, we give a summary of our results. We also discuss the possible generalisation of our results to higher half-odd-integer spins, as well as to other vacuum spacetimes with positive cosmological constant.  

There are six Appendices. In Appendix~\ref{Appendix_spin5/2_solving_SN}, we construct the STSSH's of rank 2 on $S^{N}$ by making use of the method of separation of variables. In this method, the STSSH's of rank 2 on $S^{N}$ are expressed in terms of STSSH's of rank $\tilde{r}$ ($0 \leq \tilde{r} \leq 2$) on $S^{N-1}$. In Appendix~\ref{Appendix_transfrmn_proeprties_norm_fac}, we present technical details omitted in Section~\ref{section_transf_SN}. To be specific, we first give a detailed derivation of the formulae for the spin$(N+1)$ transformation of the rank-1 STSSH's and we determine their normalisation factors. Then, we discuss briefly the derivation of the transformation formulae and the normalisation factors for the rank-2 STSSH's on $S^{N}$. The rest of the Appendices concern other technical details that were omitted in the main text. 

\textbf{Notation and conventions.} We denote a point on $S^{N-1}$ as $\bm{\theta}_{N-1}=(\theta_{N-1},...,\theta_{1})$ and the line element of $S^{N-1}$ as $ds^{2}_{N-1}$ (see also Section~\ref{Section_geometry_N_sphere}). We use the mostly plus metric sign convention for $dS_{N}$. Lowercase Greek tensor indices refer to components with
respect to the “coordinate basis”. Lowercase Latin tensor indices refer to components with respect to the vielbein basis. Summation over repeated indices is
understood. We denote the symmetrisation of a pair of indices as $A_{(\mu \nu)} \equiv  (A_{\mu \nu}+A_{\nu \mu})/2$ and the anti-symmetrisation as $A_{[\mu \nu]} \equiv  (A_{\mu \nu}-A_{\nu \mu})/2$.
 Spinor indices are always suppressed throughout this paper. We use the term strictly/partially massless field of spin $s \in \set{3/2,5/2}$ to refer to either one of the following three cases (unless otherwise stated): the strictly massless spin-3/2 field ($r=\uptau=1$), the strictly massless spin-5/2 field ($r=\uptau+1=2$), the partially massless spin-5/2 field ($r=\uptau=2$). The complex conjugate of the complex number $z$ is denoted as $z^{*}$. The notation concerning the representation-theoretic labels of the eigenmodes is slightly different than the notation used in our previous article~\cite{Letsios_announce}. However, we ensure a clear explanation of the representation-theoretic meaning to prevent any potential confusion. 

\section{Geometry of the \texorpdfstring{${N}\textbf{-}$}{N}sphere and tensor-spinor fields}\label{Section_geometry_N_sphere}

\subsection{Coordinate system, Christoffel symbols and spin connection}\label{Geometry_of_N_sphere}
The $N$-sphere ($S^{N}$) embedded in the Euclidean space $\mathbb{R}^{N+1}$ is described by
\begin{align}
    \delta_{ab}X^{a} X^{b}=1,
\end{align}
$(a,b=1,2,...,N+1)$ where $\delta_{ab}$ is the Kronecker delta symbol
and $X^{1},X^{2},...,X^{N+1}$ are the standard coordinates for $\mathbb{R}^{N+1}$. The `geodesic polar coordinates'\footnote{The geodesic polar coordinates are also known as hyperspherical coordinates. They correspond to the straightforward generalisation of the standard spherical coordinates on $S^{2}$. The North Pole of $S^{N}$ is located at $\theta_{N} = 0$. The geodesic distance, $\mu_{S^{N}}$, between two points $\bm{\theta}_{N-1}= (\theta_{N},...,\theta_{1})$ and $\bm{\theta}'_{N-1}= (\theta_{N}',...,\theta_{1}')$ on $S^{N}$ is given by $\cos{\mu_{S^{N}}} = \cos{\theta_{N}} \cos{\theta_{N}'}+ \sin{\theta_{N}} \sin{\theta_{N}'}\cos{\mu_{S^{N-1}}}$. If we fix $\bm{\theta}_{N}'$ to be at the North Pole, then the geodesic distance is given as $\cos{\mu_{S^{N}}} =\cos{ \theta_{N}}$.} are given by
\begin{align}\label{global_coordinates}
    &X^{N+1}=X^{N+1}{(\theta_{N})} =\cos{\theta_{N}} \nonumber \\
   & X^{i}=X^{i}{(\theta_{N},\bm{\theta}_{N-1})}=\sin{\theta_{N}} \,\, \tilde{X}^{i}(\bm{\theta}_{N-1}), \hspace{4mm} i=1,...,N,
\end{align}
where $0 \leq \theta_{N} \leq \pi$ is the geodesic distance from the North Pole and $\bm{\theta}_{N-1}=(\theta_{N-1},...,\theta_{1})$ (where $0 \leq \theta_{1} <2 \pi$ and $0\leq \theta_{i} \leq  \pi$ for $i=2,3,...,N-1$). The $\tilde{X}^{i}$'s in eq.~(\ref{global_coordinates}) are the geodesic polar coordinates for $S^{N-1}$ in $N$-dimensional Euclidean space. 

The line element for $S^{N}$ is expressed in coordinates~(\ref{global_coordinates}) as
\begin{align}\label{line_element_SN}
        ds^{2}_{N}=d\theta_{N}^{2}+\sin^{2}{\theta_{N}}ds^{2}_{N-1},
    \end{align}
where $ds^{2}_{N-1}$ is the line element for $S^{N-1}$. (Note that we define $ds^{2}_{1}\equiv d\theta^{2}_{1}$.) The non-zero Christoffel symbols in geodesic polar coordinates are 
\begin{align}\label{Christoffels_SN}
    &\Gamma^{\theta_{N}}_{\hspace{0.2mm}\theta_{i} \theta_{j}}=-\sin{\theta_{N}} \cos{\theta_{N}} \hspace{1mm}\tilde{g}_{\theta_{i} \theta_{j}}, \hspace{2mm} \Gamma^{\theta_{i}}_{\hspace{0.2mm}\theta_{j} \theta_{N}} =\cot{\theta_{N}}  \hspace{1mm}\tilde{g}^{\theta_{i}}_{\theta_{j}}, \nonumber \\ 
& \Gamma^{\theta_{k}}_{\hspace{0.2mm}\theta_{i} \theta_{j}}=\tilde{\Gamma}^{\theta_{k}}_{\hspace{0.2mm}\theta_{i} \theta_{j}},
\end{align}
where $\tilde{g}_{\theta_{i} \theta_{j} }$ and $\tilde{\Gamma}^{\theta_{k}}_{\hspace{0.2mm}\theta_{i} \theta_{j}}$ are the metric tensor and the Christoffel symbols, respectively, on $S^{N-1}$. The vielbein fields $\bm{e}_{a}=e^{\mu}{\hspace{0.2mm}}_{a}\partial_{\mu}$ (where $a=1,...,N$ and $\mu=\theta_{1},...,\theta_{N}$), determining an orthonormal frame, satisfy
\begin{align}
    e_{\mu}{\hspace{0.2mm}}^{a} \, e_{\nu}{\hspace{0.2mm}}^{b}\delta_{ab}=g_{\mu \nu}, \hspace{4mm}e^{\mu}{\hspace{0.2mm}}_{a}\,e_{\mu}{\hspace{0.2mm}}^{b}=\delta^{b}_{a},
\end{align}
where the co-vielbein fields $\bm{e}^{a}=e_{\mu}{\hspace{0.2mm}}^{a}\,dx^{\mu}$ define the dual coframe. The co-vielbein transforms under local rotations $\Lambda: S^{N} \rightarrow$ SO($N$) as
\begin{align}\label{local rotation of covielbein}
    \bm{e}^{a} \rightarrow \Lambda(x)^{a}_{\hspace{2mm}b}\,\bm{e}^{b}.
\end{align}
In geodesic polar coordinates the non-zero components of the vielbein fields are given by
\begin{equation}\label{vielbeins}
    e^{\theta_{N}}{\hspace{0.2mm}}_{N}=1, \hspace{5mm}  e^{\theta_{i}}{\hspace{0.2mm}}_{i}=\frac{1}{\sin{\theta_{N}}} \tilde{e}^{\theta_{i}}{\hspace{0.2mm}}_{i} , \hspace{5mm}i=1,...,N-1,
\end{equation}
 where $\tilde{e}^{\theta_{i}}{\hspace{0.2mm}}_{i}$ are the vielbein fields on $S^{N-1}$.
  The spin connection ${\omega}_{abc}={\omega}_{a[bc]} \equiv ({\omega}_{abc}-{\omega}_{acb})/2 $ is given by
 \begin{equation}
           {\omega}_{abc} = - e^{\mu}\hspace{0.2mm}_{a} \Big( \partial_{\mu}  e^{\lambda}\hspace{0.2mm}_{b} + {\Gamma}^{\lambda}_{\mu \nu} e^{\nu}\hspace{0.2mm}_{b} \Big) e_{\lambda c}      
       \end{equation}
       and its only non-zero components are
\begin{equation}\label{spin_connection_components}
   \omega_{ijk} = \frac{  \tilde{\omega}_{ijk}}{ \sin{\theta_{N}}  } , \hspace{4mm}  \omega_{iNk} = -\omega_{ikN} =- \cot{\theta_{N}}\hspace{1mm} \delta_{ik}, \hspace{5mm} i,j,k=1,...,N-1,
        \end{equation} 
where $\tilde{\omega}_{ijk}$ are the spin connection components on $S^{N-1}$. (Note that the sign convention we use for the
spin connection is the opposite of the one used in Refs.~\cite{Camporesi, Letsios}.)

\subsection{Gamma matrices and tensor-spinor fields on the \texorpdfstring{$N$}{N}-sphere}\label{Subsection_spinors_and_clifford}

A Clifford algebra representation in $N$ dimensions is generated by $N$ gamma matrices. These are matrices of dimension $2^{[N/2]}$ - where $[N/2]=N/2$ if $N$ is even and $[N/2]=(N-1)/2$ if $N$ is odd - satisfying the anti-commutation relations
\begin{equation}\label{anticommutation_relations_gamma}
   \{\gamma^{a}, \gamma^{b}\}  = 2 \delta^{ab} \bm{1}, \hspace{10mm} a,b=1,2,...,N,
\end{equation}
where $\bm{1}$ is the identity matrix.
We adopt the representation of gamma matrices used in Ref.~\cite{Camporesi}, where gamma matrices in $N$ dimensions are expressed in terms of gamma matrices in $N-1$ dimensions ($\widetilde{\gamma}^{i}$) as follows:

\begin{itemize}
    \item For $N$ even

\begin{equation}\label{even_gammas}
 \gamma^{N}= \begin{pmatrix}  
   0 & \bm{1} \\
   \bm{1} & 0
    \end{pmatrix} , \hspace{5mm}
    \gamma^{j}=\begin{pmatrix}  
   0 & i\widetilde{\gamma}^{j} \\
   -i\widetilde{ \gamma}^{j} & 0
    \end{pmatrix} ,
     \end{equation} 
($ j=1,..., N-1$) where the lower-dimensional gamma matrices satisfy the Euclidean Clifford algebra anti-commutation relations
\begin{equation} \label{Euclidean_Clifford_relns}
    \{ \widetilde{\gamma}^{j}, \widetilde{\gamma}^{k}\} = 2 \delta^{jk} \bm{1}, \hspace{5mm}j,k=1,...,N-1.
\end{equation}
By using the vielbein fields~(\ref{vielbeins}) we can express the gamma matrices~(\ref{even_gammas}) in the ``coordinate basis'' as $\gamma^{\mu}(x)= e^{\mu}{\hspace{0.2mm}}_{a}(x)\,\gamma^{a}$. Note that one can construct the extra gamma matrix $\gamma^{N+1}$, which is given by the product $\gamma^{N+1} \equiv \epsilon \,\gamma^{1}\gamma^{2}...\gamma^{N}$, where $\epsilon$ is a phase factor. The matrix $\gamma^{N+1}$ anti-commutes with each of the $\gamma^{a}$'s in eq.~(\ref{even_gammas}). As in Ref.~\cite{Camporesi}, we choose the phase factor $\epsilon$ such that
\begin{align}\label{gamma(N+1)}
    \gamma^{N+1}=\begin{pmatrix}
    \bm{1} & 0\\
    0      & -\bm{1}
    \end{pmatrix}.
\end{align}
\item For $N$ odd

\begin{equation*}
   \gamma^{N}= \begin{pmatrix}  
    \bm{1} & 0 \\
 0& -\bm{1} 
    \end{pmatrix} ,\hspace{5mm}\gamma^{N-1}=\tilde{\gamma}^{N-1}= \begin{pmatrix}  
   0 &  \bm{1} \\
   \bm{1} & 0
    \end{pmatrix} , 
    \end{equation*}
\begin{equation}\label{odd_gammas}
    \gamma^{j}= \widetilde{\gamma}^{j}=\begin{pmatrix}  
   0 & i\widetilde{\widetilde{ \gamma}}^{j} \\
   -i\widetilde{\widetilde{ \gamma}}^{j} & 0
    \end{pmatrix},\hspace{5mm}j=1,...,N-2.
\end{equation}
The tilde is used to denote gamma matrices in $N-1$ dimensions, while the double-tilde is used to denote gamma matrices in $N-2$ dimensions. In $N=1$ dimension the only (one-dimensional) gamma matrix is equal to $1$. The gamma matrices~(\ref{odd_gammas}) are expressed in the ``coordinate basis'' by using the vielbein fields~(\ref{vielbeins}), as in the case with $N$ even.
\end{itemize}
Note that all gamma matrices in eqs.~(\ref{even_gammas})-(\ref{odd_gammas}) are hermitian.

 The tensor-spinor fields $\psi_{\mu_{1}...\mu_{r}}$ of rank $r$ are defined as $r^{th}$-rank tensors where each one of the tensorial components transforms as a $2^{[N/2]}$-dimensional spinor under Spin$(N)$ (double cover of SO$(N)$). In other words, tensor-spinors transform under the local rotation of the co-vielbein in eq.~(\ref{local rotation of covielbein}) as
\begin{equation}
    \psi_{\mu_{1}...\mu_{r}}{(x)} \rightarrow  \Lambda{(x)}_{\mu_{1}}^{\hspace{1mm}\nu_{1}}...\Lambda{(x)}_{\mu_{r}}^{\hspace{1mm}\nu_{r}}\hspace{1mm} S(\Lambda{(x)})\,\psi_{\nu_{1}...\nu_{r}}{(x)},
\end{equation}
where the matrix $\Lambda(x) \in$ SO$(N)$ acts on the tensor indices of $\psi_{\mu_{1}...\mu_{r}}$, while the matrix $S(\Lambda{(x)}) \in$ Spin$(N)$ acts on the spinor indices of $\psi_{\mu_{1}...\mu_{r}}$ (the spinor indices have been suppressed for convenience). For any $\Lambda(x) \in$ SO$(N)$ we have~\cite{osti_5650644}
\begin{align}
    S(\Lambda(x))^{-1}\, \gamma^{a}\, S(\Lambda(x))=\Lambda(x)^{a}_{\hspace{2mm}b}\gamma^{b},
\end{align}
where $S(\Lambda(x))$ is either one of the two matrices in Spin$(N)$ that correspond to $\Lambda(x)$. (See, e.g., Ref.~\cite{Camporesi} and Appendix D of Ref.~\cite{osti_5650644} for more detailed discussions on spinor representations of orthogonal groups.) 

 The covariant derivative for a vector-spinor field is given by
  \begin{equation}\label{covariant_deriv_vector_spinor}
      \nabla_{\nu} \psi_{\mu} = \partial_{\nu}  \psi_{\mu}  + \frac{1}{2} \omega_{\nu bc} \Sigma^{bc} \psi_{\mu}-\Gamma^{\lambda}_{\hspace{1mm}\nu \mu} \psi_{\lambda},
  \end{equation}
while the covariant derivative for a rank-2 tensor-spinor field is given by
\begin{align}\label{covariant_deriv_tensor_spinor}
      \nabla_{\nu} \psi_{\mu_{1}\mu_{2}} =& \partial_{\nu}  \psi_{\mu_{1}\mu_{2}}  + \frac{1}{2} \omega_{\nu bc} \Sigma^{bc} \psi_{\mu_{1}\mu_{2}}-\Gamma^{\lambda}_{\hspace{1mm}\nu \mu_{1}} \psi_{\lambda \mu_{2}}-\Gamma^{\lambda}_{\hspace{1mm}\nu \mu_{2}} \psi_{\mu_{1}\lambda},
  \end{align}
  where $\omega_{\nu  bc}= e_{\nu}^{\hspace{2mm}d}\, \omega_{d bc}$ [see eq.~(\ref{spin_connection_components})]. The matrices $\Sigma^{ab}$ are the generators of the $2^{[N/2]}$-dimensional spinor representation of Spin$(N)$ and they are given by
\begin{align}\label{Spin(N)_generators}
    \Sigma^{ab}&=\frac{1}{4}[\gamma^{a},\gamma^{b}] \\
    &=\frac{1}{2}\gamma^{a}\, \gamma^{b} -\frac{1}{2}\delta^{ab},\hspace{5mm} a,b=1,...,N.
\end{align}
They satisfy the Spin$(N)$ algebra commutation relations
\begin{equation}
       [\Sigma^{ab},\Sigma^{cd}]= \delta^{bc} \Sigma^{ad}-\delta^{ac} \Sigma^{bd} + \delta^{ad} \Sigma^{bc} - \delta^{bd} \Sigma^{ac}.
   \end{equation}
(The gamma matrices are covariantly constant, i.e. $\nabla_{a}\gamma^{b}=0$ - see e.g. Appendix D of Ref.~\cite{osti_5650644}.)

\noindent \textbf{Eigenspinors on $\bm{S^{N-1}}$}. For later convenience, let us introduce the spinor eigenmodes $\chi_{\pm \ell \tilde{\rho}}(\bm{\theta}_{N-1})$ of the Dirac operator on $S^{N-1}$ (see also Ref.~\cite{Camporesi} and Appendix~\ref{appendix_spinor_eigenmodes_SN-1} of the present paper). These spinor eigenmodes satisfy~\cite{Camporesi}
 \begin{align}\label{eigenspinors on S_(N-1)}
     \tilde{\slashed{\nabla}}\chi_{\pm \ell \tilde{\rho}}=\pm i \left(\ell+\frac{N-1}{2}\right)\chi_{\pm \ell \tilde{\rho}},
 \end{align}
 where $\tilde{\slashed{\nabla}}= \gamma^{a} \tilde{\nabla}_{a}$ is the Dirac operator on $S^{N-1}$, $\tilde{\nabla}_{a}$ is the spinor covariant derivative on $S^{N-1}$ and $\ell$ is the angular momentum quantum number on $S^{N-1}$. The symbol $\tilde{\rho}$ represents labels other than $\ell$. The requirement for regularity of the spinor eigenmodes~(\ref{eigenspinors on S_(N-1)}) on $S^{N-1}$ restricts $\ell$ to take the values $\ell=0,1,2,...$~\cite{Camporesi}. We suppose that the spinor eigenmodes~(\ref{eigenspinors on S_(N-1)}) are normalised as
 \begin{align}\label{normalization_SN-1_spinors}
     \int_{S^{N-1}}\sqrt{\tilde{g}}\,d \bm{\theta}_{N-1}\,\chi_{\pm \ell \tilde{\rho}}(\bm{\theta}_{N-1})^{\dagger}\,\chi_{\pm \ell' \tilde{\rho}'}(\bm{\theta}_{N-1})=\delta_{\ell \ell '} \delta_{\tilde{\rho}  \tilde{\rho}'},
 \end{align}
 where $d \bm{\theta}_{N-1}=d\theta_{N-1}\,d\theta_{N-2}...d\theta_{1}$. The square root of the determinant of the metric on $S^{N-1}$ is
 \begin{align}\label{determinant_metric_S_N-1}
   \sqrt{\tilde{g}}&=\sin^{N-2}{\theta_{N-1}}\,\sin^{N-3}{\theta_{N-2}}\,...\,\sin{\theta_{2}} \\
   &= \sin^{N-2}{\theta_{N-1}}\sqrt{\tilde{\tilde{g}}},
 \end{align}
 where $\tilde{\tilde{g}}$ is the determinant of the metric on $S^{N-2}$. All the $\chi_{+}$ eigenspinors are orthogonal to all the $\chi_{-}$ eigenspinors in eq.~(\ref{normalization_SN-1_spinors})~\cite{Camporesi}. For each allowed value of $\ell$, the eigenspinors $\chi_{+ \ell \tilde{\rho}}$ and $\chi_{- \ell \tilde{\rho}}$ separately form irreducible representations of spin$(N)$~\cite{Homma}.  For odd $N=2p+1$, the spinors $\chi_{+ \ell \tilde{\rho}}$ (or $\chi_{- \ell \tilde{\rho}}$) form a spin$(2p+1)$ representation with the ($p$-component) highest weight~\cite{Camporesi} 
 $$ \vec{f}_{0} = \left(\ell+ \frac{1}{2}, \frac{1}{2},...,\frac{1}{2}  \right).   $$
 For even $N=2p$, the spinors $\chi_{\pm \ell \tilde{\rho}}$ form a spin$(2p)$ representation with the ($p$-component) highest weight~\cite{Camporesi} 
 $$ \vec{f}^{\pm}_{0} = \left(\ell+ \frac{1}{2}, \frac{1}{2},...,\frac{1}{2}, \pm\frac{1}{2}  \right).   $$



  \section{Technical details for the functions describing the dependence of STSSH's on \texorpdfstring{${\theta_{N}}$}{theta{N}}} \label{section_functions describing the dependence on theta_N}
  Before writing down the explicit form of the STSSH's of rank $r~(=1,2)$ on $S^{N}$, it is useful to introduce the functions $\phi^{(a)}_{n \ell}(\theta_{N})$ [eq.~(\ref{phi_a})] and $\psi^{(a)}_{n \ell}(\theta_{N})$ [eq.~(\ref{psi_a})] that describe the dependence of the STSSH's on $\theta_{N}$, since they are going to be used extensively in the rest of the paper. The properties of these functions play a crucial role in the normalisation of the STSSH's and in the derivation of the formulae for the spin$(N+1)$ transformation of the STSSH's (see Section~\ref{section_transf_SN} and Appendix~\ref{Appendix_transfrmn_proeprties_norm_fac}). Most importantly, in view of the analytic continuation of our STSSH's to $dS_{N}$, the properties of the functions $\phi^{(a)}_{n \ell}(\theta_{N})$ and $\psi^{(a)}_{n \ell}(\theta_{N})$ will play a very important role in studying the unitarity/non-unitarity of the spin$(N,1)$ representations formed by the analytically continued STSSH's.
  
As we will see in Sections~\ref{sectn_spin3/2_solving_SN} and \ref{sectn_spin5/2_solving_SN}, the $\theta_{N}$-dependence of the STSSH's on $S^{N}$ is described by functions of the following form:
  \begin{align}
    \phi^{(a)}_{n \ell}(\theta_{N})=&~\kappa_{\phi}(n,\ell)\,\left(\cos{\frac{\theta_{N}}{2}}\right)^{\ell+1-a}\left(\sin{\frac{\theta_{N}}{2}}\right)^{\ell-a}\nonumber \\
    &\times F\left(-n+\ell,n+\ell+N;\ell+\frac{N}{2};\sin^{2}\frac{\theta_{N}}{2}\right),\label{phi_a} 
\end{align}
\begin{align}
    \psi^{(a)}_{n \ell}(\theta_{N})=&~\kappa_{\phi}(n,\ell)\frac{n
    +\frac{N}{2}}{\ell+\frac{N}{2}}  \left(\cos{\frac{\theta_{N}}{2}}\right)^{\ell-a} \left(\sin{\frac{\theta_{N}}{2}} \right)^{\ell+1-a} \nonumber\\ &\times F\left(-n+\ell,n+\ell+N;\ell+\frac{N+2}{2};\sin^{2}\frac{\theta_{N}}{2}\right),\label{psi_a}
\end{align}
where the normalisation factor $\kappa_{\phi}(n,\ell)$ is given by
  \begin{align}\label{normlsn_fac_of_Jacobi}
   \kappa_{\phi}(n,\ell)= \frac{\Gamma(n+N/2)}{\Gamma(n-\ell+1)\Gamma{(\ell+N/2)}},
  \end{align}
while $F(A, B; C; z)$ is the Gauss hypergeometric function \cite{gradshteyn2007}. The number $a$ in eqs.~(\ref{phi_a}) and (\ref{psi_a}) is taken to be an integer for the purposes of this paper. The functions in eqs.~(\ref{phi_a}) and (\ref{psi_a}) can be expressed in terms of the Jacobi polynomials~\cite{gradshteyn2007}, where $\kappa_{\phi}(n, \ell)$ plays the role of the conventional normalisation factor for the Jacobi polynomials~\cite{gradshteyn2007}. (These functions with $a=0$ were used to describe spinors on $S^{N}$~\cite{Camporesi}.) As we will discuss in Section~\ref{sectn_spin3/2_solving_SN} and \ref{sectn_spin5/2_solving_SN}, the integer $n$ is the angular momentum quantum number of the STSSH's on $S^{N}$ and it labels the representation of spin$(N+1)$ formed by the STSSH's. The angular momentum quantum number on $S^{N-1}$, $\ell$, is initially assumed to be a positive integer or zero\footnote{This requirement on $\ell$ is motivated naturally in the recursive construction of the STSSH's on $S^{N}$ in terms of STSSH's on $S^{N-1}$ - see Appendix~\ref{Appendix_spin5/2_solving_SN}.}. Furthermore, the requirement for the absence of singularity in the STSSH's on $S^{N}$ will give rise to the condition
\begin{align}
     &n-\ell \,\in \mathbb{N}_{ 0} \label{restriction on n-ell}
\end{align}
or equivalently $n \geq \ell$, where $\mathbb{N}_{0}$ is the set of positive integers including zero. (This condition also arises from the branching rules for spin$(N+1) \supset$ spin$(N)$, as we will see below.) In particular, eq.~(\ref{restriction on n-ell}) is obtained in Appendix~\ref{Appendix_spin5/2_solving_SN}, by requiring the regularity of $\phi^{(a)}_{n \ell}(\theta_{N})$ and $\psi^{(a)}_{n \ell}(\theta_{N})$ in the limit $\theta_{N} \rightarrow \pi$. 

The functions
  $\phi^{(a)}_{n \ell}(\theta_{N})$ and $\psi^{(a)}_{n \ell}(\theta_{N})$ are related to each other by the following formulae:
 \begin{align}
    \left(\frac{d}{d\theta_{N}}+\frac{N+2a-1}{2}\cot{\theta_{N}}+\frac{\ell+(N-1)/2}{\sin{\theta_{N}}}\,\right)\psi^{(a)}_{n \ell}(\theta_{N})&=\left(n+\frac{N}{2}\right)\phi^{(a)}_{n \ell}(\theta_{N}) \label{psi_to_phi_sphere}\\
  \left(\frac{d}{d \theta_{N}}+\frac{N+2a-1}{2}\cot{\theta_{N}}-\frac{\ell+(N-1)/2}{\sin{\theta_{N}}} \,\right)\phi^{(a)}_{n \ell}(\theta_{N})&=-\left(n+\frac{N}{2}\right)\psi^{(a)}_{n \ell}(\theta_{N}) \label{phi_to_psi_sphere}.
  \end{align}
  Equations~(\ref{psi_to_phi_sphere}) and (\ref{phi_to_psi_sphere}) are proved using the raising and lowering operators for the Gauss hypergeometric function in Appendix~\ref{appendix_raising_lowering_hypergeom}. Note also the relation
\begin{align}
    \psi^{(a)}_{n \ell}(\theta_{N})=(-1)^{n-\ell}\phi^{(a)}_{n \ell}(\pi-\theta_{N}).
\end{align}

  \section{The STSSH's of rank 1 on the \texorpdfstring{${N}$}{N}-sphere}\label{sectn_spin3/2_solving_SN}
In this Section, we write down explicitly the unnormalised STSSH's of rank 1 [i.e. the TT vector-spinor eigenmodes of eq.~(\ref{Dirac_eqn_fermion_SN})], by following Ref.~\cite{CHH} where these eigenmodes have been originally constructed. However, we will present the results of Ref.~\cite{CHH} in a slightly modified manner that is more suitable for studying the group-theoretic properties of the eigenmodes. We also recommend that the readers refer to our previous article~\cite{Letsios_announce}, in which the steps in the method of separation of variables are discussed in greater detail for spin-3/2 eigenmodes on $dS_{N}$.

 \subsection{STSSH's of rank 1 for \texorpdfstring{${N}$}{N} even} \label{subsectn_spin3/2_solving_SN_even}

 \noindent \textbf{Representation-theoretic background.}
 The equations (\ref{Dirac_eqn_fermion_SN}) and (\ref{TT_conditions_fermions_SN}) for the TT vector-spinor eigenmodes on $S^{N}$ ($N\geq 4$) are written as
  \begin{align}
      &\slashed{\nabla}  {\psi}^{(A;\sigma;n \ell;\tilde{\rho})}_{\pm \mu}=\pm i\left(n +\frac{N}{2}\right)\, {\psi}^{(A;\sigma;n\ell;\tilde{\rho})}_{\pm \mu},\label{eigenvalue_Equation_SN_3/2}\\
      & \nabla^{\alpha}{\psi}^{(A;\sigma;n \ell; {\tilde{\rho}})}_{\pm \alpha}=\gamma^{\alpha}{\psi}^{(A;\sigma;n \ell;\tilde{\rho})}_{\pm \alpha}=0  \label{TT_SN_3/2}.
  \end{align}
  We have denoted the TT vector-spinor eigenmodes with eigenvalue $\pm i (n+\frac{N}{2})$ as ${\psi}^{(A;\sigma;n \ell;\tilde{\rho})}_{\pm \mu}$,
 where $n=1,2,...$ and $\ell=1,...,n$ are the angular momentum quantum numbers on $S^{N}$ and $S^{N-1}$, respectively.\footnote{The angular momentum quantum numbers for our STSSH's of rank $r \in \set{1,2}$ on $S^{N}$ satisfy $n \geq \ell \geq r$. The condition $n \geq \ell$ was discussed in the previous Section - see eq.~(\ref{restriction on n-ell}). However, as we will see below, the condition $\ell \geq r$ is obtained by using the explicit expressions of the STSSH's.} The representation-theoretic meaning of the labels $A, \sigma, n, \ell$ and $\tilde{\rho}$ is explained below.

 For each value of $n$ we have a representation of spin$(N+1)$ (i.e. algebra of Spin$(N+1)$) acting on the space of the eigenmodes ${\psi}^{(A;\sigma;n\ell;\tilde{\rho})}_{+ \mu}$ (or~${\psi}^{(A;\sigma;n\ell;\tilde{\rho})}_{-\mu}$) with highest weight $\vec{\lambda} = (\lambda_{1},...,\lambda_{N/2})$ given by~\cite{Homma}
 \begin{align}
     \vec{\lambda}=\left(n+\frac{1}{2},\frac{3}{2},\frac{1}{2},...,\frac{1}{2}\right), \hspace{6mm}(n=1,2,...).
 \end{align}
 Note that for $N=4$ we have $\vec{\lambda}=(n+1/2,\,3/2)$ ($n=1,2,...$). The two sets of eigenmodes,  $\set{{\psi}^{(A;\sigma;n\ell;\tilde{\rho})}_{+\mu}}$ and  $\set{{\psi}^{(A;\sigma;n\ell;\tilde{\rho})}_{-\mu}}$, form equivalent representations and they are related to each other by ${\psi}^{(A;\sigma;n\ell;\tilde{\rho})}_{+ \mu}= \gamma^{N+1}{\psi}^{(A;\sigma;n\ell;\tilde{\rho})}_{- \mu}$.

 From a representation-theoretic viewpoint, the construction of eigenmodes on $S^{N}$ using the method of separation of variables corresponds to specifying the basis vectors of a spin$(N+1)$ representation space in the decomposition spin$(N+1) \supset $ spin$(N)$. For a spin$(N+1)$ representation $\vec{\lambda} = (\lambda_{1},...,\lambda_{N/2})$ ($N$ even), the spin$(N)$ content corresponds to highest weights $\vec{f} = (f_{1},...,f_{N/2})$ with~\cite{Camporesi,barut_group,Dobrev:1977qv}
 \begin{align} \label{branch rules spin(N+1)>spin(N)}
  \lambda_{1} \geq f_{1} \geq \lambda_{2} \geq ... \geq \lambda_{N/2} \geq |f_{N/2}|,   
 \end{align}
 where $f_{N/2}$ can be negative.
 In the case of TT vector-spinor eigenmodes on $S^{N}$, $\vec{\lambda}=\left(n+\frac{1}{2},\frac{3}{2},\frac{1}{2},...,\frac{1}{2}\right)$, the spin$(N)$ content corresponds to representations with highest weights: $\vec{f}_{0}^{\sigma}=\left(\ell+\frac{1}{2},\frac{1}{2},...,\frac{1}{2}, \sigma \frac{1}{2}\right)$ and $\vec{f}_{1}^{\sigma}=\left(\ell+\frac{1}{2},\frac{3}{2},\frac{1}{2},...,\frac{1}{2}, \sigma \frac{1}{2}\right)$ with $\sigma = \pm$. We call the index $\sigma$ `the spin projection index' on $S^{N}$. The symbol $\tilde{\rho}$ stands for the representation-theoretic labels concerning the chain of subalgebras spin$(N-1) \supset$ spin$(N-2) \supset ... \supset$ spin$(2)$.

Depending on the `spin' of the spin$(N)$ representations included in our spin$(N+1)$ representation of interest, the solutions of equations~(\ref{eigenvalue_Equation_SN_3/2}) and (\ref{TT_SN_3/2}) are separated into two different types, namely, the \textbf{type}-$\bm{I}$ \textbf{modes} and the \textbf{type}-$\bm{I\!I}$ \textbf{modes}~\cite{CHH}. In particular, the type-$I$ modes correspond to the spinor representation of spin$(N)$ $$\vec{f}_{0}^{\pm}=\left(\ell+\frac{1}{2},\frac{1}{2},...,\frac{1}{2}, \pm \frac{1}{2}\right),$$while the type-$I\!I$ modes correspond to the TT vector-spinor representation $$\vec{f}_{1}^{\pm}=\left(\ell+\frac{1}{2},\frac{3}{2},\frac{1}{2},...,\frac{1}{2}, \pm \frac{1}{2}\right).$$We assign to the label $A$ the value `$I$' in order to indicate the type-$I$ modes (${\psi}^{(I;\sigma;n\ell;\tilde{\rho})}_{\pm \mu}$) and the value `$I\!I\text{-}\tilde{A}$' in order to indicate the type-$I\!I$ modes (${\psi}^{(I\!I\text{-}\tilde{A};\sigma;n\ell;\tilde{\rho})}_{\pm \mu}$), where the label $\tilde{A}$ on $S^{N-1}$ corresponds to $A$ on $S^{N}$ (the label $\tilde{A}$ is discussed further in the passage after eq.~(\ref{eigenvectorspinors_SN-1_TT})).
 

\noindent \textbf{Type-}$\bm{I}$ \textbf{modes.} The type-$I$ modes are expressed in their vector components as
\begin{align}\label{introducing_typeI_spin3/2}
  {\psi}^{(I;\sigma;n \ell;\tilde{\rho})}_{\pm \mu}=\left({\psi}^{(I;\sigma;n \ell;\tilde{\rho})}_{\pm \theta_{N}},  {\psi}^{(I;\sigma;n \ell;\tilde{\rho})}_{\pm \theta_{j}} \right) 
\end{align}
($j=1,...,N-1$), where ${\psi}^{(I;\sigma;n \ell;\tilde{\rho})}_{\pm \theta_{N}}$ is a spinor on $S^{N-1}$, while ${\psi}^{(I;\sigma;n \ell;\tilde{\rho})}_{\pm \theta_{j}}$ is a vector-spinor on $S^{N-1}$~\cite{CHH}. The type-$I$ modes with negative spin projection ($\sigma=-$) on $S^{N}$ are given by~\cite{CHH}
 \begin{equation}\label{TYPE_I_thetaN_negative_spin_3/2}
    {\psi}^{(I;-;n\ell;\tilde{\rho})}_{\pm  \theta_{N}}(\theta_{N},\bm{\theta}_{N-1})= \begin{pmatrix}\phi^{(1)}_{n \ell}(\theta_{N})  \chi_{- \ell \tilde{\rho}}(\bm{\theta}_{N-1})  \\\pm i\psi^{(1)}_{n \ell}(\theta_{N}) \chi_{- \ell \tilde{\rho}}(\bm{\theta}_{N-1})  \end{pmatrix}
\end{equation}
 \begin{align}\label{TYPE_I_theta_i_negative_spin_3/2}
   {\psi}^{(I;-;n \ell;\tilde{\rho})}_{\pm  \theta_{j}}(\theta_{N},\bm{\theta}_{N-1}) 
    = \begin{pmatrix} C_{ n \ell}^{(\uparrow)(1)}(\theta_{N}) \,\tilde{\nabla}_{\theta_{j}}\chi_{- \ell \tilde{\rho}}(\bm{\theta}_{N-1})+ D_{ n \ell}^{(\uparrow)(1)}(\theta_{N}) \,\tilde{\gamma}_{\theta_{j}}\chi_{- \ell \tilde{\rho}}(\bm{\theta}_{N-1}) \\ \\ \pm i C_{ n \ell}^{(\downarrow)(1)}(\theta_{N})\,\tilde{\nabla}_{\theta_{j}}\chi_{- \ell \tilde{\rho}}(\bm{\theta}_{N-1}) \pm i D_{ n \ell}^{(\downarrow)(1)}(\theta_{N})\,\tilde{\gamma}_{\theta_{j}}\chi_{- \ell \tilde{\rho}}(\bm{\theta}_{N-1}) \end{pmatrix} .
\end{align}
 The type-$I$ modes with positive spin projection ($\sigma=+$) on $S^{N}$ are given by~\cite{CHH}
 \begin{equation}\label{TYPE_I_thetaN_positive_spin_3/2}
   {\psi}^{(I;+;n\ell;\rho)}_{\pm  \theta_{N}}(\theta_{N},\bm{\theta}_{N-1})= \begin{pmatrix}i\psi^{(1)}_{n \ell}(\theta_{N})  \chi_{+ \ell \tilde{\rho}}(\bm{\theta}_{N-1})  \\\pm \phi^{(1)}_{n \ell}(\theta_{N}) \chi_{+ \ell \tilde{\rho}}(\bm{\theta}_{N-1})  \end{pmatrix}
\end{equation}
 \begin{align}\label{TYPE_I_theta_i_positive_spin_3/2}
  {\psi}^{(I;+;n \ell;\tilde{\rho})}_{\pm  \theta_{j}}(\theta_{N},\bm{\theta}_{N-1}) 
    = \begin{pmatrix} iC_{ n \ell}^{(\downarrow)(1)}(\theta_{N}) \,\tilde{\nabla}_{\theta_{j}}\chi_{+ \ell \tilde{\rho}}(\bm{\theta}_{N-1})-i D_{ n \ell}^{(\downarrow)(1)}(\theta_{N}) \,\tilde{\gamma}_{\theta_{j}}\chi_{+ \ell \tilde{\rho}}(\bm{\theta}_{N-1}) \\ \\ \pm  C_{ n \ell}^{(\uparrow)(1)}(\theta_{N})\,\tilde{\nabla}_{\theta_{j}}\chi_{+ \ell \tilde{\rho}}(\bm{\theta}_{N-1}) \mp i D_{ n \ell}^{(\uparrow)(1)}(\theta_{N})\,\tilde{\gamma}_{\theta_{j}}\chi_{+ \ell \tilde{\rho}}(\bm{\theta}_{N-1}) \end{pmatrix} .
\end{align}
The eigenspinors on $S^{N-1}$, $\chi_{\pm \ell \tilde{\rho}}$, satisfy eq.~(\ref{eigenspinors on S_(N-1)}) and they are written down explicitly in Appendix~\ref{appendix_spinor_eigenmodes_SN-1}.
The functions $\phi^{(1)}_{n \ell}$ and $\psi^{(1)}_{n \ell}$ are given by eqs.~(\ref{phi_a}) and (\ref{psi_a}), respectively. The functions $C_{ n\ell}^{(\uparrow)(a)}, C_{ n\ell}^{(\downarrow)(a)}$ are expressed in terms of $\phi^{(a)}_{n \ell}$ and $\psi^{(a)}_{n \ell}$ as follows~\cite{CHH}:
 \begin{align}\label{C1_a_function}
    C_{ n\ell}^{(\uparrow)(a)}(\theta_{N})=&~  \frac{1}{\ell(\ell+N-1)}\Bigg\{ \sin{\theta_{N}}\left[\frac{N-1}{2} \cos{\theta_{N}}+ \ell+\frac{N-1}{2} \right]   \phi^{(a)}_{n \ell}(\theta_{N}) \nonumber\\
    &-\frac{N-1}{N-2} (n+\frac{N}{2}) \sin^{2}{\theta_{N}} \,\psi^{(a)}_{n \ell}(\theta_{N}) \Bigg\},
\end{align}
 \begin{align}\label{C2_a_function}
    C_{ n\ell}^{(\downarrow)(a)}(\theta_{N})=&~  \frac{1}{\ell(\ell+N-1)}  \times\Bigg\{ \sin{\theta_{N}}\left[\frac{N-1}{2} \cos{\theta_{N}}- \ell-\frac{N-1}{2} \right]   \psi^{(a)}_{n \ell}(\theta_{N})\nonumber\\
    &+\frac{N-1}{N-2} (n+\frac{N}{2}) \sin^{2}{\theta_{N}} \phi^{(a)}_{n \ell} (\theta_{N})\Bigg\},
\end{align}
 while the functions $D_{n\ell}^{(\uparrow)(a)}$ and $D_{ n\ell}^{(\downarrow)(a)}$ are given by:
 \begin{align}
    D^{(\uparrow)(a)}_{n \ell}(\theta_{N})=&~\frac{-i}{N-1}\left[-\left(\ell+\frac{N-1}{2}\right)\,C^{(\uparrow)(a)}_{n \ell}(\theta_{N})+\sin{\theta_{N}}\, \phi^{(a)}_{n \ell}(\theta_{N})\right]\label{D1_a_function}\end{align}
    and
    \begin{align}
     D^{(\downarrow)(a)}_{n \ell}(\theta_{N})=&~\frac{-i}{N-1}\left[-\left(\ell+\frac{N-1}{2}\right)\,C^{(\downarrow)(a)}_{n \ell}(\theta_{N})-\sin{\theta_{N}}\, \psi^{(a)}_{n \ell}(\theta_{N})\right],\label{D2_a_function}
\end{align}
respectively.~\textbf{Allowed values for angular momentum quantum numbers:} The appearance of $\ell$ in the denominator in eqs.~(\ref{C1_a_function}) and (\ref{C2_a_function}) reflects the fact that there is no type-$I$ eigenmode if the $\theta_{N}$-component~(\ref{TYPE_I_thetaN_negative_spin_3/2}) [or (\ref{TYPE_I_thetaN_positive_spin_3/2})] has $\ell=0$ (i.e. $\ell$ has to satisfy $\ell \geq r=1$). The condition $n \geq \ell$ and the quantisation of the eigenvalue in eq.~(\ref{eigenvalue_Equation_SN_3/2}) follow from the requirement of regularity of the functions $\phi^{(a)}_{n \ell}(\theta_{N})$ and $\psi^{(a)}_{n \ell}(\theta_{N})$ (see Appendix~\ref{Appendix_spin5/2_solving_SN}). Thus, we have verified that the allowed values for the angular momentum quantum numbers are $n=1,2,...$ and $\ell=1,...,n$.
 
\noindent \textbf{Type-}$\bm{I\!I}$ \textbf{modes.} The vector components of the type-$I\!I$ modes are expressed as~\cite{CHH}
\begin{align}\label{introducing_typeII_spin3/2}
  {\psi}^{(I\!I \text{-}\tilde{A};\sigma;n \ell;\tilde{\rho})}_{\pm \mu}=\left(0,  {\psi}^{(I\!I \text{-}\tilde{A};\sigma;n \ell;\tilde{\rho})}_{\pm \theta_{j}} \right) ,
\end{align}
($j=1,...,N-1$) where ${\psi}^{(I\!I \text{-}\tilde{A};\sigma;n \ell;\tilde{\rho})}_{\pm \theta_{N}}=0$. The type-$I\!I$ modes~(\ref{introducing_typeII_spin3/2}) are TT vector-spinors on $S^{N-1}$. Thus, they can be constructed in terms of TT vector-spinor eigenmodes $\tilde{\psi}^{(\tilde{A}; \ell \tilde{\rho})}_{\pm \theta_{j}}(\bm{\theta}_{N-1})$ on $S^{N-1}$ that satisfy
\begin{align}
   & \tilde{\slashed{\nabla}} \tilde{\psi}^{(\tilde{A}; \ell \tilde{\rho})}_{\pm \theta_{j}}= \pm i \left(\ell+\frac{N-1}{2}\right)  \tilde{\psi}^{(\tilde{A}; \ell \tilde{\rho})}_{\pm \theta_{j}} \label{eigenvectorspinors_SN-1} \\
   &\tilde{\gamma}^{\theta_{i}}\tilde{\psi}^{(\tilde{A}; \ell \tilde{\rho})}_{\pm \theta_{i}}=\tilde{\nabla}^{\theta_{i}}\tilde{\psi}^{(\tilde{A}; \ell \tilde{\rho})}_{\pm \theta_{i}}=0  \label{eigenvectorspinors_SN-1_TT},
\end{align}
where the label $\tilde{A}$ indicates the type of the eigenmode 
$\tilde{\psi}^{(\tilde{A}; \ell \tilde{\rho})}_{\pm \theta_{j}}$~\footnote{As in the case of the label $A$ for eigenmodes on $S^{N}$, the label $\tilde{A}$ in $\tilde{\psi}^{(\tilde{A}; \ell \tilde{\rho})}_{\pm \theta_{j}}$ refers to the `spin' of the spin$(N-1)$ representations appearing in the spin$(N-1)$ content of the spin$(N)$ representations formed by $\{ \tilde{\psi}^{(\tilde{A}; \ell \tilde{\rho})}_{\pm \theta_{j}}\}$.}. (The TT vector-spinor eigenmodes and the corresponding types of modes on odd-dimensional spheres are presented in Subsection~\ref{subsectn_spin3/2_solving_SN_odd}.) The requirement for regularity of $\tilde{\psi}^{(\tilde{A}; \ell \tilde{\rho})}_{\pm \theta_{j}}$ on $S^{N-1}$ gives the allowed values for $\ell$, i.e. $\ell=1,2,...\,$. This requirement for $\ell$ follows naturally from the recursive construction of the STSSH's of rank 1 in Ref.~\cite{CHH}. We suppose that the eigenmodes $\tilde{\psi}^{(\tilde{A}; \ell \tilde{\rho})}_{\pm \theta_{j}}$ are normalised on $S^{N-1}$ as
\begin{align}\label{normalization_SN-1_vectorspinors}
     \int_{S^{N-1}}&\sqrt{\tilde{g}}\,d \bm{\theta}_{N-1}\, \tilde{\psi}^{(\tilde{A}; \ell \tilde{\rho})}_{\pm \theta_{i}}(\bm{\theta}_{N-1})^{\dagger}\, \tilde{\psi}^{(\tilde{A}'; \ell' \tilde{\rho}')\theta_{i}}_{\pm}(\bm{\theta}_{N-1}) =\delta_{\ell \ell '} \delta_{\tilde{\rho}  \tilde{\rho}'} \delta_{\tilde{A} \tilde{A}'},
 \end{align}
where $\sqrt{\tilde{g}}$ is given by eq.~(\ref{determinant_metric_S_N-1}). For each allowed value of $\ell$, the set of eigenmodes $\{\tilde{\psi}^{(\tilde{A}; \ell \tilde{\rho})}_{\pm \theta_{j}}  \}$ forms a spin$(N)$ representation with highest weight $\vec{f}_{1}^{\pm}=\left(\ell+\frac{1}{2},\frac{3}{2},\frac{1}{2},...,\frac{1}{2}, \pm \frac{1}{2}\right)$~\cite{Homma}.

The type-$I\!I$ modes ${\psi}^{(I\!I \text{-}\tilde{A};\sigma;n \ell;\tilde{\rho})}_{\pm \mu}$ on $S^{N}$ with negative ($\sigma=-$) and positive ($\sigma=+$) spin projections are given by~\cite{CHH}
 \begin{align}\label{TYPE_II_thetai_negative_spin_3/2}
   {\psi}^{(I\!I \text{-}\tilde{A};-;n\ell;\tilde{\rho})}_{\pm  \theta_{N}}(\theta_{N},\bm{\theta}_{N-1})=&~0 \nonumber \\
    {\psi}^{(I\!I \text{-}\tilde{A};-;n\ell;\tilde{\rho})}_{\pm  \theta_{j}}(\theta_{N},\bm{\theta}_{N-1})=& \begin{pmatrix}\phi^{(-1)}_{n \ell}(\theta_{N})  \,\tilde{\psi}^{(\tilde{A}; \ell \tilde{\rho})}_{-\theta_{j}}(\bm{\theta}_{N-1})  \\\pm i\psi^{(-1)}_{n \ell}(\theta_{N})\, \tilde{\psi}^{(\tilde{A}; \ell \tilde{\rho})}_{-\theta_{j}}(\bm{\theta}_{N-1})\end{pmatrix}
\end{align}
and
 \begin{align}\label{TYPE_II_thetai_positive_spin_3/2}
   {\psi}^{({I\!I}\text{-}\tilde{A};+;n\ell;\tilde{\rho})}_{\pm  \theta_{N}}(\theta_{N},\bm{\theta}_{N-1})&=\, 0 \nonumber \\ {\psi}^{(I\!I \text{-}\tilde{A};+;n\ell;\tilde{\rho})}_{\pm  \theta_{j}}(\theta_{N},\bm{\theta}_{N-1})&= \begin{pmatrix}i\psi^{(-1)}_{n \ell}(\theta_{N})  \tilde{\psi}^{(\tilde{A}; \ell \tilde{\rho})}_{+\theta_{i}}(\bm{\theta}_{N-1})  \\\pm \phi^{(-1)}_{n \ell}(\theta_{N}) \tilde{\psi}^{(\tilde{A}; \ell \tilde{\rho})}_{+\theta_{j}}(\bm{\theta}_{N-1})\end{pmatrix},
\end{align}
($j=1,...,N-1$) respectively. The functions
$\phi^{(-1)}_{n \ell}$ and $\psi^{(-1)}_{n \ell}$ are given by eqs.~(\ref{phi_a}) and (\ref{psi_a}), respectively. As in the case of type-$I$ modes, we find the allowed values $n=1,2,...$ and $\ell=1,...,n$.


 \subsection{STSSH's of rank 1 for \texorpdfstring{${N}$}{N} {odd}} \label{subsectn_spin3/2_solving_SN_odd}
\noindent \textbf{Representation-theoretic background.}
The eigenvalue equation and the TT conditions are given again by eqs.~(\ref{eigenvalue_Equation_SN_3/2}) and (\ref{TT_SN_3/2}), respectively, while the gamma matrices are now given by eq.~(\ref{odd_gammas}). The TT eigenmodes on $S^{N}$ are denoted as ${\psi}^{(A;n\ell;\tilde{\rho})}_{\pm  \mu}$. The allowed values for the angular momentum quantum numbers are $n=1,2,...$ and $\ell=1,...,n$. As we can see, the labeling of the eigenmodes is slightly different than in the case with $N$ even. However, the representation-theoretic meaning of the labels $A, n, \ell$ and $\tilde{\rho}$ is similar to the case with $N$ even, as will be explained below.

For each allowed value of $n$ we have a representation of spin$(N+1)$ acting on the space of the eigenmodes ${\psi}^{(A;n\ell;\tilde{\rho})}_{\pm \mu}$. The highest weights $\vec{\lambda} = (\lambda_{1},...,\lambda_{(N+1)/2})$ for these representations are given by~\cite{Homma}
 \begin{align}\label{highest_weight_3/2_SN_odd}
    \vec{\lambda}^{\pm}=\left(n+\frac{1}{2},\frac{3}{2},\frac{1}{2},...,\frac{1}{2},\pm\frac{1}{2}\right), \hspace{6mm}(n=1,2,...).
 \end{align}
 Unlike the case with $N$ even, for $N$ odd there does not exist any spinorial matrix that relates ${\psi}^{(A;n\ell;\tilde{\rho})}_{+ \mu}$ and ${\psi}^{(A;n\ell;\tilde{\rho})}_{- \mu}$, since the two sets of modes form inequivalent representations of spin$(N+1)$\footnote{In general, for $N$ odd there does not exist any spinorial matrix that relates two STSSH's of arbitrary rank $r$ with different sign for the eigenvalue.}. Note that for $N=3$ we have $\vec{\lambda}^{\pm}=(n+1/2,\,\pm3/2)$ ($n=1,2,...$).

As in the case with $N$ even, the construction of eigenmodes on $S^{N}$ using the method of separation of variables corresponds to specifying the basis vectors of a spin$(N+1)$ representation space in the decomposition spin$(N+1) \supset $ spin$(N)$. For a spin$(N+1)$ representation $\vec{\lambda} = (\lambda_{1},...,\lambda_{(N+1)/2})$ ($N$ odd), where $\lambda_{(N+1)/2}$ can be negative, the spin$(N)$ content corresponds to highest weights $\vec{f} = (f_{1},...,f_{(N-1)/2})$ with~\cite{Camporesi,barut_group,Dobrev:1977qv}
 \begin{align} \label{branch rules spin(N+1)>spin(N) Nodd}
  \lambda_{1} \geq f_{1} \geq \lambda_{2} \geq ... \geq \lambda_{(N-1)/2} \geq f_{(N-1)/2} \geq |\lambda_{(N+1)/2}|.  
 \end{align}
 In the case of TT vector-spinor eigenmodes on $S^{N}$, $\vec{\lambda}^{\pm}=\left(n+\frac{1}{2},\frac{3}{2},\frac{1}{2},...,\frac{1}{2}, \pm \frac{1}{2}\right)$, the spin$(N)$ content corresponds to representations with highest weights: $\vec{f}_{0}=\left(\ell+\frac{1}{2},\frac{1}{2},..., \frac{1}{2}\right)$ and $\vec{f}_{1}=\left(\ell+\frac{1}{2},\frac{3}{2},\frac{1}{2},...,\frac{1}{2}\right)$. As the representation $\vec{f}_{\tilde{r}}=\left(\ell+\frac{1}{2},\tilde{r}+\frac{1}{2},..., \frac{1}{2}, \frac{1}{2}\right)$ is equivalent to $\vec{f}^{~'}_{\tilde{r}}=\left(\ell+\frac{1}{2},\tilde{r}+\frac{1}{2},...,\frac{1}{2}, -\frac{1}{2}\right)$~\cite{Dobrev:1977qv}, there is no need to introduce the notion of the `spin projection index' for eigenmodes on odd-dimensional $S^{N}$. As in the case with $N$ even, the symbol $\tilde{\rho}$ stands for representation-theoretic labels concerning the chain of subalgebras spin$(N-1) \supset$ spin$(N-2) \supset ... \supset$ spin$(2)$.


As in the even-dimensional case, the label $A$ denotes the type of the mode, i.e. the `spin' of the corresponding spin$(N)$ representation. In particular, the type-$I$ modes (${\psi}^{(I;n\ell;\tilde{\rho})}_{\pm  \mu}$) on $S^{N}$ are constructed in terms of eigenspinors on $S^{N-1}$, which correspond to the spin$(N)$ highest weight $\vec{f}_{0}=\left(\ell+\frac{1}{2},\frac{1}{2},..., \frac{1}{2}\right)$. The type-$I\!I$ modes (${\psi}^{(I\!I \text{-}\tilde{A};n\ell;\tilde{\rho})}_{\pm  \mu}$) on $S^{N}$ are constructed in terms of TT eigenvector-spinors of type-$\tilde{A}$ on $S^{N-1}$, which correspond to the spin$(N)$ highest weight $\vec{f}_{1}=\left(\ell+\frac{1}{2},\frac{3}{2},\frac{1}{2},..., \frac{1}{2}\right)$. {Note that TT eigenvector-spinor modes of any type on $S^{N}$ (with arbitrary $N$) exist only for $N \geq 3$, while type-$I\!I$ modes exist only for $N \geq 4$~\cite{CHH}.}

\noindent \textbf{Type-}$\bm{I}$ \textbf{modes.} The type-$I$ modes are given by~\cite{CHH} 
\begin{align}\label{TYPE_I_thetaN_Nodd_spin_3/2}
    {\psi}^{(I;n \ell ; \tilde{\rho})}_{\pm \theta_{N}}(\theta_{N},\bm{\theta}_{N-1})
    &=\frac{1}{\sqrt{2}}(1+i \gamma^{N}) \left\{\phi^{(1)}_{n \ell}(\theta_{N}) \pm i  \psi^{(1)}_{n \ell}(\theta_{N})\gamma^{N}\right\}  \chi_{-\ell \tilde{\rho}}(\bm{\theta}_{N-1})
  \end{align}
  \begin{align}\label{TYPE_I_thetaj_Nodd_spin_3/2}
    {\psi}^{(I;n \ell ; \tilde{\rho})}_{\pm \theta_{j}}(\theta_{N},\bm{\theta}_{N-1})=&~\frac{1}{\sqrt{2}}(1+i \gamma^{N})\Bigg\{ \left(C^{(\uparrow)(1)}_{n \ell}(\theta_{N})\pm  iC^{(\downarrow)(1)}_{n \ell}(\theta_{N})\gamma^{N}\right) \tilde{\nabla}_{\theta_{j}}  \chi_{-\ell \tilde{\rho}}(\bm{\theta}_{N-1})\nonumber \\ &+\left(D^{(\uparrow)(1)}_{n \ell}(\theta_{N})\pm  iD^{(\downarrow)(1)}_{n \ell}(\theta_{N})\gamma^{N}\right) \tilde{\gamma}_{\theta_{j}}  \chi_{-\ell \tilde{\rho}}(\bm{\theta}_{N-1})\Bigg\} ,
  \end{align}
($j=1,...,N-1$) where $\chi_{- \ell \tilde{\rho}}$ are the eigenspinors on $S^{N-1}$ satisfying eq.~(\ref{eigenspinors on S_(N-1)}). (Since $\gamma^{N}$ anti-commutes with $\tilde{\slashed{\nabla}}$ we have $\gamma^{N}\chi_{- \ell \tilde{\rho}}=\chi_{+ \ell \tilde{\rho}}$~\cite{Camporesi}.) As in the case with $N$ even, the functions $\phi^{(1)}_{n \ell}$ and $\psi^{(1)}_{n \ell}$ are given by eqs.~(\ref{phi_a}) and (\ref{psi_a}), respectively, while the functions
$C^{(\uparrow)(1)}_{n \ell}, C^{(\downarrow)(1)}_{n \ell}, D^{(\uparrow)(1)}_{n \ell}$ and $D^{(\downarrow)(1)}_{n \ell}$ are given by eqs.~(\ref{C1_a_function}), (\ref{C2_a_function}), (\ref{D1_a_function}) and (\ref{D2_a_function}), respectively. As in the even-dimensional case, one finds that the angular momentum quantum numbers are allowed to take the values $n=1,2,...$ and $\ell=1,...,n$.

\noindent \textbf{Type-}$\bm{I\!I}$ \textbf{modes.} The type-$I\!I$ modes are given by~\cite{CHH}
\begin{align}
    {\psi}^{(I\!I \text{-}\tilde{A};n \ell ; \tilde{\rho})}_{\pm \theta_{N}}(\theta_{N},\bm{\theta}_{N-1})=&0~ \nonumber \\
    {\psi}^{(I\!I \text{-}\tilde{A};n \ell ; \tilde{\rho})}_{\pm \theta_{j}}(\theta_{N},\bm{\theta}_{N-1})=&~\frac{1}{\sqrt{2}}(1+i \gamma^{N})\left\{\phi^{(-1)}_{n \ell}(\theta_{N}) \pm i  \psi^{(-1)}_{n \ell}(\theta_{N})\gamma^{N}\right\}  \tilde{\psi}^{(\tilde{A}; \ell \tilde{\rho})}_{-\theta_{j}}(\bm{\theta}_{N-1})\label{TYPE_II_thetai_Nodd_spin_3/2},
  \end{align}
where the functions $\phi^{(-1)}_{n \ell}$ and $\psi^{(-1)}_{n \ell}$ are given by eqs.~(\ref{phi_a}) and (\ref{psi_a}), respectively, while the rank-1 STSSH's of type-$\tilde{A}$ on $S^{N-1}$, $\tilde{\psi}^{(\tilde{A}; \ell \tilde{\rho})}_{-\theta_{j}}$, satisfy eqs.~(\ref{eigenvectorspinors_SN-1})-(\ref{normalization_SN-1_vectorspinors}) (where $\gamma^{N}\tilde{\psi}^{(\tilde{A}; \ell \tilde{\rho})}_{-\theta_{j}}= \tilde{\psi}^{(\tilde{A}; \ell \tilde{\rho})}_{+\theta_{j}}$). As in the case with $N$ even, we find that the angular momentum quantum numbers are allowed to take the values: $n=1,2,...$ and $\ell=1,...,n$. 
 \section{The STSSH's of rank 2 on the \texorpdfstring{$N$}{N}-sphere}\label{sectn_spin5/2_solving_SN}
  In this Section we write down explicitly the STSSH's of rank 2 on $S^{N}$ by using the method of separation of variables. In this method the STSSH's of rank 2 on $S^{N}$ are expressed in terms of STSSH's of rank $\tilde{r}$ (where $\tilde{r} \leq r$) on $S^{N-1}$. (The $0$th rank STSSH's are the eigenspinors of the Dirac operator constructed in Ref.~\cite{Camporesi}.) We present the details of the calculations in Appendix~\ref{Appendix_spin5/2_solving_SN}. 
  
  The representation-theoretic background concerning the STSSH's of rank 2 is very similar to the case of STSSH's of rank 1 presented in the previous Section. Therefore, we are not going to discuss the corresponding representation-theoretic details here; we will just focus on the explicit expressions for the STSSH's of rank 2 on $S^{N}$. Let us recall the main idea: the construction of eigenmodes on $S^{N}$ using the method of separation of variables corresponds to specifying the basis vectors of a spin$(N+1)$ representation space in the decomposition spin$(N+1) \supset $ spin$(N)$.

\subsection{STSSH's of rank 2 for \texorpdfstring{${N}$}{N} even}\label{subsectn_spin5/2_solving_SNeven}
 The equations for the STSSH's of rank 2 are given by:
  \begin{align}
      &\slashed{\nabla} {\psi}^{(B;\sigma;n \ell;\tilde{\rho})}_{\pm \mu \nu}=\pm i |\zeta_{n,N}| \,{\psi}^{(B;\sigma;\ell_{N}\ell;\tilde{\rho})}_{\pm \mu \nu},\label{eigenvalue_Equation_SN_5/2}\\
      & \nabla^{\alpha}{\psi}^{(B;\sigma;n \ell; {\tilde{\rho}})}_{\pm \alpha \nu}=\gamma^{\alpha}{\psi}^{(B;\sigma;n \ell;\tilde{\rho})}_{\pm \alpha \nu}=0 , \label{TT_SN_5/2} \\
      & g^{\alpha \beta}{\psi}^{(B;\sigma;n \ell;\tilde{\rho})}_{\pm \alpha \beta}=0, \label{traceless_SN_5/2}
  \end{align}
 [see eqs.~(\ref{Dirac_eqn_fermion_SN}) and (\ref{TT_conditions_fermions_SN})] where the labels $\sigma,n,\ell,\tilde{\rho}$ have the same meaning as in the case of STSSH's of rank 1 [see the discussion after eqs.~(\ref{eigenvalue_Equation_SN_3/2}) and (\ref{TT_SN_3/2})]. Note that eq.~(\ref{traceless_SN_5/2}) is not independent of the gamma-tracelessness condition, as it arises by contracting eq.~(\ref{TT_SN_5/2}) with $\gamma^{\nu}$. As demonstrated in Appendix~\ref{Appendix_spin5/2_solving_SN}, by requiring our eigenmodes to be non-singular, we find the quantisation condition for the eigenvalue in eq.~(\ref{eigenvalue_Equation_SN_5/2}),
 \begin{align}\label{eigenvalue_quantz_5/2}
     |\zeta_{n,N}|=n+\frac{N}{2},\hspace{5mm} n \in \mathbb{N}_{0},
 \end{align}
($\mathbb{N}_{0}$ is the set of positive integers including zero), while the allowed values for the angular momentum quantum numbers are found to be $n=2,3,...$ and $\ell=2,...,n$.

  For each value of $n$ we have a representation of spin$(N+1)$ acting on the space of the eigenmodes ${\psi}^{(B;\sigma;n \ell;\tilde{\rho})}_{+ \mu \nu}$ (or ${\psi}^{(B;\sigma;n \ell;\tilde{\rho})}_{- \mu \nu}$). The highest weight $\vec{\lambda} = (\lambda_{1},...,\lambda_{N/2})$ for this representation is given by~\cite{Homma}
 \begin{align}
     \vec{\lambda}=\left(n+\frac{1}{2},\frac{5}{2},\frac{1}{2},...,\frac{1}{2} \right), \hspace{6mm}(n=2,3,...).
 \end{align}
Note that for $N=4$ we have $\vec{\lambda}= (n+1/2,  5/2)$. As in the case of STSSH's of rank 1, the two sets of eigenmodes, $\set{{\psi}^{(B;\sigma;n\ell;\tilde{\rho})}_{+\mu \nu}}$ and $\set{{\psi}^{(B;\sigma;n\ell;\tilde{\rho})}_{-\mu \nu}}$, form equivalent representations and they are related to each other by ${\psi}^{(B;\sigma;n\ell;\tilde{\rho})}_{+ \mu \nu}= \gamma^{N+1}{\psi}^{(B;\sigma;n\ell;\tilde{\rho})}_{- \mu \nu}$.

 \noindent \textbf{spin$\bm{(N)}$ content and types of eigenmodes.}~Equations~(\ref{eigenvalue_Equation_SN_5/2})-(\ref{traceless_SN_5/2}) have three different types of mode solutions, namely, the \textbf{type-}$\bm{I}$ \textbf{modes}, the \textbf{type-}$\bm{I\!I}$ \textbf{modes} and the \textbf{type-}$\bm{I\!I\!I}$ \textbf{modes}. The label $B$ is used in order to indicate the type of the STSSH ${\psi}^{(B;\sigma;n \ell;\tilde{\rho})}_{\pm \mu \nu}$ on $S^{N}$.
  In analogy with the rank-1 STSSH's discussed in Section~\ref{sectn_spin3/2_solving_SN}, the rank-2 type-$I$ modes are constructed using the eigenspinors $\chi_{\pm \ell \tilde{\rho}}$ on $S^{N-1}$ [eq.~(\ref{eigenspinors on S_(N-1)})], while the type-$I\!I$ modes are constructed using the TT eigenvector-spinors $\tilde{\psi}^{(\tilde{A}; \ell \tilde{\rho})}_{\pm \theta_{i}}$ on $S^{N-1}$ [eqs.~(\ref{eigenvectorspinors_SN-1}) and (\ref{eigenvectorspinors_SN-1_TT})].
  
  The rank-2 type-$I\!I\!I$ modes are constructed using the STSSH's of rank 2 on $S^{N-1}$ ($\tilde{\psi}^{(\tilde{B}; \ell \tilde{\rho})}_{\pm \theta_{i} \theta_{j}}$), satisfying
  \begin{align}
   & \tilde{\slashed{\nabla}} \tilde{\psi}^{(\tilde{B}; \ell \tilde{\rho})}_{\pm \theta_{i} \theta_{j}}= \pm i \left(\ell+\frac{N-1}{2}\right)  \tilde{\psi}^{(\tilde{B}; \ell \tilde{\rho})}_{\pm \theta_{i} \theta_{j}} \label{eigentensorspinors_SN-1} \\
   &\tilde{\gamma}^{\theta_{i}}\tilde{\psi}^{(\tilde{B}; \ell \tilde{\rho})}_{\pm \theta_{i} \theta_{j}}=\tilde{\nabla}^{\theta_{i}}\tilde{\psi}^{(\tilde{B}; \ell \tilde{\rho})}_{\pm \theta_{i} \theta_{j}}=0 , \label{eigentensorspinors_SN-1_TT} \\
  &\tilde{g}^{\theta_{i} \theta_{j}} \tilde{\psi}^{(\tilde{B}; \ell \tilde{\rho})}_{\pm \theta_{i} \theta_{j}}=0  \label{eigentensorspinors_SN-1_trace}  ,
\end{align}
  where the label $\tilde{B}$ indicates the type of the STSSH $\tilde{\psi}^{(\tilde{B}; \ell \tilde{\rho})}_{\pm \theta_{i} \theta_{j}}$ on $S^{N-1}$. (The rank-2 STSSH's on odd-dimensional spheres are presented in Subsection~\ref{subsectn_spin5/2_solving_SNodd}.) We require $\ell=2,3,...$ in order for $\tilde{\psi}^{(\tilde{B}; \ell \tilde{\rho})}_{\pm \theta_{i} \theta_{j}}$ to be non-singular on $S^{N-1}$.\footnote{This requirement for $\ell$ is motivated naturally in the recursive construction of the STSSH's of rank 2 in Appendix~\ref{Appendix_spin5/2_solving_SN}.} 
  We suppose that the STSSH's on $S^{N-1}$, $\tilde{\psi}^{(\tilde{B}; \ell \tilde{\rho})}_{\pm \theta_{i} \theta_{j}} $, are normalised as
  \begin{align}\label{normalization_SN-1_tensorspinors}
     \int_{S^{N-1}}&\sqrt{\tilde{g}}\,d \bm{\theta}_{N-1}\, \tilde{\psi}^{(\tilde{B}; \ell \tilde{\rho})}_{\pm \theta_{i} \theta_{j}}(\bm{\theta}_{N-1})^{\dagger}\, \tilde{\psi}^{(\tilde{B}'; \ell' \tilde{\rho}')\theta_{i} \theta_{j}}_{\pm}(\bm{\theta}_{N-1})=\delta_{\ell \ell '} \delta_{\tilde{\rho}  \tilde{\rho}'} \delta_{\tilde{B} \tilde{B}'},
 \end{align}
  where all the $\tilde{\psi}_{+ \theta_{i} \theta_{j}}$ modes are orthogonal to all the $\tilde{\psi}_{- \theta_{i} \theta_{j}}$ modes.
  For each value of $\ell$, the set of eigenmodes $\{\tilde{\psi}^{(\tilde{B}; \ell \tilde{\rho})}_{\pm \theta_{i} \theta_{j}}  \}$ forms a spin$(N)$ representation with highest weight~\cite{Homma}:
  $$\vec{f}^{\pm}_{2}=\left(\ell+\frac{1}{2},\frac{5}{2},\frac{1}{2},...,\frac{1}{2}, \pm \frac{1}{2}\right)$$
  for $N$ even, and
  $$\vec{f}_{2}=\left(\ell+\frac{1}{2},\frac{5}{2},\frac{1}{2},...,  \frac{1}{2}\right)$$for $N$ odd.
  
  Now let us present the explicit form of the STSSH's of rank 2 on $S^{N}$ (see Appendix~\ref{Appendix_spin5/2_solving_SN} for the derivation).
  
\noindent\textbf{Type-}$\bm{I}$ \textbf{modes.} The type-$I$ modes with negative spin projection ($\sigma=-$) on $S^{N}$ are given by
 \begin{equation}\label{TYPE_I_thetaNthetaN_negative_spin_5/2}
    {\psi}^{(I;-;n\ell;\tilde{\rho})}_{\pm  \theta_{N} \theta_{N}}(\theta_{N},\bm{\theta}_{N-1})= \begin{pmatrix}\phi^{(2)}_{n \ell}(\theta_{N})  \chi_{- \ell \tilde{\rho}}(\bm{\theta}_{N-1})  \\\pm i\psi^{(2)}_{n \ell}(\theta_{N}) \chi_{- \ell \tilde{\rho}}(\bm{\theta}_{N-1})  \end{pmatrix}
\end{equation} 
 \begin{align}\label{TYPE_I_thetaNthetaj_negative_spin_5/2}
   {\psi}^{(I;-;n \ell;\tilde{\rho})}_{\pm  \theta_{N}\theta_{j}}(\theta_{N},\bm{\theta}_{N-1}) 
    = \begin{pmatrix} C_{ n \ell}^{(\uparrow)(2)}(\theta_{N}) \,\tilde{\nabla}_{\theta_{j}}\chi_{- \ell \tilde{\rho}}(\bm{\theta}_{N-1})+ D_{ n \ell}^{(\uparrow)(2)}(\theta_{N}) \,\tilde{\gamma}_{\theta_{j}}\chi_{- \ell \tilde{\rho}}(\bm{\theta}_{N-1}) \\ \\ \pm i C_{ n \ell}^{(\downarrow)(2)}(\theta_{N})\,\tilde{\nabla}_{\theta_{j}}\chi_{- \ell \tilde{\rho}}(\bm{\theta}_{N-1}) \pm i D_{ n \ell}^{(\downarrow)(2)}(\theta_{N})\,\tilde{\gamma}_{\theta_{j}}\chi_{- \ell \tilde{\rho}}(\bm{\theta}_{N-1}) \end{pmatrix}  
\end{align}
\begin{align}\label{TYPE_I_thetajthetak_negative_spin_5/2}
 {\psi}^{(I;-;n \ell;\tilde{\rho})}_{\pm  \theta_{j}\theta_{k}}&(\theta_{N},\bm{\theta}_{N-1}) \nonumber\\
    =& \begin{pmatrix}  K^{(\uparrow)}_{n \ell}(\theta_{N})\,\tilde{g}_{\theta_{j} \theta_{k}}\chi_{-\ell \tilde{\rho}}(\bm{\theta}_{N-1}) \\ \\
 \pm i K^{(\downarrow)}_{n \ell}(\theta_{N})\,\tilde{g}_{\theta_{j} \theta_{k}}\chi_{-\ell \tilde{\rho}}(\bm{\theta}_{N-1})
 \end{pmatrix} \nonumber\\
 &+\begin{pmatrix}
W^{(\uparrow)}_{n \ell}(\theta_{N})\, \tilde{H}_{\theta_{j}\theta_{k}}\chi_{-\ell \tilde{\rho}}(\bm{\theta}_{N-1})+ T^{(\uparrow)}_{n \ell}(\theta_{N})\, \tilde{H}'_{\theta_{j}\theta_{k}}\chi_{-\ell \tilde{\rho}}(\bm{\theta}_{N-1}) \\ \\
 \pm i W^{(\downarrow)}_{n \ell}(\theta_{N}) \,\tilde{H}_{\theta_{j}\theta_{k}}\chi_{-\ell \tilde{\rho}}(\bm{\theta}_{N-1})\pm i T^{(\downarrow)}_{n \ell}(\theta_{N}) \,\tilde{H}'_{\theta_{j}\theta_{k}}\chi_{-\ell \tilde{\rho}}(\bm{\theta}_{N-1}) 
 \end{pmatrix},
\end{align}
($j,k=1,...,N-1$) where $\chi_{\pm \ell \tilde{\rho}}$ are the eigenspinors on $S^{N-1}$ [see eq.~(\ref{eigentensorspinors_SN-1_trace})] and we have defined
\begin{align}
    \tilde{H}_{\theta_{j}\theta_{k}}&\equiv \tilde{\nabla}_{(\theta_{j}}\tilde{\nabla}_{\theta_{k})}-\tilde{g}_{\theta_{j} \theta_{k}}\frac{\tilde{\Box}}{N-1},\label{traceless_tensor_SN-1}\\
    \tilde{H}'_{\theta_{j}\theta_{k}}&\equiv \tilde{\gamma}_{(\theta_{j}}\tilde{\nabla}_{\theta_{k})}-\tilde{g}_{\theta_{j} \theta_{k}}\frac{\tilde{\slashed{\nabla}}}{N-1}.\label{traceless_tensor'_SN-1}
\end{align}
These differential operators satisfy $\tilde{g}^{\theta_{j} \theta_{k}} \tilde{H}_{\theta_{j}\theta_{k}}=\tilde{g}^{\theta_{j} \theta_{k}} \tilde{H}'_{\theta_{j}\theta_{k}}=0$. Note that $\tilde{\slashed{\nabla}}\chi_{\pm\ell \tilde{\rho}}=\pm i\left(\ell+\frac{N-1}{2}\right)\chi_{\pm \ell \tilde{\rho}}$ [eq.~(\ref{eigenspinors on S_(N-1)})], while $\tilde{\Box}\chi_{\pm \ell \tilde{\rho}}\equiv \tilde{\nabla}^{\theta_{k}}\tilde{\nabla}_{\theta_{k}}\chi_{\pm \ell \tilde{\rho}}$ is given by eq.~(\ref{LB_op_eigenspin_SN-1}). The function $\phi^{(2)}_{n \ell}$ is given by eq.~(\ref{phi_a}), the function $\psi^{(2)}_{n \ell}$ is given by eq.~(\ref{psi_a}), the functions $C^{(\uparrow)(2)}_{n \ell}$ and $C^{(\downarrow)(2)}_{n \ell}$ are given by eqs.~(\ref{C1_a_function}) and (\ref{C2_a_function}), respectively, while the functions $D^{(\uparrow)(2)}_{n \ell}$ and $D^{(\downarrow)(2)}_{n \ell}$ are given by eqs.~(\ref{D1_a_function}) and (\ref{D2_a_function}), respectively. The functions describing the dependence on $\theta_{N}$ in eq.~(\ref{TYPE_I_thetajthetak_negative_spin_5/2}) are given by 
\begin{align}
    K^{(\uparrow)}_{n \ell}(\theta_{N})&=-\frac{\sin^{2}{\theta_{N}}}{N-1} \phi^{(2)}_{n \ell}(\theta_{N}),\label{Kup_function}\\
      K^{(\downarrow)}_{n \ell}(\theta_{N})&=-\frac{\sin^{2}{\theta_{N}}}{N-1} \psi^{(2)}_{n \ell}(\theta_{N}) \label{Kdown_function}, \\
    T^{(\uparrow)}_{n \ell}(\theta_{N})&=\frac{-2i}{N+1}\left\{\sin{\theta_{N}}\, C^{(\uparrow)(2)}_{n\ell}(\theta_{N})-\left(\ell+\frac{N-1}{2} \right) W^{(\uparrow)}_{n \ell}(\theta_{N})\, \right\}, \label{Tup_function} \\
     T^{(\downarrow)}_{n \ell}(\theta_{N})&=\frac{-2i}{N+1} \left\{-\sin{\theta_{N}}\, C^{(\downarrow)(2)}_{n\ell}(\theta_{N})-\left(\ell+\frac{N-1}{2} \right) W^{(\downarrow)}_{n \ell}(\theta_{N})\, \right \},\label{Tdown_function}
\end{align} 
\begin{align}\label{Wup_function}
 W^{(\uparrow)}_{n \ell}(\theta_{N})=~&\frac{\sin{\theta_{N}}}{(\ell-1)(\ell+N)(N-1)} \nonumber\\
 &\times \Bigg\{\, \,\Big[\,\,\frac{N(N-3)\left(\ell+\frac{N-1}{2}\right)}{N-1}+\frac{N(N+1)}{2}\cos{\theta_{N}}\Big]\,C^{(\uparrow)(2)}_{n\ell}(\theta_{N}) \nonumber\\ &-(n+\frac{N}{2})(N+1)\sin{\theta_{N}}\,C^{(\downarrow)(2)}_{n\ell}(\theta_{N})+\frac{N+1}{N-1}\sin{\theta_{N}}\,\phi^{(2)}_{n \ell}(\theta_{N})      \Bigg\}
\end{align}
and
\begin{align}\label{Wdown_function}
 W^{(\downarrow)}_{n \ell}(\theta_{N})=&~\frac{\sin{\theta_{N}}}{(\ell-1)(\ell+N)(N-1)} \nonumber \\
 &\times \Bigg\{\, \,\Big[\,\,-\frac{N(N-3)\left(\ell+\frac{N-1}{2}\right)}{N-1}+\frac{N(N+1)}{2}\cos{\theta_{N}}\Big]\,C^{(\downarrow)(2)}_{n\ell}(\theta_{N}) \nonumber\\ &+(n+\frac{N}{2})(N+1)\sin{\theta_{N}}\,C^{(\uparrow)(2)}_{n\ell}(\theta_{N})+\frac{N+1}{N-1}\sin{\theta_{N}}\,\psi^{(2)}_{n \ell}(\theta_{N})      \Bigg\}.
\end{align} 
The type-$I$ modes with positive spin projection, ${\psi}^{(I;+;n\ell;\tilde{\rho})}_{\pm  \mu \nu}$, are given by expressions similar to the expressions for ${\psi}^{(I;-;n\ell;\tilde{\rho})}_{\pm  \mu \nu}$. To be specific, the expression for ${\psi}^{(I;+;n\ell;\tilde{\rho})}_{\pm  \theta_{N} \theta_{N}}$ is found by exchanging $\phi^{(2)}_{n \ell}$ and $i \psi^{(2)}_{n \ell}$ and replacing $\chi_{- \ell \tilde{\rho}}$ by $\chi_{+ \ell \tilde{\rho}}$ in eq.~(\ref{TYPE_I_thetaNthetaN_negative_spin_5/2}) and the expression for the component ${\psi}^{(I;+;n\ell;\tilde{\rho})}_{\pm  \theta_{N} \theta_{j}}$ is found using eq.~(\ref{TYPE_I_thetaNthetaj_negative_spin_5/2}) as follows: we exchange $C^{(\uparrow)(2)}_{n \ell}$ and $i C^{(\downarrow)(2)}_{n \ell}$; we also exchange $D^{(\uparrow)(2)}_{n \ell}$ and $i D^{(\downarrow)(2)}_{n \ell}$ and we make the replacements $\tilde{\nabla}_{\theta_{j}}\chi_{- \ell \tilde{\rho}} \rightarrow \tilde{\nabla}_{\theta_{j}}\chi_{+ \ell \tilde{\rho}}$
and $\tilde{\gamma}_{\theta_{j}}\chi_{- \ell \tilde{\rho}} \rightarrow -\tilde{\gamma}_{\theta_{j}}\chi_{+ \ell \tilde{\rho}}$. Similarly, ${\psi}^{(I;+;n\ell;\tilde{\rho})}_{\pm  \theta_{j} \theta_{k}}$ is found using the expression for ${\psi}^{(I;-;n\ell;\tilde{\rho})}_{\pm  \theta_{j} \theta_{k}}$ [eq.~(\ref{TYPE_I_thetajthetak_negative_spin_5/2})] as follows: we exchange the functions with superscript `$(\uparrow)$' and the functions with superscript `$(\downarrow)$', i.e., $K^{(\uparrow)}_{n \ell} \leftrightarrow iK^{(\downarrow)}_{n \ell}$, $W^{(\uparrow)}_{n \ell} \leftrightarrow iW^{(\downarrow)}_{n \ell}$ and $T^{(\uparrow)}_{n \ell} \leftrightarrow iT^{(\downarrow)}_{n \ell}$ (the symbol $\leftrightarrow$ denotes the exchange of the functions  appearing in the two sides of the `left-right' arrow) and we also make the replacements 
$\chi_{- \ell \tilde{\rho}} \rightarrow \chi_{+ \ell \tilde{\rho}}$
and $\tilde{H}'_{\theta_{j} \theta_{k}} \rightarrow -\tilde{H}'_{\theta_{j} \theta_{k}}$ in eq.~(\ref{TYPE_I_thetajthetak_negative_spin_5/2}).

\noindent \textbf{Allowed values for angular momentum quantum numbers.} Let us now verify that the allowed values for the angular momentum quantum numbers $n$ and $\ell$ for the type-$I$ modes satisfy $n \geq \ell \geq r=2$. As in the case of STSSH's of rank 1 (see Subsection~\ref{subsectn_spin3/2_solving_SN_even}), the appearance of $\ell$ in the denominator in eqs.~(\ref{C1_a_function}) and (\ref{C2_a_function}) implies that there is no type-$I$ mode if the $\theta_{N}\theta_{N}$-component (\ref{TYPE_I_thetaNthetaN_negative_spin_5/2}) has $\ell=0$. Similarly, as eqs.~(\ref{Wup_function}) and (\ref{Wdown_function}) indicate, there is no type-$I$ mode with $\theta_{N}\theta_{N}$-component given by eq.~(\ref{TYPE_I_thetaNthetaN_negative_spin_5/2}) with $\ell=1$. Also, as demonstrated in Appendix~\ref{Appendix_spin5/2_solving_SN}, the quantisation condition~(\ref{eigenvalue_quantz_5/2}) for the eigenvalue, as well as the condition $n - \ell \geq 0$, arise as the requirement for the absence of singularity in the functions $\phi^{(2)}_{n \ell}$ and $\psi^{(2)}_{n \ell}$. Thus, the allowed values for $n$ and $\ell$ are $n=2,3,...$ and $\ell=2,...,n$, respectively.

\noindent\textbf{Type-}$\bm{I\!I}$ \textbf{modes.} The type-$I\!I$ modes with negative spin projection ($\sigma=-$) on $S^{N}$ are given by 
 \begin{align}
    {\psi}^{(I\!I \text{-}\tilde{A};-;n\ell;\tilde{\rho})}_{\pm  \theta_{N}\theta_{N}}(\theta_{N},\bm{\theta}_{N-1})=&~ 0 \label{TYPE_II_thetaNthetaN_negative_spin_5/2} \\ 
    {\psi}^{(I\!I \text{-}\tilde{A};-;n\ell;\tilde{\rho})}_{\pm  \theta_{N}\theta_{j}}(\theta_{N},\bm{\theta}_{N-1})=& \begin{pmatrix}\phi^{(0)}_{n \ell}(\theta_{N})  \,\tilde{\psi}^{(\tilde{A}; \ell \tilde{\rho})}_{-\theta_{j}}(\bm{\theta}_{N-1})  \\\pm i\psi^{(0)}_{n \ell}(\theta_{N})\, \tilde{\psi}^{(\tilde{A}; \ell \tilde{\rho})}_{-\theta_{j}}(\bm{\theta}_{N-1})\end{pmatrix}\label{TYPE_II_thetaNthetj_negative_spin_5/2}
\end{align}
\begin{align}\label{TYPE_II_thetajthetk_negative_spin_5/2}
  &{\psi}^{(I\!I \text{-}\tilde{A};-;n\ell;\tilde{\rho})}_{\pm  \theta_{j}\theta_{k}}(\theta_{N},\bm{\theta}_{N-1})= \begin{pmatrix} \Gamma_{ n \ell}^{(\uparrow)}(\theta_{N}) \,\tilde{\nabla}_{(\theta_{j}}\tilde{\psi}^{(\tilde{A}; \ell \tilde{\rho})}_{- \theta_{k})}(\bm{\theta}_{N-1})+ \Delta_{ n \ell}^{(\uparrow)}(\theta_{N}) \,\tilde{\gamma}_{(\theta_{j}}\tilde{\psi}^{(\tilde{A}; \ell \tilde{\rho})}_{- \theta_{k})}(\bm{\theta}_{N-1}) \\ \\
  \pm i\Gamma_{ n \ell}^{(\downarrow)}(\theta_{N}) \,\tilde{\nabla}_{(\theta_{j}}\tilde{\psi}^{(\tilde{A}; \ell \tilde{\rho})}_{- \theta_{k})}(\bm{\theta}_{N-1})\pm i \Delta_{ n \ell}^{(\downarrow)}(\theta_{N}) \,\tilde{\gamma}_{(\theta_{j}}\tilde{\psi}^{(\tilde{A}; \ell \tilde{\rho})}_{- \theta_{k})}(\bm{\theta}_{N-1})
  \end{pmatrix}
\end{align}
($j,k=1,...,N-1$), where $\phi^{(0)}_{n \ell}$ is given by eq.~(\ref{phi_a}) and $\psi^{(0)}_{n \ell}$ is given by eq.~(\ref{psi_a}). The type-$\tilde{A}$ TT vector-spinor eigenmodes $\tilde{\psi}^{(\tilde{A}; \ell \tilde{\rho})}_{\pm \theta_{k}}$ on $S^{N-1}$ satisfy eqs.~(\ref{eigenvectorspinors_SN-1})-(\ref{normalization_SN-1_vectorspinors}) and they are non-singular on $S^{N-1}$ for $\ell=1,2,...\,$ (see Section~\ref{sectn_spin3/2_solving_SN}). The functions describing the dependence on $\theta_{N}$ in eq.~(\ref{TYPE_II_thetajthetk_negative_spin_5/2}) are given by
\begin{align}
\frac{\Delta^{(\uparrow)}_{n \ell}(\theta_{N})}{2}&= \frac{-i}{N+1} \left[-  \frac{\ell+\frac{N-1}{2}}{2}\Gamma^{(\uparrow)}_{n \ell}(\theta_{N}) +\sin{\theta_{N}}\, \phi^{(0)}_{n \ell}(\theta_{N})\right], \label{Deltaup_function}\\
\frac{\Delta^{(\downarrow)}_{n \ell}(\theta_{N})}{2}&= \frac{-i}{N+1} \left[ -\frac{\ell+\frac{N-1}{2}}{2}\Gamma^{(\downarrow)}_{n \ell}(\theta_{N}) -\sin{\theta_{N}}\, \psi^{(0)}_{n \ell}(\theta_{N})\right]\label{Deltadown_function}
\end{align}
 and 
\begin{align}
    \frac{\Gamma_{ n\ell}^{(\uparrow)}(\theta_{N})}{2}=&~ \frac{1}{(\ell-1)(\ell+N)}\Bigg\{ \sin{\theta_{N}}\left[\frac{N+1}{2} \cos{\theta_{N}}+ \ell+\frac{N-1}{2}  \right]   \phi^{(0)}_{n \ell}(\theta_{N}) \nonumber\\
    &-\frac{N+1}{N} (n+\frac{N}{2}) \sin^{2}{\theta_{N}} \psi^{(0)}_{n \ell}(\theta_{N}) \Bigg\}, \label{Gammaup_function}\\
      \frac{\Gamma_{ n\ell}^{(\downarrow)}(\theta_{N})}{2}=&~ \frac{1}{(\ell-1)(\ell+N)} \Bigg\{ \sin{\theta_{N}}\left[\frac{N+1}{2} \cos{\theta_{N}}- \ell-\frac{N-1}{2}  \right]   \psi^{(0)}_{n \ell}(\theta_{N}) \nonumber\\
    &+\frac{N+1}{N} (n+\frac{N}{2}) \sin^{2}{\theta_{N}} \phi^{(0)}_{n \ell}(\theta_{N}) \Bigg\}. \label{Gammadown_function}
\end{align}
The expressions for the type-$I\!I$ modes with positive spin projection, ${\psi}^{(I\!I \text{-}\tilde{A};+;n\ell;\tilde{\rho})}_{\pm  \mu \nu}$, are similar to the expressions for ${\psi}^{(I\!I \text{-}\tilde{A};-;n\ell;\tilde{\rho})}_{\pm  \mu \nu}$ presented above. More specifically, the expression for ${\psi}^{(I\!I \text{-}\tilde{A};+;n\ell;\tilde{\rho})}_{\pm  \theta_{N}\theta_{j}}$ is found by exchanging $\phi^{(0)}_{n \ell}$
 and $i \psi^{(0)}_{n \ell}$ and making the replacement $\tilde{\psi}^{(\tilde{A}; \ell \tilde{\rho})}_{-\theta_{j}} \rightarrow \tilde{\psi}^{(\tilde{A}; \ell \tilde{\rho})}_{+\theta_{j}}$ in eq.~(\ref{TYPE_II_thetaNthetj_negative_spin_5/2}). The steps required in order to find the expression for ${\psi}^{(I\!I \text{-}\tilde{A};+;n\ell;\tilde{\rho})}_{\pm  \theta_{j}\theta_{k}}$ by using eq.~(\ref{TYPE_II_thetajthetk_negative_spin_5/2}) are: we exchange $\Gamma^{(\uparrow)}_{n \ell}$ and $i \Gamma^{(\downarrow)}_{n \ell}$, as well as $\Delta^{(\uparrow)}_{n \ell}$ and $i \Delta^{(\downarrow)}_{n \ell}$, and we make the replacements $\tilde{\nabla}_{(\theta_{j}}\tilde{\psi}^{(\tilde{A}; \ell \tilde{\rho})}_{-\theta_{k})} \rightarrow \tilde{\nabla}_{(\theta_{j}}\tilde{\psi}^{(\tilde{A}; \ell \tilde{\rho})}_{+\theta_{k)}}$ and $\tilde{\gamma}_{(\theta_{j}}\tilde{\psi}^{(\tilde{A}; \ell \tilde{\rho})}_{-\theta_{k})} \rightarrow -\tilde{\gamma}_{(\theta_{j}}\tilde{\psi}^{(\tilde{A}; \ell \tilde{\rho})}_{+\theta_{k)}}$ in eq.~(\ref{TYPE_II_thetajthetk_negative_spin_5/2}).

\noindent \textbf{Allowed values for angular momentum quantum numbers.} Let us now verify that the allowed values for the angular momentum quantum numbers $n$ and $\ell$ for the type-$I\!I$ modes satisfy $n \geq \ell \geq r=2$. As mentioned in Section~\ref{sectn_spin3/2_solving_SN}, the eigenvector-spinors on $S^{N-1}$ ($\tilde{\psi}^{(\tilde{A}; \ell \tilde{\rho})}_{-\theta_{j}}$) are non-singular for $\ell \geq 1$. Also, since $\ell-1$ appears in the denominator in eqs.~(\ref{Gammaup_function}) and (\ref{Gammadown_function}), there is no type-$I\!I$ mode with $\theta_{N}\theta_{j}$-component given by eq.~(\ref{TYPE_II_thetaNthetj_negative_spin_5/2}) with $\ell=1$.  As in the case of the type-$I$ modes, the quantisation condition~(\ref{eigenvalue_quantz_5/2}) and the condition $n-\ell \geq 0$ arise by demanding $\phi^{(0)}_{n \ell}$ and $\psi^{(0)}_{n \ell}$ to be non-singular. Hence, the allowed values for the angular momentum quantum numbers are $n=2,3,...$ and $\ell=2,...,n$. 

\noindent\textbf{Type-}$\bm{I\!I\!I}$ \textbf{modes.} The type-$I\!I\!I$ modes with negative ($\sigma=-$) and positive ($\sigma=+$) spin projections on $S^{N}$ are given by 
 \begin{align}
  {\psi}^{(I\!I\!I \text{-}\tilde{B};-;n\ell;\tilde{\rho})}_{\pm  \theta_{N}\theta_{N}}(\theta_{N},\bm{\theta}_{N-1})=&~ 0  \label{TYPE_III_thetaNthetaN_negative_spin_5/2}\\
    {\psi}^{(I\!I\!I \text{-}\tilde{B};-;n\ell;\tilde{\rho})}_{\pm  \theta_{N}\theta_{j}}(\theta_{N},\bm{\theta}_{N-1})=&~ 0  \\ 
  {\psi}^{(I\!I\!I \text{-}\tilde{B};-;n\ell;\tilde{\rho})}_{\pm  \theta_{j}\theta_{k}}(\theta_{N},\bm{\theta}_{N-1})
  =& \begin{pmatrix}\phi^{(-2)}_{n \ell}(\theta_{N})  \,\tilde{\psi}^{(\tilde{B}; \ell \tilde{\rho})}_{-\theta_{j} \theta_{k}}(\bm{\theta}_{N-1})  \\\pm i\psi^{(-2)}_{n \ell}(\theta_{N})\, \tilde{\psi}^{(\tilde{B}; \ell \tilde{\rho})}_{-\theta_{j}\theta_{k}}(\bm{\theta}_{N-1})\end{pmatrix}\label{TYPE_III_thetajthetak_negative_spin_5/2}
\end{align}
and
\begin{align}
   {\psi}^{(I\!I\!I \text{-}\tilde{B};+;n\ell;\tilde{\rho})}_{\pm  \theta_{N}\theta_{N}}(\theta_{N},\bm{\theta}_{N-1})=&~ 0  \\
    {\psi}^{(I\!I\!I \text{-}\tilde{B};+;n\ell;\tilde{\rho})}_{\pm  \theta_{N}\theta_{j}}(\theta_{N},\bm{\theta}_{N-1})=&~ 0  \\
  {\psi}^{(I\!I\!I \text{-}\tilde{B};+;n\ell;\tilde{\rho})}_{\pm  \theta_{j}\theta_{k}}(\theta_{N},\bm{\theta}_{N-1})=& \begin{pmatrix}i\psi^{(-2)}_{n \ell}(\theta_{N})  \,\tilde{\psi}^{(\tilde{B}; \ell \tilde{\rho})}_{+\theta_{j} \theta_{k}}(\bm{\theta}_{N-1})  \\\pm \phi^{(-2)}_{n \ell}(\theta_{N})\, \tilde{\psi}^{(\tilde{B}; \ell \tilde{\rho})}_{+\theta_{j}\theta_{k}}(\bm{\theta}_{N-1})\end{pmatrix}\label{TYPE_III_thetajthetak_positive_spin_5/2},
\end{align}
($j,k=1,...,N-1$) respectively, where $\phi^{(-2)}_{n \ell}$ is given by eq.~(\ref{phi_a}) and $\psi^{(-2)}_{n \ell}$ is given by eq.~(\ref{psi_a}). The STSSH's of rank 2 on $S^{N-1}$, $\tilde{\psi}^{(\tilde{B}; \ell \tilde{\rho})}_{+\theta_{j}\theta_{k}}$, satisfy eqs.~(\ref{eigentensorspinors_SN-1})-(\ref{normalization_SN-1_tensorspinors}) and they are non-singular for $\ell=2,3,...\,$ (see the next Subsection).
By working as in the case of type-$I$ and type-$I\!I$ modes discussed above, we find again that the allowed values for the angular momentum quantum numbers are $n=2,3,...$ and $\ell=2,...,n$.
 \subsection{STSSH's of rank 2 for \texorpdfstring{${N}$}{N} odd}\label{subsectn_spin5/2_solving_SNodd}
The equations for the STSSH's of rank 2 are given by eqs.~(\ref{eigenvalue_Equation_SN_5/2})-(\ref{traceless_SN_5/2}), where the gamma matrices are given by eq.~(\ref{odd_gammas}). We denote the STSSH's of rank 2 as ${\psi}^{(B;n\ell;\tilde{\rho})}_{\pm  \mu \nu}$ (with $n=2,...$ and $\ell=2,...,n$), where  the label $B$ denotes the type of the mode. Note that for $N$ odd there is no spin projection index on $S^{N}$ [see also the discussion after eq.~(\ref{highest_weight_3/2_SN_odd})]. The labels $n,\ell$ and $\tilde{\rho}$ have the same meaning as in the case of the STSSH's of rank 1 in Subsection~\ref{subsectn_spin3/2_solving_SN_odd}.

For each value of $n$ we have a representation of spin$(N+1)$ acting on the space of the eigenmodes ${\psi}^{(B;n\ell;\tilde{\rho})}_{\pm \mu \nu}$. The highest weights $\vec{\lambda} = (\lambda_{1},...,\lambda_{(N+1)/2})$ for these representations are~\cite{Homma} 
 \begin{align}
    \vec{\lambda}^{\pm}=\left(n+\frac{1}{2},\frac{5}{2},\frac{1}{2},...,\frac{1}{2},\pm\frac{1}{2}\right), \hspace{6mm}(n=2,3,...).
 \end{align}
  Note that for $N=3$ we have $\vec{\lambda}^{\pm}= (n+1/2, \pm 5/2)$.

\noindent\textbf{Type-}$\bm{I}$ \textbf{modes.} The type-$I$ modes on $S^{N}$ are given by 
\begin{align}
   {\psi}^{(I;n \ell ; \tilde{\rho})}_{\pm \theta_{N} \theta_{N}}(\theta_{N},\bm{\theta}_{N-1})=&\frac{1}{\sqrt{2}}(\bm{1}+i \gamma^{N})\Big\{\phi^{(2)}_{n \ell}(\theta_{N}) \pm i  \psi^{(2)}_{n \ell}(\theta_{N})\gamma^{N}\Big\} \chi_{-\ell \tilde{\rho}}(\bm{\theta}_{N-1})\label{TYPE_I_thetaNthetaN_Nodd_spin_5/2}\\
    {\psi}^{(I;n \ell ; \tilde{\rho})}_{\pm \theta_{N}\theta_{j}}(\theta_{N},\bm{\theta}_{N-1})=
     &\frac{1}{\sqrt{2}}(\bm{1}+i \gamma^{N})\Bigg\{ \left(C^{(\uparrow)(2)}_{n \ell}(\theta_{N})\pm  iC^{(\downarrow)(2)}_{n \ell}(\theta_{N})\gamma^{N}\right) \tilde{\nabla}_{\theta_{j}}\chi_{-\ell \tilde{\rho}}(\bm{\theta}_{N-1}) \nonumber \\ &+\left(D^{(\uparrow)(2)}_{n \ell}(\theta_{N})\pm  iD^{(\downarrow)(2)}_{n \ell}(\theta_{N})\gamma^{N}\right) \tilde{\gamma}_{\theta_{j}}\chi_{-\ell \tilde{\rho}}(\bm{\theta}_{N-1}) \Bigg\}\label{TYPE_I_thetaNthetaj_Nodd_spin_5/2}\\
  {\psi}^{(I;n \ell ; \tilde{\rho})}_{\pm \theta_{j}\theta_{k}}(\theta_{N}, \bm{\theta}_{N-1})=&\frac{1}{\sqrt{2}}(\bm{1}+i \gamma^{N})\Bigg\{ \left(K^{(\uparrow)}_{n \ell}(\theta_{N})\pm  iK^{(\downarrow)}_{n \ell}(\theta_{N})\gamma^{N}\right) \tilde{g}_{\theta_{j}\theta_{k}}\chi_{-\ell \tilde{\rho}}(\bm{\theta}_{N-1}) \nonumber\\
    &+  \left(W^{(\uparrow)}_{n\ell}(\theta_{N})\pm  iW^{(\downarrow)}_{n \ell}(\theta_{N})\gamma^{N}\right)  \tilde{H}_{\theta_{j} \theta_{k}}\chi_{-\ell \tilde{\rho}}(\bm{\theta}_{N-1})\nonumber\\
    &+  \left(T^{(\uparrow)}_{n \ell}(\theta_{N})\pm  iT^{(\downarrow)}_{n \ell}(\theta_{N})\gamma^{N}\right) \tilde{H}'_{\theta_{j} \theta_{k}}\chi_{-\ell \tilde{\rho}}(\bm{\theta}_{N-1}) \Bigg\}\label{TYPE_I_thetajthetak_Nodd_spin_5/2}
  \end{align}
  ($j,k=1,...,N-1$) where the eigenspinors $\chi_{- \ell \tilde{\rho}}$ on $S^{N-1}$ satisfy eq.~(\ref{eigenspinors on S_(N-1)}). The functions $\phi^{(2)}_{n \ell}, \psi^{(2)}_{n \ell}, C^{(b)(2)}_{n \ell}, D^{(b)(2)}_{n \ell},K^{(b)}_{n \ell}, W^{(b)}_{n \ell}$ and $T^{(b)}_{n \ell}$ (where $b= \uparrow, \downarrow$), describing the dependence on $\theta_{N}$, are the same as in the even-dimensional case [see eqs.~(\ref{TYPE_I_thetaNthetaN_negative_spin_5/2})-(\ref{TYPE_I_thetajthetak_negative_spin_5/2})], while $\tilde{H}_{\theta_{j} \theta_{k}}$ and $\tilde{H}'_{\theta_{j} \theta_{k}}$ are given again by eqs.~(\ref{traceless_tensor_SN-1}) and (\ref{traceless_tensor'_SN-1}), respectively.
 
\noindent\textbf{Type-}$\bm{I\!I}$ \textbf{modes.} The type-$I\!I$ modes on $S^{N}$ are given by
 \begin{align}
 {\psi}^{(I\!I \text{-}\tilde{A};n \ell ; \tilde{\rho})}_{\pm \theta_{N} \theta_{N}}(\theta_{N},\bm{\theta}_{N-1})=&~0\label{TYPE_II_thetaNthetaN_Nodd_spin_5/2}\\
{\psi}^{(I\!I \text{-}\tilde{A};n \ell ; \tilde{\rho})}_{\pm \theta_{N} \theta_{j}}(\theta_{N},\bm{\theta}_{N-1})=&~\frac{1}{\sqrt{2}}(\bm{1}+i \gamma^{N})\Big\{\phi^{(0)}_{n \ell}(\theta_{N}) \pm i  \psi^{(0)}_{n \ell}(\theta_{N})\gamma^{N}\Big\}  \tilde{\psi}^{(\tilde{A};\ell \tilde{\rho})}_{-\theta_{j}}(\bm{\theta}_{N-1}) \label{TYPE_II_thetaNthetaj_Nodd_spin_5/2}\\
   {\psi}^{(I\!I \text{-}\tilde{A};n \ell ;\tilde{\rho})}_{\pm \theta_{j}\theta_{k}}(\theta_{N},\bm{\theta}_{N-1})  =&~\frac{1}{\sqrt{2}}(\bm{1}+i \gamma^{N})\Bigg\{ 
       \left(\Gamma^{(\uparrow)}_{n \ell}(\theta_{N})\pm  i\Gamma^{(\downarrow)}_{n \ell}(\theta_{N})\gamma^{N}\right)  \tilde{\nabla}_{(\theta_{j}} \tilde{\psi}^{(\tilde{A};\ell \tilde{\rho})}_{-\theta_{k})}(\bm{\theta}_{N-1}) \nonumber\\
      &+ \left(\Delta^{(\uparrow)}_{n \ell}(\theta_{N})\pm  i\Delta^{(\downarrow)}_{n \ell}(\theta_{N})\gamma^{N}\right)  \tilde{\gamma}_{(\theta_{j}} \tilde{\psi}^{(\tilde{A};\ell \tilde{\rho})}_{-\theta_{k})}(\bm{\theta}_{N-1}) \Bigg\},\label{TYPE_II_thetajthetak_Nodd_spin_5/2}
  \end{align}
($j,k=1,...,N-1$) where the TT eigenvector-spinors $\tilde{\psi}^{(\tilde{A};\ell \tilde{\rho})}_{-\theta_{k}}$ on $S^{N-1}$ satisfy eqs.~(\ref{eigenvectorspinors_SN-1})-(\ref{normalization_SN-1_vectorspinors}). As in the even-dimensional case, the functions $\phi^{(0)}_{n \ell}$ and $\psi^{(0)}_{n \ell}$ are given by eqs.~(\ref{phi_a}) and (\ref{psi_a}), respectively. The functions $\Delta^{(\uparrow)}_{n \ell},\Delta^{(\downarrow)}_{n \ell}, \Gamma^{(\uparrow)}_{n \ell} $ and $\Gamma^{(\downarrow)}_{n \ell}$ are given by eqs.~(\ref{Deltaup_function}) , (\ref{Deltadown_function}), (\ref{Gammaup_function}) and (\ref{Gammadown_function}), respectively.

\noindent\textbf{Type-}$\bm{I\!I\!I}$ \textbf{modes.} The type-$I\!I\!I$ modes on $S^{N}$ are given by
  \begin{align}
 {\psi}^{(I\!I\!I \text{-}\tilde{B};n \ell ; \tilde{\rho})}_{\pm \theta_{N} \theta_{N}}(\theta_{N},\bm{\theta}_{N-1})&=0\label{TYPE_III_thetaNthetaN_Nodd_spin_5/2}\\
{\psi}^{(I\!I\!I \text{-}\tilde{B};n \ell ; \tilde{\rho})}_{\pm \theta_{N} \theta_{j}}(\theta_{N},\bm{\theta}_{N-1})&=0 \\
     {\psi}^{(I\!I\!I \text{-}\tilde{B};n \ell ;\tilde{\rho})}_{\pm \theta_{j}\theta_{k}}(\theta_{N},\bm{\theta}_{N-1})  &=\frac{1}{\sqrt{2}}(\bm{1}+i \gamma^{N})\Big\{\phi^{(-2)}_{n \ell}(\theta_{N}) \pm i  \psi^{(-2)}_{n \ell}(\theta_{N})\gamma^{N}\Big\}  \tilde{\psi}^{(\tilde{B};\ell \tilde{\rho})}_{-\theta_{j} \theta_{k}}(\bm{\theta}_{N-1}),\label{TYPE_III_thetajthetak_Nodd_spin_5/2}
  \end{align}
 ($j,k=1,...,N-1$) where the rank-2 STSSH's on $S^{N-1}$ ($\tilde{\psi}^{(\tilde{B};\ell \tilde{\rho})}_{-\theta_{j} \theta_{k}}$) satisfy eqs.~(\ref{eigentensorspinors_SN-1})-(\ref{normalization_SN-1_tensorspinors}), while the functions $\phi^{(-2)}_{n \ell}$ and $\psi^{(-2)}_{n \ell}$ are given by eqs.~(\ref{phi_a}) and (\ref{psi_a}), respectively. 
 
 As in the case with $N$ even, by requiring that the rank-2 STSSH's of all types (i.e. type-$I$, type-$I\!I$ and type-$I\!I\!I$) on $S^{N}$ are non-singular, we obtain the quantisation condition~(\ref{eigenvalue_quantz_5/2}) for the eigenvalue, while the allowed values for the angular momentum quantum numbers are found to be $n=2,3,...$ and $\ell=2,...,n$.


 \section{Normalisation factors and transformation properties under spin(\texorpdfstring{{$N+1$}}{Nplus1}) of rank-1 and rank-2 STSSH's}\label{section_transf_SN}
 In this Section, we study the transformation properties of a specific class of STSSH's of ranks 1 and 2 on $S^{N}$ under a spin$(N+1)$~transformation. (This class will be specified by determining the spin$(N)$, as well as the spin$(N-1)$, content in the decomposition spin$(N+1) \supset$ spin$(N) \supset$ spin$(N-1)$.) We also write down explicitly the normalisation factors for all STSSH's of ranks 1 and 2 and we make a conjecture for the normalisation factors for STSSH's of arbitrary rank $r$. 
 
 In order to derive the transformation formulae and determine the normalisation factors for STSSH's of ranks 1 and 2, we introduce an inner product on the solution space of eqs.~(\ref{Dirac_eqn_fermion_SN}) and~(\ref{TT_conditions_fermions_SN}) and we also exploit the spin$(N+1)$ invariance of this inner product. The transformation properties and the normalisation factors that we present in this Section have been obtained after long and tedious calculations. For this reason, in this Section, we simply present the results of our lengthy calculations and provide the necessary mathematical background (for example, we discuss the Lie-Lorentz derivative~(\ref{Lie_Lorentz})~\cite{Ortin}). We refer the reader to Appendix~\ref{Appendix_transfrmn_proeprties_norm_fac} for details of the calculations.
\subsection{Lie-Lorentz derivative and spin\texorpdfstring{${(N+1)}$}{(N+1)} invariant inner product}
 Let $\psi_{\mu_{1}...\mu_{r}}$ be any tensor-spinor of rank $r$ and $\xi$ be any Killing vector on $S^{N}$. The infinitesimal change $\delta_{\xi} \psi_{\mu_{1}...\mu_{r}}$ due to the spin$(N+1)$ transformation generated by $\xi$ is conveniently described by the \textbf{Lie-Lorentz derivative}~\cite{Ortin} 
 \begin{align}\label{Lie_Lorentz}
 \mathbb{L}_{{\xi}}~{\psi}_{\mu_{1}...\mu_{r}}  =~&  \xi^{\nu} \nabla_{\nu} {\psi}_{ \mu_{1}...\mu_{r}} +{\psi}_{ \nu \mu_{2}...\mu_{r}} \nabla_{\mu_{1}}\xi^{\nu}+{\psi}_{ \mu_{1} \nu \mu_{3}...\mu_{r}} \nabla_{\mu_{2}}\xi^{\nu}+...+ {\psi}_{ \mu_{1} ...\mu_{r-1}\nu}\nabla_{\mu_{r}}\xi^{\nu}\nonumber \\
 &+ \frac{1}{4}  \nabla_{\kappa} \xi_{\lambda}  \gamma^{\kappa} \gamma^{\lambda }   {\psi}_{\mu_{1}...\mu_{r}}.
\end{align}
The Lie-Lorentz derivative satisfies~\cite{Ortin}
\begin{subequations}
\begin{align}
&\mathbb{L}_{\xi}~e_{\mu}^{\hspace{2mm}a}=0 ,\\
& \mathbb{L}_{\xi}~\gamma^{a}=0  
\end{align}
\end{subequations}
 and - after a straightforward calculation - one can verify that
 \begin{align}\label{commutator_cov_LieLorentz}
     \left( \mathbb{L}_{{\xi}} \nabla_{\mu}-\nabla_{\mu}\mathbb{L}_{\xi}\right)~\psi_{\mu_{1}...\mu_{r}}=0.
 \end{align}
Thus, if $\psi_{\mu_{1}...\mu_{r}}$ satisfies eqs.~(\ref{Dirac_eqn_fermion_SN}) and (\ref{TT_conditions_fermions_SN}) (i.e., if $\psi_{\mu_{1}...\mu_{r}}$ is a STSSH of rank $r$), then $\mathbb{L}_{\mathcal{\xi}}~{\psi}_{ \mu_{1}...\mu_{r}}$ also satisfies eqs.~(\ref{Dirac_eqn_fermion_SN}) and (\ref{TT_conditions_fermions_SN}). Also, the Lie-Lorentz derivative preserves the Lie bracket between two Killing vectors $ [ \mathbb{L}_{\xi} , \mathbb{L}_{\xi'} ] = \mathbb{L}_{[\xi , \xi ']}$, and, thus, generates a spin$(N+1)$ representation in the space of eigenmodes.

Let us introduce the following inner product on the solution space of eqs.~(\ref{Dirac_eqn_fermion_SN}) and (\ref{TT_conditions_fermions_SN}):
 \begin{align}\label{inner_prod_SN}
     \left({\psi}^{(1)}, {\psi}^{(2)}  \right)_{(r)} =~ \int_{S^{N}}\sqrt{g}\, d\bm{\theta}_{N}\,  {\psi}^{(1)\dagger}_{ \mu_{1}...\mu_{r}}\,\, {\psi}^{(2)\mu_{1}...\mu_{r}},
 \end{align}
where $d \bm{\theta}_{N}$ stands for $d \theta_{N}...d\theta_{2}\,d\theta_{1}$, while $\psi^{(1)}_{\mu_{1}...\mu_{r}}$ and $\psi^{(2)}_{\mu_{1}...\mu_{r}}$ are any two STSSH's of rank $r$ with the same angular momentum $n$ on $S^{N}$.\footnote{Any two STSSH's with different signs for the eigenvalue in eq.~(\ref{Dirac_eqn_fermion_SN}) and/or with different $n$  are orthogonal to each other, since $i\slashed{\nabla}$ is hermitian with respect to the inner product~(\ref{inner_prod_SN}).} Since the inner product~(\ref{inner_prod_SN}) is invariant under spin$(N+1)$, we have
 \begin{align}\label{Spin(N+1)-invariance-inner_prod}
    \left(\mathbb{L}_{\xi}{\psi}^{(1)}, {\psi}^{(2)}   \right)_{(r)}  +   \left({\psi}^{(1)}, \mathbb{L}_{\xi}{\psi}^{(2)}  \right)_{(r)}=0
 \end{align}
for any Killing vector $\xi$ on $S^{N}$.

We will study the transformation properties of a certain class of STSSH's of ranks 1 and 2 under spin$(N+1)$, by specialising to the case where the Killing vector in eq.~(\ref{Lie_Lorentz}) is given by $\xi = \mathcal{S}$
\begin{equation}\label{Killing_vector_sphere}
    \mathcal{S}=\mathcal{S}^{\mu}\partial_{\mu}=\cos{\theta_{N-1}} \,\frac{\partial}{\partial {\theta_{N}}} - \cot{\theta_{N}} \, \sin{\theta_{N-1}}\, \frac{\partial}{\partial \theta_{N-1}}.
\end{equation}

Now, let us discuss the certain class of STSSH's of ranks 1 and 2 on $S^{N}$ ($N \geq 3$), the transformation properties of which we are interested in.
\begin{itemize}
\item  \textbf{STSSH's of rank $r=1$}. We will study the transformation properties of the class of STSSH's which comprises: the type-$I$ modes and a certain kind of type-$I\!I$ modes, called type-$I\!I\text{-}I$ modes. The type-$I\!I\text{-}I$ modes on $S^{N}$ are defined for $N \geq 4$ and they are constructed in terms of type-$I$ eigenvector-spinors on $S^{N-1}$. Thus, the type-$I\!I\text{-}I$ modes on $S^{N}$ are given by letting $\tilde{A}=I$ in eqs.~(\ref{TYPE_II_thetai_negative_spin_3/2}) and (\ref{TYPE_II_thetai_positive_spin_3/2}) (for $N$ even) and in eq.~(\ref{TYPE_II_thetai_Nodd_spin_3/2}) (for $N$ odd).

From a representation-theoretic viewpoint, the type-$I$ modes correspond to the following spin$(N)$ and spin$(N-1)$ highest weights: \newline
\noindent \textbf{Type-$\bm{I}$ modes for $\bm{N}$ even:} $$\vec{f}^{\sigma}_{0} = \left(\ell+\frac{1}{2}, \frac{1}{2},..., \frac{1}{2}, \sigma \frac{1}{2} \right) \hspace{2mm}\text{for spin}(N) $$ 
(with $\sigma = \pm$) and
$$\vec{l}_{0} = \left(m+\frac{1}{2}, \frac{1}{2},..., \frac{1}{2} \right) \hspace{2mm}\text{for spin}(N-1), $$ 
where $\vec{f}^{\pm}_{0}$ has $N/2$ components, while $\vec{l}_{0}$ has $N/2-1$ components. These highest weights satisfy the branching rules for spin$(N+1)  \supset $ spin$(N)  \supset $ spin$(N-1)$~\cite{Homma, Dobrev:1977qv}, which imply $n \geq \ell  \geq 1$ and $\ell \geq m \geq 0$.
\newline
\noindent \textbf{Type-$\bm{I}$ modes for $\bm{N}$ odd:} $$\vec{f}_{0} = \left(\ell+\frac{1}{2}, \frac{1}{2},..., \frac{1}{2},  \frac{1}{2} \right) \hspace{2mm}\text{for spin}(N) $$ 
and
$$\vec{l}^{~\sigma_{N-1}}_{0} = \left(m+\frac{1}{2}, \frac{1}{2},..., \frac{1}{2}, \sigma_{N-1} \frac{1}{2} \right) \hspace{2mm}\text{for spin}(N-1), $$ 
where $\sigma_{N-1} = \pm$, $\vec{f}_{0}$ has $(N-1)/2$ components, while $\vec{l}^{\pm}_{0}$ also has $(N-1)/2$ components. We will call the index $\sigma_{N-1}$ the `spin projection index' on $S^{N-1}$ ($N$ odd). Again, these highest weights satisfy the branching rules for spin$(N+1)  \supset $ spin$(N)  \supset $ spin$(N-1)$~\cite{Homma, Dobrev:1977qv}, which imply $n \geq \ell  \geq 1$ and $\ell \geq m \geq 0$.

Similarly, the type-$I\! I$-$I$ modes correspond to the following spin$(N)$ and spin$(N-1)$ highest weights: \newline
\noindent \textbf{Type-$\bm{I\!I}$-$\bm{I}$ modes for $\bm{N}$ even:} $$\vec{f}^{\sigma}_{1} = \left(\ell+\frac{1}{2}, \frac{3}{2},\frac{1}{2},..., \frac{1}{2}, \sigma \frac{1}{2} \right) \hspace{2mm}\text{for spin}(N) $$ 
(with $\sigma = \pm $) and
$$\vec{l}_{0} = \left(m+\frac{1}{2}, \frac{1}{2},..., \frac{1}{2} \right) \hspace{2mm}\text{for spin}(N-1).$$ 
The weight $\vec{f}^{\pm}_{1}$ has $N/2$ components, while $\vec{l}_{0}$ has $N/2-1$ components. These highest weights satisfy the branching rules for spin$(N+1)  \supset $ spin$(N)  \supset $ spin$(N-1)$~\cite{Homma, Dobrev:1977qv}, which imply $n \geq \ell  \geq 1$ and $\ell \geq m \geq 1$.
\newline
\noindent \textbf{Type-$\bm{I\!I}$-$\bm{I}$ modes for $\bm{N}$ odd:} $$\vec{f}_{1} = \left(\ell+\frac{1}{2}, \frac{3}{2},\frac{1}{2},...,\frac{1}{2} \right) \hspace{2mm}\text{for spin}(N) $$ 
and
$$\vec{l}^{~\sigma_{N-1}}_{0} = \left(m+\frac{1}{2}, \frac{1}{2},..., \frac{1}{2}, \sigma_{N-1} \frac{1}{2} \right) \hspace{2mm}\text{for spin}(N-1), $$ 
where $\sigma_{N-1} = \pm$, while $\vec{f}_{1}$ has $(N-1)/2$ components, and $\vec{l}_{0}^{\pm}$ also has $(N-1)/2$ components. Again, these highest weights satisfy the branching rules for spin$(N+1)  \supset $ spin$(N)  \supset $ spin$(N-1)$~\cite{Homma, Dobrev:1977qv}, which imply $n \geq \ell  \geq 1$ and $\ell \geq m \geq 1$.

\item \textbf{STSSH's of rank $r=2$.} We will study the class of STSSH's which comprises: the  type-$I$ modes, the type-$I\!I\text{-}I$ modes and the type-$I\!I\!I\text{-}I$ modes. As in the case of rank-1 STSSH's, the type-$I\!I$-$I$ modes on $S^{N}$ are defined for $N \geq 4$ and they are constructed in terms of type-$I$ eigenvector-spinors on $S^{N-1}$. Thus, these modes are given by letting $\tilde{A}=I$ in eqs.~(\ref{TYPE_II_thetaNthetaN_negative_spin_5/2})-(\ref{TYPE_II_thetajthetk_negative_spin_5/2}) (for $N$ even) and in eqs.~(\ref{TYPE_II_thetaNthetaN_Nodd_spin_5/2})-(\ref{TYPE_II_thetajthetak_Nodd_spin_5/2}) (for $N$ odd). The type-$I\!I\!I$-$I$ modes on $S^{N}$ are defined for $N \geq 4$ and they are constructed in terms of type-$I$ STSSH's of rank 2 on $S^{N-1}$. Thus, the type-$I\!I\!I$-$I$ modes on $S^{N}$ are given by letting $\tilde{B}=I$ in eqs.~(\ref{TYPE_III_thetajthetak_negative_spin_5/2}) and (\ref{TYPE_III_thetajthetak_positive_spin_5/2}) (for $N$ even) and in eq.~(\ref{TYPE_III_thetajthetak_Nodd_spin_5/2}) (for $N$ odd). 

The representation-theoretic content of this class of eigenmodes concerning the decomposition spin$(N+1)  \supset $ spin$(N)  \supset $ spin$(N-1)$ can be found as in the case of STSSH's of rank 1 discussed above.
\end{itemize}
 \subsection{Normalisation factors and transformation properties under spin\texorpdfstring{${(N+1)}$}{(N+1)} of STSSH's of ranks 1 and 2}\label{subsection_transf_SN}
\noindent \textbf{Case 1: \texorpdfstring{$N$}{N} even.} Using the inner product~(\ref{inner_prod_SN}), we define 
the normalisation factors $c^{(B;r)}_{N}(n, \ell)$ for the STSSH's of arbitrary rank $r$ and type $B$ on $S^{N}$, ${\psi}^{(B;\sigma;n\ell;\tilde{\rho})}_{\pm \mu_{1}...\mu_{r}}$, as
\begin{align}\label{define_normn_factors_SN_r}
    & \left({\psi}^{(B;\sigma;n\ell;\tilde{\rho})}_{\pm}, {\psi}^{(B';\sigma';n\ell';\tilde{\rho}')}_{\pm}   \right)_{(r)}\equiv  \left|\frac{c^{(B;r)}_{N}(n, \ell)}{\sqrt{2}}\right|^{-2}\,\delta_{BB'} \delta_{\sigma \sigma'} \delta_{\ell \ell'} \delta_{\tilde{\rho}  \tilde{\rho}'} .
  \end{align}
  The normalised STSSH's are ${c^{(B;r)}_{N}(n, \ell)}/{\sqrt{2}}\, \,{\psi}^{(B;\sigma;n\ell;\tilde{\rho})}_{\pm \mu_{1}...\mu_{r}}$.
  
  As discussed in Sections~\ref{sectn_spin3/2_solving_SN} and \ref{sectn_spin5/2_solving_SN} (for $r=1$ and $r=2$, respectively), the STSSH's of rank $r$ on $S^{N}$, ${\psi}^{(B;\sigma;n\ell;\tilde{\rho})}_{\pm \mu_{1}...\mu_{r}}$, are constructed in terms of STSSH's of rank $\tilde{r} \leq r$ on $S^{N-1}$, using the method of separation of variables. 
  The type of the mode ${\psi}^{(B;\sigma;n\ell;\tilde{\rho})}_{\pm \mu_{1}...\mu_{r}}$ (i.e. the value assigned to the label $B$) depends on the choice of $\tilde{r}$. For convenience, instead of using the symbol $\tilde{r}$, let us denote the rank of the STSSH's on $S^{N-1}$ as $\tilde{r}_{(B)}$, where the type-$I$ STSSH's ($\psi^{(I;\sigma;n\ell;\tilde{\rho})}_{\pm \mu_{1}...\mu_{r}}$) have $\tilde{r}_{(I)}=0$, the type-$I\!I$ STSSH's ($\psi^{(I\!I\text{-}\tilde{A};\sigma;n\ell;\tilde{\rho})}_{\pm \mu_{1}...\mu_{r}}$) have $\tilde{r}_{(I\!I)}=1$, the type-$I\!I\!I$ STSSH's ($\psi^{(I\!I\!I\text{-}\tilde{B};\sigma;n\ell;\tilde{\rho})}_{\pm \mu_{1}...\mu_{r}}$) have $\tilde{r}_{(I\!I\!I)}=~2$ and so forth. As shown in Appendix~\ref{Appendix_transfrmn_proeprties_norm_fac}, the normalisation factors for STSSH's of rank $r\in \set{1,2}$ are given by 
 \begin{align}\label{normln_fac_SN_TYPEB_r}
   \left| \frac{c_{N}^{(B;r)}( n,\ell)}{\sqrt{2}}\right|^{2}=&~  \frac{2^{-N-2r+1+4\tilde{r}_{(B)}}}{\binom{r}{\tilde{r}_{(B)}}}\, \frac{\Gamma(n-\ell+1)  \Gamma(n+\ell+N)}{|\Gamma(n+\frac{N}{2})|^{2}} \nonumber\\
   &\times \left(\prod_{j=\tilde{r}_{(B)}}^{r-1} \frac{N+j+\tilde{r}_{(B)}-2}{N+2j-1} \right) \left(\prod_{j=\tilde{r}_{(B)}}^{r-1} {(\ell-j)(\ell+N-1+j)} \right) \nonumber \\ 
      &\times\prod_{j=1}^{r-\tilde{r}_{(B)}} \frac{1}{{\left(n+\frac{N}{2}\right)^{2}-\left(r-j+\frac{N-2}{2}\right)^{2}}} 
 \end{align}
($\tilde{r}_{(B)} \leq r$) where $\binom{r}{\tilde{r}_{(B)}}$ is the binomial coefficient. The normalisation factors~(\ref{normln_fac_SN_TYPEB_r}) have been determined by either direct calculation or by exploiting the spin$(N+1)$ transformation properties of the STSSH's (the most convenient way to determine the normalisation factors in each case depends on the type $B$ of the eigenmode) - see Appendix \ref{Appendix_transfrmn_proeprties_norm_fac}. Here, if $\nu_{1} > \nu_{2}$, then $\prod_{j=\nu_{1}}^{\nu_{2}}=1$. We have proved eq.~(\ref{normln_fac_SN_TYPEB_r}) only for $r=1$ (where ${B}=I, I\!I$) and for $r=2$ (where ${B}=I, I\!I, I\!I\!I$). 
However, we make the following conjecture, which is true for $r=1$ and $r=2$:
 
 \noindent \textbf{Conjecture}: The normalisation factors for all types of STSSH's (i.e. STSSH's with all possible values of $B$) of arbitrary rank $r \geq 1$ on $S^{N}$ are given by eq.~(\ref{normln_fac_SN_TYPEB_r}), where $n \geq \ell \geq r \geq \tilde{r}_{(B)}$ and $\tilde{r}_{(B)} \in \set{0,1,...,r}$. (This conjecture will be useful in future attempts to extend our present study to the case with spin $s \geq 7/2$.)

\noindent  \textbf{Useful shorthand notation.} Before presenting the transformation properties of our STSSH's of rank $r=1,2$ under spin$(N+1)$, let us introduce the shorthand notation $\psi^{(B;\sigma;n \ell m;\rho)}_{\pm {\bm{N}_{r}}}$ for the STSSH's of ranks 1 and 2, defined as follows:
\begin{subequations}\label{shorthand_not_Neven}
\begin{align}
  \psi^{(B;\sigma;n \ell m;\rho)}_{\pm {\bm{N}_{1}}}&=    
      \psi^{(B;\sigma;n \ell m;\rho)}_{\pm \mu_{1}}\hspace{4mm} (B=I,\, I\!I\text{-}I),\\ 
    \psi^{(I\!I\!I\text{-}I;\sigma;n \ell m;\rho)}_{\pm {\bm{N}_{1}}}&=~0, \\
    \psi^{(B;\sigma;n \ell m;\rho)}_{\pm {\bm{N}_{2}}}&=  \psi^{(B;\sigma;n \ell m;\rho)}_{\pm \mu_{1} \mu_{2}}\hspace{4mm}(B=I, I\!I\text{-}I, I\!I\!I\text{-}I),
\end{align}
\end{subequations}
where we have also written out explicitly the dependence on the angular momentum quantum number on $S^{N-2}$, $m$, which corresponds to $\ell$ on $S^{N-1}$. The symbol $\rho$ represents labels other than $\sigma, n, \ell$ and $m$. For the type-$I$ modes we have $m=0,1,...,\ell$, for the type-$I\!I\text{-}I$ modes we have $m=1,2,...,\ell$ and for the type-$I\!I\!I\text{-}I$ modes we have $m=2,3,...,\ell$. (In other words $\ell \geq m \geq \tilde{r}_{(B)}$.)

\noindent \textbf{Transformation formulae for type-$I$ modes}. As demonstrated in Appendix~\ref{Appendix_transfrmn_proeprties_norm_fac}, the spin$(N+1)$ transformation of the type-$I$ modes is expressed as
\begin{align}\label{transfrmn_type_I_unnormalsd_r_SN_even}
     \mathbb{L}_{\mathcal{S}}{\psi}^{(I;\sigma;n \ell m;\rho)}_{\pm \bm{N}_{r} } =&~  \mathcal{A}^{(I)}\, {\psi}^{(I;\sigma;n \,(\ell+1)\,m;\rho)}_{\pm \bm{N}_{r} } + \mathcal{B}^{(I)}\, {\psi}^{(I;\sigma;n \,(\ell-1)\,m;\rho)}_{\pm \bm{N}_{r} }- i \varkappa^{(I)}\, {\psi}^{(I;-\sigma;n \ell m ;\rho)}_{\pm \bm{N}_{r} }\nonumber\\
     &+ \mathcal{K}^{(I\rightarrow I\!I)}\,{\psi}^{(I\!I\text{-}I;\sigma;n \ell m ;\rho)}_{\pm\bm{N}_{r}},
\end{align}
where the coefficients on the right-hand side of eq.~(\ref{transfrmn_type_I_unnormalsd_r_SN_even}) are
\begin{align}
    \mathcal{A}^{(I)}=&-\frac{(n+\ell+N)(\ell +N+r-1)}{2(\ell+\frac{N}{2})(\ell+N-1)} \times \sqrt{(\ell-m+1)(\ell+N-1+m)}, \label{trnsf_coffcient_raise_unnormtypeI_r}\\
     \mathcal{B}^{(I)}=&~\frac{(n-\ell+1)(\ell-r)}{2(\ell+\frac{N-2}{2})\ell}\times\sqrt{(\ell-m)(\ell+m+N-2)},\label{trnsf_coffcient_lower_unnormtypeI_r} \end{align}
     \begin{align}
   \varkappa^{(I)}=&-\frac{(n+\frac{N}{2})(m+\frac{N-2}{2})(N+2r-2)}{2(\ell+\frac{N-2}{2})(\ell+\frac{N}{2})(N-2)} ,\label{varkap_I_coeff_r} \end{align} 
   and
   \begin{align}
     \mathcal{K}^{(I\rightarrow I\!I)}=&-\frac{4\left[ \left(n+\frac{N}{2}\right)^{2}-{(N-2)^{2}}/{4}\right](N+r-2)}{\ell(\ell+N-1)(N-2)}\times\sqrt{\frac{N-3}{N-2}\frac{m(m+N-2)}{(\ell+1)(\ell+N-2)}}\label{trnsf_coffcient_mixI->II_unnorm_r}.
\end{align} 
Equations~(\ref{transfrmn_type_I_unnormalsd_r_SN_even})-(\ref{trnsf_coffcient_mixI->II_unnorm_r}) hold for $r=1,2$. Note that the sign of the spin projection index $\sigma$ is flipped in the third term of the linear combination in eq.~(\ref{transfrmn_type_I_unnormalsd_r_SN_even}), while $i\varkappa^{(I)}$ is the only imaginary coefficient on the right-hand side of this equation. Also, note that $\mathcal{K}^{(I \rightarrow I\!I)}$ vanishes for $m=0$, i.e. for $m=0$ there is no mixing between type-$I$ and type-$I\!I\text{-}I$ modes in eq.~(\ref{transfrmn_type_I_unnormalsd_r_SN_even}). This is consistent with the fact that type-$I\!I \text{-}I$ modes are defined only for $m=1,2,...,\ell$. 
 
 \noindent \textbf{Transformation formulae for type-$I\!I\text{-}I$ modes.} The spin$(N+1)$ transformation of the type-$I\!I\text{-}I$ modes is expressed as
\begin{align}\label{transfrmn_type_II-I_unnormalsd_r_SN_even}
     \mathbb{L}_{\mathcal{S}} {\psi}^{(I\!I\text{-}I;\sigma;n \ell m;\rho)}_{\pm  \bm{N}_{r} }=&~  \mathcal{A}^{(I\!I)} \,{\psi}^{(I\!I\text{-}I;\sigma;n \,(\ell+1)\,m;\rho)}_{\pm  \bm{N}_{r} } + \mathcal{B}^{(I\!I)} \,{\psi}^{(I\!I\text{-}I;\sigma;n \,(\ell-1)\,m;\rho)}_{\pm \bm{N}_{r} }\nonumber\\
     &- i \varkappa^{(I\!I)} {\psi}^{(I\!I\text{-}I;-\sigma;n \ell m ;\rho)}_{\pm \bm{N}_{r}  }\nonumber\\
     &+ \mathcal{K}^{(I\!I\rightarrow I)}\,{\psi}^{(I;\sigma;n \ell m ;\rho)}_{\pm \bm{N}_{r}}+ \mathcal{K}^{(I\!I\rightarrow I\!I\!I)}\,{\psi}^{(I\!I\!I\text{-}I;\sigma;n \ell m ;\rho)}_{\pm  \bm{N}_{r}} 
\end{align}
 where
 \begin{align}
\mathcal{A}^{(I\!I)}=&-\frac{(n+\ell+N)(\ell+N+r-1)}{2(\ell+\frac{N}{2})(\ell+N)}\nonumber\\
&\times \sqrt{\frac{(\ell+2)(\ell+N-2)}{(\ell+1)(\ell+N-1)}(\ell-m+1)(\ell+m+N-1)},\label{trnsf_coffcient_raise_unnormtypeII_r}\end{align}
\begin{align}
   \mathcal{B}^{(I\!I)} =&~\frac{(n-\ell+1)(\ell-r)}{2(\ell+\frac{N-2}{2})(\ell-1)}\times\sqrt{\frac{(\ell+1)(\ell+N-3)}{\ell(\ell+N-2)}(\ell-m)(\ell+m+N-2)},\label{trnsf_coffcient_lower_unnormtypeII_r} \end{align}
   \begin{align}
    \varkappa^{(I\!I)}=&\frac{-(n+\frac{N}{2})(m+\frac{N-2}{2})(N-4)}{2(\ell+\frac{N-2}{2})(\ell+\frac{N}{2})(N-2)}\times \left( \frac{N+2}{N} \right)^{r-1}\label{varkap_II_coeff_r}
 \end{align}
 \begin{align}\label{trnsf_coffcient_mixII->I_unnorm_r}
  \mathcal{K}^{(I\!I \rightarrow I)} =&\frac{r}{4}\times\sqrt{\frac{(N-3)m(m+N-2)}{(N-2)(\ell+1)(\ell+N-2)}},
 \end{align} 
 where $r=1,2$ and
  \begin{align}
 \mathcal{K}^{(I\!I \rightarrow I\!I\!I)} =&-{2}^{3}\frac{\left[ \left(n+\frac{N}{2}\right)^{2}-{N^{2}}/{4}\right](N+1)}{(\ell-1)(\ell+N)N} \times\sqrt{\frac{N-2}{N}\,\frac{(m-1)(m+N-1)}{\ell(\ell+N-1)}}\label{trnsf_coffcient_mixII->III_unnorm_r=2}
 \end{align}
 [eq.~(\ref{trnsf_coffcient_mixII->III_unnorm_r=2}) is defined only for $r=2$].
 The sign of the spin projection index is flipped in the third term of the linear combination in eq.~(\ref{transfrmn_type_II-I_unnormalsd_r_SN_even}), while $i\varkappa^{(I\!I)}$ is the only imaginary coefficient on the right-hand side of this equation. \textbf{Note} that $\varkappa^{(I\!I)}$ vanishes for $N=4$ and thus type-$I\!I\text{-}I$ modes with different spin projections on $S^{4}$ do not mix with each other under the transformation~(\ref{transfrmn_type_II-I_unnormalsd_r_SN_even}).

 \noindent \textbf{Transformation formulae for type-$I\!I\!I\text{-}I$ modes.} The spin$(N+1)$ transformation of the rank-2 type-$I\!I\!I\text{-}I$ modes is expressed as a linear combination of other STSSH's of rank 2, as follows:
\begin{align}\label{transfrmn_type_III_unnormalsd_5/2_SN_even}
     \mathbb{L}_{\mathcal{S}}{\psi}^{(I\!I\!I\text{-}I;\sigma;n \ell m;\rho)}_{\pm \mu_{1} \mu_{2} } =~&  \mathcal{A}^{(I\!I\!I)}\, {\psi}^{(I\!I\!I\text{-}I;\sigma;n \,(\ell+1)\,m;\rho)}_{\pm \mu_{1} \mu_{2}} + \mathcal{B}^{(I\!I\!I)}\, {\psi}^{(I\!I\!I\text{-}I;\sigma;n \,(\ell-1)\,m;\rho)}_{\pm \mu_{1} \mu_{2}}\nonumber\\
     &- i \varkappa^{(I\!I\!I)}\, {\psi}^{(I\!I\!I\text{-}I;-\sigma;n \ell m ;\rho)}_{\pm \mu_{1} \mu_{2}}+ \mathcal{K}^{(I\!I\!I \rightarrow I\!I)}\,{\psi}^{(I\!I\text{-}I;\sigma;n \ell m ;\rho)}_{\pm \mu_{1} \mu_{2}},
\end{align}
where
 \begin{align}
\mathcal{A}^{(I\!I\!I)}=&-\frac{(n+\ell+N)}{2(\ell+\frac{N}{2})}\times \sqrt{\frac{(\ell+2)(\ell+N-2)}{\ell(\ell+N)}(\ell-m+1)(\ell+m+N-1)},\label{trnsf_coffcient_raise_unnormtypeIII_5/2}\\
   \mathcal{B}^{(I\!I\!I)} =&~\frac{(n-\ell+1)}{2(\ell+\frac{N-2}{2})} \times\sqrt{\frac{(\ell+1)(\ell+N-3)}{(\ell-1)(\ell+N-1)}(\ell-m)(\ell+m+N-2)},\label{trnsf_coffcient_lower_unnormtypeIII_5/2}\end{align}
   \begin{align}
    \varkappa^{(I\!I\!I)}&=-\frac{(n+\frac{N}{2})(m+\frac{N-2}{2})(N-4)}{2(\ell+\frac{N-2}{2})(\ell+\frac{N}{2})N}\label{varkap_III_coeff_spin5/2}
 \end{align}
 and
  \begin{align}
  \mathcal{K}^{(I\!I\!I \rightarrow I\!I)} =&~\frac{1}{4}\sqrt{\frac{(N-2)(m-1)(m+N-1)}{N\,\ell(\ell+N-1)}}\label{trnsf_coffcient_mixIII->II_unnorm_5/2}.
  \end{align}
 As in eqs.~(\ref{transfrmn_type_I_unnormalsd_r_SN_even}) and (\ref{transfrmn_type_II-I_unnormalsd_r_SN_even}), the spin projection index $\sigma$ has flipped sign in the third term of the linear combination in eq.~(\ref{transfrmn_type_III_unnormalsd_5/2_SN_even}). \textbf{Note} that the STSSH's ${\psi}^{(I\!I\!I\text{-}I;-;n \ell m;\rho)}_{\pm \mu \nu }$ and ${\psi}^{(I\!I\!I\text{-}I;+;n \ell m;\rho)}_{\pm \mu \nu }$ do not mix with each other for $N=4$ since the coefficient $\varkappa^{(I\!I\!I)}$ [eq.~(\ref{varkap_III_coeff_spin5/2})] vanishes for this value of $N$.

 \noindent \textbf{Case 2: \texorpdfstring{$N$}{N} odd.}
 As in the case with $N$ even, the normalisation factors for the STSSH's ${\psi}^{(B;n\ell;\tilde{\rho})}_{\pm \mu_{1}...\mu_{r}}$ are defined using the inner product~(\ref{inner_prod_SN}), as
 \begin{align}\label{define_normn_factors_SN_r_Nodd}
     &\left({\psi}^{(B;n\ell;\tilde{\rho})}_{\pm }, {\psi}^{(B';n\ell';\tilde{\rho}')}_{\pm}   \right)_{(r)} \equiv  \left|\frac{c^{(B;r)}_{N}(n, \ell)}{\sqrt{2}}\right|^{-2}\,\delta_{BB'}  \delta_{\ell \ell'} \delta_{\tilde{\rho}  \tilde{\rho}'} .
  \end{align}
As demonstrated in Appendix~\ref{Appendix_transfrmn_proeprties_norm_fac}, the normalisation factors for $N$ odd are given again by eq.~(\ref{normln_fac_SN_TYPEB_r}). The conjecture for the normalisation factors of the STSSH's in the passage below eq.~(\ref{normln_fac_SN_TYPEB_r}) is made for both $N$ odd and $N$ even.
  
As in the case with $N$ even, we introduce the shorthand notation $\psi^{(B;n \ell;\sigma_{N-1};m;\rho)}_{\pm {\bm{N}_{r}}}$ for the STSSH's of ranks 1 and 2, as
\begin{subequations}\label{shorthand_not_Nodd}
\begin{align}
  \psi^{(B;n \ell ;\sigma_{N-1};m;\rho)}_{\pm {\bm{N}_{1}}}&=    
      \psi^{(B;n \ell ;\sigma_{N-1};m;\rho)}_{\pm \mu_{1}}\hspace{4mm} (B=I,\, I\!I\text{-}I),\\ 
    \psi^{(I\!I\!I\text{-}I;n \ell ;\sigma_{N-1};m;\rho)}_{\pm {\bm{N}_{1}}}&=~0, \\
    \psi^{(B;n \ell ;\sigma_{N-1};m;\rho)}_{\pm {\bm{N}_{2}}}&=  \psi^{(B;n \ell ;\sigma_{N-1};m;\rho)}_{\pm \mu_{1} \mu_{2}}\hspace{4mm}(B=I, I\!I\text{-}I, I\!I\!I\text{-}I),
\end{align}
\end{subequations}
where we have also written out explicitly the dependence on the angular momentum quantum number on $S^{N-2}$, $m$, which corresponds to $\ell$ on $S^{N-1}$, as well as the dependence on the spin projection index on $S^{N-1}$ ($\sigma_{N-1}=\pm$). The symbol $\rho$ represents labels other than $n,\ell,\sigma_{N-1}$ and $m$.

\noindent \textbf{Transformation formulae.} As shown in Appendix~\ref{Appendix_transfrmn_proeprties_norm_fac}, the spin$(N+1)$ ($N$ odd) transformation of the type-$I$, type-$I\!I\text{-}I$ and type-$I\!I\!I\text{-}I$ modes are expressed as
 \begin{align}\label{transfrmn_type_I_unnormalsd_r_SN_odd}
     \mathbb{L}_{\mathcal{S}} \psi^{(I;n \ell;\sigma_{N-1}; m;\rho)}_{\pm \bm{N}_{r} } =&~\mathcal{A}^{(I)} \, \psi^{(I;n \,(\ell+1);\sigma_{N-1};m;\rho)}_{\pm \bm{N}_{r} } +\mathcal{B}^{(I)}\,\psi^{(I;n \,(\ell-1);\sigma_{N-1};m;\rho)}_{\pm \bm{N}_{r} }\nonumber\\
     &\pm i\,\sigma_{N-1}\, \varkappa^{(I)}\,\psi^{(I;n \ell;\sigma_{N-1} ;m ;\rho)}_{\pm \bm{N}_{r} }+\mathcal{K}^{(I\rightarrow I\!I)}\,\psi^{(I\!I\text{-}I;n \ell;\sigma_{N-1} ;m ;\rho)}_{\pm \bm{N}_{r} },
\end{align}
\begin{align}\label{transfrmn_type_II-I_unnormalsd_r_SN_odd}
     \mathbb{L}_{\mathcal{S}} \psi^{(I\!I\text{-}I;n \ell;\sigma_{N-1}; m;\rho)}_{\pm \bm{N}_{r} } =&~\mathcal{A}^{(I\!I)} \, \psi^{(I\!I\text{-}I;n \,(\ell+1);\sigma_{N-1};m;\rho)}_{\pm \bm{N}_{r} } +\mathcal{B}^{(I\!I)}\,\psi^{(I\!I\text{-}I;n \,(\ell-1);\sigma_{N-1};m;\rho)}_{\pm \bm{N}_{r} }\nonumber\\
     &\pm i\,\sigma_{N-1}\, \varkappa^{(I\!I)}\,\psi^{(I\!I\text{-}I;n \ell;\sigma_{N-1} ;m ;\rho)}_{\pm \bm{N}_{r} }+\mathcal{K}^{(I\!I \rightarrow I)}\,\psi^{(I;\sigma;n \ell;\sigma_{N-1} ;m ;\rho)}_{\pm \bm{N}_{r} } \nonumber\\
    &+\mathcal{K}^{(I\!I \rightarrow I\!I\!I)}\,\psi^{(I\!I\!I\text{-}I;n \ell;\sigma_{N-1} ;m ;\rho)}_{\pm \bm{N}_{r} } ,
\end{align}
and
\begin{align}\label{transfrmn_type_III-I_unnormalsd_5/2_SN_odd}
     \mathbb{L}_{\mathcal{S}} \psi^{(I\!I\!I\text{-}I;n \ell;\sigma_{N-1}; m;\rho)}_{\pm \mu_{1}  \mu_{2} } =&~\mathcal{A}^{(I\!I\!I)} \, \psi^{(I\!I\!I\text{-}I;n \,(\ell+1);\sigma_{N-1};m;\rho)}_{\pm \mu_{1} \mu_{2} } +\mathcal{B}^{(I\!I\!I)}\,\psi^{(I\!I\!I\text{-}I;n \,(\ell-1);\sigma_{N-1};m;\rho)}_{\pm \mu_{1} \mu_{2} }\nonumber\\
     &\pm i\,\sigma_{N-1}\, \varkappa^{(I\!I\!I)}\,\psi^{(I\!I\!I\text{-}I;n \ell;\sigma_{N-1} ;m ;\rho)}_{\pm \mu_{1} \mu_{2} }+\mathcal{K}^{(I\!I\!I \rightarrow I\!I)}\,\psi^{(I\!I\text{-}I;n \ell;\sigma_{N-1} ;m ;\rho)}_{\pm \mu_{1}\mu_{2} } ,
\end{align}
respectively. [In eqs.~(\ref{transfrmn_type_I_unnormalsd_r_SN_odd}) and (\ref{transfrmn_type_II-I_unnormalsd_r_SN_odd}) we have $r\in \set{1,2}$, while eq.~(\ref{transfrmn_type_III-I_unnormalsd_5/2_SN_odd}) is relevant only for $r=2$.] All coefficients in eqs.~(\ref{transfrmn_type_I_unnormalsd_r_SN_odd})-(\ref{transfrmn_type_III-I_unnormalsd_5/2_SN_odd}) are given by the same expressions as the coefficients in the case with $N$ even [see eqs.~(\ref{transfrmn_type_I_unnormalsd_r_SN_even}), (\ref{transfrmn_type_II-I_unnormalsd_r_SN_even}) and (\ref{transfrmn_type_III_unnormalsd_5/2_SN_even})]. Unlike the even-dimensional case, the two spin projections $\sigma_{N-1}=\pm$ do not mix with each other in eqs.~(\ref{transfrmn_type_I_unnormalsd_r_SN_odd})-(\ref{transfrmn_type_III-I_unnormalsd_5/2_SN_odd}). However, the two spin projections $\sigma_{N-1}=\pm$ mix with each other under spin$(N)$ transformations.
 Note that the transformation formulae~(\ref{transfrmn_type_II-I_unnormalsd_r_SN_odd}) and (\ref{transfrmn_type_III-I_unnormalsd_5/2_SN_odd}) are defined only for $N \geq 5$ ($N$ odd), since type-$I\!I$ and type-$I\!I\!I$ modes on $S^{N}$ do not exist\footnote{This is consistent with the fact that the coefficient $\mathcal{K}^{(I \rightarrow I\!I)}$, given by eq.~(\ref{trnsf_coffcient_mixI->II_unnorm_r}), vanishes for $N=3$.} for $N=3$.

We are now ready to analytically continue our rank-1 and rank-2 STSSH's to $dS_{N}$ and study the group representation properties of the analytically continued STSSH's.
  \section{Obtaining spin-3/2 and spin-5/2 mode solutions on \texorpdfstring{${N}$}{N}-dimensional de Sitter spacetime by the analytic continuation of STSSH's} \label{section_analytic_cont}
  \subsection{Analytic continuation techniques}
  In this Section, we begin by discussing our analytic continuation techniques for STSSH's of arbitrary rank $r$ and then we specialise to the cases with $r=1$ and $r=2$.
  
 It is well known that $dS_{N}$ can be obtained by an “analytic continuation” of $S^{N}$~(see, e.g., Ref.~\cite{STSHS}). By replacing the angle $\theta_{N}$ in the line element of $S^{N}$~(\ref{line_element_SN}) as:
 \begin{align}\label{coord_change_analytic_cont}
     \theta_{N} \rightarrow x(t)\equiv \frac{\pi}{2} - i t,
 \end{align}
 ($t \in \mathbb{R}$) we find the line element for global $dS_{N}$: 
\begin{equation} \label{dS_metric}
    ds^{2}=-dt^{2}+\cosh^{2}{t} \,ds^{2}_{N-1}.
\end{equation}
 Motivated by this observation, we can obtain the field equations (\ref{Dirac_eqn_fermion_dS}) and (\ref{TT_conditions_fermions_dS}) on $dS_{N}$ by analytically continuing eqs.~(\ref{Dirac_eqn_fermion_SN}) and (\ref{TT_conditions_fermions_SN}), respectively, for the STSSH's on $S^{N}$. For convenience, let us give here again eqs.~(\ref{Dirac_eqn_fermion_SN}) and (\ref{TT_conditions_fermions_SN}) for STSSH's on $S^{N}$:
 \begin{align}
   &\slashed{\nabla}\psi_{\pm\mu_{1}...\mu_{r}}=\pm i \left(n+\frac{N}{2} \right) \psi_{\pm\mu_{1}...\mu_{r}} ,\hspace{5mm}(n=r,\,r+1,...) \label{Dirac_eqn_fermion_SN_minus}\\
   & \nabla^{\alpha}\psi_{\pm\alpha \mu_{2}...\mu_{r}}=0, \hspace{4mm}  \gamma^{\alpha}\psi_{\pm\alpha \mu_{2}...\mu_{r}}=0 \label{TT_conditions_fermions_SN_minus}.
\end{align}
Without loss of generality, we can choose to analytically continue the STSSH's with either one of the two signs for the eigenvalue in eq.~(\ref{Dirac_eqn_fermion_SN_minus}), since each of the two sets of modes, $\{\psi_{+\mu_{1}...\mu_{r}}\}$ and $\{\psi_{-\mu_{1}...\mu_{r}}\}$, forms independently a unitary representation of spin$(N+1)$ labelled by $n$ (see the beginning of Sections~\ref{sectn_spin3/2_solving_SN} and \ref{sectn_spin5/2_solving_SN}). Here we choose to analytically continue the STSSH's $\psi_{-\mu_{1}...\mu_{r}}$. By making the following replacements in eqs.~(\ref{Dirac_eqn_fermion_SN_minus}) and (\ref{TT_conditions_fermions_SN_minus}):
\begin{equation}\label{replacements}
        \theta_{N} \rightarrow x(t)\equiv\frac{\pi}{2} -it , \hspace{10mm} n \rightarrow \tilde{M}  - \frac{N}{2}\hspace{10mm}(t\in \mathbb{R},~ \tilde{M} \in \mathbb{R}\setminus \{ 0\})
    \end{equation}
we obtain eqs.~(\ref{Dirac_eqn_fermion_dS}) and (\ref{TT_conditions_fermions_dS}), respectively, with imaginary mass parameter $M= i \tilde{M}$ ($\tilde{M} \neq 0$) on $dS_{N}$. Recall that we are mainly interested in field equations with imaginary mass parameter because our aim is to study strictly and partially massless representations of spin$(N,1)$, where the mass parameter takes the imaginary values~(\ref{values_mass_parameter_masslessness_fermion}). Note that the gamma matrices on $S^{N}$ [eqs.~(\ref{even_gammas}) and (\ref{odd_gammas})] transform under the replacement~(\ref{coord_change_analytic_cont}) as: $\gamma^{N} \rightarrow i\gamma^{N}=\gamma^{0}$, while the $\gamma^{j}$'s ($j=1,...,N-1$) remain unchanged.\footnote{Alternatively, we could analytically continue the STSSH's on $S^{N}$ by making the replacement $\theta_{N} \rightarrow \pi/2 + it$ instead of the replacement~(\ref{coord_change_analytic_cont}). The analytically continued STSSH's with $\theta_{N} \rightarrow \pi/2 - it$ and the ones with $\theta_{N} \rightarrow \pi/2 + it$ are related to each other by charge conjugation. However, these two cases of analytically continued STSSH's form equivalent representations of spin$(N,1)$.}


\noindent  {\textbf{Note.} Each of the sets $\{ \psi_{+\mu_{1}...\mu_{r}} \}$ and $\{ \psi_{-\mu_{1}...\mu_{r}} \}$ in eq.~(\ref{Dirac_eqn_fermion_SN_minus}) separately forms a representation space for UIR's of spin$(N+1)$. Thus, as we mentioned earlier, we can analytically continue either $\{ \psi_{+\mu_{1}...\mu_{r}} \}$ or $ \{\psi_{-\mu_{1} ... \mu_{r}}\}$  to obtain TT tensor-spinor eigenmodes satisfying eq.~(\ref{Dirac_eqn_fermion_dS}) on $dS_{N}$. Then, one might wonder whether we can analytically continue some eigenmodes from the set $\{ \psi_{+\mu_{1}...\mu_{r}} \}$ and some from the set $\{ \psi_{-\mu_{1}...\mu_{r}} \}$ in order to obtain a spin$(N,1)$ representation space. The answer is yes, but apart from complicating things by doing so, there are also some caveats. Let us elaborate on this a bit more.} 
\begin{itemize} 
\item Each of the eigenmodes in $\{ \psi_{+ \mu_{1}...\mu_{r}} \}$ is a basis vector in a spin$(N+1)$ representation space, and so is each of the eigenmodes in $\{ \psi_{- \mu_{1}...\mu_{r}} \}$.  Each basis vector is uniquely labeled by its own set of quantum numbers.

\item Each set of quantum numbers (and, thus, each basis vector) essentially corresponds to a unique ``chain'' of highest weights concerning the chain of subalgebras of spin$(N+1)$: spin$(N) \supset$ spin$(N-1) \supset ... \supset$ spin$(2)$. (Recall that a spin$(N+1)$ representation includes many representations of its subalgebras and each one of them is determined by its own highest weight~\cite{barut_group}).

\item It is also known that for a given irreducible spin$(D+1)$ representation (with $D$ arbitrary) in the decomposition spin$(D+1)   \supset$ spin$(D)$, a spin$(D)$ representation appears with multiplicity one or it does not appear at all (this statement is also true for irreducible spin$(D,1)$ representations in the decomposition spin$(D,1) \supset$  spin$(D)$)~\cite{barut_group}. This explains why each basis vector/eigenmode forming the representation space for an irreducible spin$(N+1)$ representation is uniquely labeled by its own quantum numbers
\end{itemize}
{The situation is the same for irreducible representations of the dS algebra spin$(N,1)$ in the decomposition spin$(N,1) \supset$ spin$(N) \supset ... \supset$ spin$(2)$: each basis vector/analytically continued eigenmode is uniquely labeled by its own quantum numbers. \textbf{Thus,} if one decides to analytically continue some eigenmodes from the set $\{ \psi_{+\mu_{1}...\mu_{r}} \}$ and some from the set $\{ \psi_{-\mu_{1}...\mu_{r}} \}$ to form an irreducible spin$(N,1)$ representation, then care must be taken not to include basis vectors labeled by the same quantum numbers concerning the chain of subalgebras spin$(N) \supset ... \supset$ spin$(2)$ more than once. }


$$\textbf{Analytic continuation technicalities}$$

\noindent Let us now give a prescription for obtaining the explicit form of the spin-3/2 and spin-5/2 TT mode functions with mass parameter $M=i \tilde{M}$ on $dS_{N}$ by analytically continuing the STSSH's of rank 1 and 2, respectively.

\noindent\textbf{Functions describing the time-dependence.} The functions describing the time-dependence of the analytically continued STSSH's are found by making the replacements~(\ref{replacements}) in the (unnormalised) functions $\phi^{(a)}_{n \ell}(\theta_{N})$ [eq.~(\ref{phi_a})] and $\psi^{(a)}_{n \ell}(\theta_{N})$ [eq.~(\ref{psi_a})], as
\begin{align}
    \hat{\phi}^{(a)}_{\tilde{M} \ell}(t)\equiv&~\left[\kappa_{\phi}\left(\tilde{M}-\frac{N}{2},\ell\right)\right]^{-1}{\phi}^{(a)}_{(\tilde{M}-\frac{N}{2})\, \ell}(x(t))\label{phi_aM_t_initial} \\
    =&~\left(\cos{\frac{x(t)}{2}}\right)^{\ell+1-a}\left(\sin{\frac{x(t)}{2}}\right)^{\ell-a}\nonumber\\
    & \times F\left(-\tilde{M}+\frac{N}{2}+\ell,\tilde{M}+\ell+\frac{N}{2};\ell+\frac{N}{2};\sin^{2}\frac{x(t)}{2}\right),\label{phi_aM_t} 
\end{align}
\begin{align}
   \hat{\psi}^{(a)}_{\tilde{M} \ell}(t)\equiv &~\left[\kappa_{\phi}\left(\tilde{M}-\frac{N}{2},\ell\right)\right]^{-1}{\psi}^{(a)}_{(\tilde{M}-\frac{N}{2})\, \ell}(x(t))\label{psi_aM_t_initial}\\
    =&~\frac{\tilde{M}}{\ell+\frac{N}{2}}  \left(\cos{\frac{x(t)}{2}}\right)^{\ell-a} \left(\sin{\frac{x(t)}{2}} \right)^{\ell+1-a} \nonumber\\ &\times F\left(-\tilde{M}+\frac{N}{2}+\ell,\tilde{M}+\ell+\frac{N}{2};\ell+\frac{N+2}{2};\sin^{2}\frac{x(t)}{2}\right),\label{psi_aM_t}
\end{align}
where $\kappa_{\phi}(\tilde{M}-\frac{N}{2}, \ell)$ is given by eq.~(\ref{normlsn_fac_of_Jacobi}) with $n$ replaced by $\tilde{M}-\frac{N}{2}$, while
 \begin{align}
   &\cos{\frac{x(t)}{2}}=\left(  \sin{\frac{x(t)}{2}} \right)^{*}=\frac{\sqrt{2}}{2}\,\left(\cosh{\frac{t}{2}} + i \sinh{\frac{t}{2}} \right)    \label{cosx/2}.
       \end{align}
       Note that $\hat{\phi}^{(a)}_{(-\tilde{M}) \ell}= \hat{\phi}^{(a)}_{\tilde{M} \ell}$ and $\hat{\psi}^{(a)}_{(-\tilde{M}) \ell}= -\hat{\psi}^{(a)}_{\tilde{M} \ell}$. The condition $\ell \leq n$ does not hold for $dS_{N}$. Now $\ell$ can be any positive integer with $\ell \geq r$.\footnote{In our previous article~\cite{Letsios_announce}, the function $\hat{\phi}^{(a)}_{\tilde{M} \ell}(t) = \hat{\phi}^{(a)}_{(-i {M}) \ell}(t)$ is denoted as ${\varPhi}^{(a)}_{M \ell}(t)$ (where $M = i\tilde{M}$). Similarly, the function $\hat{\psi}^{(a)}_{\tilde{M} \ell}(t) = \hat{\psi}^{(a)}_{(-i {M}) \ell}(t)$ is denoted as ${\varPsi}^{(a)}_{M \ell}(t)$.}

\noindent \textbf{Analytic continuation of eigenmodes.}   For brevity, let us use again the shorthand notation introduced in eqs.~(\ref{shorthand_not_Neven}) (for $N$ even) and (\ref{shorthand_not_Nodd}) (for $N$ odd).  For $N$ even, we denote the analytically continued STSSH's as ${\Psi}^{(B;\sigma;\tilde{M}\ell m;{\rho})}_{ \bm{N}_{r}}(t, \bm{\theta}_{N-1})$ (where $\sigma= \pm$ is the spin projection index on $dS_{N}$, while $ m \leq \ell$ and $\ell=r,\,r+1,...\,$). We define the modes ${\Psi}^{(B;\sigma;\tilde{M}\ell m;{\rho})}_{ \bm{N}_{r}}$  by making the replacements~(\ref{replacements}) in the STSSH's ${\psi}^{\left(B;\sigma;n\ell m;{\rho}\right)}_{- \bm{N}_{r}}$ on $S^{N}$, as
\begin{align}\label{define_anal_continued_STSSH's}
    \Psi^{(B;\sigma;\tilde{M}\ell m;{\rho})}_{\bm{N}_{r}}(t, \bm{\theta}_{N-1})=\left[\kappa_{\phi}\left(\tilde{M}-\frac{N}{2},\ell\right)\right]^{-1}\,{\psi}^{\left(B;\sigma;(\tilde{M}-N/2)\,\ell m;{\rho}\right)}_{- \bm{N}_{r}}({\pi}/{2}-it, \bm{\theta}_{N-1})
\end{align}
where $\left[\kappa_{\phi}\left(\tilde{M}-\frac{N}{2},\ell\right)\right]^{-1}$ is essentially the factor used in eqs.~(\ref{phi_aM_t_initial}) and (\ref{psi_aM_t_initial}) [it is used in order to cancel the normalisation factor~(\ref{normlsn_fac_of_Jacobi}) of the Jacobi polynomials]. Note that, by viewing the replacement $\theta_{N}\rightarrow \frac{\pi}{2} - it$ as a coordinate change, we find that ${\psi}^{\left(B;\sigma;n\ell m;{\rho}\right)}_{- \theta_{N}}$ transforms as
$${\psi}^{\left(B;\sigma;n\ell m;{\rho}\right)}_{- \theta_{N}} \rightarrow i\, {\psi}^{\left(B;\sigma;(\tilde{M}-N/2)\,\ell m;{\rho}\right)}_{- t}.$$ Similarly, ${\psi}^{\left(B;\sigma;n\ell m;{\rho}\right)}_{- \theta_{N} \theta_{N}}$ and ${\psi}^{\left(B;\sigma;n\ell m;{\rho}\right)}_{- \theta_{N} \theta_{j}}$ transform as
$${\psi}^{\left(B;\sigma;n\ell m;{\rho}\right)}_{- \theta_{N} \theta_{N}}\rightarrow-{\psi}^{\left(B;\sigma;(\tilde{M}-N/2)\,\ell m;{\rho}\right)}_{- t\,t}$$ and $${\psi}^{\left(B;\sigma;n\ell m;{\rho}\right)}_{- \theta_{N} \theta_{j}}\rightarrow i\,{\psi}^{\left(B;\sigma;(\tilde{M}-N/2)\,\ell m;{\rho}\right)}_{- t\,\theta_{j}},$$ respectively.

For $N$ odd, the analytically continued STSSH's are denoted as ${\Psi}^{(B;\tilde{M}\ell; \sigma_{N-1};m;{\rho})}_{ \bm{N}_{r}}$ (where $\sigma_{N-1}=\pm$, $m \leq \ell$ and $\ell=r,\,r+1,...\,$). They are obtained by analytically continuing the STSSH's ${\psi}^{\left(B;n\ell;\sigma_{N-1};m;{\rho}\right)}_{- \bm{N}_{r}}(\theta_{N}, \bm{\theta}_{N-1})$ on $S^{N}$, as
\begin{align}\label{define_anal_continued_STSSH's_Nodd}
    \Psi^{(B;\tilde{M}\ell;\sigma_{N-1};m;{\rho})}_{\bm{N}_{r}}(t, \bm{\theta}_{N-1})=\left[\kappa_{\phi}\left(\tilde{M}-\frac{N}{2},\ell\right)\right]^{-1}{\psi}^{\left(B;(\tilde{M}-N/2)\,\ell;\sigma_{N-1};m;{\rho}\right)}_{- \bm{N}_{r}}({\pi}/{2}-it, \bm{\theta}_{N-1}).
\end{align}
Note that, unlike the case with $N$ even [eq.~(\ref{define_anal_continued_STSSH's})], the analytically continued STSSH's (\ref{define_anal_continued_STSSH's_Nodd}) have a spin projection index ($\sigma_{N-1}$) on $S^{N-1}$ instead of a spin projection index on $dS_{N}$.

The aforementioned analytically continued eigenmodes have been also constructed directly on $dS_{N}$ using the method of separation of variables in our previous article~\cite{Letsios_announce}, where representation-theoretic details concerning the decomposition spin$(N,1) \supset$ spin$(N)$ can also be found.
\subsection{Pure gauge modes for the strictly/partially massless spin-3/2 and spin-5/2 theories}\label{subsection_PG}
As in Minkowski spacetime, (strictly and partially) massless field theories in $dS_{N}$ enjoy gauge symmetry~\cite{Deser_Waldron_phases}. In terms of mode solutions of the corresponding field equations, gauge symmetry manifests itself through the appearance of `pure gauge' modes in the set of mode solutions. The `pure gauge' modes do not describe propagating degrees of freedom of the field theory and - assuming that there exists an invariant inner product for the mode solutions - these modes have zero norm (see, e.g. Ref.~\cite{STSHS}).

For later convenience, let us present the `pure gauge' modes that appear among the analytically continued STSSH's of rank $r$ ($r=1,2$) when we tune the imaginary mass parameter ($M=i \tilde{M}$) to the strictly/partially massless values $\tilde{M}=\pm \left[r-\uptau+ (N-2)/2\right]$, where $\uptau=1,..,r$ [see eq.~(\ref{values_mass_parameter_masslessness_fermion})]. For each strictly/partially massless value of $\tilde{M}$, the analytically continued STSSH's of rank $r$ with $r- \uptau \geq \tilde{r} \geq 0$ are `pure gauge' modes, where $\tilde{r}$ is the rank of the STSSH on $S^{N-1}$ used in the method of separation of variables (see Sections~\ref{sectn_spin3/2_solving_SN} and \ref{sectn_spin5/2_solving_SN}). In Section~\ref{section_(non)unitarity} we will verify that our `pure gauge' modes have zero norm associated to a spin$(N,1)$ invariant scalar product for $N$ even. We will also demonstrate that for $N$ odd there does not exist any spin$(N,1)$ invariant scalar product for the analytically continued STSSH's with imaginary mass parameter. Thus, for $N$ odd the norm of the `pure gauge' modes cannot be calculated in a meaningful way, as there is no de Sitter invariant notion of norm.

\noindent \textbf{Strictly massless spin-3/2 field.} The mass parameter for the strictly massless spin-3/2 field is given by $M=i \tilde{M}=\pm i (N-2)/2 $ [this is found by letting $r=\uptau=1$ in eq.~(\ref{values_mass_parameter_masslessness_fermion})]. The analytically continued STSSH's of type-$I$ ($\tilde{r}=0$) are `pure gauge' modes. As demonstrated in Appendix~\ref{appendix_pure_gauge_modes}, the analytically continued rank-1 STSSH's~(\ref{define_anal_continued_STSSH's}) of type-$I$ with $\tilde{M}=\pm  (N-2)/2$ are expressed in a `pure gauge' form as follows:
\begin{align}\label{PG_TYPEI_spin3/2}
    \Psi^{\left(I;(\pm \frac{N-2}{2});\tilde{\ell}\right)}_{\mu}(t, \bm{\theta}_{N-1})=\left( \nabla_{\mu}\pm \frac{i}{2}\gamma_{\mu}  \right) \Lambda_{\pm}^{\left(\tilde{\ell}\right)}(t, \bm{\theta}_{N-1}),
\end{align}
where for brevity we use the symbol $\tilde{\ell}$ to represent all the labels of the analytically continued STSSH's which have not been written down explicitly. The Dirac spinors $\Lambda_{\pm}^{\left(\tilde{\ell}\right)}(t, \bm{\theta}_{N-1})$ satisfy
\begin{align}
    \slashed{\nabla}\Lambda_{\pm}^{\left(\tilde{\ell} \right)}=\mp i \,\frac{N}{2}\,\Lambda_{\pm}^{\left( \tilde{\ell}\right)}.
\end{align}
The `pure gauge' expression~(\ref{PG_TYPEI_spin3/2}) for the type-$I$ modes coincides with the form of the infinitesimal gauge transformation~\cite{Deser_Waldron_phases} (with a specific gauge condition) that leaves invariant the action for the strictly massless spin-3/2 field in $dS_{4}$. In Section~\ref{section_(non)unitarity} we show that the `pure gauge' modes~(\ref{PG_TYPEI_spin3/2}) have vanishing dS invariant norm for even $N \geq 4$.

   \noindent \textbf{Strictly massless spin-5/2 field.} The mass parameter for the strictly massless spin-5/2 field is given by $M=i \tilde{M}=\pm i N/2 $ [this is found by letting $r=2$ and $\uptau=1$ in eq.~(\ref{values_mass_parameter_masslessness_fermion})]. There are two types of `pure gauge' modes, namely the analytically continued STSSH's of type-$I$ ($\tilde{r}=0$) and type-$I\!I$ ($\tilde{r}=1$). As demonstrated in Appendix~\ref{appendix_pure_gauge_modes}, the analytically continued rank-2 STSSH's~(\ref{define_anal_continued_STSSH's}) of type-$I$ and type-$I\!I$ with $\tilde{M}=\pm  N/2$ are expressed in the following `pure gauge' form: 
   \begin{align}\label{PG_TYPEB_spin5/2}
    {\Psi}^{\left(B;(\pm \frac{N}{2});\tilde{\ell}\right)}_{ \mu \nu}(t, \bm{\theta}_{N-1}) = \left(\nabla_{(\mu}\pm \frac{i}{2} \gamma_{(\mu} \right) \lambda^{\left(B;\tilde{\ell}\right)}_{\pm \nu)}(t, \bm{\theta}_{N-1}),\hspace{9mm}B=I,I\!I,
\end{align}
where the gauge functions $\lambda^{\left(B;\tilde{\ell}\right)}_{\pm \mu}(t, \bm{\theta}_{N-1})$ ($B=I,I\!I$) are vector-spinor fields satisfying
\begin{align}
   & \slashed{\nabla}\lambda^{\left(B;\tilde{\ell}\right)}_{\pm \mu}= \mp i \,\frac{N+2}{2} \,\lambda^{\left(B;\tilde{\ell}\right)}_{\pm \mu}\\
   & \gamma^{\mu} \lambda^{\left(B;\tilde{\ell}\right)}_{\pm \mu}= \nabla^{\mu}  \lambda^{\left(B;\tilde{\ell}\right)}_{\pm \mu}=0.
\end{align} 
The vector-spinors $\lambda^{\left(B;\tilde{\ell}\right)}_{\pm \mu}(t, \bm{\theta}_{N-1})$ are given by the analytic continuation of rank-1 STSSH's of type-$B$ ($B= I, I\!I$) - see Appendix~\ref{appendix_pure_gauge_modes}.
Note that the `pure gauge' expressions~(\ref{PG_TYPEB_spin5/2}) for the type-$I$ and type-$I\!I$ modes coincide with the form of the infinitesimal gauge transformation~\cite{Deser_Waldron_phases} (with a specific gauge condition) for the gauge-invariant action for the strictly massless spin-5/2 field in $dS_{4}$. In Section~\ref{section_(non)unitarity} we show that the `pure gauge' modes~(\ref{PG_TYPEB_spin5/2}) have zero (dS invariant) norm for even $N \geq 4$.

 \noindent \textbf{Partially massless spin-5/2 field.} The mass parameter for the partially massless spin-5/2 field is given by $M=i \tilde{M}=\pm i (N-2)/2 $ [this is found by letting $r=2$ and $\uptau=2$ in eq.~(\ref{values_mass_parameter_masslessness_fermion})]. The analytically continued STSSH's of type-$I$ ($\tilde{r}=0$) are `pure gauge' modes. As demonstrated in Appendix~\ref{appendix_pure_gauge_modes}, the analytically continued rank-2 STSSH's~(\ref{define_anal_continued_STSSH's}) of type-$I$ with $\tilde{M}=\pm  (N-2)/2$ are expressed in a `pure gauge' form as follows:
   \begin{align}\label{PG_TYPEI_spin5/2_partially}
   {\Psi}^{\left(I;(\pm \frac{N-2}{2});\tilde{\ell}\right)}_{ \mu \nu}(t, \bm{\theta}_{N-1})= \left( \nabla_{(\mu}\nabla_{\nu)} \pm i  \gamma_{(\mu}\nabla_{\nu)} +\frac{3}{4} g_{\mu \nu} \right) \varphi_{\pm}^{\left(\tilde{\ell}\right)}(t, \bm{\theta}_{N-1}),
\end{align} 
where the spinor modes $\varphi_{\pm}^{\left(\tilde{\ell}\right)}(t, \bm{\theta}_{N-1})$ satisfy
    \begin{align}
    \slashed{\nabla}\varphi_{\pm}^{\left(\tilde{\ell} \right)}=\mp i\, \frac{N+2}{2}\,\varphi_{\pm}^{\left( \tilde{\ell}\right)}.
\end{align}
In Section~\ref{section_(non)unitarity} we show that the `pure gauge' modes~(\ref{PG_TYPEI_spin5/2_partially}) have zero (dS invariant) norm for even $N \geq 4$.
We note that we have not constructed a gauge-invariant action for the partially massless spin-5/2 field in $dS_{N}$ with infinitesimal gauge transformation of the form~(\ref{PG_TYPEI_spin5/2_partially}). However, we call the modes~(\ref{PG_TYPEI_spin5/2_partially}) `pure gauge' modes because we expect that such an action exists and that the expression~(\ref{PG_TYPEI_spin5/2_partially}) describes infinitesimal gauge transformations (satisfying a specific gauge condition) for this action. 

In Appendix~\ref{appendix_pure_gauge_modes}, we discuss the relation between our `pure gauge' modes~(\ref{PG_TYPEI_spin5/2_partially}) and the gauge transformation of the partially massless spin-5/2 field in $dS_{4}$ given in Ref.~\cite{Deser_Waldron_phases}. More specifically, we observe the following intriguing fact: for a specific choice for the spinor gauge function in the gauge transformation used in Ref.~\cite{Deser_Waldron_phases}, the gamma-traceless part of this gauge transformation can be expressed in our `pure gauge' form~(\ref{PG_TYPEI_spin5/2_partially}).
 \section{(Non)unitarity of the strictly/partially massless representations of spin\texorpdfstring{${(N,1)}$}{(N,1)} formed by the analytically continued rank-1 and rank-2 STSSH's}\label{section_(non)unitarity}
 For each value of the imaginary mass parameter $M=i\tilde{M}$ in eq.~(\ref{Dirac_eqn_fermion_dS}), the TT tensor-spinor mode solutions (i.e. the analytically continued STSSH's) form a representation of spin$(N,1)$. If one introduces a dS invariant scalar product among the analytically continued STSSH's, then the unitarity of the representation is equivalent to the positive-definiteness of the associated norm. If there is no dS invariant scalar product, then the corresponding representation of spin$(N,1)$ is, by definition, not unitary.
 
 In this Section, we prove statements \ref{basic_result_Neven}, \ref{basic_result_N=4} and \ref{basic_result_Nodd} presented in the Introduction, which give the technical explanation of the main result of our paper (which we mention here again for convenience): the strictly massless spin-3/2 field theory and the strictly and partially massless spin-5/2 field theories on $dS_{N}$ ($N \geq 3$) are unitary only for $N=4$. 
 
 
\subsection{The strictly/partially massless spin-3/2 and spin-5/2 representations of spin\texorpdfstring{${(N,1)}$}{(N,1)} are non-unitary for even \texorpdfstring{$N>4$}{N>4}} \label{Subsection_reps_Neven_part1}
In this Subsection, we show that the representations of spin$(N,1)$ with even $N>4$ formed by the spin-3/2 and spin-5/2 TT mode solutions of eq.~(\ref{Dirac_eqn_fermion_dS}) with arbitrary imaginary mass parameter $M=i\tilde{M}$ ($\tilde{M} \neq 0$) are non-unitary (i.e. we prove statement~\ref{basic_result_Neven}). In order to arrive at this result we study the transformation properties of our analytically continued STSSH's under a spin$(N,1)$ boost and then we demonstrate the indefiniteness of the norm associated to a dS invariant scalar product for even $N >4$. (In this Subsection we work without specifying the form of the dS invariant scalar product. Thus, our results hold for any dS invariant scalar product.) We also find that for $N=4$ the requirement for dS invariance of the scalar product does not imply the indefiniteness of the norm if and only if the mass parameter $\tilde{M}$ is tuned to the strictly/partially massless values~(\ref{values_mass_parameter_masslessness_fermion}). Also, for even $N \geq 4$, we show that the `pure gauge' modes in the strictly/partially massless theories with spin $s \in \set{3/2, 5/2}$ have zero norm with respect to any dS invariant scalar product. Furthermore, for $N=4$ and $\tilde{M}$ given by eq.~(\ref{values_mass_parameter_masslessness_fermion}), we show that the TT modes in the strictly/partially massless theories are divided into two spin$(4,1)$ invariant subspaces, denoted as $\mathcal{H}^{-}$ and $\mathcal{H}^{+}$ (where each subspace contains modes with definite helicity). The positivity of the norm in each of these subspaces is shown in Subsection~\ref{Subsection_explicit_calculationnorms} by calculating explicitly the norms of the eigenmodes with respect to a specific dS invariant scalar product. (In Subsection~\ref{Subsection_explicit_calculationnorms} we also verify the results obtained in the present Subsection for even $N>4$ by explicit calculation of the norms of the eigenmodes with arbitrary imaginary mass parameter $M=i \tilde{M} \neq 0$.)

The analytic continuation techniques introduced in Section~\ref{section_analytic_cont} can also be applied to the transformation properties of the STSSH's under spin$(N+1)$. By doing so, one obtains the transformation properties of the analytically continued STSSH's on $dS_{N}$ under spin$(N,1)$. Let us make the replacement~(\ref{coord_change_analytic_cont}) in the Killing vector $\mathcal{S}^{\mu} $ [eq.~(\ref{Killing_vector_sphere})] on $S^{N}$. One finds that the analytically continued version of $\mathcal{S}^{\mu}$ is expressed as $i X^{\mu}$, where $X^{\mu}$ is the following boost generator of spin$(N,1)$:
\begin{equation}\label{Killing_vector_boost}
    X^{\mu}\partial_{\mu}=\cos{\theta_{N-1}} \,\frac{\partial}{\partial t} - \tanh{t} \, \sin{\theta_{N-1}}\, \frac{\partial}{\partial \theta_{N-1}}.
\end{equation}
The de Sitter algebra spin$(N,1)$ is generated by the de Sitter boost~(\ref{Killing_vector_boost}) and the generators of spin$(N)$.

\noindent \textbf{spin$\bm{(N,1)}$ transformation formulae.} By making the replacements~(\ref{replacements}) in the spin$(N+1)$ transformation formulae (\ref{transfrmn_type_I_unnormalsd_r_SN_even}), (\ref{transfrmn_type_II-I_unnormalsd_r_SN_even}) and (\ref{transfrmn_type_III_unnormalsd_5/2_SN_even}) [and using eq.~(\ref{define_anal_continued_STSSH's})], we find
\begin{align}\label{transfrmn_type_I_unnormalsd_r_boost_even}
     \mathbb{L}_{X}{\Psi}^{(I;\sigma;\tilde{M} \ell m;\rho)}_{ \bm{N}_{r} } =&~ -i\,c_{(\ell)}\, \mathcal{A}^{(I)}\, {\Psi}^{(I;\sigma;\tilde{M} \,(\ell+1)\,m;\rho)}_{ \bm{N}_{r} } -\frac{i}{c_{(\ell-1)}} \mathcal{B}^{(I)}\, {\Psi}^{(I;\sigma;\tilde{M} \,(\ell-1)\,m;\rho)}_{ \bm{N}_{r} }\nonumber\\
     &-  \varkappa^{(I)}\, {\Psi}^{(I;-\sigma;\tilde{M} \ell m ;\rho)}_{ \bm{N}_{r} }-i \mathcal{K}^{(I\rightarrow I\!I)}\,{\Psi}^{(I\!I\text{-}I;\sigma;\tilde{M} \ell m ;\rho)}_{\bm{N}_{r}},
\end{align}
\begin{align}\label{transfrmn_type_II-I_unnormalsd_r_boost_even}
     \mathbb{L}_{X} {\Psi}^{(I\!I\text{-}I;\sigma;\tilde{M} \ell m;\rho)}_{\  \bm{N}_{r} }=&~-i\,c_{(\ell)} \mathcal{A}^{(I\!I)} \,{\Psi}^{(I\!I\text{-}I;\sigma;\tilde{M} \,(\ell+1)\,m;\rho)}_{  \bm{N}_{r} } -\frac{i}{c_{(\ell-1)}} \mathcal{B}^{(I\!I)} \,{\Psi}^{(I\!I\text{-}I;\sigma;\tilde{M} \,(\ell-1)\,m;\rho)}_{ \bm{N}_{r} } \nonumber\\
     &-  \varkappa^{(I\!I)} {\Psi}^{(I\!I\text{-}I;-\sigma;\tilde{M}\ell m ;\rho)}_{ \bm{N}_{r}  }-i \mathcal{K}^{(I\!I\rightarrow I)}\,{\Psi}^{(I;\sigma;\tilde{M} \ell m ;\rho)}_{ \bm{N}_{r}}\nonumber\\
     &-i \mathcal{K}^{(I\!I\rightarrow I\!I\!I)}\,{\Psi}^{(I\!I\!I\text{-}I;\sigma;\tilde{M} \ell m ;\rho)}_{  \bm{N}_{r}} 
\end{align}
($r=1,2,$) and
\begin{align}\label{transfrmn_type_III-I_unnormalsd_5/2_boost_even}
     \mathbb{L}_{X} {\Psi}^{(I\!I\!I\text{-}I;\sigma;\tilde{M} \ell m;\rho)}_{\mu_{1}\mu_{2} }=&~-i\,c_{(\ell)} \mathcal{A}^{(I\!I\!I)} \,{\Psi}^{(I\!I\!I\text{-}I;\sigma;\tilde{M} \,(\ell+1)\,m;\rho)}_{ \mu_{1} \mu_{2} } -\frac{i}{c_{(\ell-1)}} \mathcal{B}^{(I\!I\!I)} \,{\Psi}^{(I\!I\!I\text{-}I;\sigma;\tilde{M} \,(\ell-1)\,m;\rho)}_{ \mu_{1} \mu_{2} } \nonumber\\
     &-  \varkappa^{(I\!I\!I)} {\Psi}^{(I\!I\!I\text{-}I;-\sigma;\tilde{M}\ell m ;\rho)}_{ \mu_{1} \mu_{2}  }-i \mathcal{K}^{(I\!I\!I\rightarrow I\!I)}\,{\Psi}^{(I\!I\text{-}I;\sigma;\tilde{M} \ell m ;\rho)}_{ \mu_{1}\mu_{2}},
\end{align}
respectively, with
\begin{align}
    c_{(\ell)}= \frac{\kappa_{\phi}(\tilde{M}-\frac{N}{2},\ell+1)}{\kappa_{\phi}(\tilde{M}-\frac{N}{2},\ell)}=\frac{\tilde{M}-\ell-\frac{N}{2}}{\ell+N/2},
\end{align}
where $\kappa_{\phi}(\tilde{M}-N/2,\ell)$ is found by eq.~(\ref{normlsn_fac_of_Jacobi}) and $\mathbb{L}_{X}$ is the Lie-Lorentz derivative~(\ref{Lie_Lorentz}) on $dS_{N}$. The coefficients $\mathcal{A}^{(B)},\mathcal{B}^{(B)}, \varkappa^{(B)}$ (with $B=I,\, I\!I,\, I\!I\!I$), $\mathcal{K}^{(I \rightarrow I\!I)}, \mathcal{K}^{(I\!I \rightarrow I)},\mathcal{K}^{(I\!I \rightarrow I\!I\!I)} $ and $\mathcal{K}^{(I\!I\!I \rightarrow I\!I)}$ are found by making the replacement $n \rightarrow \tilde{M} - N/2$ in the corresponding expressions for the coefficients of STSSH's on $S^{N}$ [see eqs.~(\ref{transfrmn_type_I_unnormalsd_r_SN_even}), (\ref{transfrmn_type_II-I_unnormalsd_r_SN_even}) and (\ref{transfrmn_type_III_unnormalsd_5/2_SN_even})]. Note that we use the same symbols to represent the coefficients in the transformation formulae on $S^{N}$ and the analytically continued coefficients on $dS_{N}$.


 $$\textbf{Investigating the (non-)existence of positive-definite, dS invariant scalar products.}$$ 
 
 \noindent Let $\braket{\Psi^{(1)}, \Psi^{(2)} }_{(r)}$ be a spin$(N,1)$ invariant scalar product for any two analytically continued rank-$r$ STSSH's $\Psi^{(1)}_{\bm{N}_{r}}, \Psi^{(2)}_{\bm{N}_{r}}$ ($r=1,2$) with imaginary mass parameter $M= i\tilde{M}$ ($\tilde{M} \neq 0$). Due to the spin$(N,1)$ invariance of the scalar product we have
\begin{align}\label{dS_invar_inner_prod}
 \braket{\mathbb{L}_{\xi}\Psi^{(1)},  \Psi^{(2)}}_{(r)}  +\braket{\Psi^{(1)},  \mathbb{L}_{\xi}\Psi^{(2)}}_{(r)}=0
\end{align} 
for any Killing vector $\xi$ on $dS_{N}$. Then, by letting $\Psi^{(1)}_{\bm{N}_{r}}={\Psi}^{(B;-;\tilde{M} \ell m;\rho)}_{ \bm{N}_{r} }$ and $\Psi^{(2)}_{\bm{N}_{r}}={\Psi}^{(B;+;\tilde{M} \ell m;\rho)}_{ \bm{N}_{r} }$ (with $B=I,I\!I\text{-}I,I\!I\!I\text{-}I $) in eq.~(\ref{dS_invar_inner_prod}) with $\xi=X$ and using the transformation formulae~(\ref{transfrmn_type_I_unnormalsd_r_boost_even})-(\ref{transfrmn_type_III-I_unnormalsd_5/2_boost_even}), we find that the norms of eigenmodes with opposite spin projections must satisfy:
    \begin{align}\label{proof_negnorms_typeI}
    \varkappa^{(I)}\times \left( \braket{{\Psi}^{(I;-;\tilde{M} \ell m;\rho)},{\Psi}^{(I;-;\tilde{M} \ell m;\rho)} }_{(r)}  +\braket{{\Psi}^{(I;+;\tilde{M} \ell m;\rho)},{\Psi}^{(I;+;\tilde{M} \ell m;\rho)} }_{(r)} \right)=0, \end{align}
    \begin{align}\label{proof_negnorms_typeII-I}
    \varkappa^{(I\!I)}\times \Big(& \braket{{\Psi}^{(I\!I\text{-}I;-;\tilde{M} \ell m;\rho)},{\Psi}^{(I\!I\text{-}I;-;\tilde{M} \ell m;\rho)} }_{(r)} \nonumber\\
    &+\braket{{\Psi}^{(I\!I\text{-}I;+;\tilde{M} \ell m;\rho)},{\Psi}^{(I\!I\text{-}I;+;\tilde{M} \ell m;\rho)} }_{(r)} \Big)=0,\end{align}
    \begin{align}\label{proof_negnorms_typeIII-I}
 \varkappa^{(I\!I\!I)}\times \Big(& \braket{{\Psi}^{(I\!I\!I\text{-}I;-;\tilde{M} \ell m;\rho)},{\Psi}^{(I\!I\!I\text{-}I;-;\tilde{M} \ell m;\rho)} }_{(r=2)} \nonumber\\
 &+\braket{\Psi^{(I\!I\!I\text{-}I;+;\tilde{M} \ell m;\rho)},\Psi^{(I\!I\!I\text{-}I;+;\tilde{M} \ell m;\rho)} }_{(r=2)} \Big) =0. \end{align}
 Note that, since the scalar product is also spin$(N)$ invariant, analytically continued STSSH's of different type or/and with different values for $\ell$ are orthogonal to each other because they correspond to inequivalent irreducible representations of spin$(N)$ in the decomposition spin$(N,1)$ $\supset$ spin$(N)$.
For convenience, we give here the explicit form of the analytically continued coefficients $\varkappa^{(I)}$ [eq.~(\ref{varkap_I_coeff_r})], $\varkappa^{(I\!I)}$ [eq.~(\ref{varkap_II_coeff_r})] and $\varkappa^{(I\!I\!I)}$ [eq.~(\ref{varkap_III_coeff_spin5/2})]:
 \begin{align}
 \varkappa^{(I)}=&-\frac{\tilde{M}(m+\frac{N-2}{2})(N+2r-2)}{2(\ell+\frac{N-2}{2})(\ell+\frac{N}{2})(N-2)} \hspace{14mm}(r=1,2) ,\label{varkap_I_coeff_r_anal_con} \end{align}
 \begin{align}
\varkappa^{(I\!I)}=&-\frac{\tilde{M}(m+\frac{N-2}{2})(N-4)}{2(\ell+\frac{N-2}{2})(\ell+\frac{N}{2})(N-2)}\times \left( \frac{N+2}{N} \right)^{r-1} \hspace{10mm}(r=1,2),\label{varkap_II_coeff_r_anal_con}
 \end{align} 
   \begin{align}
    \varkappa^{(I\!I\!I)}&=-\frac{\tilde{M}(m+\frac{N-2}{2})(N-4)}{2(\ell+\frac{N-2}{2})(\ell+\frac{N}{2})N}\label{varkap_III_coeff_spin5/2_anal_con}
 \end{align}
 [eq.~(\ref{varkap_III_coeff_spin5/2_anal_con}) is relevant only for spin-5/2 modes, i.e. only for $r=2$]. We also give the explicit form of the analytically continued coefficients $\mathcal{K}^{(I \rightarrow I\!I)}$ [eq.~(\ref{trnsf_coffcient_mixI->II_unnorm_r})] and $\mathcal{K}^{(I\!I \rightarrow I\!I\!I)}$ [eq.~(\ref{trnsf_coffcient_mixII->III_unnorm_r=2})]:
  \begin{align}
     \mathcal{K}^{(I\rightarrow I\!I)}=&-\frac{4\left( \tilde{M}^{2}-{(N-2)^{2}}/{4}\right)(N+r-2)}{\ell(\ell+N-1)(N-2)}\times\sqrt{\frac{N-3}{N-2}\frac{m(m+N-2)}{(\ell+1)(\ell+N-2)}} \hspace{4mm}(r=1,2)\label{trnsf_coffcient_mixI->II_unnorm_r_anal_con},
\end{align} 
\begin{align}
 \mathcal{K}^{(I\!I \rightarrow I\!I\!I)} =&-{2}^{3}\frac{\left( \tilde{M}^{2}-{N^{2}}/{4}\right)(N+1)}{(\ell-1)(\ell+N)N} \times\sqrt{\frac{N-2}{N}\,\frac{(m-1)(m+N-1)}{\ell(\ell+N-1)}}\label{trnsf_coffcient_mixII->III_unnorm_r=2_anal_cont},
 \end{align}
 where eq.~(\ref{trnsf_coffcient_mixII->III_unnorm_r=2_anal_cont}) is relevant only for $r=2$. The analytically continued coefficients $\mathcal{K}^{(I\!I \rightarrow I)}$ and $ \mathcal{K}^{(I\!I\!I \rightarrow I\!I)}$ are given by the same expressions as the coefficients on $S^{N}$, i.e. eqs.~(\ref{trnsf_coffcient_mixII->I_unnorm_r}) and (\ref{trnsf_coffcient_mixIII->II_unnorm_5/2}), respectively.
 
\noindent \textbf{$\bullet$ Cases with even $\bm{N >4}$.} Let us first discuss the case with even $N>4$, where $\varkappa^{(I)}, \varkappa^{(I\!I)}$ and $\varkappa^{(I\!I\!I)}$ are all non-zero (for all $\tilde{M} \neq 0$). The representation can be unitary only if eqs.~(\ref{proof_negnorms_typeI})-(\ref{proof_negnorms_typeIII-I}) are consistent with the positive-definiteness of the norm. However, it is clear from eqs.~(\ref{proof_negnorms_typeI})-(\ref{proof_negnorms_typeIII-I}) that the norm of the modes ${\Psi}^{(B;-;\tilde{M} \ell m;\rho)}_{\bm{N}_{r}}$ is opposite of the norm of the modes ${\Psi}^{(B;+;\tilde{M} \ell m;\rho)}_{\bm{N}_{r}}$ ($B=I, I\!I\text{-}I, I\!I\!I \text{-}I$) for all $\tilde{M} \neq 0$. Hence, for even $N>4$, there are negative-norm modes for all values of $\tilde{M} \neq 0$, unless all modes have zero norm. (Not all modes could have zero norm if the field were to describe a physical particle.) Thus, we have proved statement~\ref{basic_result_Neven}.

\noindent \textbf{$\bullet$ dS invariance requires the norm of `pure gauge' modes to be zero.} Before discussing the case with $N=4$, we can show that the `pure gauge' modes (discussed in Subsection~\ref{subsection_PG}), which appear among the TT mode solutions in the strictly/partially massless theories, have zero norm with respect to any dS invariant scalar product for even $N \geq 4$, as follows~\cite{Higuchi_private_com}. For the strictly massless spin-3/2 theory ($r=\uptau=1$), as well as for the partially massless spin-5/2 theory ($r=\uptau=2$), the mass parameter is $\tilde{M}^{2}=(N-2)^{2}/4$ [see eq.~(\ref{values_mass_parameter_masslessness_fermion})], while the type-$I$ modes are `pure gauge' modes. We observe that the coefficient $\mathcal{K}^{(I \rightarrow I\!I)}$ [eq.~(\ref{trnsf_coffcient_mixI->II_unnorm_r_anal_con})] vanishes for $\tilde{M}^{2}=(N-2)^{2}/4$ (with $r=1,2$). Then, by letting $\Psi^{(1)}_{\bm{N}_{r}}={\Psi}^{\left(I;\sigma;(\pm  \frac{N-2}{2}) \ell m;\rho\right)}_{ \bm{N}_{r} }$ and $\Psi^{(2)}_{\bm{N}_{r}}={\Psi}^{\left(I\!I\text{-}I;\sigma;(\pm  \frac{N-2}{2})\ell m;\rho\right)}_{ \bm{N}_{r} }$ in eq.~(\ref{dS_invar_inner_prod}) with $\xi=X$ and using the transformation formulae~(\ref{transfrmn_type_I_unnormalsd_r_boost_even}) and (\ref{transfrmn_type_II-I_unnormalsd_r_boost_even}), we straightforwardly find $\braket{{\Psi}^{\left(I;\sigma;(\pm  \frac{N-2}{2})\ell m;\rho\right)},{\Psi}^{\left(I;\sigma;(\pm  \frac{N-2}{2})\ell m;\rho\right)} }_{(r)}=0$ (with $r=1,2$), i.e. the type-$I$ modes have zero norm for even $N \geq 4$.
For the strictly massless spin-5/2 theory ($r=\uptau+1=2$) the mass parameter is $\tilde{M}^{2}=N^{2}/4$ [see eq.~(\ref{values_mass_parameter_masslessness_fermion})], while both type-$I$ and type-$I\!I$ modes are `pure gauge' modes. For this value of $\tilde{M}^{2}$ the coefficient $\mathcal{K}^{(I\!I \rightarrow I\!I\!I)}$ [eq.~(\ref{trnsf_coffcient_mixII->III_unnorm_r=2_anal_cont})] vanishes. By letting $\Psi^{(1)}_{\bm{N}_{r}}={\Psi}^{\left(I\!I\text{-}I;\sigma;(\pm  \frac{N}{2}) \ell m;\rho\right)}_{\mu_{1} \mu_{2} }$ and $\Psi^{(2)}_{\bm{N}_{r}}={\Psi}^{\left(I\!I\!I\text{-}I;\sigma;(\pm  \frac{N}{2})\ell m;\rho\right)}_{ \mu_{1}  \mu_{2} }$ in eq.~(\ref{dS_invar_inner_prod}) with $\xi=X$ and using the transformation formulae~(\ref{transfrmn_type_II-I_unnormalsd_r_boost_even}) (with $r=2$) and (\ref{transfrmn_type_III-I_unnormalsd_5/2_boost_even}), we find $\braket{{\Psi}^{\left(I\!I\text{-}I;\sigma;(\pm  \frac{N}{2})\ell m;\rho\right)},{\Psi}^{\left(I\!I\text{-}I;\sigma;(\pm  \frac{N}{2})\ell m;\rho\right)} }_{(r=2)}=0$. Then, by letting $\Psi^{(1)}_{\bm{N}_{r}}={\Psi}^{\left(I;\sigma;(\pm  \frac{N}{2}) \ell m;\rho\right)}_{\mu_{1} \mu_{2} }$ and $\Psi^{(2)}_{\bm{N}_{r}}={\Psi}^{\left(I\!I\text{-}I;\sigma;(\pm  \frac{N}{2})\ell m;\rho\right)}_{ \mu_{1}  \mu_{2} }$ in eq.~(\ref{dS_invar_inner_prod}) with $\xi=X$ and using the transformation formulae~(\ref{transfrmn_type_I_unnormalsd_r_boost_even}) (with $r=2$) and (\ref{transfrmn_type_II-I_unnormalsd_r_boost_even}) (with $r=2$), we find $\braket{{\Psi}^{\left(I;\sigma;(\pm  \frac{N}{2})\ell m;\rho\right)},{\Psi}^{\left(I;\sigma;(\pm  \frac{N}{2})\ell m;\rho\right)} }_{(r=2)}=0$. Thus, in the strictly massless spin-5/2 theory the `pure gauge' modes have zero norm for even $N \geq 4$.

\noindent $\bullet$ \textbf{The special case $\bm{N=4}$.} Let us now discuss the case with $N=4$. First, we show that if $N=4$, then the dS invariance of the scalar product~(\ref{dS_invar_inner_prod}) (with $\xi=X$) for the analytically continued STSSH's with imaginary mass parameter $M=i \tilde{M} \neq 0$ does not require indefiniteness of the norm if and only if $\tilde{M}$ is tuned to the strictly/partially massless values~(\ref{values_mass_parameter_masslessness_fermion}). This can be shown as follows. For $N=4$ eqs.~(\ref{proof_negnorms_typeII-I}) and (\ref{proof_negnorms_typeIII-I}) are trivial due to the vanishing of $\varkappa^{(I\!I)}$ [eq.~(\ref{varkap_II_coeff_r_anal_con})] and $\varkappa^{(I\!I\!I)}$ [eq.~(\ref{varkap_III_coeff_spin5/2_anal_con})], respectively. It is clear that if eq.~(\ref{proof_negnorms_typeI}) is not trivial, then the indefiniteness of the norm cannot be avoided. Equation~(\ref{proof_negnorms_typeI}) becomes trivial if we tune $\tilde{M}$ to the strictly/partially massless values~(\ref{values_mass_parameter_masslessness_fermion}) because for this value of $\tilde{M}$ the type-$I$ modes are pure gauge (i.e. zero-norm modes). Hence, for $N=4$ the dS invariance of the scalar product does not require the indefiniteness of the norm for the strictly/partially massless theories with spin $s \in \set{3/2,5/2}$. Note that, since $\varkappa^{(I\!I)}$ and $\varkappa^{(I\!I\!I)}$ are zero, the (non-zero-norm) eigenmodes with negative spin projection do not mix with the eigenmodes with positive spin projection under the spin$(4,1)$ boost in eqs.~(\ref{transfrmn_type_II-I_unnormalsd_r_boost_even}) and (\ref{transfrmn_type_III-I_unnormalsd_5/2_boost_even}). We have also verified that (non-zero-norm) eigenmodes with different spin projections on $dS_{4}$ do not mix each other under spin$(4)$. 

According to our analysis in the previous paragraph, in the case of strictly/partially massless theories with spin $s=r+1/2$ ($r \in \set{1,2}$) on $dS_{4}$, we conclude the following:
\begin{itemize}
    \item The set $\mathcal{H}^{-}=\set{ {\Psi}^{(B;-;\tilde{M} \ell ;\tilde{\rho})}_{\bm{N}_{r}} }$ of (non-zero-norm) TT eigenmodes with negative spin projection forms an irreducible representation of spin$(4,1)$.
    \item The set $\mathcal{H}^{+}=\set{ {\Psi}^{(B;+;\tilde{M} \ell ;\tilde{\rho})}_{\bm{N}_{r}} }$ of (non-zero-norm) TT eigenmodes with positive spin projection forms separately an irreducible representation of spin$(4,1)$.\footnote{This situation is similar to the case of the strictly massless spin-2 field in $dS_{4}$~\cite{HiguchiLinearised}, where self-dual and anti-self-dual modes correspond to different irreducible representations of SO$(4,1)$.}
\end{itemize}
\textbf{Conclusion concerning the irreducibility of strictly/partially massless representations in $\bm{N=4}$ dimensions.} The two sets of eigenmodes, $\mathcal{H}^{+}$ and $\mathcal{H}^{-}$, form a direct sum of irreducible representations of spin$(4,1)$. In Subsection~\ref{Subsection_explicit_calculationnorms} we are going to show that these irreducible representations are unitary by demonstrating the positivity of the norm in each subspace. [As we demonstrated in our previous article~\cite{Letsios_announce}, this is a direct sum of Discrete Series representations of spin$(4,1)$.] 

\noindent \textbf{Note.} {Note that zero-norm modes (i.e. `pure gauge' modes) transform only into zero-norm modes under spin$(4,1)$ and they can be identified with zero, since, as we discussed above, the coefficient~(\ref{trnsf_coffcient_mixI->II_unnorm_r_anal_con}) (in the transformation formula~(\ref{transfrmn_type_I_unnormalsd_r_boost_even}) with $r \in \set{1,2}$) vanishes for $\tilde{M}^{2}=(N-2)^{2}/4$, while the coefficient~(\ref{trnsf_coffcient_mixII->III_unnorm_r=2_anal_cont}) (in the transformation formula~(\ref{transfrmn_type_II-I_unnormalsd_r_boost_even}) with $r=2$) vanishes for $\tilde{M}^{2}=N^{2}/4$.} For the strictly massless spin-3/2 theory ($r=\uptau=1$, $\tilde{M}^{2}=(N-2)^{2}/4$) and the partially massless spin-5/2 theory ($r=\uptau=2$, $\tilde{M}^{2}=(N-2)^{2}/4$), where the type-$I$ modes have zero norm, the action of spin$(4,1)$ is defined on equivalence classes of the TT modes contained in $\mathcal{H}^{\sigma}$ ($\sigma=\pm$) with the equivalence relation 
$${\Psi}^{\left(B;\sigma;(\pm \frac{N-2}{2})\ell ;\tilde{\rho}\right)}_{\bm{N}_{r}} \sim {\Psi}^{\left(B;\sigma;(\pm \frac{N-2}{2})\ell ;\tilde{\rho}\right)}_{\bm{N}_{r}}+ {\Psi}^{\left(I;\sigma';(\pm \frac{N-2}{2})\ell' ;\tilde{\rho}'\right)}_{\bm{N}_{r}}$$
(with $B=I\!I\text{-}I$ for $r=1$ and $B=I\!I\text{-}I, I\!I\!I\text{-}I$ for $r=2$), where ${\Psi}^{\left(I;\sigma';(\pm \frac{N-2}{2})\ell' ;\tilde{\rho}'\right)}_{\bm{N}_{r}}$ is any type-$I$ mode, i.e. the labels $\sigma', \ell'$ and $\tilde{\rho}'$ are no necessarily equal to $\sigma, \ell$ and $\tilde{\rho}$, respectively.
For the strictly massless spin-5/2 theory ($r=\uptau+1=2$, $\tilde{M}^{2}=N^{2}/4$), where both type-$I$ and type-$I\!I\text{-}I$ modes have zero norm, the action of spin$(4,1)$ is defined on equivalence classes of type-$I\!I\!I\text{-}I$ modes in $\mathcal{H}^{\sigma}$ ($\sigma=\pm$) with the equivalence relation
$${\Psi}^{\left(I\!I\!I\text{-}I;\sigma;(\pm \frac{N}{2})\ell ;\tilde{\rho}\right)}_{\mu_{1} \mu_{2}} \sim {\Psi}^{\left(I\!I\!I\text{-}I;\sigma;(\pm \frac{N}{2})\ell ;\tilde{\rho}\right)}_{\mu_{1} \mu_{2}}+ {\Psi}^{(PG)}_{\mu_{1} \mu_{2}},$$
where ${\Psi}^{(PG)}_{\mu_{1} \mu_{2}}$ is any (finite or infinite) linear combination of type-$I$ and type-$I\!I$ modes.

$$ \textbf{Eigenmodes and helicity for strictly/partially massless theories on}~\bm{dS_{4}}   $$
  For the strictly massless theories with spin $s \in \set{3/2, 5/2}$ on $dS_{4}$, the set $\mathcal{H}^{-}$ is identified with the set of states with `negative helicity' ($-s$), while the set $\mathcal{H}^{+}$ is identified with the set of states with `positive helicity' ($+s$). This can be understood as follows. As in Ref.~\cite{HiguchiLinearised}, let us introduce the helicity operator $\tilde{\epsilon}_{\theta_{i}}^{\hspace{2mm} \theta_{j} \theta_{k}} \tilde{\nabla}_{\theta_{j}}$, where $\tilde{\epsilon}_{\theta_{i} \theta_{j} \theta_{k}}$ is the invariant 3-form on $S^{3}$ ($i,j,k \in \set{1,2,3}$). For the strictly massless spin-3/2 theory on $dS_{4}$, where $$\mathcal{H}^{\sigma}=\set{{\Psi}^{(B;\sigma;\tilde{M} \ell ;\tilde{\rho})}_{\bm{N}_{1}}}=\set{ {\Psi}^{(I\!I\text{-}I;\sigma;(\pm 1) \ell ;\tilde{\rho})}_{\mu}},$$ it can readily be shown that eigenmodes with different spin projections belong to different eigenspaces of the helicity operator, as
\begin{align}
    \tilde{\epsilon}_{\theta_{i}}^{\hspace{2mm} \theta_{j} \theta_{k}} \tilde{\nabla}_{\theta_{j}} {\Psi}^{(I\!I\text{-}I;\sigma; (\pm 1) \ell ;\tilde{\rho})}_{\theta_{k}}  \propto \tilde{\slashed{\nabla}}{\Psi}^{(I\!I\text{-}I;\sigma;(\pm 1) \ell ;\tilde{\rho})}_{\theta_{i}}=i\sigma\,\left(\ell+\frac{3}{2}\right)\,{\Psi}^{(I\!I\text{-}I;\sigma;(\pm 1) \ell ;\tilde{\rho})}_{\theta_{i}}.
\end{align}
(This equation can be readily proved using the fact that $\tilde{\epsilon}_{\theta_{i} \theta_{j} \theta_{k}} \propto  \tilde{\gamma}_{\theta_{i} \theta_{j} \theta_{k}}$, where $\tilde{\gamma}_{\theta_{i} \theta_{j} \theta_{k}}$ is the third-rank gamma matrix on $S^{3}$ which is given by the anti-symmetrised product of three gamma matrices $\tilde{\gamma}_{\theta_{i} \theta_{j} \theta_{k}}=\tilde{\gamma}_{[\theta_{i}} \tilde{\gamma}_{\theta_{j}}\tilde{\gamma}_{\theta_{k}]}$ - see e.g. Ref.~\cite{Freedman}.)
Similarly, for the strictly massless spin-5/2 theory on $dS_{4}$, where $$\mathcal{H}^{\sigma}=\set{{\Psi}^{(B;\sigma;\tilde{M} \ell ;\tilde{\rho})}_{\bm{N}_{2}}}=\set{ {\Psi}^{(I\!I\!I\text{-}I;\sigma;(\pm 2) \ell ;\tilde{\rho})}_{\mu \nu}},$$ it can readily be shown that 
\begin{align}
    \tilde{\epsilon}_{\theta_{i}}^{\hspace{2mm} \theta_{j} \theta_{k} } \tilde{\nabla}_{\theta_{j}} {\Psi}^{(I\!I\!I\text{-}I;\sigma; (\pm 2) \ell ;\tilde{\rho})}_{\theta_{k}\theta_{l}}  \propto \tilde{\slashed{\nabla}}{\Psi}^{(I\!I\!I\text{-}I;\sigma;(\pm 2) \ell ;\tilde{\rho})}_{\theta_{i} \theta_{l}}=i\sigma\,\left(\ell+\frac{3}{2}\right)\,{\Psi}^{(I\!I\!I\text{-}I;\sigma;(\pm 2) \ell ;\tilde{\rho})}_{\theta_{i} \theta_{l}}.
\end{align}
In the case of the partially massless spin-5/2 field on $dS_{4}$, where $$\mathcal{H}^{\sigma}=\set{{\Psi}^{(B;\sigma;\tilde{M} \ell ;\tilde{\rho})}_{\bm{N}_{2}}}=\set{ {\Psi}^{(I\!I\text{-}I;\sigma;(\pm 1) \ell ;\tilde{\rho})}_{\mu \nu}, {\Psi}^{(I\!I\!I\text{-}I;\sigma;(\pm 1) \ell ;\tilde{\rho})}_{\mu \nu}},$$
the helicity operator cannot be defined in the same way. However, it is natural to identify $\mathcal{H}^{-}$ with the set of states with helicities $(-5/2, -3/2)$ and $\mathcal{H}^{+}$ with the set of states with helicities $(+5/2, +3/2)$.


Below we choose a specific dS invariant scalar product for the analytically continued STSSH's with imaginary mass parameter. By calculating the associated norms of the modes we will verify the non-unitarity of the spin$(N,1)$ representations for even $N >4$ for arbitrary imaginary mass parameter $M=i \tilde{M}$ ($\tilde{M} \neq 0$). Also, in the case of strictly/partially massless theories on $dS_{4}$, we will show that each of the spin$(4,1)$ invariant subspaces, $\mathcal{H}^{-}$ and $\mathcal{H}^{+}$, separately forms a unitary representation of spin$(4,1)$ (and, thus, we have a direct sum of UIR's of spin$(4,1)$).

\subsection{Strictly/partially massless spin-3/2 and spin-5/2 representations of spin\texorpdfstring{${(N,1)}$}{(N,1)} for \texorpdfstring{${N}$}{N} even: norms of the eigenmodes} \label{Subsection_explicit_calculationnorms}
In this Subsection, by calculating the norms of the analytically continued STSSH's explicitly, we show that the representations of spin$(N,1)$ (even $N \geq 4$) formed by the spin-3/2 and spin-5/2 TT mode solutions of eq.~(\ref{Dirac_eqn_fermion_dS}) with arbitrary imaginary mass parameter $M=i\tilde{M}$ ($\tilde{M} \neq 0$) are non-unitary, unless the following two conditions hold at the same time: i) $N=4$ and ii) $\tilde{M}$ is tuned to the strictly/partially massless values~(\ref{values_mass_parameter_masslessness_fermion}). For $N=4$, we show that the TT modes in the strictly/partially massless theories form a direct sum of UIR's of spin$(4,1)$. In other words, in the present Subsection we verify the results of Subsection~\ref{Subsection_reps_Neven_part1} for even $N>4$ and we prove statement~\ref{basic_result_N=4}.

{Let $\Psi^{(1)}_{\mu_{1}...\mu_{r}}$ and $\Psi^{(2)}_{\mu_{1}...\mu_{r}}$ be any two analytically continued STSSH's [satisfying eqs.~(\ref{Dirac_eqn_fermion_dS}) and (\ref{TT_conditions_fermions_dS})] with the same imaginary mass parameter $M= i\tilde{M}$ ($\tilde{M} \neq 0$) on $dS_{N}$ ($N$ even). For a spin$(N,1)$ representation to be unitary we need a notion of a dS invariant and positive-definite scalar product. (Recall that dS invariance of the scalar product means that the spin$(N,1)$ generators, which in our case correspond to the Lie-Lorentz derivatives with respect to dS Killing vectors, are anti-hermitian.) In the case of half-odd-integer-spin fields, the standard choice for a scalar product is the Dirac scalar product (see, e.g., \cite{Letsios}) \begin{align*}
   \braket{{\Psi}^{(1)}, {\Psi}^{(2)}}_{Dirac}   =& \,\int_{S^{N-1}}\sqrt{-{g}}\, d\bm{\theta}_{N-1}\,  {\Psi}^{(1)\dagger}_{ \mu_{1}...\mu_{r}} \,{\Psi}^{(2)\mu_{1}...\mu_{r}} = \int_{S^{N-1}}\sqrt{-{g}}\, d\bm{\theta}_{N-1} \,J^{0}_{Dirac},
 \end{align*}
 where $d \bm{\theta}_{N-1}$ stands for $d \theta_{1} \,d\theta_{2}...d\theta_{N-1}$, $g$ is the determinant of the de Sitter metric, while $J^{\mu}_{Dirac}$ is the Dirac current
 $$J^{\mu}_{Dirac}= i\,\overline{\Psi}^{(1)}_{\mu_{1}...\mu_{r}} \gamma^{\mu} \Psi^{(2)\mu_{1}...\mu_{r}}$$
  with $\overline{\Psi}^{(1)}_{\mu_{1}...\mu_{r}}= i {\Psi}^{(1)\dagger}_{\mu_{1}...\mu_{r}} \gamma^{0}$. Interestingly, unlike the case of a real mass parameter, in the case of an imaginary mass parameter the Dirac current is \textbf{not} covariantly conserved. This means that the Dirac scalar product is \textbf{not} de Sitter invariant~\cite{Letsios}, and thus, it cannot be used to study the unitarity of the spin$(N,1)$ representations. (The non-conservation of the Dirac current can be understood using the equations satisfied by $\overline{\Psi}^{(1)}_{\mu_{1}...\mu_{r}}$ and $\Psi^{(2)}_{\mu_{1}...\mu_{r}}$:  $ \slashed{\nabla}\Psi^{(2)}_{\mu_{1}...\mu_{r}}=-i\, \tilde{M}\,\Psi^{(2)}_{\mu_{1}...\mu_{r}}  $ and $\nabla_{\nu}\overline{\Psi}^{(1)}_{\mu_{1}...\mu_{r}} \gamma^{\nu} =- i \tilde{M} \,\overline{\Psi}^{(1)}_{\mu_{1}...\mu_{r}}$. Note that the mass parameter has the same sign in both equations. If the mass parameter were real, the signs in the two equations would be opposite. A straightforward calculation reveals that $\nabla_{\mu} J^{\mu}_{Dirac} \neq 0$.)}
  
  {On the other hand}, the axial current
 \begin{equation}\label{current}
     J^{\mu}= i\,\overline{\Psi}^{(1)}_{\mu_{1}...\mu_{r}} \gamma^{\mu}\gamma^{N+1} \Psi^{(2)\mu_{1}...\mu_{r}}
    \end{equation}
is covariantly conserved~\cite{Higuchi_private_com}. Then, it follows that the scalar product
\begin{align}
     \braket{{\Psi}^{(1)}, {\Psi}^{(2)}}_{(r)} =& \int_{S^{N-1}}\sqrt{-g}\, d\bm{\theta}_{N-1} \,J^{0}\end{align}
    is time-independent.  This scalar product is equivalently written as
     \begin{align}\label{inner_prod_dSN_gamma5}
   \braket{{\Psi}^{(1)}, {\Psi}^{(2)}}_{(r)}   =& \cosh^{N-1}{t}\,\int_{S^{N-1}}\sqrt{\tilde{g}}\, d\bm{\theta}_{N-1}\,  {\Psi}^{(1)\dagger}_{ \mu_{1}...\mu_{r}}\gamma^{N+1}{\Psi}^{(2)\mu_{1}...\mu_{r}},
 \end{align}
where we used $(\gamma^{0})^{2}=-\bm{1}$, as well as
\begin{align}
    \sqrt{-g}=\cosh^{N-1}{t}\, \sqrt{\tilde{g}},
\end{align}
while $\sqrt{\tilde{g}}$ is given by eq.~(\ref{determinant_metric_S_N-1}).

Now let us show that the scalar product~(\ref{inner_prod_dSN_gamma5}) is de Sitter invariant. 
Let $\xi^{\mu}$ be a Killing vector of $dS_{N}$ satisfying
\begin{equation} \label{Killing_equation}
    \nabla_{\mu}\xi_{\nu}+ \nabla_{\nu}\xi_{\mu}=0.
\end{equation}
The infinitesimal change  $\delta_{\xi} J^{\mu}$ of the current~(\ref{current}) under the spin$(N,1)$ transformation generated by $\xi^{\mu}$ is described by the Lie derivative
\begin{align}
    \delta_{\xi} J ^{\mu}=\mathcal{L}_{\xi} J^{\mu}&= \xi^{\nu} \nabla_{\nu} J^{\mu}- J^{\nu}  \nabla_{\nu} \xi^{\mu} \nonumber \\&=\nabla_{\nu} (  \xi^{\nu} J^{\mu}- J^{\nu}   \xi^{\mu}),
\end{align}
where we used $\nabla_{\mu}J^{\mu}=\nabla_{\mu}\xi^{\mu}=0$.
Then, it is straightforward to find
\begin{equation}\label{delta_J_0}
      \delta_{\xi} J ^{0}=  \frac{1}{\sqrt{-g}}\partial_{\theta_{\kappa}}\big[ \sqrt{-g} (  \xi^{\theta_{\kappa}}  J^{0}- J^{\theta_{\kappa}}   \xi^{0}) \big],
\end{equation}
where $\kappa={1},...,{N-1}$. By integrating eq.~(\ref{delta_J_0}) over $S^{N-1}$ we find that the scalar product~(\ref{inner_prod_dSN_gamma5}) is de Sitter invariant, as
\begin{equation}
  \delta_{\xi} \braket{\Psi^{(1)}, \Psi^{(2)}}_{(r)}=  \int_{S^{N-1}} d \bm{\theta}_{N-1}\, \sqrt{-g}\, \delta_{\xi} J^{0}= 0.
\end{equation}
 
It is possible to calculate the norms of the analytically continued STSSH's of ranks 1 and 2 [the analytically continued STSSH's are defined by eq.~(\ref{define_anal_continued_STSSH's})] using the de Sitter invariant scalar product~(\ref{inner_prod_dSN_gamma5}). We find in this manner
 \begin{align}\label{normln^(-1)_fac_dSN_TYPEB_r}
   \braket{{\Psi}^{(B;\sigma;\tilde{M} \ell ;\tilde{\rho})},{\Psi}^{(B';\sigma';\tilde{M} \ell'; \tilde{\rho}')} }_{(r)} =&~(-\sigma)\times {\binom{r}{\tilde{r}_{(B)}}}\,2^{N+2r-1-4\tilde{r}_{(B)}}\,\nonumber\\
   &\times \frac{|\Gamma(\ell+\frac{N}{2})|^{2}}{\Gamma(\ell+\frac{N}{2}+\tilde{M})  \Gamma(\ell+\frac{N}{2}-\tilde{M})} \nonumber\\
   &\times \left(\prod_{j=\tilde{r}_{(B)}}^{r-1} \frac{N+2j-1}{N+j+\tilde{r}_{(B)}-2}\right) \nonumber\\
   &\times \left(\prod_{j=\tilde{r}_{(B)}}^{r-1} \frac{1}{(\ell-j)(\ell+N-1+j)} \right) \nonumber \\ 
      &\times \left( \prod_{j=1}^{r-\tilde{r}_{(B)}} \left\{-\tilde{M}^{2}+\left(r-j+\frac{N-2}{2}\right)^{2}  \right\}\right) \delta_{\sigma \sigma'} \delta_{\ell \ell'} \delta_{\tilde{\rho} \tilde{\rho}'}
 \end{align}
 for $r\in \set{1,2}$ and $B=I,I\!I, I\!I\!I$ (where $\sigma=\pm$, $\tilde{M} \in \mathbb{R} \setminus \{ 0\}$, $\tilde{r}_{(B)} \leq r$, while $\tilde{r}_{(I)}=0$, $\tilde{r}_{(I\!I)}=1$ and $\tilde{r}_{(I\!I\!I)}=2$). The norms of type-$I$ and type-$I\!I$ spin-3/2 modes, as well as the norms of type-$I\!I$ and type-$I\!I\!I$ spin-5/2 modes, can be determined by direct calculation using the time-independence of the scalar product~(\ref{inner_prod_dSN_gamma5}). The calculations are simplified by using
 \begin{align}\label{phiaM^2-psiaM^2}
  \left| \hat{\phi}^{(a)}_{\tilde{M} \ell}(t=0)\right|^{2}-\left|  \hat{\psi}^{(a)}_{\tilde{M} \ell}(t=0)\right|^{2}=\frac{2^{N+2a-1}\,|\Gamma(\ell+\frac{N}{2})|^{2}}{\Gamma(\ell+\frac{N}{2}+\tilde{M})  \Gamma(\ell+\frac{N}{2}-\tilde{M})} .
 \end{align}
[This equation can readily be proved using eqs.~(\ref{Fab_(a+b)/2}) and (\ref{Fab_(a+b)/2+1}).] Once the norms of type-$I\!I$ and type-$I\!I\!I$ spin-5/2 modes have been calculated, the norm of the type-$I$ spin-5/2 modes is readily found using the dS invariance~(\ref{dS_invar_inner_prod}) of the inner product between type-$I$ and type-$I\!I$ modes (by making use of the transformation formulae~(\ref{transfrmn_type_I_unnormalsd_r_boost_even}) and (\ref{transfrmn_type_II-I_unnormalsd_r_boost_even})).
 
\noindent \textbf{Consistency check.} As a consistency check, by using our result for the norms~(\ref{normln^(-1)_fac_dSN_TYPEB_r}) of the eigenmodes with spin $s \in \set{3/2, 5/2}$, we can reproduce the strictly/partially massless tunings~(\ref{values_mass_parameter_masslessness_fermion}) for the imaginary mass parameter as follows.
 For $r=1$ (spin-3/2 field), we find that the norm~(\ref{normln^(-1)_fac_dSN_TYPEB_r}) of type-$I$ modes ($\tilde{r}_{(I)}=0$) becomes zero if $\tilde{M}^{2}=(N-2)^{2}/4$, corresponding to the strictly massless spin-3/2 theory. For $r=2$ (spin-5/2 field), we find that both type-$I$ ($\tilde{r}_{(I)}=0$) and type-$I\!I$ ($\tilde{r}_{(I\!I)}=1$) modes have zero norm~(\ref{normln^(-1)_fac_dSN_TYPEB_r}) for $\tilde{M}^{2}=N^{2}/4$, corresponding to the strictly massless spin-5/2 theory. Finally, for $r=2$, we find that type-$I$ ($\tilde{r}_{(I)}=0$) modes have zero norm~(\ref{normln^(-1)_fac_dSN_TYPEB_r}) for $\tilde{M}^{2}=(N-2)^{2}/4$, corresponding to the partially massless spin-5/2 theory.
 
\noindent \textbf{Note.} We observe that the sign of the norm~(\ref{normln^(-1)_fac_dSN_TYPEB_r}) depends on the sign of the spin projection index $\sigma=\pm$, as expected from the dS invariance of the scalar product~(\ref{proof_negnorms_typeI})-(\ref{proof_negnorms_typeIII-I}). Thus, it is easy to understand that representations of spin$(N,1)$ with spin $s \in \set{3/2,5/2}$ and arbitrary imaginary mass parameter $M=i \tilde{M} \neq 0$ are non-unitary for even $N >4$, since positive-norm and negative-norm modes mix with each other under spin$(N,1)$ [see the transformation formulae~(\ref{transfrmn_type_II-I_unnormalsd_r_boost_even}) and (\ref{transfrmn_type_III-I_unnormalsd_5/2_boost_even})]. Similarly, we find that for $N=4$ the representations of spin$(4,1)$ are not unitary if $\tilde{M}$ is not given by the strictly/partially massless values in eq.~(\ref{values_mass_parameter_masslessness_fermion}).

\noindent \textbf{Striclty/partially massless theories and direct sum of spin$\bm(4,1)$ UIRs.} Now, let us suppose that the following two conditions are satisfied at the same time: i) $N=4$ and ii) the imaginary mass parameter is tuned to the strictly/partially massless values~(\ref{values_mass_parameter_masslessness_fermion}). According to our discussion for the $N=4$ case in Subsection~\ref{Subsection_reps_Neven_part1}, each of the solution subspaces, $\mathcal{H}^{-}$ and $\mathcal{H}^{+}$, forms separately an irreducible representation of spin$(4,1)$ with spin $s=r+1/2$ ($r \in \set{1,2}$). (The `pure gauge' modes are identified with zero in each subspace.) We can show that the subspaces $\mathcal{H}^{-}$ and $\mathcal{H}^{+}$ form a direct sum of UIR's of spin$(4,1)$ as follows\footnote{These UIR's have been identified with Discrete Series UIR's of spin$(4,1)$ in our previous article~\cite{Letsios_announce}.}. By observing that the norms~(\ref{normln^(-1)_fac_dSN_TYPEB_r}) of the eigenmodes depend on the spin projection, we have: 
 \begin{itemize}
     \item   For the set of eigenmodes with negative spin projection (or negative helicity), $\mathcal{H}^{-}=\set{ {\Psi}^{(B;-;\tilde{M} \ell ;\tilde{\rho})}_{\bm{N}_{r}} }$, the positive-definite inner product is
     \begin{align*}
        &\braket{{\Psi}^{(B;-;\tilde{M} \ell ;\tilde{\rho})},{\Psi}^{(B';-;\tilde{M} \ell' ;\tilde{\rho}')} }_{(r)}\nonumber\\
        &=\cosh^{3}{t}\,\int_{S^{3}}\sqrt{\tilde{g}}\, d\bm{\theta}_{3}\,  {\Psi}^{(B;-;\tilde{M} \ell ;\tilde{\rho})\dagger}_{ \mu_{1}...\mu_{r}}\,\gamma^{5}\,{\Psi}^{(B';-;\tilde{M} \ell' ;\tilde{\rho}')\mu_{1}...\mu_{r}} 
     \end{align*}
     The explicit expression for the positive-definite norm is given by eq.~(\ref{normln^(-1)_fac_dSN_TYPEB_r}). 
     \item  For the set of eigenmodes with positive spin projection (or positive helicity), $\mathcal{H}^{+}=\set{ {\Psi}^{(B;+;\tilde{M} \ell ;\tilde{\rho})}_{\bm{N}_{r}} }$, the positive-definite inner product is $$- \braket{{\Psi}^{(B;+;\tilde{M} \ell ;\tilde{\rho})},{\Psi}^{(B';+;\tilde{M} \ell' ;\tilde{\rho}')} }_{(r)}.$$  
     The explicit expression for the positive-definite norm is given by the negative of eq.~(\ref{normln^(-1)_fac_dSN_TYPEB_r}).
 \end{itemize}
 \subsection{The strictly/partially massless spin-3/2 and spin-5/2 representations of spin\texorpdfstring{${(N,1)}$}{(N,1)} are non-unitary for \texorpdfstring{${N}$}{N} odd} \label{Subsection_reps_Nodd}

In this Subsection, we show that the strictly massless spin-3/2 field theory, as well as the strictly and partially massless spin-5/2 field theories, on $dS_{N}$ ($N$ odd) are not unitary (i.e. we prove statement~\ref{basic_result_Nodd}).

\noindent \textbf{spin$\bm{(N,1)}$ transformation formulae.} As in the case with $N$ even, we study the transformation properties of the analytically continued STSSH's under the de Sitter boost~(\ref{Killing_vector_boost}).
By making the replacements~(\ref{replacements}) in the spin$(N+1)$ transformation formulae (\ref{transfrmn_type_I_unnormalsd_r_SN_odd}), (\ref{transfrmn_type_II-I_unnormalsd_r_SN_odd}) and (\ref{transfrmn_type_III-I_unnormalsd_5/2_SN_odd}) [and using eq.~(\ref{define_anal_continued_STSSH's_Nodd})], we find
\begin{align}\label{transfrmn_type_I_unnormalsd_r_boost_odd}
     \mathbb{L}_{X}{\Psi}^{(I;\tilde{M} \ell;\sigma_{N-1}; m;\rho)}_{ \bm{N}_{r} } =&~ -i\,c_{(\ell)}\, \mathcal{A}^{(I)}\, {\Psi}^{(I;\tilde{M} \,(\ell+1);\sigma_{N-1}; m;\rho)}_{ \bm{N}_{r} } -\frac{i}{c_{(\ell-1)}} \mathcal{B}^{(I)}\, {\Psi}^{(I;\tilde{M} \,(\ell-1);\sigma_{N-1};m;\rho)}_{ \bm{N}_{r} }\nonumber\\
     &-\sigma_{N-1} \, \varkappa^{(I)}\, {\Psi}^{(I;\tilde{M} \ell;\sigma_{N-1}; m ;\rho)}_{ \bm{N}_{r} }-i \mathcal{K}^{(I\rightarrow I\!I)}\,{\Psi}^{(I\!I\text{-}I;\tilde{M} \ell;\sigma_{N-1}; m ;\rho)}_{\bm{N}_{r}},
\end{align}
\begin{align}\label{transfrmn_type_II-I_unnormalsd_r_boost_odd}
     \mathbb{L}_{X} {\Psi}^{(I\!I\text{-}I;\tilde{M} \ell ; \sigma_{N-1};m;\rho)}_{\  \bm{N}_{r} }=&~-i\,c_{(\ell)} \mathcal{A}^{(I\!I)} \,{\Psi}^{(I\!I\text{-}I;\tilde{M} \,(\ell+1);\sigma_{N-1};m;\rho)}_{  \bm{N}_{r} }\nonumber\\
     &-\frac{i}{c_{(\ell-1)}} \mathcal{B}^{(I\!I)} \,{\Psi}^{(I\!I\text{-}I;\tilde{M} \,(\ell-1);\sigma_{N-1};m;\rho)}_{ \bm{N}_{r} } \nonumber\\
     &-\sigma_{N-1} \, \varkappa^{(I\!I)} {\Psi}^{(I\!I\text{-}I;\tilde{M}\ell;\sigma_{N-1}; m ;\rho)}_{ \bm{N}_{r}  }-i \mathcal{K}^{(I\!I\rightarrow I)}\,{\Psi}^{(I;\tilde{M} \ell;\sigma_{N-1}; m ;\rho)}_{ \bm{N}_{r}}\nonumber\\
     &-i \mathcal{K}^{(I\!I\rightarrow I\!I\!I)}\,{\Psi}^{(I\!I\!I\text{-}I;\tilde{M} \ell;\sigma_{N-1}; m ;\rho)}_{  \bm{N}_{r}} 
\end{align}
($r=1,2,$) and
\begin{align}\label{transfrmn_type_III-I_unnormalsd_5/2_boost_odd}
     \mathbb{L}_{X} &{\Psi}^{(I\!I\!I\text{-}I;\tilde{M} \ell;\sigma_{N-1}; m;\rho)}_{\mu_{1}\mu_{2} }\nonumber\\
     =&~-i\,c_{(\ell)} \mathcal{A}^{(I\!I\!I)} \,{\Psi}^{(I\!I\!I\text{-}I;\tilde{M} \,(\ell+1);\sigma_{N-1};m;\rho)}_{ \mu_{1} \mu_{2} } -\frac{i}{c_{(\ell-1)}} \mathcal{B}^{(I\!I\!I)} \,{\Psi}^{(I\!I\!I\text{-}I;\tilde{M} \,(\ell-1);\sigma_{N-1};m;\rho)}_{ \mu_{1} \mu_{2} } \nonumber\\
     &-\sigma_{N-1}\,  \varkappa^{(I\!I\!I)} {\Psi}^{(I\!I\!I\text{-}I;\tilde{M} \ell;\sigma_{N-1}; m ;\rho)}_{ \mu_{1} \mu_{2}  }-i \mathcal{K}^{(I\!I\!I\rightarrow I\!I)}\,{\Psi}^{(I\!I\text{-}I;\tilde{M} \ell;\sigma_{N-1}; m ;\rho)}_{ \mu_{1}\mu_{2}},
\end{align}
respectively, where all the coefficients on the right-hand sides of eqs.~(\ref{transfrmn_type_I_unnormalsd_r_boost_odd})-(\ref{transfrmn_type_III-I_unnormalsd_5/2_boost_odd}) are the same as the coefficients used in the case with $N$ even [see eqs.~(\ref{transfrmn_type_I_unnormalsd_r_boost_even})-(\ref{transfrmn_type_III-I_unnormalsd_5/2_boost_even})]. 
 
 $$\textbf{Non-existence of positive-definite, dS invariant scalar products}$$ Now, we will show that the representations of spin$(N,1)$ ($N$ odd) formed by the spin-3/2 and spin-5/2 TT mode solutions of eq.~(\ref{Dirac_eqn_fermion_dS}) are non-unitary for all values of the imaginary mass parameter $M=i \tilde{M}$ ($\tilde{M} \neq 0$), including the strictly/partially massless tunings~(\ref{values_mass_parameter_masslessness_fermion}). Let $\braket{\Psi^{(1)}, \Psi^{(2)}}$ be a dS invariant scalar product for any two analytically continued STSSH's $\Psi^{(1)}, \Psi^{(2)}$ [satisfying eqs.~(\ref{Dirac_eqn_fermion_dS}) and (\ref{TT_conditions_fermions_dS})] with $M = i \tilde{M}$ and  $\tilde{M}\neq 0$. We will show that this scalar product must vanish for all eigenmodes. First, let us make the following observation. The infinitesimal transformations $\mathbb{L}_{X} {\Psi}^{(B;\tilde{M} \ell;\sigma_{N-1}; m;\rho)}_{\bm{N}_{r} }$ given by eqs.~(\ref{transfrmn_type_I_unnormalsd_r_boost_odd})-(\ref{transfrmn_type_III-I_unnormalsd_5/2_boost_odd}), always give rise to a term of the form $\varkappa^{(B)}\,{\Psi}^{(B;\tilde{M} \ell;\sigma_{N-1}; m;\rho)}_{\bm{N}_{r}}$ in the linear combination on the right-hand sides of each of~eqs.~(\ref{transfrmn_type_I_unnormalsd_r_boost_odd})-(\ref{transfrmn_type_III-I_unnormalsd_5/2_boost_odd}). The coefficients $\varkappa^{(I)}, \varkappa^{(I\!I)}$ and $\varkappa^{(I\!I\!I)}$ are given by eqs.~(\ref{varkap_I_coeff_r_anal_con}), (\ref{varkap_II_coeff_r_anal_con}) and (\ref{varkap_III_coeff_spin5/2_anal_con}), respectively, and they are all non-zero for $N$ odd. Thus, by combining the dS invariance of the scalar product:
 \begin{align}
    \braket{\mathbb{L}_{X} {\Psi}^{(B;\tilde{M} \ell;\sigma_{N-1}; m;\rho)}, {\Psi}^{(B;\tilde{M} \ell;\sigma_{N-1}; m;\rho)}}+\braket{ {\Psi}^{(B;\tilde{M} \ell;\sigma_{N-1}; m;\rho)}, \mathbb{L}_{X}{\Psi}^{(B;\tilde{M} \ell;\sigma_{N-1}; m;\rho)}}=0
 \end{align}
 with the transformation formulae~(\ref{transfrmn_type_I_unnormalsd_r_boost_odd})-(\ref{transfrmn_type_III-I_unnormalsd_5/2_boost_odd}), we find
 \begin{align}
    \braket{ {\Psi}^{(B;\tilde{M} \ell;\sigma_{N-1}; m;\rho)}, {\Psi}^{(B;\tilde{M} \ell;\sigma_{N-1}; m;\rho)}}=0
 \end{align}
 for $B=I, I\!I\text{-}I, I\!I\!I\text{-}I$ and for all $\tilde{M}\neq   0$. Then, since the eigenmodes with different labels are orthogonal, we conclude that there is no dS invariant scalar product (which is not identically zero).

  
\section{Summary and discussions}
\label{summary and discussions}
\noindent \textbf{Summary.} In this paper, we provided a technical explanation of the results of our previous article~\cite{Letsios_announce}. In particular, we showed that the strictly massless spin-3/2 field (i.e. gravitino field) theory, as well as the strictly and partially massless spin-5/2  field theories on $dS_{N}$ ($N \geq 3$) are unitary only in $N=4$ dimensions. In order to arrive at this result, we studied the group-theoretic properties of the eigenmodes for the following field theories with imaginary mass parameter on $dS_{N}$ ($N \geq 3$): the vector-spinor field and the symmetric rank-2 tensor-spinor field. The corresponding eigenmodes satisfy eq.~(\ref{Dirac_eqn_fermion_dS}) with $M = i \tilde{M}$ ($\tilde{M} \neq 0$)
and the TT conditions~(\ref{TT_conditions_fermions_dS}). These eigenmodes were obtained by analytically continuing STSSH's on $S^{N}$. The transformation properties of these eigenmodes under a spin$(N,1)$ boost were studied. By using these transformation properties, we showed that all dS invariant scalar products for even $N>4$ are indefinite. We also showed that all dS invariant scalar products must vanish identically for odd $N$. It was found that dS invariant scalar products that are positive-definite are allowed only for strictly and partially massless theories in $N=4$ dimensions (and, thus, these theories are unitary). Also, for these unitary spin-$s$ ($s \in \set{ 3/2, 5/2}$) strictly/partially massless theories on $dS_{4}$, we showed that eigenmodes with positive helicity and the ones with negative helicity separately form UIR's of spin$(4,1)$. (In particular, we have a direct sum of Discrete Series UIR's of spin$(4,1)$~\cite{Letsios_announce}.) All the results mentioned in this paragraph are summarised as statements \ref{basic_result_Neven}, \ref{basic_result_N=4} and \ref{basic_result_Nodd} in the Introduction.


\noindent \textbf{Towards future work.} The analysis of our previous article~\cite{Letsios_announce} implies that our main result extends to all strictly and partially massless fields with half-odd-integer spin $ s \geq 7/2$ on $dS_{N}$ ($N \geq 3$). The technical proof for this is something that we will leave for future work. It would also be interesting to investigate whether our result about the non-unitarity of the strictly/partially spin-3/2 and spin-5/2 theories on $dS_{N}$ for $N \neq 4$ could be extended to other $N$-dimensional vacuum spacetimes with positive cosmological constant. As an argument indicating that this generalisation of our result is indeed possible, we would like to mention the forbidden mass range for the symmetric spin-2 field on $dS_{N}$~\cite{STSHS, Higuchiforb}. More specifically, the forbidden mass range for the spin-2 field on $dS_{N}$ was explained group-theoretically in Ref.~\cite{STSHS}, while it was first observed for $dS_{4}$ in Ref.~\cite{Higuchiforb}. However, it was later shown that the forbidden mass range exists in all 4-dimensional vacuum spacetimes with positive cosmological constant~\cite{HiguchiMassivespin2}.

\acknowledgments

\noindent \textit{This work is dedicated to the memory of Roberto Camporesi.} \\

 \noindent The author is grateful to Atsushi Higuchi for guidance, suggestions and encouragement, as well as for invaluable discussions and ideas that inspired the majority of the material presented in this paper and helpful comments on earlier versions of this paper. Also, it is a pleasure to thank Stanley Deser for communications and Andrew Waldron for useful correspondence. The author would also like to thank Lasse Schmieding, Nikolaos Koutsonikos-Kouloumpis and F. F. John for useful discussions. I would like to thank Alex for her encouragement. This work was supported by a studentship from the Department of Mathematics, University of York.

\appendix

\section{Raising and lowering operators for the Gauss hypergeometric function and other useful formulae}\label{appendix_raising_lowering_hypergeom}
The Gauss hypergeometric function $F(a,b;c;z)$ satisfies~\cite{NIST:DLMF}
\begin{align}
&\frac{d}{dz}F(a,b;c;z)= \frac{ab}{c}F(a+1,b+1;c+1;z), \label{hyper_raise_all}\\
&(z \frac{d}{dz}+c-1 ) F(a,b;c;z)=(c-1) F(a,b;c-1;z), \label{hyper_lower_c}\\
&(z \frac{d}{dz}+a ) F(a,b;c;z)= a F(a+1,b;c;z). \label{hyper_raise_a}
\end{align}
By combining eq.~(\ref{hyper_raise_a}) with the following relation~\cite{hyper}:
\begin{align}
   (c-b)& F(a+1,b-1;c;z)+  (b-a-1)(1-z)  F(a+1,b;c;z)
   = (c-a-1) F(a,b;c;z),
\end{align}
we find
\begin{align}
    \Big( a(b-c)&+a(-b+a+1)z-(-b+a+1)z(1-z)\frac{d}{dz}\Big)F(a,b;c;z)\nonumber\\
    &= a (b-c) F(a+1,b-1;c;z) \label{hyper_raise_a_lower_b}.
\end{align}
Using eqs.~(\ref{hyper_raise_all}) and (\ref{hyper_lower_c}) we can show the ladder relations (\ref{raising_phi_psi}) and (\ref{lowering_phi_psi}), while using eq.~(\ref{hyper_raise_a_lower_b}) we can show the ladder relations (\ref{raising_tilde}) and (\ref{lowering_tilde}).

The behaviour of the functions~(\ref{phi_a}) and (\ref{psi_a}) in the limit $\theta_{N} \rightarrow \pi$ is studied by using the transformation formula~\cite{gradshteyn2007}
\begin{align}\label{transformation_hypergeom}
      F(\alpha,\beta;\gamma;z)&=\frac{\Gamma(\gamma) \Gamma(\gamma-\alpha-\beta)}{\Gamma(\gamma-\alpha) \Gamma(\gamma-\beta)}\  F(\alpha,\beta;\alpha+\beta-\gamma+1;1-z)\nonumber\\
     &+(1-z)^{\gamma-\alpha-\beta}\frac{\Gamma(\gamma) \Gamma(-\gamma+\alpha+\beta)}{\Gamma(\alpha) \Gamma(\beta)}  F(\gamma-\alpha,\gamma-\beta;\gamma-\alpha-\beta+1;1-z).
  \end{align}

  Equation~(\ref{phiaM^2-psiaM^2}) is proved using~\cite{wolf1}
  \begin{align}\label{Fab_(a+b)/2}
   F\left(a, b,  \frac{a+b}{2} ; \frac{1}{2}\right)=&~\sqrt{\pi}\,\Gamma\left(\frac{a+b}{2}\right)\left[ \frac{1}{\Gamma((a+1)/2)   \Gamma(b/2)}   + \frac{1}{\Gamma((b+1)/2)   \Gamma(a/2)}   \right]
\end{align}
and \cite{wolf2} 
\begin{align}\label{Fab_(a+b)/2+1}
    F\left(a, b,  \frac{a+b}{2}+1 ; \frac{1}{2}\right)=&~\frac{2\sqrt{\pi}}{a-b} \Gamma\left(\frac{a+b}{2}+1\right)\nonumber\\
    &\times \left[ \frac{1}{\Gamma((b+1)/2)} \frac{1}{   \Gamma(a/2)}- \frac{1}{\Gamma((a+1)/2)   \Gamma(b/2)} \right].
\end{align}

\section{Spinor eigenmodes of the Dirac operator on the \texorpdfstring{${({N-1})}$}{N-1}-sphere} \label{appendix_spinor_eigenmodes_SN-1}
The spinor eigenmodes of the Dirac operator (i.e. the STSSH's of rank 0) on spheres of arbitrary dimension have been computed in Ref.~\cite{Camporesi}. Here we write down explicitly the eigenspinors on $S^{N-1}$ that satisfy eq.~(\ref{eigenspinors on S_(N-1)}). These eigenspinors play an important role in the derivation of the formulae for the spin$(N+1)$ transformation of the STSSH's in Appendix~\ref{Appendix_transfrmn_proeprties_norm_fac}.

\noindent \textbf{Case 1:} $\bm{N-1}$ \textbf{odd.} We denote the eigenspinors on $S^{N-1}$ as $\chi_{\pm \ell m \rho}(\theta_{N-1}, \bm{\theta}_{N-2})$, where $\rho$ stands for labels other than $\ell$ and $m$. These eigenspinors are given by
\begin{align}\label{form_of_eigenspinors_SN-1odd}
\chi_{\pm \ell m \rho}(\theta_{N-1}, \bm{\theta}_{N-2})    =~\frac{\tilde{c}_{N-1}(\ell,m)}{\sqrt{2}}\,\Big\{\tilde{\phi}^{(0)}_{\ell m}(\theta_{N-1})\hat{\tilde{\chi}}_{-m {\rho}}(\bm{\theta}_{N-2}) \pm i \tilde{\psi}^{(0)}_{ \ell m}(\theta_{N-1})\hat{\tilde{\chi}}_{+m {\rho}}(\bm{\theta}_{N-2})\Big\} ,
\end{align}
where $\tilde{\phi}^{(0)}_{\ell m}(\theta_{N-1})$ and $\tilde{\psi}^{(0)}_{\ell m}(\theta_{N-1})$ are given by eqs.~(\ref{phitilde_a}) and (\ref{psitilde_a}), respectively, and
\begin{align}
\hat{\tilde{\chi}}_{\pm m {\rho}}(\bm{\theta}_{N-2})= &   \frac{1}{\sqrt{2}}(1+i \tilde{\gamma}^{N-1}){\tilde{\chi}}_{\pm m {\rho}}(\bm{\theta}_{N-2}),\label{tildechihat_in_terms_tildechi_SN-2}\\
\hat{\tilde{\chi}}_{+m {\rho}}(\bm{\theta}_{N-2})=& \tilde{\gamma}^{N-1} {\tilde{\chi}}_{-m {\rho}}(\bm{\theta}_{N-2}),
\end{align}\label{tildechi+_becomes_tildechi-_SN-2}
where the spinors $\tilde{\chi}_{\pm m {\rho}}(\bm{\theta}_{N-2})$ are the eigenspinors of the Dirac operator on $S^{N-2}$. [The gamma matrices on $S^{N-1}$ are denoted as $\tilde{\gamma}^{a}$ - see eq.~(\ref{even_gammas}).]
In order for the eigenspinors~(\ref{form_of_eigenspinors_SN-1odd}) to be non-singular we require $\ell \geq m$ and $\ell=0,1,...$ \cite{Camporesi}. The eigenspinors~(\ref{form_of_eigenspinors_SN-1odd}) satisfy the normalisation condition~(\ref{normalization_SN-1_spinors}), while the normalisation factor is given by~\cite{Camporesi}
\begin{equation}\label{S_N-1_normlzn_fac_spinor}
     \left|\frac{\tilde{c}_{N-1}( \ell, m)}{\sqrt{2}}\right|^{2}=\frac{{\Gamma(\ell-m +1) \Gamma(\ell+N-1+m)}}{2^{N-2}|\Gamma{(\frac{N-1}{2}+\ell)}|^{2}}.
 \end{equation}

\noindent \textbf{Case 2:} $\bm{N-1}$ \textbf{even.} We denote the eigenspinors on $S^{N-1}$ as $\chi^{(\sigma_{N-1})}_{\pm \ell m \rho}(\theta_{N-1}, \bm{\theta}_{N-2})$, where $\sigma_{N-1}=\pm$ is the spin projection index on $S^{N-1}$ and $\rho$ stands for labels other than $\sigma_{N-1},\ell$ and $m$. The eigenspinors with negative spin projection are given by
\begin{align}\label{form_of_eigenspinors_SN-1even_negspin}
&\chi^{(-)}_{\pm \ell m \rho}(\theta_{N-1}, \bm{\theta}_{N-2})   =~ \frac{\tilde{c}_{N-1}( \ell, m)}{\sqrt{2}}\begin{pmatrix}\tilde{\phi}^{(0)}_{ \ell m}(\theta_{N-1})  \,\tilde{\chi}_{- m {\rho}}(\bm{\theta}_{N-2}) \\\pm i \tilde{\psi}^{(0)}_{ \ell m}(\theta_{N-1})\,\tilde{\chi}_{- m {\rho}}(\bm{\theta}_{N-2})\end{pmatrix} 
\end{align}
and those with positive spin projection are given by
\begin{align}\label{form_of_eigenspinors_SN-1even_posspin}
&\chi^{(+)}_{\pm \ell m \rho}(\theta_{N-1}, \bm{\theta}_{N-2})   =~ \frac{\tilde{c}_{N-1}( \ell, m)}{\sqrt{2}}\begin{pmatrix}i\tilde{\psi}^{(0)}_{ \ell m}(\theta_{N-1})  \,\tilde{\chi}_{+ m {\rho}}(\bm{\theta}_{N-2}) \\\pm  \tilde{\phi}^{(0)}_{ \ell m}(\theta_{N-1})\,\tilde{\chi}_{+ m {\rho}}(\bm{\theta}_{N-2})\end{pmatrix} 
\end{align}
and they both satisfy eq.~(\ref{eigenspinors on S_(N-1)}). The normalisation factors ${\tilde{c}_{N-1}( \ell, m)}$, as well as the functions $\tilde{\phi}^{(0)}_{ \ell m}(\theta_{N-1})$ and $\tilde{\psi}^{(0)}_{ \ell m}(\theta_{N-1})$, have the same  expressions as in the case with $N-1$ odd.


\section{Some useful formulae on \texorpdfstring{$\bm{S^{N-1}}$}{SN-1}}
Let $\tilde{g}_{\mu \nu}$ be the metric tensor on $S^{N-1}$. The Riemann tensor on $S^{N-1}$ is
\begin{align}\label{Riemann_tens}
    \tilde{R}_{\mu \nu \kappa \lambda}=\tilde{g}_{\mu \kappa} \tilde{g}_{\nu \lambda}-\tilde{g}_{\nu \kappa} \tilde{g}_{\mu \lambda}.
\end{align}
Let $\tilde{\psi}, \tilde{\psi}_{\mu}$ and $\tilde{\psi}_{\mu \nu }$ be any spinor, vector-spinor and rank-2 tensor-spinor field, respectively, on $S^{N-1}$. The commutator of covariant derivatives acting on these fields is given by
\begin{align}
    [\tilde{\nabla}_{\mu}, \tilde{\nabla}_{\nu}]\tilde{\psi}&=\frac{1}{4}\tilde{R}_{\mu \nu \kappa \lambda} \tilde{\gamma}^{\kappa}\tilde{\gamma}^{\lambda}\tilde{\psi}\\
    &=\frac{1}{2}(\tilde{\gamma}_{\mu}\tilde{\gamma}_{\nu}-\tilde{g}_{\mu \nu})\tilde{\psi},\label{commutr_deriv_spinor}\\
    [\tilde{\nabla}_{\mu}, \tilde{\nabla}_{\nu}]\tilde{\psi}_{\alpha}&=\frac{1}{4}\tilde{R}_{\mu \nu \kappa \lambda} \tilde{\gamma}^{\kappa}\tilde{\gamma}^{\lambda}\tilde{\psi}_{\alpha}+\tilde{R}^{\lambda}_{\hspace{1mm}\alpha \nu \mu}\tilde{\psi}_{\lambda}\\
    &=\frac{1}{2}(\tilde{\gamma}_{\mu}\tilde{\gamma}_{\nu}-\tilde{g}_{\mu \nu})\tilde{\psi}_{\alpha}+2\tilde{g}_{\alpha[\mu}\tilde{\psi}_{\nu]},\label{commutr_deriv_vecspinor}\\
     [\tilde{\nabla}_{\mu}, \tilde{\nabla}_{\nu}]\tilde{\psi}_{\alpha \beta}&=\frac{1}{2}(\tilde{\gamma}_{\mu}\tilde{\gamma}_{\nu}-\tilde{g}_{\mu \nu})\tilde{\psi}_{\alpha \beta}+2\tilde{g}_{\alpha[\mu}\tilde{\psi}_{\nu]\beta}+2\tilde{\psi}_{\alpha [\nu} \tilde{g}_{\mu]\beta}.\label{commutr_deriv_tensspinor_rank2}
\end{align}

The Laplace-Beltrami operator on $S^{N-1}$ is defined as $\tilde{\Box} \equiv \tilde{g}^{\kappa \lambda}\tilde{\nabla}_{\kappa} \tilde{\nabla}_{\lambda}$. The eigenspinors on $S^{N-1}$ [see eq.~(\ref{eigenspinors on S_(N-1)})] satisfy \cite{Camporesi}
\begin{align}\label{LB_op_eigenspin_SN-1}
    \tilde{\Box} \chi_{\pm \ell \tilde{\rho}}&=\left[  \tilde{\slashed{\nabla}}^{2}+\frac{(N-1)(N-2)}{4}  \right]\chi_{\pm \ell \tilde{\rho}} \nonumber\\
   &=\left[-  \left( \ell +\frac{N-1}{2}\right)^{2}+\frac{(N-1)(N-2)}{4}  \right]\chi_{\pm \ell \tilde{\rho}} .
\end{align}
Note also the following relations:
\begin{align}
   & \tilde{\gamma}^{\theta_{i}} \tilde{\nabla}_{( \theta_{i}}\tilde{\nabla}_{ \theta_{j})}\chi_{\pm \ell \tilde{\rho}}=\pm i \left(\ell+\frac{N-1}{2} \right)\tilde{\nabla}_{\theta_{j}}\chi_{\pm \ell \rho}+\frac{N-2}{4}\tilde{\gamma}_{\theta_{j}}\chi_{\pm \ell \tilde{\rho}} ,\label{gamma-trace_(d-d)-chi_N-1}\\
    & \tilde{\gamma}^{\theta_{i}} \tilde{\gamma}_{( \theta_{i}}\tilde{\nabla}_{ \theta_{j})}\chi_{\pm \ell \tilde{\rho}}=\frac{N+1}{2} \tilde{\nabla}_{\theta_{j}}\chi_{\pm \ell \rho}\mp i \frac{\ell+\frac{N-1}{2}}{2}\tilde{\gamma}_{\theta_{j}}\chi_{\pm \ell \tilde{\rho}}, \label{gamma-trace_(gam-d)-chi_N-1} \\
    &\tilde{\nabla}^{\theta_{i}} \tilde{\nabla}_{( \theta_{i}}\tilde{\nabla}_{ \theta_{j})}\chi_{\pm \ell \tilde{\rho}}=\tilde{\nabla}_{\theta_{j}}\left(\tilde{\Box}+N-\frac{5}{4}\right)\chi_{\pm \ell \tilde{\rho}} \mp  \frac{3}{4}i\left(\ell+\frac{N-1}{2} \right)\tilde{\gamma}_{\theta_{j}}\chi_{\pm \ell \tilde{\rho}}\label{div_(d-d)-chi_N-1} ,\\
    &\tilde{\nabla}^{\theta_{i}} \tilde{\gamma}_{( \theta_{i}}\tilde{\nabla}_{ \theta_{j})}\chi_{\pm \ell \tilde{\rho}}=\pm i \frac{\ell+\frac{N-1}{2}}{2} \tilde{\nabla}_{\theta_{j}}\chi_{\pm \ell \tilde{\rho}}+\frac{1}{2}\tilde{\gamma}_{\theta_{j}}\left(\tilde{\Box}+\frac{N-2}{2}\right)\chi_{\pm \ell \tilde{\rho}}\label{div_(gam-d)-chi_N-1},
\end{align}
where in order to prove eqs.~(\ref{gamma-trace_(d-d)-chi_N-1}) and (\ref{div_(gam-d)-chi_N-1}) we have to use eq.~(\ref{commutr_deriv_spinor}), while in order to prove eq.~(\ref{div_(d-d)-chi_N-1}) we have to use eqs.~(\ref{commutr_deriv_spinor}) and (\ref{commutr_deriv_vecspinor}).

The TT vector-spinor eigenmodes [see eqs.~(\ref{eigenvectorspinors_SN-1})-(\ref{eigenvectorspinors_SN-1_TT})] satisfy
\begin{align}\label{LB_op_eigenvecspin_SN-1}
     \tilde{\Box} \tilde{\psi}^{(\tilde{A}; \ell \tilde{\rho})}_{\pm \theta_{j}}&=\left[-  \left( \ell +\frac{N-1}{2}\right)^{2}+\frac{(N-1)(N-2)}{4} 
  +1\right]\tilde{\psi}^{(\tilde{A}; \ell \tilde{\rho})}_{\pm \theta_{j}}
\end{align}
($j=1,...,N-1$).
By combining this equation with eq.~(\ref{commutr_deriv_vecspinor}) we can prove the following relation:
\begin{align}
    \tilde{\nabla}^{\theta_{i}} \tilde{\nabla}_{(\theta_{i}}\tilde{\psi}^{(\tilde{A};\ell \tilde{\rho})}_{\pm \theta_{k})}=&~\frac{1}{2}\left(\tilde{\Box}+N-\frac{3}{2} \right)\tilde{\psi}^{(\tilde{A};\ell \tilde{\rho})}_{\pm \theta_{k}}\nonumber\\
    =&~\frac{1}{2}\left(\tilde{\slashed{\nabla}}^{2}+\frac{N(N+1)}{4} \right)\tilde{\psi}^{(\tilde{A};\ell \tilde{\rho})}_{\pm \theta_{k}} .\label{div_(d-psi)-vecspin_N-1}
\end{align}

The rank-2 STSSH's on $S^{N-1}$ [see eqs.~(\ref{eigentensorspinors_SN-1})-
(\ref{eigentensorspinors_SN-1_trace})] satisfy
\begin{align}\label{LB_op_STSSH_SN-1}
     \tilde{\Box} \tilde{\psi}^{(\tilde{B}; \ell \tilde{\rho})}_{\pm \theta_{j} \theta_{k}}&=\left[-  \left( \ell +\frac{N-1}{2}\right)^{2}+\frac{(N-1)(N-2)}{4} 
  +2\right]\tilde{\psi}^{(\tilde{B}; \ell \tilde{\rho})}_{\pm \theta_{j} \theta_{k}}
\end{align}
($j,k=1,...,N-1$).
\section{Constructing the STSSH's of rank 2 on the \texorpdfstring{${N}$}{N}-sphere}\label{Appendix_spin5/2_solving_SN}
  In this Appendix, we construct the STSSH's of rank 2 on $S^{N}$. These STSSH's satisfy eqs.~(\ref{eigenvalue_Equation_SN_5/2})-(\ref{traceless_SN_5/2}) and we construct them explicitly by using the method of separation of variables in geodesic polar coordinates~(\ref{global_coordinates}), as in Refs.~\cite{Camporesi, CHH}. In the method of separation of variables, the STSSH's of rank 2 on $S^{N}$ are expressed in terms of STSSH's of rank $\tilde{r}$ (with $\tilde{r}=0,1,2$) on $S^{N-1}$.
  
  For later convenience, note that the functions~$\phi^{(a)}_{n \ell} (\theta_{N})$ [eq.~(\ref{phi_a})] satisfy the following differential equation:
  \begin{align}\label{diff_equation_for_phi_a}
  D_{(a)} \phi^{(a)}_{n \ell}(\theta_{N}) =-\zeta^{2}_{n,N} \phi^{(a)}_{n \ell}(\theta_{N}),
  \end{align}
  where $\zeta_{n,N}^{2}\equiv \zeta^{2}=(n+\frac{N}{2})^{2}$ is the eigenvalue of the STSSH in eq.~(\ref{Dirac_eqn_fermion_SN}), while the differential operator is given by 
   \begin{align}\label{diff_operator}
      D_{(a)} =&~\frac{\partial^{2}}{\partial \theta_{N}^{2}}+(N+2a-1)\cot{\theta_{N}}\frac{\partial}{\partial \theta_{N}}+\left(\ell+\frac{N-1}{2}\right)\frac{\cos{\theta_{N}}}{\sin^{2}{\theta_{N}}} \nonumber\\
      &-\frac{(\ell+\frac{N-1}{2})^{2}-\frac{1}{4}{(N+2a-1)(N+2a-3)}}{\sin^{2}{\theta_{N}}} -\frac{(N+2a-1)^{2}}{4}.
  \end{align}
One can readily verify that the functions $\phi^{(a)}_{n \ell}(\theta_{N})$ [eq.~(\ref{phi_a})] are the unique regular solutions (up to a normalisation constant) of the differential equation~(\ref{diff_equation_for_phi_a}) by using the results of Ref.~\cite{Camporesi}, as follows. By expressing $\phi^{(a)}_{n \ell}$ as
  \begin{align}\label{phi_a_interms_phi0}
      \phi^{(a)}_{n \ell}(\theta_{N})= \left( \sin{\frac{\theta_{N}}{2}} \cos{\frac{\theta_{N}}{2}}\right)^{-a} \phi^{(0)}_{n \ell}(\theta_{N})
  \end{align}
  [see eq.~(\ref{phi_a})] we rewrite eq.~(\ref{diff_equation_for_phi_a}) as $D_{(0)} \phi^{(0)}_{n \ell} =-\zeta^{2}_{n,N} \phi^{(0)}_{n \ell}$. The latter has been solved in Ref.~\cite{Camporesi} and it was found that the unique regular solutions $\phi^{(0)}_{n \ell}$ are the ones given by eq.~(\ref{phi_a}) (with $a=0$). For the rank-$1$ STSSH's on $S^{N}$ the integer $a$ takes the values $a=-1,1$ (see Section~\ref{sectn_spin3/2_solving_SN}), while for rank-2 STSSH's $a$ takes the values $a=-2,0,2$ (see Section~\ref{sectn_spin5/2_solving_SN}). The functions $\phi^{(a)}_{n \ell}(\theta_{N})$ are regular for $a=1$ and $a=2$ despite the factor $\left( \sin{\tfrac{\theta_{N}}{2}} \cos{\tfrac{\theta_{N}}{2}}\right)^{-a}$ in eq.~(\ref{phi_a_interms_phi0}) because of the restriction $\ell \geq r$ (this restriction on $\ell$ is proved in Section~\ref{sectn_spin3/2_solving_SN} for $r=1$ and in Section~\ref{sectn_spin5/2_solving_SN} for $r=2$).
  
  The differential equation satisfied by the functions $\psi^{(a)}_{n \ell}(\theta_{N})$ [eq.~(\ref{psi_a})] is obtained from eq.~(\ref{diff_equation_for_phi_a}) by making the replacement $\theta_{N} \rightarrow \pi - \theta_{N}$ in the expression~(\ref{diff_operator}) for the differential operator $D_{(a)}$.
  
 Let us also briefly explain how to obtain the condition $n \geq \ell $ [eq.~(\ref{restriction on n-ell})]. By taking the limit $\theta_{N} \rightarrow \pi$ for $\phi^{(a)}_{n \ell}(\theta_{N})$ and using the transformation formula~(\ref{transformation_hypergeom}) for the Gauss hypergeometric function, we readily find that the requirement for absence of singularity in $\phi^{(a)}_{n \ell}(\theta_{N})$ gives rise to the condition $n \geq \ell$, as well as to the quantisation condition
  \begin{align}
     & |\zeta_{n,N}|=n+\frac{N}{2},\hspace{4mm}n \in \mathbb{N}_{0}. \label{regularity_thetaN_quantiz}
  \end{align}

\subsection{Constructing the STSSH's of rank 2 for \texorpdfstring{${N}$}{N} even}
Our aim is to obtain the STSSH's ${\psi}^{(B;\sigma;n \ell;\tilde{\rho})}_{\mu \nu}$ that satisfy eqs.~(\ref{eigenvalue_Equation_SN_5/2})-(\ref{traceless_SN_5/2}), where the gamma matrices for $N$ even are given by eq.~(\ref{even_gammas}). As in Ref.~\cite{Camporesi}, we write  ${\psi}^{(B;\sigma;n \ell;\tilde{\rho})}_{\mu \nu}$ in terms of upper and lower $2^{N/2-1}$-dimensional spinor components
\begin{align}\label{upper_lower_comp}
    {\psi}^{(B;\sigma;n \ell;\tilde{\rho})}_{\pm \mu \nu} (\theta_{N}, \bm{\theta}_{N-1})=\begin{pmatrix}  {}^{(\uparrow)}{\psi}^{(B;\sigma;n \ell;\tilde{\rho})}_{\pm \mu \nu}(\theta_{N}, \bm{\theta}_{N-1})\\ \\
  {}^{(\downarrow)}{\psi}^{(B;\sigma;n \ell;\tilde{\rho})}_{\pm \mu \nu} (\theta_{N}, \bm{\theta}_{N-1}) \end{pmatrix}.
\end{align}  
 It is clear that eqs.~(\ref{eigenvalue_Equation_SN_5/2})-(\ref{traceless_SN_5/2}) - which determine the form of our STSSH's - reduce to a system of equations for the upper and lower components. We will now obtain the system of equations for the upper and lower components.
 By using eqs.~(\ref{Christoffels_SN}), (\ref{spin_connection_components}), (\ref{even_gammas}), (\ref{covariant_deriv_tensor_spinor}), (\ref{Spin(N)_generators}), (\ref{eigenvalue_Equation_SN_5/2}) and (\ref{TT_SN_5/2}) and by expressing ${\psi}^{(B;\sigma;n \ell;\tilde{\rho})}_{\pm \mu \nu}$ in terms of the upper and lower components as in~(\ref{upper_lower_comp}), we find that the eigenvalue equation $\slashed{\nabla}{\psi}^{(B;\sigma;n \ell;\tilde{\rho})}_{\pm \theta_{N}  \theta_{N}}= \pm i |\zeta_{n,N}|{\psi}^{(B;\sigma;n \ell;\tilde{\rho})}_{\pm \theta_{N}  \theta_{N}}$ is written as
\begin{subequations}\label{Dirac_op_thetaN_thetaN_5/2_system}
     \begin{align}
    \left( \frac{\partial}{\partial \theta_{N}}+\frac{N+3}{2}\cot{\theta_{N}}    +\frac{i}{\sin{\theta_{N}}} \tilde{\slashed{\nabla}}\right){}^{(\downarrow)}{\psi}^{(B;\sigma;n \ell;\tilde{\rho})}_{\pm \theta_{N} \theta_{N}} &= \pm  i|\zeta_{n,N}|\, {}^{(\uparrow)}{\psi}^{(B;\sigma;n \ell;\tilde{\rho})}_{\pm \theta_{N} \theta_{N}},\\ 
   \left( \frac{\partial}{\partial \theta_{N}}+\frac{N+3}{2}\cot{\theta_{N}}    -\frac{i}{\sin{\theta_{N}}} \tilde{\slashed{\nabla}}\right){}^{(\uparrow)}{\psi}^{(B;\sigma;n \ell;\tilde{\rho})}_{\pm \theta_{N} \theta_{N}} &= \pm  i|\zeta_{n,N}|\, {}^{(\downarrow)}{\psi}^{(B;\sigma;n \ell;\tilde{\rho})}_{\pm \theta_{N} \theta_{N}}.
    \end{align}
\end{subequations}
Similarly, we find that the eigenvalue equation $\slashed{\nabla}{\psi}^{(B;\sigma;n \ell;\tilde{\rho})}_{\pm \theta_{N}  \theta_{j}}= \pm i |\zeta_{n,N}|{\psi}^{(B;\sigma;n \ell;\tilde{\rho})}_{\pm \theta_{N}  \theta_{j}}$ ($j=1,...,N-1$) is written as
\begin{subequations}\label{Dirac_op_thetaN_thetai_5/2_system}
     \begin{align}
    \Big( \frac{\partial}{\partial \theta_{N}}&+\frac{N-1}{2}\cot{\theta_{N}}    +\frac{i}{\sin{\theta_{N}}} \tilde{\slashed{\nabla}}\Big){}^{(\downarrow)}{\psi}^{(B;\sigma;n \ell;\tilde{\rho})}_{\pm \theta_{N} \theta_{j}}+i\cos{\theta_{N}}\tilde{\gamma}_{\theta_{j}}{}^{(\downarrow)}{\psi}^{(B;\sigma;n \ell;\tilde{\rho})}_{\pm \theta_{N} \theta_{N}}\nonumber\\
    &= \pm  i|\zeta_{n,N}|\, {}^{(\uparrow)}{\psi}^{(B;\sigma;n \ell;\tilde{\rho})}_{\pm \theta_{N} \theta_{j}},\\ 
   \Big( \frac{\partial}{\partial \theta_{N}}&+\frac{N-1}{2}\cot{\theta_{N}}    -\frac{i}{\sin{\theta_{N}}} \tilde{\slashed{\nabla}}\Big){}^{(\uparrow)}{\psi}^{(B;\sigma;n \ell;\tilde{\rho})}_{\pm \theta_{N} \theta_{j}}-i\cos{\theta_{N}}\tilde{\gamma}_{\theta_{j}}{}^{(\uparrow)}{\psi}^{(B;\sigma;n \ell;\tilde{\rho})}_{\pm \theta_{N} \theta_{N}}\nonumber\\
   &= \pm  i|\zeta_{n,N}|\, {}^{(\downarrow)}{\psi}^{(B;\sigma;n \ell;\tilde{\rho})}_{\pm \theta_{N} \theta_{j}},
    \end{align}
\end{subequations}
 while $\slashed{\nabla}{\psi}^{(B;\sigma;n \ell;\tilde{\rho})}_{\pm \theta_{j} \theta_{k}}= \pm i |\zeta_{n,N}|{\psi}^{(B;\sigma;n \ell;\tilde{\rho})}_{\pm  \theta_{j} \theta_{k}}$ ($j,k=1,...,N-1$) is written as
\begin{subequations}\label{Dirac_op_thetai_thetak_5/2_system}
     \begin{align}
    \Big( \frac{\partial}{\partial \theta_{N}}&+\frac{N-5}{2}\cot{\theta_{N}}    +\frac{i}{\sin{\theta_{N}}} \tilde{\slashed{\nabla}}\Big){}^{(\downarrow)}{\psi}^{(B;\sigma;n \ell;\tilde{\rho})}_{\pm  \theta_{j} \theta_{k}}+2i\cos{\theta_{N}}\tilde{\gamma}_{(\theta_{j}}{}^{(\downarrow)}{\psi}^{(B;\sigma;n \ell;\tilde{\rho})}_{\pm \theta_{k})\theta_{N} }\nonumber\\
    &= \pm  i|\zeta_{n,N}|\, {}^{(\uparrow)}{\psi}^{(B;\sigma;n \ell;\tilde{\rho})}_{\pm  \theta_{j} \theta_{k}},\\ 
   \Big( \frac{\partial}{\partial \theta_{N}}&+\frac{N-5}{2}\cot{\theta_{N}}    -\frac{i}{\sin{\theta_{N}}} \tilde{\slashed{\nabla}}\Big){}^{(\uparrow)}{\psi}^{(B;\sigma;n \ell;\tilde{\rho})}_{\pm  \theta_{j} \theta_{k}}-2i\cos{\theta_{N}} \tilde{\gamma}_{(\theta_{j}}{}^{(\uparrow)}{\psi}^{(B;\sigma;n \ell;\tilde{\rho})}_{\pm \theta_{k}) \theta_{N}} \nonumber\\
   &= \pm  i|\zeta_{n,N}|\, {}^{(\downarrow)}{\psi}^{(B;\sigma;n \ell;\tilde{\rho})}_{\pm  \theta_{j} \theta_{k}}.
    \end{align}
\end{subequations}
 By making use of eq.~(\ref{upper_lower_comp}), we express the gamma-tracelessness condition~(\ref{TT_SN_5/2}) as
 \begin{align}\label{gamma_traceless_5/2_upper_lower}
 \begin{cases} 
  &{}^{(\downarrow)}{\psi}^{(B;\sigma;n \ell;\tilde{\rho})}_{\pm  \theta_{N} \mu}+\dfrac{i}{\sin{\theta_{N}}}\tilde{\gamma}^{\theta_{i}}\,\,{}^{(\downarrow)}{\psi}^{(B;\sigma;n \ell;\tilde{\rho})}_{\pm  \theta_{i} \mu}=0, \\ \\
  &{}^{(\uparrow)}{\psi}^{(B;\sigma;n \ell;\tilde{\rho})}_{\pm  \theta_{N} \mu}-\dfrac{i}{\sin{\theta_{N}}}\tilde{\gamma}^{\theta_{i}}\,\,{}^{(\uparrow)}{\psi}^{(B;\sigma;n \ell;\tilde{\rho})}_{\pm  \theta_{i} \mu}=0,\hspace{8mm}(\mu=\theta_{1},...,\theta_{N}\hspace{3mm}\text{and}\hspace{2mm} \theta_{i}=\theta_{1},...,\theta_{N-1})
 \end{cases}
 \end{align}
 and the tracelessness condition~(\ref{traceless_SN_5/2}) as
 \begin{align}\label{traceless_5/2_upper_lower}
 \begin{cases} 
  &{}^{(\downarrow)}{\psi}^{(B;\sigma;n \ell;\tilde{\rho})}_{\pm  \theta_{N} \theta_{N}}+\dfrac{1}{\sin^{2}{\theta_{N}}}\tilde{g}^{\theta_{i} \theta_{j}}\,\,{}^{(\downarrow)}{\psi}^{(B;\sigma;n \ell;\tilde{\rho})}_{\pm  \theta_{i} \theta_{j}}=0, \\ \\
  &{}^{(\uparrow)}{\psi}^{(B;\sigma;n \ell;\tilde{\rho})}_{\pm  \theta_{N} \theta_{N}}+\dfrac{1}{\sin^{2}{\theta_{N}} }\tilde{g}^{\theta_{i} \theta_{j}}\,\,{}^{(\uparrow)}{\psi}^{(B;\sigma;n \ell;\tilde{\rho})}_{\pm \theta_{i}\theta_{j}}=0.
 \end{cases}
 \end{align}
 Similarly, by substituting eq.~(\ref{upper_lower_comp}) into the divergence-free condition~(\ref{TT_SN_5/2}), we may express the condition $\nabla^{\alpha} {\psi}^{(B;\sigma;n \ell;\tilde{\rho})}_{\pm  \alpha \theta_{N}}=0$ as 
 \begin{align}\label{transeverse_thetaN_5/2_upper_lower}
     \begin{cases}  
    &\left[\frac{\partial}{\partial \theta_{N}}+(N+\frac{1}{2})\cot{\theta_{N}} \right] {}^{(\uparrow)}{\psi}^{(B;\sigma;n \ell;\tilde{\rho})}_{\pm  \theta_{N} \theta_{N}}+\dfrac{1}{\sin^{2}{\theta_{N}}}\tilde{\nabla}^{\theta_{i}}\,{}^{(\uparrow)}{\psi}^{(B;\sigma;n \ell;\tilde{\rho})}_{\pm  \theta_{i} \theta_{N}} =0,\\ \\
     &\left[\frac{\partial}{\partial \theta_{N}}+(N+\frac{1}{2})\cot{\theta_{N}} \right] {}^{(\downarrow)}{\psi}^{(B;\sigma;n \ell;\tilde{\rho})}_{\pm  \theta_{N} \theta_{N}}+\dfrac{1}{\sin^{2}{\theta_{N}}}\tilde{\nabla}^{\theta_{i}}\,{}^{(\downarrow)}{\psi}^{(B;\sigma;n \ell;\tilde{\rho})}_{\pm  \theta_{i} \theta_{N}}=0,\end{cases}
 \end{align}
while the condition $\nabla^{\alpha} {\psi}^{(B;\sigma;n \ell;\tilde{\rho})}_{\pm  \alpha \theta_{j}}=0$ ($j=1,...,N-1$) is expressed as 
  \begin{align}\label{transeverse_thetaj_5/2_upper_lower}
     \begin{cases}  
    &\left[\frac{\partial}{\partial \theta_{N}}+(N-\frac{1}{2})\cot{\theta_{N}} \right] {}^{(\uparrow)}{\psi}^{(B;\sigma;n \ell;\tilde{\rho})}_{\pm  \theta_{N} \theta_{j}}+\dfrac{1}{\sin^{2}{\theta_{N}}}\tilde{\nabla}^{\theta_{i}}\,{}^{(\uparrow)}{\psi}^{(B;\sigma;n \ell;\tilde{\rho})}_{\pm  \theta_{i} \theta_{j}} =0,\\ \\
     &\left[\frac{\partial}{\partial \theta_{N}}+(N-\frac{1}{2})\cot{\theta_{N}} \right] {}^{(\downarrow)}{\psi}^{(B;\sigma;n \ell;\tilde{\rho})}_{\pm  \theta_{N} \theta_{j}}+\dfrac{1}{\sin^{2}{\theta_{N}}}\tilde{\nabla}^{\theta_{i}}\,{}^{(\downarrow)}{\psi}^{(B;\sigma;n \ell;\tilde{\rho})}_{\pm  \theta_{i} \theta_{j}}=0.\end{cases}
 \end{align}
 
\noindent \textbf{Type-}$\bm{I}$ \textbf{STSSH's of rank 2 for} $\bm{N}$ \textbf{even.} Let us start by describing how to obtain the type-$I$ modes, given by eqs.~(\ref{TYPE_I_thetaNthetaN_negative_spin_5/2})-(\ref{TYPE_I_thetajthetak_negative_spin_5/2}). The component ${\psi}^{(I;\sigma;n \ell;\tilde{\rho})}_{\pm \theta_{N} \theta_{N}}$ is a spinor on $S^{N-1}$. Thus, in order to solve the system of equations~(\ref{Dirac_op_thetaN_thetaN_5/2_system}) we separate variables as in the case of spinor eigenmodes in Ref.~\cite{Camporesi}, i.e.
\begin{align}
  &  {}^{(\uparrow)}{\psi}^{(I;-;n \ell;\tilde{\rho})}_{\pm \theta_{N} \theta_{N}}(\theta_{N}, \bm{\theta}_{N-1})= \phi^{(2)}_{ n \ell}(\theta_{N}) \,\chi_{- \ell \tilde{\rho}}(\bm{\theta}_{N-1}),\nonumber\\ &{}^{(\downarrow)}{\psi}^{(I;-;n \ell;\tilde{\rho})}_{\pm \theta_{N} \theta_{N}}(\theta_{N}, \bm{\theta}_{N-1})= \pm i\psi^{(2)}_{ n \ell}(\theta_{N}) \,\chi_{- \ell \tilde{\rho}}(\bm{\theta}_{N-1})\label{separ_type_I_thetaNthetaN_-}\\
  &   {}^{(\uparrow)}{\psi}^{(I;+;n \ell;\tilde{\rho})}_{\pm \theta_{N} \theta_{N}}(\theta_{N}, \bm{\theta}_{N-1})= i\psi^{(2)}_{ n \ell}(\theta_{N}) \,\chi_{+ \ell \tilde{\rho}}(\bm{\theta}_{N-1}),\nonumber\\
  &{}^{(\downarrow)}{\psi}^{(I;+;n \ell;\tilde{\rho})}_{\pm \theta_{N} \theta_{N}}(\theta_{N}, \bm{\theta}_{N-1})= \pm \phi^{(2)}_{n \ell}(\theta_{N}) \,\chi_{+ \ell \tilde{\rho}}(\bm{\theta}_{N-1}),\label{separ_type_I_thetaNthetaN_+}
\end{align}
where $\chi_{\pm \ell \tilde{\rho}}$ are the eigenspinors on $S^{N-1}$ (see eq.~(\ref{eigenspinors on S_(N-1)})). By substituting eq.~(\ref{separ_type_I_thetaNthetaN_-}) [or eq.~(\ref{separ_type_I_thetaNthetaN_+})] into the system of equations~(\ref{Dirac_op_thetaN_thetaN_5/2_system}) and eliminating ${}^{(\downarrow)}{\psi}^{(I;-;n \ell;\tilde{\rho})}_{\pm \theta_{N} \theta_{N}}$ (or ${}^{(\downarrow)}{\psi}^{(I;+;n \ell;\tilde{\rho})}_{\pm \theta_{N} \theta_{N}}$) we find that $\phi^{(2)}_{n \ell}$ has to satisfy the differential equation~(\ref{diff_equation_for_phi_a}) (with $a=2$), while $\psi^{(2)}_{n \ell}$ has to satisfy the differential equation~(\ref{diff_equation_for_phi_a}) ($a=2$) with $\theta_{N}$ replaced by $\pi-\theta_{N}$ in the differential operator $D_{(2)}$ [eq.~(\ref{diff_operator})].
Thus, we find that $\phi^{(2)}_{n \ell}$ and $\psi^{(2)}_{n \ell}$ are given by eqs.~(\ref{phi_a}) and (\ref{psi_a}), respectively.
As a check, one readily finds that the components defined by eqs.~(\ref{separ_type_I_thetaNthetaN_-}) and (\ref{separ_type_I_thetaNthetaN_+}) satisfy the system of equations~(\ref{Dirac_op_thetaN_thetaN_5/2_system}) by making use of the formulae~(\ref{psi_to_phi_sphere}) and (\ref{phi_to_psi_sphere}).

The components ${\psi}^{(I;\sigma;n \ell;\tilde{\rho})}_{\pm \theta_{N} \theta_{j}}$ ($j=1,...,N-1$) are vector-spinors on $S^{N-1}$ and thus we may separate variables analogously to eqs.~(\ref{TYPE_I_theta_i_negative_spin_3/2}) and (\ref{TYPE_I_theta_i_positive_spin_3/2}). Thus, for STSSH's with negative spin projection ($\sigma=-$) we separate variables as 
 \begin{align}
  {}^{(\uparrow)}{\psi}^{(I;-;n \ell;\tilde{\rho})}_{\pm \theta_{N} \theta_{j}}(\theta_{N}, \bm{\theta}_{N-1})=&  \,  C_{ n \ell}^{(\uparrow)(2)}(\theta_{N}) \,\tilde{\nabla}_{\theta_{j}}\chi_{- \ell \tilde{\rho}}(\bm{\theta}_{N-1})+ D_{ n \ell}^{(\uparrow)(2)}(\theta_{N}) \,\tilde{\gamma}_{\theta_{j}}\chi_{- \ell \tilde{\rho}}(\bm{\theta}_{N-1}), \nonumber \\
   {}^{(\downarrow)}{\psi}^{(I;-;n \ell;\tilde{\rho})}_{\pm \theta_{N} \theta_{j}}(\theta_{N}, \bm{\theta}_{N-1})= &\pm i   C_{ n \ell}^{(\downarrow)(2)}(\theta_{N}) \,\tilde{\nabla}_{\theta_{j}}\chi_{- \ell \tilde{\rho}}(\bm{\theta}_{N-1})\pm i D_{ n \ell}^{(\downarrow)(2)}(\theta_{N}) \,\tilde{\gamma}_{\theta_{j}}\chi_{- \ell \tilde{\rho}}(\bm{\theta}_{N-1})\label{separ_type_I_thetaNthetaj_-},
 \end{align}
 while for STSSH's with positive spin projection ($\sigma=+$) we separate variables as
  \begin{align}
  {}^{(\uparrow)}{\psi}^{(I;+;n \ell;\tilde{\rho})}_{\pm \theta_{N} \theta_{j}}(\theta_{N}, \bm{\theta}_{N-1})=&  \, i C_{ n \ell}^{(\downarrow)(2)}(\theta_{N}) \,\tilde{\nabla}_{\theta_{j}}\chi_{+ \ell \tilde{\rho}}(\bm{\theta}_{N-1})- iD_{ n \ell}^{(\downarrow)(2)}(\theta_{N}) \,\tilde{\gamma}_{\theta_{j}}\chi_{+ \ell \tilde{\rho}}(\bm{\theta}_{N-1}), \nonumber \\
   {}^{(\downarrow)}{\psi}^{(I;+;n \ell;\tilde{\rho})}_{\pm \theta_{N} \theta_{j}}(\theta_{N}, \bm{\theta}_{N-1})= &\pm   C_{ n \ell}^{(\uparrow)(2)}(\theta_{N}) \,\tilde{\nabla}_{\theta_{j}}\chi_{+ \ell \tilde{\rho}}(\bm{\theta}_{N-1})\mp  D_{ n \ell}^{(\uparrow)(2)}(\theta_{N}) \,\tilde{\gamma}_{\theta_{j}}\chi_{+ \ell \tilde{\rho}}(\bm{\theta}_{N-1})\label{separ_type_I_thetaNthetaj_+}.
 \end{align}
 By using the gamma-tracelessness condition~(\ref{gamma_traceless_5/2_upper_lower}) we readily find that the functions
 $D_{ n \ell}^{(b)(2)}$ and $C_{ n \ell}^{(b)(2)}$ ($b=\uparrow,\downarrow$) are related to each other by eqs.~(\ref{D1_a_function}) and (\ref{D2_a_function}). Then, using the divergence-free condition~(\ref{transeverse_thetaN_5/2_upper_lower}), we find that
 $C_{ n \ell}^{(\uparrow)(2)}$ is given by eq.~(\ref{C1_a_function}) and $C_{ n \ell}^{(\downarrow)(2)}$ is given by eq.~(\ref{C2_a_function}), where we also have used eqs.~(\ref{psi_to_phi_sphere}), (\ref{phi_to_psi_sphere}) and eq.~(\ref{LB_op_eigenspin_SN-1}). One can straightforwardly verify that the components defined by eqs.~(\ref{separ_type_I_thetaNthetaj_-}) and (\ref{separ_type_I_thetaNthetaj_+}) are solutions of the system of equations~(\ref{Dirac_op_thetaN_thetai_5/2_system}), where the calculations are significantly simplified by using the following formulae:
 \begin{align}
    \Bigg( \frac{\partial}{\partial{\theta_{N}}}&+\frac{N-1}{2}\cot{\theta_{N}} -\frac{\ell+\frac{N-1}{2}}{\sin{\theta_{N}}} \Bigg)C^{(\uparrow)(2)}_{n \ell}(\theta_{N})-
    \frac{2i}{\sin{\theta_{N}}}D^{(\uparrow)(2)}_{n \ell}(\theta_{N})\nonumber\\
    &=-\left(n+\frac{N}{2}\right)~C^{(\downarrow)(2)}_{n \ell}(\theta_{N}), \label{Cup_to_Cdown_a=2}\\
   \Bigg( \frac{\partial}{\partial{\theta_{N}}}&+\frac{N-1}{2}\cot{\theta_{N}} +\frac{\ell+\frac{N-1}{2}}{\sin{\theta_{N}}} \Bigg)C^{(\downarrow)(2)}_{n \ell}(\theta_{N})+
    \frac{2i}{\sin{\theta_{N}}}D^{(\downarrow)(2)}_{n \ell}(\theta_{N})\nonumber\\
    &=\left(n+\frac{N}{2}\right)~C^{(\uparrow)(2)}_{n \ell} (\theta_{N})\label{Cdown_to_Cup_a=2},
\end{align}
which can be proved by using the formulae~(\ref{psi_to_phi_sphere}) and (\ref{phi_to_psi_sphere}).
 
 The components ${\psi}^{(I;\sigma;n \ell;\tilde{\rho})}_{\pm \theta_{j} \theta_{k}}$ ($j,k=1,...,N-1$) are rank-2 symmetric tensor-spinors on $S^{N-1}$. Let us first discuss the case with negative spin projection ($\sigma=-$). We choose to separate variables for ${\psi}^{(I;-;n \ell;\tilde{\rho})}_{\pm \theta_{j} \theta_{k}}$ as follows:
  \begin{align}
  {}^{(\uparrow)}{\psi}^{(I;-;n \ell;\tilde{\rho})}_{\pm \theta_{j} \theta_{k}}(\theta_{N}, \bm{\theta}_{N-1})=&  K^{(\uparrow)}_{n \ell}(\theta_{N})\,\tilde{g}_{\theta_{j} \theta_{k}}\chi_{-\ell \tilde{\rho}}(\bm{\theta}_{N-1})\nonumber\\
  &+W^{(\uparrow)}_{n \ell}(\theta_{N}) \left( \tilde{\nabla}_{(\theta_{j}}\tilde{\nabla}_{\theta_{k})}-\tilde{g}_{\theta_{j} \theta_{k}}\frac{\tilde{\Box}}{N-1}\right)\chi_{-\ell \tilde{\rho}}(\bm{\theta}_{N-1})\nonumber \\
 &+T^{(\uparrow)}_{n \ell}(\theta_{N}) \left( \tilde{\gamma}_{(\theta_{j}}\tilde{\nabla}_{\theta_{k})}-\tilde{g}_{\theta_{j} \theta_{k}}\frac{\tilde{\slashed{\nabla}}}{N-1}\right)\chi_{-\ell \tilde{\rho}}(\bm{\theta}_{N-1}), \nonumber\\
     {}^{(\downarrow)}{\psi}^{(I;-;n \ell;\tilde{\rho})}_{\pm \theta_{j} \theta_{k}}(\theta_{N}, \bm{\theta}_{N-1})=&  \pm i K^{(\downarrow)}_{n \ell}(\theta_{N})\,\tilde{g}_{\theta_{j} \theta_{k}}\chi_{-\ell \tilde{\rho}}(\bm{\theta}_{N-1})\nonumber\\
     &\pm iW^{(\downarrow)}_{n \ell}(\theta_{N}) \left( \tilde{\nabla}_{(\theta_{j}}\tilde{\nabla}_{\theta_{k})}-\tilde{g}_{\theta_{j} \theta_{k}}\frac{\tilde{\Box}}{N-1}\right)\chi_{-\ell \tilde{\rho}}(\bm{\theta}_{N-1})\nonumber \\
 &\pm i T^{(\downarrow)}_{n \ell}(\theta_{N}) \left( \tilde{\gamma}_{(\theta_{j}}\tilde{\nabla}_{\theta_{k})}-\tilde{g}_{\theta_{j} \theta_{k}}\frac{\tilde{\slashed{\nabla}}}{N-1}\right)\chi_{-\ell \tilde{\rho}}(\bm{\theta}_{N-1})\label{separ_type_I_thetajthetak_-},
 \end{align}
 where $\tilde{\slashed{\nabla}}\chi_{-\ell \tilde{\rho}}=-i\left(\ell+\frac{N-1}{2}\right)\chi_{-\ell \tilde{\rho}}$ (see eq.~(\ref{eigenspinors on S_(N-1)})) and $\tilde{\Box}\chi_{-\ell \tilde{\rho}}\equiv \nabla^{\theta_{k}}\nabla_{\theta_{k}}\chi_{-\ell \tilde{\rho}}$ is given by eq.~(\ref{LB_op_eigenspin_SN-1}).
 By using the tracelessness condition~(\ref{traceless_5/2_upper_lower}), we find that $K^{(\uparrow)}_{n \ell}$ and $K^{(\downarrow)}_{n \ell}$ are given by eqs.~(\ref{Kup_function}) and (\ref{Kdown_function}), respectively.
Then, by using the gamma-tracelessness condition~(\ref{gamma_traceless_5/2_upper_lower}) (and by making use of eqs.~(\ref{gamma-trace_(d-d)-chi_N-1}) and (\ref{gamma-trace_(gam-d)-chi_N-1})) we find that the function $T^{(\uparrow)}_{n \ell}$ ($T^{(\downarrow)}_{n \ell}$) is expressed in terms of $W^{(\uparrow)}_{n \ell}$ ($W^{(\downarrow)}_{n \ell}$) as in eq.~(\ref{Tup_function}) (eq.~(\ref{Tdown_function})). Then, by making use of the divergence-free condition~(\ref{transeverse_thetaj_5/2_upper_lower}) (and using eqs.~(\ref{div_(d-d)-chi_N-1}) and (\ref{div_(gam-d)-chi_N-1})) we find
\begin{align}
    \Bigg( \frac{\partial}{\partial{\theta_{N}}}&+(N-\frac{1}{2})\cot{\theta_{N}}  \Bigg)C^{(b)(2)}_{n \ell}(\theta_{N})+\frac{1}{\sin^{2}{\theta_{N}}}K^{(b)}_{n\ell}(\theta_{N}) \nonumber\\
    &+\frac{1}{\sin^{2}{\theta_{N}}} W^{(b)}_{n \ell}(\theta_{N})\left\{ -\frac{\left(\ell+\frac{N-1}{2}\right)^{2}(N-2)}{N-1}+\frac{N^{2}-1}{4} \right\} \nonumber\\
    &-i\frac{1}{2 \sin^{2}{\theta_{N}}}\frac{\left(\ell+\frac{N-1}{2} \right)(N-3)}{N-1}T^{(b)}_{n \ell}(\theta_{N})=0, \hspace{8mm}b=\uparrow, \downarrow\label{T_W_second_relation}.
\end{align}
Finally, by solving the system of equations consisting of eqs.~(\ref{Tup_function}), (\ref{Tdown_function}) and (\ref{T_W_second_relation}) (and using eqs.~(\ref{Cup_to_Cdown_a=2}) and (\ref{Cdown_to_Cup_a=2})) we find
that $W^{(\uparrow)}_{n \ell}$ is given by eq.~(\ref{Wup_function}), while $W^{(\downarrow)}_{n \ell}$ is given by eq.~(\ref{Wdown_function}). 

    By working as in the case with negative spin projection, we find that the components ${\psi}^{(I;+;n \ell;\tilde{\rho})}_{\pm \theta_{j} \theta_{k}}$ with positive spin projection are expressed in terms of upper and lower spinorial components as follows:
    \begin{align}
  {}^{(\uparrow)}{\psi}^{(I;+;n \ell;\tilde{\rho})}_{\pm \theta_{j} \theta_{k}}(\theta_{N}, \bm{\theta}_{N-1})=&  iK^{(\downarrow)}_{n \ell}(\theta_{N})\,\tilde{g}_{\theta_{j} \theta_{k}}\chi_{+\ell \tilde{\rho}}(\bm{\theta}_{N-1})\nonumber\\
  &+iW^{(\downarrow)}_{n \ell}(\theta_{N}) \left( \tilde{\nabla}_{(\theta_{j}}\tilde{\nabla}_{\theta_{k})}-\tilde{g}_{\theta_{j} \theta_{k}}\frac{\tilde{\Box}}{N-1}\right)\chi_{+\ell \tilde{\rho}}(\bm{\theta}_{N-1})\nonumber \\
 &-iT^{(\downarrow)}_{n \ell}(\theta_{N}) \left( \tilde{\gamma}_{(\theta_{j}}\tilde{\nabla}_{\theta_{k})}-\tilde{g}_{\theta_{j} \theta_{k}}\frac{\tilde{\slashed{\nabla}}}{N-1}\right)\chi_{+\ell \tilde{\rho}}(\bm{\theta}_{N-1}), \nonumber\\
     {}^{(\downarrow)}{\psi}^{(I;+;n \ell;\tilde{\rho})}_{\pm \theta_{j} \theta_{k}}(\theta_{N}, \bm{\theta}_{N-1})=&  \pm  K^{(\uparrow)}_{n \ell}(\theta_{N})\,\tilde{g}_{\theta_{j} \theta_{k}}\chi_{+\ell \tilde{\rho}}(\bm{\theta}_{N-1})\nonumber\\
     &\pm W^{(\uparrow)}_{n \ell}(\theta_{N}) \left( \tilde{\nabla}_{(\theta_{j}}\tilde{\nabla}_{\theta_{k})}-\tilde{g}_{\theta_{j} \theta_{k}}\frac{\tilde{\Box}}{N-1}\right)\chi_{+\ell \tilde{\rho}}(\bm{\theta}_{N-1})\nonumber \\
 &\mp  T^{(\uparrow)}_{n \ell}(\theta_{N}) \left( \tilde{\gamma}_{(\theta_{j}}\tilde{\nabla}_{\theta_{k})}-\tilde{g}_{\theta_{j} \theta_{k}}\frac{\tilde{\slashed{\nabla}}}{N-1}\right)\chi_{+\ell \tilde{\rho}}(\bm{\theta}_{N-1})\label{separ_type_I_thetajthetak_+}.
 \end{align}
We have verified using Mathematica 11.2 that the components defined by eqs.~(\ref{separ_type_I_thetajthetak_-}) and (\ref{separ_type_I_thetajthetak_+}) satisfy the system of equations~(\ref{Dirac_op_thetai_thetak_5/2_system}).
 
 \noindent \textbf{Type-}$\bm{I\!I}$ \textbf{STSSH's of rank 2 for} $\bm{N}$ \textbf{even.} Now let us describe how to obtain the type-$I\!I$ modes given by eqs.~(\ref{TYPE_II_thetaNthetj_negative_spin_5/2}) and (\ref{TYPE_II_thetajthetk_negative_spin_5/2}). The type-$I\!I$ modes satisfy ${\psi}^{(I\!I \text{-}\tilde{A};\sigma;n \ell;\tilde{\rho})}_{\pm \theta_{N} \theta_{N}}=0$ by definition. The components ${\psi}^{(I\!I \text{-}\tilde{A};\sigma;n \ell;\tilde{\rho})}_{\pm \theta_{N} \theta_{j}}$ ($j=1,...,N-1$) may be expressed as
 \begin{align}
  &  {}^{(\uparrow)}{\psi}^{(I\!I \text{-}\tilde{A};-;n \ell;\tilde{\rho})}_{\pm \theta_{N} \theta_{j}}(\theta_{N}, \bm{\theta}_{N-1})= \phi^{(0)}_{ n \ell}(\theta_{N}) \,\tilde{\psi}^{(\tilde{A}; \ell \tilde{\rho})}_{- \theta_{j}}(\bm{\theta}_{N-1}), \nonumber\\
  &{}^{(\downarrow)}{\psi}^{(I\!I \text{-}\tilde{A};-;n \ell;\tilde{\rho})}_{\pm \theta_{N} \theta_{j}}(\theta_{N}, \bm{\theta}_{N-1})= \pm i\psi^{(0)}_{ n \ell}(\theta_{N}) \,\tilde{\psi}^{(\tilde{A}; \ell \tilde{\rho})}_{- \theta_{j}}(\bm{\theta}_{N-1}),\label{separ_type_II_thetaNthetaj_-}\\
  &   {}^{(\uparrow)}{\psi}^{(I\!I \text{-}\tilde{A};+;n \ell;\tilde{\rho})}_{\pm \theta_{N} \theta_{j}}(\theta_{N}, \bm{\theta}_{N-1})= i\psi^{(0)}_{ n \ell}(\theta_{N}) \,\tilde{\psi}^{(\tilde{A}; \ell \tilde{\rho})}_{+ \theta_{j}}(\bm{\theta}_{N-1}) \nonumber \\
  &{}^{(\downarrow)}{\psi}^{(I\!I \text{-}\tilde{A};+;n \ell;\tilde{\rho})}_{\pm \theta_{N} \theta_{j}}(\theta_{N}, \bm{\theta}_{N-1})= \pm \phi^{(0)}_{n \ell}(\theta_{N}) \,\tilde{\psi}^{(\tilde{A}; \ell \tilde{\rho})}_{+ \theta_{j}}(\bm{\theta}_{N-1})\label{separ_type_II_thetaNthetaj_+}.
\end{align}
The TT eigenvector-spinors $\tilde{\psi}^{(\tilde{A}; \ell \tilde{\rho})}_{\pm \theta_{j}}$ ($j=1,...,N-1$) on $S^{N-1}$ satisfy eqs.~(\ref{eigenvectorspinors_SN-1}) and (\ref{eigenvectorspinors_SN-1_TT}).  By working as in the case of type-$I$ modes presented above, we find that $\phi^{(0)}_{n \ell}$ has to satisfy the differential equation~(\ref{diff_equation_for_phi_a}) with $a=0$, while $\psi^{(0)}_{n \ell}$ has to satisfy the differential equation~(\ref{diff_equation_for_phi_a}) ($a=0$) with $\theta_{N}$ replaced by $\pi-\theta_{N}$ in the differential operator $D_{(0)}$ [eq.~(\ref{diff_operator})].
Thus, we find that $\phi^{(0)}_{n \ell}$ and $\psi^{(0)}_{n \ell}$ are given by eqs.~(\ref{phi_a}) and (\ref{psi_a}), respectively.
By making use of the formulae~(\ref{psi_to_phi_sphere}) and (\ref{phi_to_psi_sphere}), one can readily verify that the components defined by eqs.~(\ref{separ_type_II_thetaNthetaj_-}) and (\ref{separ_type_II_thetaNthetaj_+}) are solutions of the system of equations~(\ref{Dirac_op_thetaN_thetai_5/2_system}). 

The components ${\psi}^{(I\!I \text{-}\tilde{A};\sigma;n \ell;\tilde{\rho})}_{\pm \theta_{j} \theta_{k}}$ ($j,k=1,...,N-1$) are symmetric rank-2 tensor-spinors on $S^{N-1}$. Let us first discuss the case with negative spin projection ($\sigma=-$). We separate variables as
 \begin{align}
  {}^{(\uparrow)}{\psi}^{(I\!I \text{-}\tilde{A};-;n \ell;\tilde{\rho})}_{\pm \theta_{j} \theta_{k}}(\theta_{N}, \bm{\theta}_{N-1})=&  \,  \Gamma_{ n \ell}^{(\uparrow)}(\theta_{N}) \,\tilde{\nabla}_{(\theta_{j}}\tilde{\psi}^{(\tilde{A}; \ell \tilde{\rho})}_{- \theta_{k})}(\bm{\theta}_{N-1})+ \Delta_{ n \ell}^{(\uparrow)}(\theta_{N}) \,\tilde{\gamma}_{(\theta_{j}}\tilde{\psi}^{(\tilde{A}; \ell \tilde{\rho})}_{- \theta_{k})}(\bm{\theta}_{N-1}), \nonumber \\
   {}^{(\downarrow)}{\psi}^{(I\!I \text{-}\tilde{A};-;n \ell;\tilde{\rho})}_{\pm \theta_{j} \theta_{k}}(\theta_{N}, \bm{\theta}_{N-1})= &\pm i   \Gamma_{ n \ell}^{(\downarrow)}(\theta_{N}) \,\tilde{\nabla}_{(\theta_{j}}\tilde{\psi}^{(\tilde{A}; \ell \tilde{\rho})}_{- \theta_{k})}(\bm{\theta}_{N-1})\pm i \Delta_{ n \ell}^{(\downarrow)}(\theta_{N}) \,\tilde{\gamma}_{(\theta_{j}}\tilde{\psi}^{(\tilde{A}; \ell \tilde{\rho})}_{- \theta_{k})}(\bm{\theta}_{N-1})\label{separ_type_II_thetajthetak_-},
 \end{align}
 where we have to determine the functions $\Gamma^{(b)}_{n \ell}$ and $\Delta^{(b)}_{n \ell}$ (with $b=\uparrow, \downarrow$). By using the TT conditions as in the case of type-$I$ modes, we find that $\Delta^{(\uparrow)}_{n \ell}$ and $\Delta^{(\downarrow)}_{n \ell}$ are given by eqs.~(\ref{Deltaup_function}) and (\ref{Deltadown_function}), respectively, while $\Gamma^{(\uparrow)}_{n \ell}$ and $\Gamma^{(\downarrow)}_{n \ell}$ are given by eqs.~(\ref{Gammaup_function}) and (\ref{Gammadown_function}), respectively,
where we also have used eqs.~(\ref{psi_to_phi_sphere}), (\ref{phi_to_psi_sphere}) and (\ref{div_(d-psi)-vecspin_N-1}). By using the formulae~(\ref{psi_to_phi_sphere}) and (\ref{phi_to_psi_sphere}), we can also prove the following formulae:
\begin{align}
   & \left( \frac{\partial}{\partial{\theta_{N}}}+\frac{N-5}{2}\cot{\theta_{N}} -\frac{\ell+\frac{N-1}{2}}{\sin{\theta_{N}}} \right)\Gamma^{(\uparrow)}_{n \ell}(\theta_{N})-
    \frac{2i}{\sin{\theta_{N}}}\Delta^{(\uparrow)}_{n \ell}(\theta_{N})=-(n+\frac{N}{2})\Gamma^{(\downarrow)}_{n \ell}(\theta_{N}), \label{Gammaup_to_Gammadown}\\
  & \left( \frac{\partial}{\partial{\theta_{N}}}+\frac{N-5}{2}\cot{\theta_{N}} +\frac{\ell+\frac{N-1}{2}}{\sin{\theta_{N}}} \right)\Gamma^{(\downarrow)}_{n \ell}(\theta_{N})+
    \frac{2i}{\sin{\theta_{N}}}\Delta^{(\downarrow)}_{n \ell}(\theta_{N})=(n+\frac{N}{2})\Gamma^{(\uparrow)}_{n \ell} (\theta_{N})\label{Gammadown_to_Gammaup}.
\end{align}
Similarly, we find that the upper and lower components of ${\psi}^{(I\!I \text{-}\tilde{A};+;n \ell;\tilde{\rho})}_{\pm \theta_{j} \theta_{k}}$ ($j,k=1,...,N-1$) are given by
 \begin{align}
  {}^{(\uparrow)}{\psi}^{(I\!I \text{-}\tilde{A};+;n \ell;\tilde{\rho})}_{\pm \theta_{j} \theta_{k}}(\theta_{N}, \bm{\theta}_{N-1})=&  \, i \Gamma_{ n \ell}^{(\downarrow)}(\theta_{N}) \,\tilde{\nabla}_{(\theta_{j}}\tilde{\psi}^{(\tilde{A}; \ell \tilde{\rho})}_{+ \theta_{k})}(\bm{\theta}_{N-1})-i \Delta_{ n \ell}^{(\downarrow)}(\theta_{N}) \,\tilde{\gamma}_{(\theta_{j}}\tilde{\psi}^{(\tilde{A}; \ell \tilde{\rho})}_{+ \theta_{k})}(\bm{\theta}_{N-1}),\nonumber \\
   {}^{(\downarrow)}{\psi}^{(I\!I \text{-}\tilde{A};+;n \ell;\tilde{\rho})}_{\pm \theta_{j} \theta_{k}}(\theta_{N}, \bm{\theta}_{N-1})= &\pm    \Gamma_{ n \ell}^{(\uparrow)}(\theta_{N}) \,\tilde{\nabla}_{(\theta_{j}}\tilde{\psi}^{(\tilde{A}; \ell \tilde{\rho})}_{+ \theta_{k})}(\bm{\theta}_{N-1})\mp  \Delta_{ n \ell}^{(\uparrow)}(\theta_{N}) \,\tilde{\gamma}_{(\theta_{j}}\tilde{\psi}^{(\tilde{A}; \ell \tilde{\rho})}_{+ \theta_{k})}(\bm{\theta}_{N-1})\label{separ_type_II_thetajthetak_+}.
 \end{align}
By making use of the formulae (\ref{Gammaup_to_Gammadown}) and (\ref{Gammadown_to_Gammaup}), as well as eq.~(\ref{commutr_deriv_vecspinor}), one can readily verify that the system of equations~(\ref{Dirac_op_thetai_thetak_5/2_system}) is satisfied by the type-$I\!I$ modes in eqs.~(\ref{separ_type_II_thetajthetak_-}) and (\ref{separ_type_II_thetajthetak_+}).

\noindent \textbf{Type-}$\bm{I\!I\!I}$ \textbf{STSSH's of rank 2 for} $\bm{N}$ \textbf{even.} Finally, let us construct the type-$I\!I\!I$ modes, given by eqs.~(\ref{TYPE_III_thetajthetak_negative_spin_5/2}) and (\ref{TYPE_III_thetajthetak_positive_spin_5/2}). The type-$I\!I\!I$ modes satisfy ${\psi}^{(I\!I\!I \text{-}\tilde{B};\sigma;n \ell;\tilde{\rho})}_{\pm \theta_{N} \theta_{N}}=0$ and ${\psi}^{(I\!I\!I \text{-}\tilde{B};\sigma;n \ell;\tilde{\rho})}_{\pm \theta_{N} \theta_{i}}=0$ ($i=1,...,N-1$) by definition. The components ${\psi}^{(I\!I\!I\text{-}\tilde{A};\sigma;n \ell;\tilde{\rho})}_{\pm \theta_{j} \theta_{k}}$ ($j,k=1,...,N-1$) are rank-2 symmetric tensor-spinors on $S^{N-1}$. Since type-$I\!I\!I$ modes are divergence-free and gamma-traceless, we separate variables in the following way:
 \begin{align}
  &  {}^{(\uparrow)}{\psi}^{(I\!I\!I \text{-}\tilde{B};-;n \ell;\tilde{\rho})}_{\pm \theta_{j} \theta_{k}}(\theta_{N}, \bm{\theta}_{N-1})= \phi^{(-2)}_{ n \ell}(\theta_{N}) \,\tilde{\psi}^{(\tilde{B}; \ell \tilde{\rho})}_{- \theta_{j} \theta_{k}}(\bm{\theta}_{N-1}),\nonumber\\
  &{}^{(\downarrow)}{\psi}^{(I\!I\!I \text{-}\tilde{B};-;n \ell;\tilde{\rho})}_{\pm \theta_{j} \theta_{k}}(\theta_{N}, \bm{\theta}_{N-1})= \pm i\psi^{(-2)}_{ n \ell}(\theta_{N}) \,\tilde{\psi}^{(\tilde{B}; \ell \tilde{\rho})}_{- \theta_{j} \theta_{k}}(\bm{\theta}_{N-1}),\label{separ_type_III_thetajthetak_-}\\
  &   {}^{(\uparrow)}{\psi}^{(I\!I\!I \text{-}\tilde{B};+;n \ell;\tilde{\rho})}_{\pm \theta_{j} \theta_{k}}(\theta_{N}, \bm{\theta}_{N-1})= i\psi^{(-2)}_{ n \ell}(\theta_{N}) \,\tilde{\psi}^{(\tilde{B}; \ell \tilde{\rho})}_{+ \theta_{j} \theta_{k}}(\bm{\theta}_{N-1}), \nonumber \\
  &{}^{(\downarrow)}{\psi}^{(I\!I\!I\text{-}\tilde{B};+;n \ell;\tilde{\rho})}_{\pm \theta_{j} \theta_{k}}(\theta_{N}, \bm{\theta}_{N-1})= \pm \phi^{(-2)}_{n \ell}(\theta_{N}) \,\tilde{\psi}^{(\tilde{B}; \ell \tilde{\rho})}_{+ \theta_{j} \theta_{k}}(\bm{\theta}_{N-1})\label{separ_type_III_thetajthetak_+},
\end{align}
where eq.~(\ref{separ_type_III_thetajthetak_-}) describes the type-$I\!I\!I$ STSSH with negative spin projection, while eq.~(\ref{separ_type_III_thetajthetak_+}) describes the type-$I\!I\!I$ STSSH with positive spin projection. The functions $\phi^{(-2)}_{n \ell}$ and $\psi^{(-2)}_{n \ell}$ are given by eqs.~(\ref{phi_a}) and (\ref{psi_a}), respectively. It is straightforward to verify that the type-$I\!I\!I$ modes in eqs.~(\ref{separ_type_III_thetajthetak_-}) and (\ref{separ_type_III_thetajthetak_+}) are solutions of the system of equations~(\ref{Dirac_op_thetai_thetak_5/2_system}) (with the use of eqs.~(\ref{psi_to_phi_sphere}) and (\ref{phi_to_psi_sphere})).
\subsection{Constructing the STSSH's of rank 2 for \texorpdfstring{$N$}{N} odd}
Now the gamma matrices are given by eq.~(\ref{odd_gammas}). By combining eqs.~(\ref{Christoffels_SN}), (\ref{spin_connection_components}), (\ref{odd_gammas}), (\ref{covariant_deriv_tensor_spinor}) and eq.~(\ref{Spin(N)_generators}) we find 
 \begin{align}
    \slashed{\nabla}{\psi}^{(B;n \ell; \tilde{\rho})}_{ \pm \theta_{N} \theta_{N}  } &=\Bigg[\left(\frac{\partial}{\partial{\theta_{N}}}   +\frac{N+3}{2} \cot{\theta_{N}}\right)\gamma^{N}+\frac{1}{\sin{\theta_{N}}} \tilde{\slashed{\nabla}} \Bigg] {\psi}^{(B;n \ell; \tilde{\rho})}_{\pm \theta_{N} \theta_{N}}=\pm i |\zeta_{n,N}|\,{\psi}^{(B;n \ell; \tilde{\rho})}_{\pm \theta_{N} \theta_{N}} \label{Dirac_op_thetaN_thetaN_5/2},
\end{align}
 where we have used the gamma-tracelessness condition $$ \gamma^{N}{\psi}^{(B;n \ell; \tilde{\rho})}_{\pm \theta_{N} \theta_{N}}=  - \gamma^{\theta_{j}}{\psi}^{(B;n \ell; \tilde{\rho})}_{\pm \theta_{j} \theta_{N}}$$ (see eq.~(\ref{TT_SN_5/2})). Similarly, we find
   \begin{align}\label{Dirac_op_thetaN_thetai_5/2}
      \slashed{\nabla}{\psi}^{(B;n \ell; \tilde{\rho})}_{\pm \theta_{N} \theta_{j}}&=\Bigg[\left(\frac{\partial}{\partial \theta_{N}}   +\frac{N-1}{2} \cot{\theta_{N}}\right)\gamma^{N}+\frac{1}{\sin{\theta_{N}}} \tilde{\slashed{\nabla}}\Bigg]{\psi}^{(B;n \ell; \tilde{\rho})}_{\pm \theta_{N} \theta_{j}}  +\cot{\theta_{N}} \gamma_{\theta_{j}} {\psi}^{(B;n \ell; \tilde{\rho})}_{\pm\theta_{N} \theta_{N}}\\
      &=\pm i |\zeta_{n,N}|\,{\psi}^{(B;n \ell; \tilde{\rho})}_{\pm \theta_{N} \theta_{j}} \nonumber
  \end{align}
($j=1,...,N-1$) and
  \begin{align}\label{Dirac_op_thetai_thetak_5/2}
     \slashed{\nabla} {\psi}^{(B;n \ell; \tilde{\rho})}_{\pm \theta_{j} \theta_{k}}&=\Bigg[ \left( \frac{\partial}{\partial \theta_{N}}   +\frac{N-5}{2} \cot{\theta_{N}} \right)\gamma^{N}+\frac{1}{\sin{\theta_{N}}} \tilde{\slashed{\nabla}} \Bigg]{\psi}^{(B;n \ell; \tilde{\rho})}_{\pm \theta_{j} \theta_{k}} +2\cot{\theta_{N}}  \gamma_{(\theta_{j}} {\psi}^{(B;n \ell; \tilde{\rho})}_{\pm \theta_{k}) \theta_{N}} \\
     &= \pm i |\zeta_{n,N}| \,{\psi}^{(B;n \ell; \tilde{\rho})}_{\pm \theta_{j} \theta_{k}}\nonumber
  \end{align}
 ($j,k=1,...,N-1$). Note that for $N$ odd we have 
 \begin{align}
     \gamma^{N} \tilde{\slashed{\nabla}}+\tilde{\slashed{\nabla}} \gamma^{N}=0,
 \end{align}
 since $\{ \gamma^{N}, \tilde{\gamma}^{j} \} =0$ ($j=1,...,N-1$) - see eq.~(\ref{odd_gammas}). Now let us separate variables in eqs.~(\ref{Dirac_op_thetaN_thetaN_5/2})-(\ref{Dirac_op_thetai_thetak_5/2}).

\noindent \textbf{Type-}$\bm{I}$ \textbf{STSSH's of rank 2 for} $\bm{N}$ \textbf{odd.} As in Ref.~\cite{Camporesi}, since $N$ is odd we choose to express the type-$I$ modes in terms of the following spinors on $S^{N-1}$:
 \begin{align}
     \hat{\chi}_{-\ell \tilde{\rho}}(\bm{\theta}_{N-1})&\equiv \frac{1}{\sqrt{2}}(\bm{1}+i \gamma^{N})\chi_{-\ell \tilde{\rho}}(\bm{\theta}_{N-1})\label{chi-_hat_in_terms_of_chi}\\
     \hat{\chi}_{+\ell \tilde{\rho}}(\bm{\theta}_{N-1})&\equiv \gamma^{N}\hat{\chi}_{-\ell \tilde{\rho}}(\bm{\theta}_{N-1})= \frac{1}{\sqrt{2}}(\bm{1}+i \gamma^{N})\chi_{+\ell \tilde{\rho}}(\bm{\theta}_{N-1})\label{chi+_hat_in_terms_of_chi},
 \end{align}
where $\chi_{\pm \ell \tilde{\rho}}$ are the eigenspinors on $S^{N-1}$ (satisfying eq.~(\ref{eigenspinors on S_(N-1)})). Since $N$ is odd, $\chi_{+ \ell \tilde{\rho}}$ and  $\chi_{- \ell \tilde{\rho}}$ are related to each other as follows~\cite{Camporesi}: 
\begin{align}
    \chi_{+\ell \tilde{\rho}}(\bm{\theta}_{N-1})= \gamma^{N} \chi_{-\ell \tilde{\rho}}(\bm{\theta}_{N-1}).
\end{align}
The spinors $\hat{\chi}_{\pm \ell \tilde{\rho}}$ are eigenfunctions of the operator $\gamma^{N}\tilde{\slashed{\nabla}}$ (that commutes with ${\slashed{\nabla}}^{2}$) and they satisfy~\cite{Camporesi}
\begin{equation}\label{chi_hat_equation}
     \gamma^{N}\tilde{\slashed{\nabla}} \hat{\chi}_{\pm \ell \tilde{\rho}} = \pm \left(\ell+\frac{N-1}{2}\right)   \hat{\chi}_{\pm \ell \tilde{\rho}}.
 \end{equation}

In order to construct the rank-2 type-$I$ modes on $S^{N}$, we separate variables as follows:
\begin{align}
   {\psi}^{(I;n \ell ; \tilde{\rho})}_{\pm \theta_{N} \theta_{N}}(\theta_{N},\bm{\theta}_{N-1})=&\phi^{(2)}_{n \ell}(\theta_{N})  \hat{\chi}_{-\ell \tilde{\rho}}(\bm{\theta}_{N-1})\pm i  \psi^{(2)}_{n \ell}(\theta_{N}) \hat{\chi}_{+\ell \tilde{\rho}}(\bm{\theta}_{N-1})\label{TYPE_I_thetaNthetaN_hatted_5/2}\\
    {\psi}^{(I;n \ell ; \tilde{\rho})}_{\pm \theta_{N}\theta_{j}}(\theta_{N},\bm{\theta}_{N-1})=
     & C^{(\uparrow)(2)}_{n \ell}(\theta_{N})\tilde{\nabla}_{\theta_{j}}\hat{\chi}_{-\ell \tilde{\rho}}(\bm{\theta}_{N-1})\pm  iC^{(\downarrow)(2)}_{n \ell}(\theta_{N}) \tilde{\nabla}_{\theta_{j}}\hat{\chi}_{+\ell \tilde{\rho}}(\bm{\theta}_{N-1}) \nonumber \\ &-iD^{(\uparrow)(2)}_{n \ell}(\theta_{N})\tilde{\gamma}_{\theta_{j}}\hat{\chi}_{+\ell \tilde{\rho}}(\bm{\theta}_{N-1})\mp  D^{(\downarrow)(2)}_{n \ell}(\theta_{N}) \tilde{\gamma}_{\theta_{j}}\hat{\chi}_{-\ell \tilde{\rho}}(\bm{\theta}_{N-1}) \label{TYPE_I_thetaNthetaj_hatted_5/2}\\
    {\psi}^{(I;n \ell ; \tilde{\rho})}_{\pm \theta_{j}\theta_{k}}(\theta_{N}, \bm{\theta}_{N-1})=& \tilde{g}_{\theta_{j}\theta_{k}}\left(\hat{\chi}_{-\ell \tilde{\rho}}(\bm{\theta}_{N-1}) K^{(\uparrow)}_{n \ell}(\theta_{N}) \pm \hat{\chi}_{+\ell \tilde{\rho}}(\bm{\theta}_{N-1})\, iK^{(\downarrow)}_{n \ell}(\theta_{N}) \right) \nonumber\\
    &+\left[\tilde{\nabla}_{(\theta_{j}}\tilde{\nabla}_{\theta_{k})}-\frac{\tilde{g}_{\theta_{j}\theta_{k}}}{N-1}\tilde{\Box} \right]\nonumber\\
    &\times\left(\hat{\chi}_{-\ell \tilde{\rho}}(\bm{\theta}_{N-1}) W^{(\uparrow)}_{n\ell}(\theta_{N})\pm \hat{\chi}_{+\ell \tilde{\rho}}(\bm{\theta}_{N-1})\, iW^{(\downarrow)}_{n \ell}(\theta_{N}) \right) \nonumber\\
    &+ \left[\tilde{\gamma}_{(\theta_{j}}\tilde{\nabla}_{\theta_{k})}-\frac{\tilde{g}_{\theta_{j}\theta_{k}}}{N-1}\tilde{\slashed{\nabla}} \right] \nonumber\\
    & \times \left(-\hat{\chi}_{+\ell \tilde{\rho}}(\bm{\theta}_{N-1})\,iT^{(\uparrow)}_{n \ell}(\theta_{N})\mp  \hat{\chi}_{-\ell \tilde{\rho}}(\bm{\theta}_{N-1})\,T^{(\downarrow)}_{n \ell}(\theta_{N})\right) , \label{TYPE_I_thetajthetak_hatted_5/2}
  \end{align}
($j,k=1,...,N-1$). By working as in the case with $N$ even, we find that the functions $\phi^{(2)}_{n\ell},\psi^{(2)}_{n\ell},C^{(b)(2)}_{n\ell},D^{(b)(2)}_{n\ell}, K^{(b)}_{n\ell}, W^{(b)}_{n\ell}$ and $T^{(b)}_{n\ell}$ (where $b= \uparrow, \downarrow$), describing the dependence on $\theta_{N}$, are the same functions as the ones used in the even-dimensional case (see eqs.~(\ref{TYPE_I_thetaNthetaN_negative_spin_5/2})-(\ref{TYPE_I_thetajthetak_negative_spin_5/2})). By expressing $\hat{\chi}_{\pm \ell \tilde{\rho}}$ in terms of ${\chi}_{\pm \ell \tilde{\rho}}$ (by making use of eqs.~(\ref{chi-_hat_in_terms_of_chi}) and (\ref{chi+_hat_in_terms_of_chi})), it is straightforward to show that eqs.~(\ref{TYPE_I_thetaNthetaN_hatted_5/2}), (\ref{TYPE_I_thetaNthetaj_hatted_5/2}) and (\ref{TYPE_I_thetajthetak_hatted_5/2}) are equal to eqs.~(\ref{TYPE_I_thetaNthetaN_Nodd_spin_5/2}), (\ref{TYPE_I_thetaNthetaj_Nodd_spin_5/2}) and (\ref{TYPE_I_thetajthetak_Nodd_spin_5/2}), respectively, as presented in Subsection~\ref{subsectn_spin5/2_solving_SNodd}. 

\noindent \textbf{Type-}$\bm{I\!I}$ \textbf{STSSH's of rank 2 for} $\bm{N}$ \textbf{odd.} In order to construct the type-$I\!I$ STSSH's of rank 2 on $S^{N}$, we use the following vector-spinors on $S^{N-1}$:
 \begin{align}
     \hat{\tilde{\psi}}^{(\tilde{A};\ell \tilde{\rho})}_{-\theta_{j}}(\bm{\theta}_{N-1})&\equiv \frac{1}{\sqrt{2}}(\bm{1}+i \gamma^{N}){\tilde{\psi}}^{(\tilde{A};\ell \tilde{\rho})}_{-\theta_{j}}(\bm{\theta}_{N-1})\label{psi-_hat_in_terms_of_psi} \\
   \hat{\tilde{\psi}}^{(\tilde{A};\ell \tilde{\rho})}_{+\theta_{j}}(\bm{\theta}_{N-1})&\equiv \gamma^{N} \hat{\tilde{\psi}}^{(\tilde{A};\ell \tilde{\rho})}_{-\theta_{j}}(\bm{\theta}_{N-1}),\label{psi+_hat_in_terms_of_psi}
 \end{align}
where ${\tilde{\psi}}^{(\tilde{A};\ell \tilde{\rho})}_{\pm \theta_{j}}$ ($j=1,...,N-1$) are the TT eigevector-spinors on $S^{N-1}$ (satisfying eqs.~(\ref{eigenvectorspinors_SN-1}) and (\ref{eigenvectorspinors_SN-1_TT})) and ${\tilde{\psi}}^{(\tilde{A};\ell \tilde{\rho})}_{+\theta_{j}}=\gamma^{N}{\tilde{\psi}}^{(\tilde{A};\ell \tilde{\rho})}_{-\theta_{j}}$ .
The vector-spinors $\hat{\tilde{\psi}}^{(\tilde{A};\ell \tilde{\rho})}_{\pm \theta_{j}}$ satisfy
\begin{align}
    & \gamma^{N}\tilde{\slashed{\nabla}} \hat{\tilde{\psi}}^{(\tilde{A};\ell \tilde{\rho})}_{\pm \theta_{j}} = \pm \left(\ell+\frac{N-1}{2}\right)   \hat{\tilde{\psi}}^{(\tilde{A};\ell \tilde{\rho})}_{\pm \theta_{j}}\\
   &\tilde{\gamma}^{\theta_{i}} \hat{\tilde{\psi}}^{(\tilde{A};\ell \tilde{\rho})}_{\pm \theta_{i}}=\tilde{\nabla}^{\theta_{i}} \hat{\tilde{\psi}}^{(\tilde{A};\ell \tilde{\rho})}_{\pm \theta_{i}}=0 .
 \end{align}
 
 By making use of the vector-spinors $\hat{\tilde{\psi}}^{(\tilde{A};\ell \tilde{\rho})}_{\pm \theta_{j}}$, we separate variables for the type-$I\!I$ STSSH's ${\psi}^{(I\!I \text{-}\tilde{A};n \ell ; \tilde{\rho})}_{\pm \mu \nu}$ on $S^{N}$ as follows:
 \begin{align}
   {\psi}^{(I\!I \text{-}\tilde{A};n \ell ; \tilde{\rho})}_{\pm \theta_{N} \theta_{j}}(\theta_{N},\bm{\theta}_{N-1})=&\phi^{(0)}_{n \ell}(\theta_{N})  \hat{\tilde{\psi}}^{(\tilde{A};\ell \tilde{\rho})}_{- \theta_{j}}(\bm{\theta}_{N-1})\pm i  \psi^{(0)}_{n \ell}(\theta_{N}) \hat{\tilde{\psi}}^{(\tilde{A};\ell \tilde{\rho})}_{+ \theta_{j}}(\bm{\theta}_{N-1})\label{TYPE_II_thetaNthetaj_hatted_5/2}\\
    {\psi}^{(I\!I \text{-}\tilde{A};n \ell ; \tilde{\rho})}_{\pm \theta_{j}\theta_{k}}(\theta_{N},\bm{\theta}_{N-1})=
     & \Gamma^{(\uparrow)}_{n \ell}(\theta_{N})\tilde{\nabla}_{(\theta_{j}}\hat{\tilde{\psi}}^{(\tilde{A};\ell \tilde{\rho})}_{- \theta_{k})}(\bm{\theta}_{N-1})\pm  i\Gamma^{(\downarrow)}_{n \ell}(\theta_{N}) \tilde{\nabla}_{(\theta_{j}}\hat{\tilde{\psi}}^{(\tilde{A};\ell \tilde{\rho})}_{+ \theta_{k})}(\bm{\theta}_{N-1}) \nonumber \\ &-i\Delta^{(\uparrow)}_{n \ell}(\theta_{N})\tilde{\gamma}_{(\theta_{j}}\hat{\tilde{\psi}}^{(\tilde{A};\ell \tilde{\rho})}_{+ \theta_{k})}(\bm{\theta}_{N-1})\mp  \Delta^{(\downarrow)}_{n \ell}(\theta_{N}) \tilde{\gamma}_{(\theta_{j}}\hat{\tilde{\psi}}^{(\tilde{A};\ell \tilde{\rho})}_{- \theta_{k})}(\bm{\theta}_{N-1}) \label{TYPE_II_thetajthetak_hatted_5/2}
  \end{align}
($j,k=1,...,N-1$), while ${\psi}^{(I\!I \text{-}\tilde{A};n \ell ; \tilde{\rho})}_{\pm \theta_{N} \theta_{N}}=0$ by definition.
By working as in the case with $N$ even, we find that the functions $\phi^{(0)}_{n\ell},\psi^{(0)}_{n\ell},\Delta^{(b)}_{n\ell}$ and $\Gamma^{(b)}_{n\ell}$ (where $b= \uparrow, \downarrow$) are given by the same expressions as in the even-dimensional case (see eqs.~(\ref{TYPE_II_thetaNthetaj_Nodd_spin_5/2}) and (\ref{TYPE_II_thetajthetak_Nodd_spin_5/2})). By expressing $\hat{\tilde{\psi}}^{(\tilde{A};\ell \tilde{\rho})}_{\pm \theta_{j}}$ in terms of ${\tilde{\psi}}^{(\tilde{A};\ell \tilde{\rho})}_{\pm \theta_{j}}$ (with the use of eqs.~(\ref{psi-_hat_in_terms_of_psi}) and (\ref{psi+_hat_in_terms_of_psi})), we straightforwardly find that eqs.~(\ref{TYPE_II_thetaNthetaj_hatted_5/2}) and (\ref{TYPE_II_thetajthetak_hatted_5/2}) are equal to eqs.~(\ref{TYPE_II_thetaNthetaj_Nodd_spin_5/2}) and (\ref{TYPE_II_thetajthetak_Nodd_spin_5/2}), respectively.

\noindent \textbf{Type-}$\bm{I\!I\!I}$ \textbf{STSSH's of rank 2 for} $\bm{N}$ \textbf{odd.} In order to construct the type-$I\!I\!I$ STSSH's of rank 2 on $S^{N}$, we use the following rank-2 symmetric tensor-spinors on $S^{N-1}$:
 \begin{align}
     \hat{\tilde{\psi}}^{(\tilde{B};\ell \tilde{\rho})}_{-\theta_{j} \theta_{k}}(\bm{\theta}_{N-1})&\equiv \frac{1}{\sqrt{2}}(\bm{1}+i \gamma^{N}){\tilde{\psi}}^{(\tilde{B};\ell \tilde{\rho})}_{-\theta_{j} \theta_{k}}(\bm{\theta}_{N-1})\\
   \hat{\tilde{\psi}}^{(\tilde{B};\ell \tilde{\rho})}_{+\theta_{j} \theta_{k}}(\bm{\theta}_{N-1})&\equiv \gamma^{N} \hat{\tilde{\psi}}^{(\tilde{B};\ell \tilde{\rho})}_{-\theta_{j} \theta_{k}}(\bm{\theta}_{N-1}),
 \end{align}
where ${\tilde{\psi}}^{(\tilde{B};\ell \tilde{\rho})}_{\pm \theta_{j} \theta_{K}}$ ($j,k=1,...,N-1$) are the STSSH's of rank 2 on $S^{N-1}$ (satisfying eqs.~(\ref{eigentensorspinors_SN-1})-(\ref{eigentensorspinors_SN-1_trace})). Also, note that ${\tilde{\psi}}^{(\tilde{B};\ell \tilde{\rho})}_{+\theta_{j}\theta_{k}}=\gamma^{N}{\tilde{\psi}}^{(\tilde{B};\ell \tilde{\rho})}_{-\theta_{j} \theta_{k}}$ .
The tensor-spinors $\hat{\tilde{\psi}}^{(\tilde{B};\ell \tilde{\rho})}_{\pm \theta_{j} \theta_{k}}$ satisfy
\begin{align}
    & \gamma^{N}\tilde{\slashed{\nabla}} \hat{\tilde{\psi}}^{(\tilde{B};\ell \tilde{\rho})}_{\pm \theta_{j} \theta_{k}} = \pm \left(\ell+\frac{N-1}{2}\right)   \hat{\tilde{\psi}}^{(\tilde{B};\ell \tilde{\rho})}_{\pm \theta_{j} \theta_{k}}\\
   &\tilde{\gamma}^{\theta_{i}} \hat{\tilde{\psi}}^{(\tilde{B};\ell \tilde{\rho})}_{\pm \theta_{i} \theta_{k}}=\tilde{\nabla}^{\theta_{i}} \hat{\tilde{\psi}}^{(\tilde{B};\ell \tilde{\rho})}_{\pm \theta_{i} \theta_{k}}=0\\
   &\tilde{g}^{\theta_{i} \theta_{j}}\hat{\tilde{\psi}}^{(\tilde{B};\ell \tilde{\rho})}_{\pm \theta_{i} \theta_{j}}=0
 \end{align}
($i,j,k=1,...,N-1$).

By making use of the tensor-spinors $\hat{\tilde{\psi}}^{(\tilde{B};\ell \tilde{\rho})}_{\pm \theta_{j} \theta_{k}}$, we separate variables for the type-$I\!I\!I$ STSSH's ${\psi}^{(I\!I\!I \text{-}\tilde{B};n \ell ; \tilde{\rho})}_{\pm \mu \nu}$ on $S^{N}$ as follows:
 \begin{align}
   {\psi}^{(I\!I\!I \text{-}\tilde{B};n \ell ; \tilde{\rho})}_{\pm \theta_{j} \theta_{k}}(\theta_{N},\bm{\theta}_{N-1})=&\phi^{(-2)}_{n \ell}(\theta_{N})  \hat{\tilde{\psi}}^{(\tilde{B};\ell \tilde{\rho})}_{- \theta_{j} \theta_{k}}(\bm{\theta}_{N-1})\pm i  \psi^{(-2)}_{n \ell}(\theta_{N}) \hat{\tilde{\psi}}^{(\tilde{B};\ell \tilde{\rho})}_{+ \theta_{j} \theta_{k}}(\bm{\theta}_{N-1})\label{TYPE_III_thetaNthetaj_hatted_5/2}
\end{align}
($j,k=1,...,N-1$), while ${\psi}^{(I\!I\!I \text{-}\tilde{B};n \ell ; \tilde{\rho})}_{\pm \theta_{N} \theta_{N}}=0$ and ${\psi}^{(I\!I\!I \text{-}\tilde{B};n \ell ; \tilde{\rho})}_{\pm \theta_{N} \theta_{j}}=0$ (by definition). By working as in the case with $N$ even, we find that the functions $\phi^{(-2)}_{n \ell}$ and $\psi^{(-2)}_{n \ell}$ are given by eqs.~(\ref{phi_a}) and (\ref{psi_a}), respectively [and, thus, eq.~(\ref{TYPE_III_thetaNthetaj_hatted_5/2}) is equal to eq.~(\ref{TYPE_III_thetajthetak_Nodd_spin_5/2})].

\section{Deriving the formulae for the spin\texorpdfstring{$(N+1)$}{(N+1)} transformation of the STSSH's of ranks 1 and 2 on \texorpdfstring{$S^{N}$}{SN} and determining their normalisation factors}\label{Appendix_transfrmn_proeprties_norm_fac}
In Subsections~\ref{Appendix_transfrmn_proeprties_norm_fac_SUB1}-\ref{Appendix_transfrmn_proeprties_norm_fac_SUB3} of this Appendix we derive the transformation formulae~(\ref{transfrmn_type_I_unnormalsd_r_SN_even}), (\ref{transfrmn_type_II-I_unnormalsd_r_SN_even}), (\ref{transfrmn_type_I_unnormalsd_r_SN_odd}) and (\ref{transfrmn_type_II-I_unnormalsd_r_SN_odd}) for STSSH's of rank 1 on $S^{N}$ and we calculate the normalisation factors $c^{(I;r=1)}_{N}(n, \ell)$ and $c^{(I\!I;r=1)}_{N}(n, \ell)$ [eq.~(\ref{normln_fac_SN_TYPEB_r})]. The derivation of the transformation formulae and the calculation of the normalisation factors for the STSSH's of rank 2 have many similarities with the case of rank-1 STSSH's and, thus, we discuss them in less detail in Subsection~\ref{Appendix_transfrmn_proeprties_norm_fac_SUB4}.
\subsection{Calculating \texorpdfstring{$c^{(I\!I;r=1)}_{N}(n, \ell)$}{CII} and making the first step towards the calculation of \texorpdfstring{$c^{(I;r=1)}_{N}(n, \ell)$}{CI}}\label{Appendix_transfrmn_proeprties_norm_fac_SUB1}
Since it is a quite simple task, let us start by calculating directly the normalisation factor for type-$I\!I$ STSSH's of rank 1 for arbitrary $N$.  For $N$ even, we substitute the unnormalised type-$I\!I$ modes~(\ref{TYPE_II_thetai_negative_spin_3/2}) (or (\ref{TYPE_II_thetai_positive_spin_3/2})) into the inner product~(\ref{define_normn_factors_SN_r}). Then, by performing the integration over $S^{N-1}$ using eq.~(\ref{normalization_SN-1_vectorspinors}), we find
 \begin{align}
  \left|\frac{c^{(I\!I;r=1)}_{N}(n, \ell)}{\sqrt{2}}\right|^{-2}  =& ~ \int_{0}^{\pi}d\theta_{N} \sin^{N-3}{\theta_{N}}\left[\left(\phi^{(-1)}_{n\ell}(\theta_{N})\right)^{2}+\left(\psi^{(-1)}_{n\ell}(\theta_{N}) \right)^{2}\right]\nonumber\\
 =& ~\frac{1}{4} \int_{0}^{\pi}d\theta_{N} \sin^{N-1}{\theta_{N}}\left[\left(\phi^{(0)}_{n\ell}(\theta_{N})\right)^{2}+\left(\psi^{(0)}_{n\ell}(\theta_{N}) \right)^{2}\right],
  \end{align}
  where the functions $\phi^{(0)}_{n \ell}$ and $\psi^{(0)}_{n \ell}$ are given by eqs.~(\ref{phi_a}) and (\ref{psi_a}), respectively.
The integral in the last line is the same integral that appears in the normalisation of spinor eigenfunctions on $S^{N}$ in Ref.~\cite{Camporesi}. Thus, using the result of Ref.~\cite{Camporesi} we readily find
\begin{align}\label{normalization_fac_TYPE_II_spin3/2}
    \left| \frac{c_{N}^{(I\!I;r=1)}( n,\ell)}{\sqrt{2}}\right|^{2}=&~  \frac{1}{2^{N-3}}\frac{\Gamma(n-\ell+1)  \Gamma(n+\ell+N)}{|\Gamma(n+\frac{N}{2})|^{2}} ,
 \end{align}
which is a special case of eq.~(\ref{normln_fac_SN_TYPEB_r}). For $N$ odd, the calculation is similar and we find again that the normalisation factor is given by eq.~(\ref{normln_fac_SN_TYPEB_r}).

The normalisation factor of the type-$I$ modes can be found by calculating the following integral:
\begin{align}
   \left|\frac{c^{(I;r=1)}_{N}(n, \ell)}{\sqrt{2}}\right|^{-2}=& ~ \int_{0}^{\pi}d\theta_{N}\sin^{N-1}{\theta_{N}}\left[ \left(\phi^{(1)}_{n \ell}(\theta_{N})  \right)^{2}   +\left(\psi^{(1)}_{n \ell}(\theta_{N})  \right)^{2}  \right]  \nonumber\\
    &+\left[\left(\ell+\frac{N-1}{2} \right)^{2}-\frac{(N-1)(N-2)}{4}\right]\nonumber\\
    &\times \int_{0}^{\pi} d\theta_{N}\sin^{N-3}{\theta_{N}}\left[\left(C^{(\uparrow)(1)}_{n \ell}(\theta_{N})  \right)^{2}   +\left(C^{(\downarrow)(1)}_{n \ell}(\theta_{N})  \right)^{2} \right] 
 \nonumber \\
 &+(N-1)\int_{0}^{\pi} d\theta_{N}\sin^{N-3}{\theta_{N}}\left[\left|D^{(\uparrow)(1)}_{n \ell}(\theta_{N})  \right|^{2}   +\left|D^{(\downarrow)(1)}_{n \ell}(\theta_{N})  \right|^{2} \right] 
 \nonumber\\
  &+2i\left(\ell+\frac{N-1}{2} \right)\nonumber\\
  &\times \int_{0}^{\pi}d\theta_{N} \sin^{N-3}{\theta_{N}}\nonumber\\
  & \times\left[C^{(\uparrow)(1)}_{n \ell}(\theta_{N})D^{(\uparrow)(1)}_{n \ell}(\theta_{N})   +C^{(\downarrow)(1)}_{n \ell}(\theta_{N})D^{(\downarrow)(1)}_{n \ell}(\theta_{N})   \right],\label{normln_fac_SN_TYPE_I_3/2_long_int}
\end{align}
where $C^{(\uparrow)(1)}_{n \ell}, C^{(\downarrow)(1)}_{n \ell},D^{(\uparrow)(1)}_{n \ell}$ and $D^{(\downarrow)(1)}_{n \ell}$ are given by eqs.~(\ref{C1_a_function}), (\ref{C2_a_function}), (\ref{D1_a_function}) and (\ref{D2_a_function}), respectively. For $N$ even, eq.~(\ref{normln_fac_SN_TYPE_I_3/2_long_int}) is derived by substituting the expressions~(\ref{TYPE_I_thetaN_negative_spin_3/2}) and (\ref{TYPE_I_theta_i_negative_spin_3/2}) for type-$I$ modes into the inner product~(\ref{define_normn_factors_SN_r}) and then performing the integration over $S^{N-1}$ (with the use of eqs.~(\ref{normalization_SN-1_spinors}) and (\ref{LB_op_eigenspin_SN-1})). For $N$ odd, by working similarly we find again eq.~(\ref{normln_fac_SN_TYPE_I_3/2_long_int}). Since the integrals in eq.~(\ref{normln_fac_SN_TYPE_I_3/2_long_int}) are not as simple as in the case of type-$I\!I$ modes, we are going to take an indirect route. To be specific, we first obtain by direct calculation the normalisation factor of the type-$I$ modes with the highest allowed value for $\ell$, i.e. $c^{(I;r=1)}_{N}(n,\ell= n)$. Then, once we have obtained the transformation formulae of the type-$I$ modes under spin$(N+1)$, the normalisation factor $c^{(I;r=1)}_{N}(n, \ell)$ (for $\ell=1,2,....n-1$) will be constructed in terms of $c^{(I;r=1)}_{N}(n, n)$ by exploiting the spin$(N+1)$ invariance of the inner product~(\ref{inner_prod_SN}). To calculate $c^{(I;r=1)}_{N}(n, n)$ we let $\ell=n$ in eq.~(\ref{normln_fac_SN_TYPE_I_3/2_long_int}) and by calculating the integrals using Mathematica 11.2 we find
\begin{align}\label{normln_fac_SN_TYPE_I_3/2_ell=n}
 \left| \frac{c_{N}^{(I;r=1)}(n,n)}{\sqrt{2}}\right|^{2}=~   \frac{n(N-2)\Gamma(n+\frac{N}{2}+\frac{1}{2})}{4^{1-n}(1+n)(N-1) \sqrt{\pi}\Gamma(n+\frac{N}{2})}.
\end{align}
\subsection{Derivation of the transformation formulae of type-\texorpdfstring{$I$}{I} and type-\texorpdfstring{$I\!I\text{-}I$}{II-I} STSSH's of rank 1 and calculation of the normalisation factor \texorpdfstring{$c_{N}^{(I;r=1)}(n,\ell)$}{cI} for \texorpdfstring{$N$}{N} even}\label{Appendix_transfrmn_proeprties_norm_fac_SUB2}

Below we give details for the derivation of the transformation formulae~(\ref{transfrmn_type_I_unnormalsd_r_SN_even}) and~(\ref{transfrmn_type_II-I_unnormalsd_r_SN_even}) for rank-1 ($r=1$) modes with positive spin projection [these modes are given by eqs.~(\ref{TYPE_I_thetaN_positive_spin_3/2}), (\ref{TYPE_I_theta_i_positive_spin_3/2}) and (\ref{TYPE_II_thetai_positive_spin_3/2})]. The calculations for the rank-1 modes with negative spin projection are not presented here, as they can be performed in the same way. 

 In order to derive the desired transformation formulae~(\ref{transfrmn_type_I_unnormalsd_r_SN_even}) and (\ref{transfrmn_type_II-I_unnormalsd_r_SN_even}), it is sufficient to study the following two components of the Lie-Lorentz derivative~(\ref{Lie_Lorentz}): $\mathbb{L}_{\mathcal{S}}~{\psi}_{ \theta_{N}}$ and $\mathbb{L}_{\mathcal{S}}~{\psi}_{ \theta_{N-1}}$. After a straightforward calculation we find 
\begin{align}\label{spinor_lie_thetaN_initial_stage_SN_3/2}
       \mathbb{L}_{\mathcal{S}}~\psi_{\theta_{N}}  = &\left( \mathcal{S}^{\mu}\partial_{\mu} +\frac{\sin{\theta_{N-1}}}{2 \sin{\theta_{N}}}\gamma^{N} \gamma^{N-1} \right)   \psi_{\theta_{N}} + \frac{\sin{\theta_{N-1}}}{\sin^{2}{\theta_{N}}}    \psi_{\theta_{N-1}}
\end{align}
and
\begin{align}\label{spinor_lie_thetaN-1_initial_stage_SN_3/2}
       \mathbb{L}_{\mathcal{S}}~\psi_{\theta_{N-1}}  =&\left( \mathcal{S}^{\mu}\partial_{\mu} -\cot{\theta_{N}}\,\cos{\theta_{N-1}}+\frac{\sin{\theta_{N-1}}}{2 \sin{\theta_{N}}}\gamma^{N} \gamma^{N-1} \right)   \psi_{\theta_{N-1}} -{\sin{\theta_{N-1}}}~\psi_{\theta_{N}},
\end{align}
where we have substituted eqs.~(\ref{Christoffels_SN}), (\ref{spin_connection_components}), (\ref{covariant_deriv_vector_spinor}) and (\ref{Killing_vector_sphere}) into eq.~(\ref{Lie_Lorentz}).
Since $N$ is even, we express $\gamma^{N} \gamma^{N-1}$ in eqs.~(\ref{spinor_lie_thetaN_initial_stage_SN_3/2}) and (\ref{spinor_lie_thetaN-1_initial_stage_SN_3/2}) as
\begin{equation}\label{N-1_0_generators_even}
  \gamma^{N}\gamma^{N-1}  = \begin{pmatrix}  
  - i\widetilde{ \gamma}^{N-1} & 0 \\
0 & i\widetilde{ \gamma}^{N-1}
    \end{pmatrix} ,
\end{equation}
where we have used eq.~(\ref{even_gammas}).

 The partial derivatives in eqs.~(\ref{spinor_lie_thetaN_initial_stage_SN_3/2}) and (\ref{spinor_lie_thetaN-1_initial_stage_SN_3/2}) act only on the coordinates $\{ \theta_{N}, \theta_{N-1} \}$. Thus, for later convenience let us introduce the functions $\tilde{\phi}^{(\tilde{a})}_{\ell m} (\theta_{N-1})$ and $\tilde{\psi}^{(\tilde{a})}_{\ell m} (\theta_{N-1})$ describing the $\theta_{N-1}$-dependence of the STSSH's on $S^{N-1}$ .
  In analogy to eqs.~(\ref{phi_a}) and (\ref{psi_a}), these functions are given by
    \begin{align}
    \tilde{\phi}^{(\tilde{a})}_{\ell m}(\theta_{N-1}) =&~\tilde{\kappa}_{\tilde{\phi}}(\ell, m) \left(\cos{\frac{\theta_{N-1}}{2}}\right)^{m+1-\tilde{a}}\left(\sin{\frac{\theta_{N-1}}{2}}\right)^{m-\tilde{a}} \nonumber \\ &\times  F\left(-\ell+m,\ell+m+N-1;m+\frac{N-1}{2};\sin^{2}\frac{\theta_{N-1}}{2}\right),\label{phitilde_a} 
\end{align} 
and
 \begin{align}
    \tilde{\psi}^{(\tilde{a})}_{\ell m}(\theta_{N-1})
    &=\,\tilde{\kappa}_{\tilde{\phi}}(\ell, m)\,\frac{\ell+\frac{N-1}{2}}{m+\frac{N-1}{2}}\left(\cos{\frac{\theta_{N-1}}{2}}\right)^{m-\tilde{a}}\left(\sin{\frac{\theta_{N-1}}{2}}\right)^{m+1-\tilde{a}}\nonumber\\  
    &\times  F\left(-\ell+m,\ell+m+N-1;m+\frac{N+1}{2};\sin^{2}\frac{\theta_{N-1}}{2}\right),\label{psitilde_a} 
    \end{align}
 where the normalisation factor is given by
  \begin{align}
   \tilde{\kappa}_{\tilde{\phi}}(\ell,m)= \frac{\Gamma(\ell+\frac{N-1}{2})}{\Gamma(\ell-m+1)\,\Gamma{(m+\frac{N-1}{2})}}.
  \end{align}
  The number $\tilde{a}$ in eqs.~(\ref{phitilde_a}) and (\ref{psitilde_a}) is an integer and $m$ is the angular momentum quantum number on $S^{N-2}$ [with $\ell \geq m$, in analogy with eq.~(\ref{restriction on n-ell})]. The formulae analogous to eqs.~(\ref{psi_to_phi_sphere}) and (\ref{phi_to_psi_sphere}) are given by
   \begin{align}
    \Bigg(\frac{d}{d \theta_{N-1}}&+\frac{N+2\tilde{a}-2}{2}\cot{\theta_{N-1}}+\frac{m+\frac{N-2}{2}}{\sin{\theta_{N-1}}}\,\Bigg)\tilde{\psi}^{(\tilde{a})}_{ \ell m} =\left(\ell+\frac{N-1}{2}\right)\tilde{\phi}^{(\tilde{a})}_{\ell m} \label{tildepsi_to_tildephi_sphere}
    \end{align}
    and
    \begin{align}
  \Bigg(\frac{d}{d \theta_{N-1}}&+\frac{N+2\tilde{a}-2}{2}\cot{\theta_{N-1}}-\frac{m+\frac{N-2}{2}}{\sin{\theta_{N-1}}} \,\Bigg)\tilde{\phi}^{(\tilde{a})}_{ \ell m}=-\left(\ell+\frac{N-1}{2}\right)\tilde{\psi}^{(\tilde{a})}_{\ell m} \label{tildephi_to_tildepsi_sphere},
  \end{align}
  respectively.

 Motivated by the techniques used in Refs.~\cite{STSHS} and~\cite{Letsios}, in order to derive the transformation formulae of our STSSH's we introduce the ladder operators for $\ell$, sending $\ell$ to $\ell \pm 1$ when acting on the functions $\phi^{(a)}_{n \ell}(\theta_{N}), \psi^{(a)}_{n \ell}(\theta_{N}),\tilde{\phi}^{(\tilde{a})}_{ \ell m }(\theta_{N-1})$ and $\tilde{\psi}^{(\tilde{a})}_{ \ell m }(\theta_{N-1})$. 
The ladder operators are given by the following expressions:
\begin{align}
  T^{(+;a)}_{\phi}=~&  \frac{d}{d\theta_{N}}+\left(-\ell+a-\frac{1}{2}\right) \cot{\theta_{N}} + \frac{1}{2 \sin{\theta_{N}}},  \\
  T^{(+;a)}_{\psi }=~&  \frac{d}{d\theta_{N}}+\left(-\ell+a-\frac{1}{2}\right) \cot{\theta_{N}} - \frac{1}{2 \sin{\theta_{N}}},\\
   T^{(-;a)}_{\phi} =~& \frac{d}{d\theta_{N}}+\left(\ell+N+a-\frac{3}{2}\right) \cot{\theta_{N}} - \frac{1}{2 \sin{\theta_{N}}}, \\
   T^{(-;a)}_{\psi} =~& \frac{d}{d\theta_{N}}+\left(\ell+N+a-\frac{3}{2}\right) \cot{\theta_{N}} + \frac{1}{2 \sin{\theta_{N}}},
   \end{align} 
   \begin{align}
  \tilde{\Pi}^{(+;\tilde{a})}_{\tilde{\phi}} =~ &  \sin{\theta_{N-1}} \frac{d}{d\theta_{N-1}}+\left(\ell+\tilde{a}+N-\frac{3}{2}\right)\cos{\theta_{N-1}} - \frac{m+\frac{N-2}{2}}{2(\ell+\frac{N}{2})},   \\
   \tilde{\Pi}^{(+;\tilde{a})}_{\tilde{\psi}} =~&  \sin{\theta_{N-1}} \frac{d}{d\theta_{N-1}}+\left(\ell+\tilde{a}+N-\frac{3}{2}\right)\cos{\theta_{N-1}} + \frac{m+\frac{N-2}{2}}{2(\ell+\frac{N}{2})},\\
    \tilde{\Pi}^{(-;\tilde{a})}_{\tilde{\phi}}=~&  \sin{\theta_{N-1}} \frac{d}{d\theta_{N-1}}+\left(-\ell+\tilde{a}-\frac{1}{2}\right)\cos{\theta_{N-1}}+\frac{m+\frac{N-2}{2}}{2(\ell+\frac{N-2}{2})} \label{lowering_tilde_phi_operator},\\
     \tilde{\Pi}^{(-;\tilde{a})}_{\tilde{\psi}}=~&  \sin{\theta_{N-1}} \frac{d}{d\theta_{N-1}}+\left(-\ell+\tilde{a}-\frac{1}{2}\right)\cos{\theta_{N-1}}-\frac{m+\frac{N-2}{2}}{2(\ell+\frac{N-2}{2})}.\label{lowering_tilde_psi_operator}
  \end{align}
 These operators act as follows:
  \begin{align}
      &T^{(+;a)}_{f} f^{(a)}_{n \ell}(\theta_{N})= k^{(+)}f^{(a)}_{n\, \ell+1}(\theta_{N}), \label{raising_phi_psi}   \\
       &T^{(-;a)}_{f} f^{(a)}_{n \ell}(\theta_{N})= k^{(-)}f^{(a)}_{n\, \ell-1}(\theta_{N}), \label{lowering_phi_psi} \\ 
        &\tilde{\Pi}^{(+;\tilde{a})}_{\tilde{f}} \tilde{f}^{(\tilde{a})}_{ \ell  m}(\theta_{N-1})= \tilde{k}^{(+)}\tilde{f}^{(\tilde{a})}_{ \ell+1 \, m}(\theta_{N-1}), \label{raising_tilde} \\
         &\tilde{\Pi}^{(-;\tilde{a})}_{\tilde{f}} \tilde{f}^{(\tilde{a})}_{ \ell  m}(\theta_{N-1})= \tilde{k}^{(-)}\tilde{f}^{(\tilde{a})}_{ \ell-1 \, m}(\theta_{N-1}),\label{lowering_tilde}
     \end{align}
     where $f^{(a)}_{n \ell}(\theta_{N}) \in \set{ \phi^{(a)}_{n \ell}(\theta_{N}),\psi^{(a)}_{n \ell}(\theta_{N})}$,  $\tilde{f}^{(\tilde{a})}_{ \ell  m}(\theta_{N-1}) \in \set { \tilde{\phi}^{(\tilde{a})}_{\ell \,m}(\theta_{N-1}),\tilde{\psi}^{(\tilde{a})}_{\ell  m}(\theta_{N-1})}$ and
  \begin{align}
        & k^{(+)}=-(n+\ell+N), \label{kay_plus}\\
         & k^{(-)}=n-\ell+1, \label{kay_minus}\\
         & \tilde{k}^{(+)}=\frac{(\ell+N-1+m)(\ell-m+1)}{\ell+{N}/{2}  },\label{tilde_kay_plus} \\
        &\tilde{k}^{(-)}= -\frac{(\ell+\frac{N-1}{2}-1)(\ell+\frac{N-1}{2})  }{\ell+ ({N-2})/{2}}.\label{tilde_kay_minus}
     \end{align}
One can straightforwardly prove the ladder relations (\ref{raising_phi_psi})-(\ref{lowering_tilde}) using the raising and lowering operators for the parameters of the Gauss hypergeometric function given in Appendix~\ref{appendix_raising_lowering_hypergeom}. (Similar ladder relations have been obtained by the author in Ref.~\cite{Letsios} while studying the Dirac field on $dS_{N}$.)

 Let us now proceed to the derivation of the transformation formulae of the type-$I$ and type-$I\!I$-$I$ modes. 
It is clear from the expressions~(\ref{spinor_lie_thetaN_initial_stage_SN_3/2}) and (\ref{spinor_lie_thetaN-1_initial_stage_SN_3/2}) for the Lie-Lorentz derivative that we need to express the type-$I$ and type-$I\!I\text{-}I$ modes in a form where the dependence on both $\theta_{N}$ and $\theta_{N-1}$ is written out explicitly. By substituting eq.~(\ref{form_of_eigenspinors_SN-1odd}) into eqs.~(\ref{TYPE_I_thetaN_positive_spin_3/2}) and (\ref{TYPE_I_theta_i_positive_spin_3/2}), we express the type-$I$ modes with positive spin projection as
\begin{align}\label{TYPE_I_3/2_thetaN_posspin_in_terms_S^(N-2)}
 {\psi}&^{(I;+;n \ell m;\rho)}_{\pm  \theta_{N}}(\theta_{N}, \theta_{N-1},\bm{\theta}_{N-2})\nonumber\\
 &=\frac{\tilde{c}_{N-1}(\ell, m)}{\sqrt{2}}  \begin{pmatrix}i\psi^{(1)}_{n \ell}(\theta_{N})\left[  \tilde{\phi}^{(0)}_{\ell m}(\theta_{N-1})\, \hat{\tilde{\chi}}_{-m {\rho}}(\bm{\theta}_{N-2})+i\tilde{\psi}^{(0)}_{\ell m}(\theta_{N-1})\, \hat{\tilde{\chi}}_{+m{\rho}}(\bm{\theta}_{N-2}) \right]   \\ \\\pm \phi^{(1)}_{n \ell}(\theta_{N})\,\left[  \tilde{\phi}^{(0)}_{\ell m}(\theta_{N-1})\, \hat{\tilde{\chi}}_{-m {\rho}}(\bm{\theta}_{N-2})+i\tilde{\psi}^{(0)}_{\ell m}(\theta_{N-1})\, \hat{\tilde{\chi}}_{+m {\rho}} (\bm{\theta}_{N-2})\right] \end{pmatrix}
\end{align}
\begin{align}\label{TYPE_I_3/2_theta(N-1)_posspin_in_terms_S^(N-2)}
 {\psi}&^{(I;+;n \ell m;\rho)}_{\pm  \theta_{N-1}}(\theta_{N}, \theta_{N-1},\bm{\theta}_{N-2})\nonumber\\
 &=\frac{\tilde{c}_{N-1}(\ell, m)}{\sqrt{2}}  \begin{pmatrix}i\left[E^{(1)}_{n \ell m}(\theta_{N}, \theta_{N-1})\, \hat{\tilde{\chi}}_{-m {\rho}}(\bm{\theta}_{N-2})+i\Sigma^{(1)}_{n \ell m}(\theta_{N},\theta_{N-1})\, \hat{\tilde{\chi}}_{+m {\rho}}(\bm{\theta}_{N-2}) \right]   \\ \\
  \pm \left[  H^{(1)}_{n\ell m}(\theta_{N},\theta_{N-1})\, \hat{\tilde{\chi}}_{-m {\rho}}(\bm{\theta}_{N-2})+iO^{(1)}_{n\ell m}(\theta_{N},\theta_{N-1})\, \hat{\tilde{\chi}}_{+m {\rho}} (\bm{\theta}_{N-2})\right] \end{pmatrix},
\end{align}
where $\tilde{c}_{N-1}(\ell, m)$ is the normalisation factor~(\ref{S_N-1_normlzn_fac_spinor}) for the eigenspinors on $S^{N-1}$, while the spinors $\hat{\tilde{\chi}}_{\pm m {\rho}} (\bm{\theta}_{N-2})$ on $S^{N-2}$ are defined by eq.~(\ref{tildechihat_in_terms_tildechi_SN-2}). Also, we have defined
\begin{align}
& O^{(a)}_{n \ell m}(\theta_{N}, \theta_{N-1})=  C^{(\uparrow)(a)}_{ n \ell}(\theta_{N}) \,\frac{\partial }{\partial \theta_{N-1}}\tilde{\psi}^{(0)}_{\ell m}(\theta_{N-1})+i  D^{(\uparrow)(a)}_{ n \ell}(\theta_{N}) \,\tilde{\phi}^{(0)}_{\ell m}(\theta_{N-1})\label{function_Onlm}\\
 & H_{n \ell m}^{(a)}(\theta_{N}, \theta_{N-1})=  C^{(\uparrow)(a)}_{ n \ell}(\theta_{N})\, \frac{\partial }{\partial \theta_{N-1}}\tilde{\phi}^{(0)}_{\ell m}(\theta_{N-1})-i  D^{(\uparrow)(a)}_{ n \ell}(\theta_{N}) \,\tilde{\psi}^{(0)}_{\ell m} (\theta_{N-1})\label{function_Hnlm} \\
 & E_{n \ell m}^{(a)}(\theta_{N}, \theta_{N-1}) = C^{(\downarrow)(a)}_{ n \ell}(\theta_{N})\, \frac{\partial }{\partial \theta_{N-1}}\tilde{\phi}^{(0)}_{\ell m}(\theta_{N-1})-i  D^{(\downarrow)(a)}_{ n \ell}(\theta_{N}) \,\tilde{\psi}^{(0)}_{\ell m}(\theta_{N-1})\label{function_Enlm}\\
  & \Sigma_{n \ell m}^{(a)}(\theta_{N}, \theta_{N-1}) = C^{(\downarrow)(a)}_{ n \ell}(\theta_{N}) \,\frac{\partial }{\partial \theta_{N-1}}\tilde{\psi}^{(0)}_{\ell m}(\theta_{N-1})+i  D^{(\downarrow)(a)}_{ n \ell}(\theta_{N}) \,\tilde{\phi}^{(0)}_{\ell m}(\theta_{N-1}) .\label{function_Sigmanlm}
    \end{align}
(Recall that $C^{(\uparrow)(a)}_{ n \ell},C^{(\downarrow)(a)}_{ n \ell}, D^{(\uparrow)(a)}_{ n \ell}$ and $D^{(\downarrow)(a)}_{ n \ell}$ are given by eqs.~(\ref{C1_a_function}), (\ref{C2_a_function}), (\ref{D1_a_function}) and (\ref{D2_a_function}), respectively.) Similarly, the type-$I$ modes with negative spin projection are expressed as
\begin{align}\label{TYPE_I_3/2_thetaN_negspin_in_terms_S^(N-2)}
  {\psi}&^{(I;-;n \ell m;\rho)}_{\pm  \theta_{N}}(\theta_{N}, \theta_{N-1},\bm{\theta}_{N-2})\nonumber\\
  &=\frac{\tilde{c}_{N-1}(\ell, m)}{\sqrt{2}}  \begin{pmatrix}\phi^{(1)}_{n \ell}(\theta_{N})\left[  \tilde{\phi}^{(0)}_{\ell m}(\theta_{N-1})\, \hat{\tilde{\chi}}_{-m {\rho}}(\bm{\theta}_{N-2})-i\tilde{\psi}^{(0)}_{\ell m}(\theta_{N-1})\, \hat{\tilde{\chi}}_{+m{\rho}}(\bm{\theta}_{N-2}) \right]   \\ \\\pm i\psi^{(1)}_{n \ell}(\theta_{N})\,\left[  \tilde{\phi}^{(0)}_{\ell m}(\theta_{N-1})\, \hat{\tilde{\chi}}_{-m {\rho}}(\bm{\theta}_{N-2})-i\tilde{\psi}^{(0)}_{\ell m}(\theta_{N-1})\, \hat{\tilde{\chi}}_{+m {\rho}} (\bm{\theta}_{N-2})\right] \end{pmatrix}
\end{align}
\begin{align}\label{TYPE_I_3/2_theta(N-1)_negspin_in_terms_S^(N-2)}
  {\psi}&^{(I;-;n \ell m;\rho)}_{\pm  \theta_{N-1}}(\theta_{N}, \theta_{N-1},\bm{\theta}_{N-2})\nonumber\\
  &=\frac{\tilde{c}_{N-1}(\ell, m)}{\sqrt{2}}  \begin{pmatrix}H^{(1)}_{n \ell m}(\theta_{N}, \theta_{N-1})\, \hat{\tilde{\chi}}_{-m {\rho}}(\bm{\theta}_{N-2})-iO^{(1)}_{n \ell m}(\theta_{N},\theta_{N-1})\, \hat{\tilde{\chi}}_{+m {\rho}}(\bm{\theta}_{N-2})  \\ \\
  \pm i \left[  E^{(1)}_{n\ell m}(\theta_{N},\theta_{N-1})\, \hat{\tilde{\chi}}_{-m {\rho}}(\bm{\theta}_{N-2})-i\Sigma^{(1)}_{n\ell m}(\theta_{N},\theta_{N-1})\, \hat{\tilde{\chi}}_{+m {\rho}} (\bm{\theta}_{N-2})\right] \end{pmatrix}.
\end{align}
Similarly, it is straightforward to express the type-$I\!I\text{-}I$ modes with positive spin projection~(\ref{TYPE_II_thetai_positive_spin_3/2}) as follows:
\begin{align}
    {\psi}^{(I\!I\text{-}I;+;n \ell m;\rho)}_{\pm  \theta_{N}}(\theta_{N}, \theta_{N-1},\bm{\theta}_{N-2})=&0,\label{TYPE_II-I_3/2_thetaN_posspin_in_terms_S^(N-2)}\end{align}
    \begin{align}
 {\psi}&^{(I\!I\text{-}I;+;n \ell m;\rho)}_{\pm  \theta_{N-1}}(\theta_{N}, \theta_{N-1},\bm{\theta}_{N-2})\nonumber\\
&=\frac{\tilde{c}^{(I;\tilde{r}=1)}_{N-1}(\ell, m)}{\sqrt{2}} 
   \begin{pmatrix}i\psi^{(-1)}_{n \ell}(\theta_{N})\left[  \tilde{\phi}^{(1)}_{\ell m}(\theta_{N-1})\, \hat{\tilde{\chi}}_{-m {\rho}}(\bm{\theta}_{N-2})+i\tilde{\psi}^{(1)}_{\ell m}(\theta_{N-1})\, \hat{\tilde{\chi}}_{+m{\rho}}(\bm{\theta}_{N-2}) \right]   \\ \\\pm \phi^{(-1)}_{n \ell}(\theta_{N})\,\left[  \tilde{\phi}^{(1)}_{\ell m}(\theta_{N-1})\, \hat{\tilde{\chi}}_{-m {\rho}}(\bm{\theta}_{N-2})+i\tilde{\psi}^{(1)}_{\ell m}(\theta_{N-1})\, \hat{\tilde{\chi}}_{+m {\rho}} (\bm{\theta}_{N-2})\right] \end{pmatrix},\label{TYPE_II-I_3/2_theta(N-1)_posspin_in_terms_S^(N-2)}
\end{align}
where $\tilde{c}^{(I;r=1)}_{N-1}(\ell, m)$ is the normalisation factor of the STSSH's of rank 1 on $S^{N-1}$ and it will be determined later. 
The type-$I\!I\text{-}I$ modes with negative spin projection~(\ref{TYPE_II_thetai_negative_spin_3/2}) are expressed as
\begin{align}
   {\psi}&^{(I\!I\text{-}I;-;n \ell m;\rho)}_{\pm  \theta_{N}}(\theta_{N}, \theta_{N-1},\bm{\theta}_{N-2})=0,\label{TYPE_II-I_3/2_thetaN_negspin_in_terms_S^(N-2)}\end{align} \begin{align}
 {\psi}&^{(I\!I\text{-}I;-;n \ell m;\rho)}_{\pm  \theta_{N-1}}(\theta_{N}, \theta_{N-1},\bm{\theta}_{N-2})\nonumber\\
 &=\frac{\tilde{c}^{(I;\tilde{r}=1)}_{N-1}(\ell, m)}{\sqrt{2}} 
   \begin{pmatrix}\phi^{(-1)}_{n \ell}(\theta_{N})\left[  \tilde{\phi}^{(1)}_{\ell m}(\theta_{N-1})\, \hat{\tilde{\chi}}_{-m {\rho}}(\bm{\theta}_{N-2})-i\tilde{\psi}^{(1)}_{\ell m}(\theta_{N-1})\, \hat{\tilde{\chi}}_{+m{\rho}}(\bm{\theta}_{N-2}) \right]   \\ \\\pm i\psi^{(-1)}_{n \ell}(\theta_{N})\,\left[  \tilde{\phi}^{(1)}_{\ell m}(\theta_{N-1})\, \hat{\tilde{\chi}}_{-m {\rho}}(\bm{\theta}_{N-2})-i\tilde{\psi}^{(1)}_{\ell m}(\theta_{N-1})\, \hat{\tilde{\chi}}_{+m {\rho}} (\bm{\theta}_{N-2})\right] \end{pmatrix},\label{TYPE_II-I_3/2_theta(N-1)_negspin_in_terms_S^(N-2)}
\end{align}

\subsubsection{Derivation of the transformation formula~(\ref{transfrmn_type_I_unnormalsd_r_SN_even}) for type-\texorpdfstring{${I}$}{I} modes of rank 1 and calculation of the normalisation factor \texorpdfstring{${c_{N}^{(I;r=1)}(n,\ell)}$}{cIr=1}}

By using the expressions~(\ref{TYPE_I_3/2_thetaN_posspin_in_terms_S^(N-2)}) and~(\ref{TYPE_I_3/2_theta(N-1)_posspin_in_terms_S^(N-2)}) for the type-$I$ modes, we express the Lie-Lorentz derivative~(\ref{spinor_lie_thetaN_initial_stage_SN_3/2}) as
\begin{align}\label{LieLorentz_thetaN_TYPEI_interms_SN-2_3/2}
  \mathbb{L}_{\mathcal{S}}&{\psi}^{(I;+;n \ell m;\rho)}_{\pm  \theta_{N}}\nonumber\\
  =&\frac{\tilde{c}_{N-1}(\ell,m)}{\sqrt{2}}  \begin{pmatrix} i \hat{\tilde{\chi}}_{-m {\rho}}(\bm{\theta}_{N-2}) \,\mathbb{T}^{(I)}_{3}(\theta_{N},\theta_{N-1}) -\hat{\tilde{\chi}}_{+m {\rho}}(\bm{\theta}_{N-2})\, \mathbb{T}^{(I)}_{4}(\theta_{N}, \theta_{N-1})\\ \\
   \pm \hat{\tilde{\chi}}_{-m {\rho}}(\bm{\theta}_{N-2}) \,\mathbb{T}^{(I)}_{1}(\theta_{N},\theta_{N-1}) \pm i \hat{\tilde{\chi}}_{+m {\rho}}(\bm{\theta}_{N-2}) \,\mathbb{T}^{(I)}_{2}(\theta_{N}, \theta_{N-1})
  \end{pmatrix},
\end{align}
where 
\begin{align}
\mathbb{T}_{1}^{(I)} &=~   \mathcal{S}^{\mu}\partial_{\mu}\left[ \phi^{(1)}_{n \ell} \tilde{\phi}^{(0)}_{\ell m}\right]-\frac{\sin{\theta_{N-1}}}{2\sin{\theta_{N}}}\phi^{(1)}_{n\ell} \tilde{\psi}^{(0)}_{\ell m}+\frac{\sin{\theta_{N-1}}}{\sin^{2}{\theta_{N}}}H^{(1)}_{n \ell m}, \label{T1_typeI_relation_define_3/2}\\
\mathbb{T}_{2}^{(I)} &=~   \mathcal{S}^{\mu}\partial_{\mu}\left[ \phi^{(1)}_{n \ell} \tilde{\psi}^{(0)}_{\ell m}\right]+\frac{\sin{\theta_{N-1}}}{2\sin{\theta_{N}}}\phi^{(1)}_{n\ell} \tilde{\phi}^{(0)}_{\ell m}+\frac{\sin{\theta_{N-1}}}{\sin^{2}{\theta_{N}}}O^{(1)}_{n \ell m}, \label{T2_typeI_relation_define_3/2}\\
\mathbb{T}_{3}^{(I)} &=~   \mathcal{S}^{\mu}\partial_{\mu}\left[ \psi^{(1)}_{n \ell}\tilde{\phi}^{(0)}_{\ell m}\right]+\frac{\sin{\theta_{N-1}}}{2\sin{\theta_{N}}}\psi^{(1)}_{n\ell}\tilde{\psi}^{(0)}_{\ell m}+\frac{\sin{\theta_{N-1}}}{\sin^{2}{\theta_{N}}}E^{(1)}_{n \ell m}, \label{T3_typeI_relation_define_3/2}\\
\mathbb{T}_{4}^{(I)} &=~   \mathcal{S}^{\mu}\partial_{\mu}\left[ \psi^{(1)}_{n \ell}\tilde{\psi}^{(0)}_{\ell m}\right]-\frac{\sin{\theta_{N-1}}}{2\sin{\theta_{N}}}\psi^{(1)}_{n\ell} \tilde{\phi}^{(0)}_{\ell m}+\frac{\sin{\theta_{N-1}}}{\sin^{2}{\theta_{N}}}\Sigma^{(1)}_{n \ell m} .\label{T4_typeI_relation_define_3/2}
\end{align}
(Recall that $O^{(1)}_{n \ell m},H^{(1)}_{n \ell m},E^{(1)}_{n \ell m}$ and $\Sigma^{(1)}_{n \ell m}$ are given by eqs.~(\ref{function_Onlm}), (\ref{function_Hnlm}), (\ref{function_Enlm}) and (\ref{function_Sigmanlm}), respectively.)
In order to proceed we need to make use of the following relations:
\begin{align}
\mathbb{T}^{(I)}_{1} &=~\mathcal{R}^{(I)}k^{(+)}\tilde{k}^{(+)}\phi^{(1)}_{n\, \ell+1} \tilde{\phi}^{(0)}_{\ell+1\, m} +\mathcal{L}^{(I)}k^{(-)}\tilde{k}^{(-)}\phi^{(1)}_{n\, \ell-1} \tilde{\phi}^{(0)}_{\ell-1\, m}+\varkappa^{(I)}\psi^{(1)}_{n \ell} \tilde{\phi}^{(0)}_{\ell m}, \label{T1_typeI_relation_3/2}\\
\mathbb{T}^{(I)}_{2} &=~\mathcal{R}^{(I)}k^{(+)}\tilde{k}^{(+)}\phi^{(1)}_{n\, \ell+1} \tilde{\psi}^{(0)}_{\ell+1\, m} +\mathcal{L}^{(I)}k^{(-)}\tilde{k}^{(-)}\phi^{(1)}_{n\, \ell-1} \tilde{\psi}^{(0)}_{\ell-1\, m}-\varkappa^{(I)}\psi^{(1)}_{n \ell} \tilde{\psi}^{(0)}_{\ell m},\label{T2_typeI_relation_3/2} \\
\mathbb{T}^{(I)}_{3} &=~  \mathcal{R}^{(I)}k^{(+)}\tilde{k}^{(+)}\psi^{(1)}_{n\, \ell+1} \tilde{\phi}^{(0)}_{\ell+1\, m} +\mathcal{L}^{(I)}k^{(-)}\tilde{k}^{(-)}\psi^{(1)}_{n\, \ell-1} \tilde{\phi}^{(0)}_{\ell-1\, m}-\varkappa^{(I)}\phi^{(1)}_{n \ell} \tilde{\phi}^{(0)}_{\ell m}, \label{T3_typeI_relation_3/2}\\
\mathbb{T}_{4}^{(I)} &=~\mathcal{R}^{(I)}k^{(+)}\tilde{k}^{(+)}\psi^{(1)}_{n\, \ell+1} \tilde{\psi}^{(0)}_{\ell+1\, m} +\mathcal{L}^{(I)}k^{(-)}\tilde{k}^{(-)}\psi^{(1)}_{n\, \ell-1} \tilde{\psi}^{(0)}_{\ell-1\, m}+\varkappa^{(I)}\phi^{(1)}_{n \ell} \tilde{\psi}^{(0)}_{\ell m},\label{T4_typeI_relation_3/2}
\end{align}
where $k^{(+)}, k^{(-)}, \tilde{k}^{(+)}$ and $\tilde{k}^{(-)}$ are given by eqs.~(\ref{kay_plus}), (\ref{kay_minus}), (\ref{tilde_kay_plus}) and (\ref{tilde_kay_minus}), respectively, while  $\varkappa^{(I)}$ is the coefficient defined in eq.~(\ref{varkap_I_coeff_r}) (with $r=1$) and
\begin{align}
    &\mathcal{R}^{(I)}=\frac{\ell +N}{2(\ell+\frac{N-1}{2})(\ell+N-1)},\, \hspace{8mm}\mathcal{L}^{(I)}=\frac{1-\ell}{2\ell(\ell+\frac{N-1}{2})}\label{R_I,L_I_3/2}.
\end{align}
Let us outline the steps required for proving eq.~(\ref{T1_typeI_relation_3/2}). (Equations (\ref{T2_typeI_relation_3/2})-(\ref{T4_typeI_relation_3/2}) are proved similarly.) First, we express $\mathbb{T}^{(I)}_{1}$ on the left-hand side of eq.~(\ref{T1_typeI_relation_3/2}) in terms of $\phi^{(1)}_{n \ell},\, d\phi^{(1)}_{n \ell}/d \theta_{N},\, \tilde{\phi}^{(0)}_{\ell m}$ and $d\tilde{\phi}^{(0)}_{\ell m}/ d\theta_{N-1}$ by making use of eqs.~(\ref{T1_typeI_relation_define_3/2}), (\ref{function_Hnlm}), (\ref{tildephi_to_tildepsi_sphere}), (\ref{C1_a_function}), (\ref{D1_a_function}) and (\ref{phi_to_psi_sphere}). As for the right-hand side, we express $\phi^{(1)}_{n \,\ell \pm 1}$ and $\tilde{\phi}^{(0)}_{\ell \pm 1 \, m}$ in terms of $\phi^{(1)}_{n \ell},\, d\phi^{(1)}_{n \ell}/d \theta_{N}$ and $\tilde{\phi}^{(0)}_{\ell m},\, d\tilde{\phi}^{(0)}_{\ell m}/ d\theta_{N-1}$, respectively, by making use of the ladder relations~(\ref{raising_phi_psi})-(\ref{lowering_tilde}) and we also express $\psi_{n \ell}^{(1)}$ in terms of $\phi^{(1)}_{n \ell}$ and $\, d\phi^{(1)}_{n \ell}/d \theta_{N}$ by making use of eq.~(\ref{phi_to_psi_sphere}). Then, it is straightforward to show that the two sides of eq.~(\ref{T1_typeI_relation_3/2}) are equal.
We have verified the calculations using Mathematica 11.2. 

Then, by substituting eqs.~(\ref{T1_typeI_relation_3/2})-(\ref{T4_typeI_relation_3/2}) into eq.~(\ref{LieLorentz_thetaN_TYPEI_interms_SN-2_3/2}), we express the latter as
\begin{align}
    \mathbb{L}_{\mathcal{S}} {\psi}^{(I;+;n \ell m;\rho)}_{\pm  \theta_{N}}=&~\frac{\tilde{c}_{N-1}(\ell, m)}{\sqrt{2}}~ \Bigg\{ \mathcal{R}^{(I)}k^{(+)}\tilde{k}^{(+)}
    \begin{pmatrix}i\psi^{(1)}_{n\, \ell+1}\left[  \tilde{\phi}^{(0)}_{\ell+1\, m}\, \hat{\tilde{\chi}}_{-m {\rho}}+i\tilde{\psi}^{(0)}_{\ell+1\, m}\, \hat{\tilde{\chi}}_{+m{\rho}} \right]   \\ \\\pm \phi^{(1)}_{n \,\ell+1}\,\left[  \tilde{\phi}^{(0)}_{\ell+1\, m}\, \hat{\tilde{\chi}}_{-m {\rho}}+i\tilde{\psi}^{(0)}_{\ell+1\, m}\, \hat{\tilde{\chi}}_{+m {\rho}} \right] \end{pmatrix} \nonumber\\
   &~+ \mathcal{L}^{(I)}k^{(-)}\tilde{k}^{(-)}\begin{pmatrix}i\psi^{(1)}_{n\, \ell-1}\left[  \tilde{\phi}^{(0)}_{\ell-1\, m}\, \hat{\tilde{\chi}}_{-m {\rho}}+i\tilde{\psi}^{(0)}_{\ell-1\, m}\, \hat{\tilde{\chi}}_{+m{\rho}} \right]   \\ \\\pm \phi^{(1)}_{n \,\ell-1}\,\left[  \tilde{\phi}^{(0)}_{\ell-1\, m}\, \hat{\tilde{\chi}}_{-m {\rho}}+i\tilde{\psi}^{(0)}_{\ell-1\, m}\, \hat{\tilde{\chi}}_{+m {\rho}} \right] \end{pmatrix} \nonumber\\
    &~- i \varkappa^{(I)}\begin{pmatrix}\phi^{(1)}_{n\ell}\left[  \tilde{\phi}^{(0)}_{\ell m}\, \hat{\tilde{\chi}}_{-m {\rho}}-i\tilde{\psi}^{(0)}_{\ell m}\, \hat{\tilde{\chi}}_{+m{\rho}} \right]   \\ \\\pm i \psi^{(1)}_{n \ell}\,\left[  \tilde{\phi}^{(0)}_{\ell m}\, \hat{\tilde{\chi}}_{-m {\rho}}-i\tilde{\psi}^{(0)}_{\ell m}\, \hat{\tilde{\chi}}_{+m {\rho}} \right] \end{pmatrix}~ \Bigg\}
\end{align}
and we straightforwardly rewrite this as  
\begin{align}
     \mathbb{L}_{\mathcal{S}} {\psi}^{(I;+;n \ell m;\rho)}_{\pm \theta_{N} }  =~&  \mathcal{A}^{(I)} {\psi}^{(I;+;n \,(\ell+1)\,m;\rho)}_{\pm \theta_{N} } + \mathcal{B}^{(I)}  {\psi}^{(I;+;n \,(\ell-1)\,m;\rho)}_{\pm \theta_{N} }- i \varkappa^{(I)}{\psi}^{(I;-;n \ell m ;\rho)}_{\pm \theta_{N}},
\end{align}
as in eq.~(\ref{transfrmn_type_I_unnormalsd_r_SN_even}), where we have defined
\begin{align}
    \mathcal{A}^{(I)}&\equiv \mathcal{R}^{(I)} k^{(+)}\tilde{k}^{(+)}\frac{\tilde{c}_{N-1}(\ell,m)}{\tilde{c}_{N-1}(\ell+1,m)}, \\
    \mathcal{B}^{(I)}&\equiv \mathcal{L}^{(I)} k^{(-)}\tilde{k}^{(-)}\frac{\tilde{c}_{N-1}(\ell,m)}{\tilde{c}_{N-1}(\ell-1,m)} . 
\end{align}
It easy to verify that these expressions for $\mathcal{A}^{(I)}$ and $\mathcal{B}^{(I)}$ agree with the expressions given by eqs.~(\ref{trnsf_coffcient_raise_unnormtypeI_r}) (with $r=1$) and (\ref{trnsf_coffcient_lower_unnormtypeI_r}) (with $r=1$), respectively. 

Now, we can determine the normalisation factor $c_{N}^{(I;r=1)}(n ,\ell)$ for the type-$I$ modes. By using the spin$(N+1)$ invariance of the inner product~(\ref{Spin(N+1)-invariance-inner_prod}) between ${\psi}^{(I;\sigma;n \ell m;\rho)}_{\pm  \mu}$ and ${\psi}^{\left(I;\sigma;n (\ell+1) m;\rho\right)}_{\pm  \mu}$ and using the transformation formula~(\ref{transfrmn_type_I_unnormalsd_r_SN_even}) we find
\begin{align}
    \left| \frac{c_{N}^{(I;r=1)}(n ,\ell)}{c_{N}^{(I;r=1)}(n ,\ell+1)}  \right|^{2}=\frac{(n-\ell) \,\ell\, (\ell +N-1)}{(\ell+1)(\ell+N)(n+\ell+N)}.
\end{align}
By iterating this equation and using eq.~(\ref{normln_fac_SN_TYPE_I_3/2_ell=n}), one can straightforwardly find 
 \begin{align}\label{normalization_fac_TYPE_I_spin3/2}
    \left| \frac{c_{N}^{(I;r=1)}( n,\ell)}{\sqrt{2}}\right|^{2}=&  \frac{1}{2^{N+1}}\frac{\Gamma(n-\ell+1)  \Gamma(n+\ell+N)}{|\Gamma(n+\frac{N}{2})|^{2}} \nonumber\\
   &\times \frac{(N-2)\ell (\ell+N-1)}{(N-1)\left( \left[n+N/2\right]^{2}-\left[N-2\right]^{2}/4  \right)},
 \end{align}
which is eq.~(\ref{normln_fac_SN_TYPEB_r}) with $r=1$ and $\tilde{r}_{(B)}=\tilde{r}_{(I)}=0$. For later convenience, note that we can easily deduce the form of the normalisation factor for the type-$I$ STSSH's of rank $1$ on $S^{N-1}$ by making the replacements $N \rightarrow N-1,\, n \rightarrow \ell$ and $\ell \rightarrow m $ in eq.~(\ref{normalization_fac_TYPE_I_spin3/2}), as
 \begin{align}\label{normln_fac_SN-1_TYPE_I_3/2}
   \left| \frac{\tilde{c}_{N-1}^{(I;\tilde{r}=1)}( \ell,m)}{\sqrt{2}}\right|^{2}&=  \frac{1}{2^{N}}\frac{\Gamma(\ell-m+1)  \Gamma(\ell+m+N-1)}{|\Gamma(\ell+\frac{N-1}{2})|^{2}} \nonumber\\
   &\times \frac{(N-3)m(m+N-2)}{(N-2)(\ell+1)(\ell+N-2)}.
 \end{align}

Let us now discuss the mixing between type-$I$ and type-$I\!I\text{-}I$ modes under the spin$(N+1)$ transformation. By using the equation $\psi^{(I\!I\text{-}I;\sigma;n \ell m;\rho)}_{\pm  \theta_{N}}=0$ and eqs.~(\ref{TYPE_I_3/2_thetaN_posspin_in_terms_S^(N-2)}) and (\ref{TYPE_II-I_3/2_theta(N-1)_posspin_in_terms_S^(N-2)}) (or eqs.~(\ref{TYPE_I_3/2_thetaN_negspin_in_terms_S^(N-2)}) and (\ref{TYPE_II-I_3/2_theta(N-1)_negspin_in_terms_S^(N-2)})), one readily finds that the component given by~(\ref{spinor_lie_thetaN_initial_stage_SN_3/2}) of the infinitesimal transformation of a type-$I\!I\text{-}I$ mode is proportional to a type-$I$ mode, as
\begin{align}
    \mathbb{L}_{\mathcal{S}}{\psi}^{(I\!I\text{-}I;\sigma;n \ell m;\rho)}_{\pm  \theta_{N}}=&\frac{\sin{\theta_{N-1}}}{\sin^{2}{\theta_{N}}}   {\psi}^{(I\!I\text{-}I;\sigma;n \ell m;\rho)}_{\pm  \theta_{N-1}}\nonumber\\
    =&\mathcal{K}^{(I\!I \rightarrow I)}{\psi}^{(I;\sigma;n \ell m;\rho)}_{\pm  \theta_{N}},
\end{align}
in agreement with eq.~(\ref{transfrmn_type_II-I_unnormalsd_r_SN_even}), where we have defined
\begin{align}
    \mathcal{K}^{(I\!I \rightarrow I)} \equiv \frac{1}{2}\frac{\tilde{c}_{N-1}^{(I;\tilde{r}=1)}(\ell,m)}{\tilde{c}_{N-1}(\ell,m)}.
\end{align}
It is easy to show that this expression for $\mathcal{K}^{(I\!I \rightarrow I)}$ is equal to the expression given by eq.~(\ref{trnsf_coffcient_mixII->I_unnorm_r}) (with $r=1$).
Then, since type-$I\!I\text{-}I$ modes transform into type-$I$ modes under the spin$(N+1)$ transformation, the spin$(N+1)$ invariance of the inner product~(\ref{Spin(N+1)-invariance-inner_prod}) (between ${\psi}^{(I;\sigma;n \ell m;\rho)}_{\pm  \mu}$ and ${\psi}^{(I\!I \text{-}I;\sigma;n \ell m;\rho)}_{\pm  \mu}$) implies that
\begin{align}
    \mathbb{L}_{\mathcal{S}} {\psi}^{(I;\sigma;n \ell m;\rho)}_{\pm  \mu}=...+ \mathcal{K}^{(I \rightarrow I\!I)}{\psi}^{(I\!I\text{-}I;\sigma;n \ell m;\rho)}_{\pm  \mu},
\end{align}
where all the STSSH's in `...' are type-$I$ modes, while $\mathcal{K}^{(I \rightarrow I\!I)}$ is given by
\begin{align}\label{trnsf_coffcient_mixI->II_unnorm_r_Append}
   \mathcal{K}^{(I \rightarrow I\!I)}= -  \mathcal{K}^{(I\!I \rightarrow I)*}\left| \frac{c^{(I\!I;r=1)}_{N}(n,\ell)}{c^{(I;r=1)}_{N}(n,\ell)}  \right|^{2},
\end{align}
where the asterisk denotes complex conjugation. Then, by using the expression for $\mathcal{K}^{(I\!I \rightarrow I)}$ [eq.~(\ref{trnsf_coffcient_mixII->I_unnorm_r})] and the expressions for the normalisation factors [eq.~(\ref{normln_fac_SN_TYPEB_r})] we find that $\mathcal{K}^{(I \rightarrow I\!I)}$ in eq.~(\ref{trnsf_coffcient_mixI->II_unnorm_r_Append}) is equal to the expression given by eq.~(\ref{trnsf_coffcient_mixI->II_unnorm_r}) (with $r=1$).
\subsubsection{{Derivation of the transformation formula~(\ref{transfrmn_type_II-I_unnormalsd_r_SN_even}) for type-}\texorpdfstring{${I\!I\text{-}I}$}{II-I} modes of rank 1}
By substituting the type-$I\!I\text{-}I$ mode~(\ref{TYPE_II-I_3/2_theta(N-1)_posspin_in_terms_S^(N-2)}) into the Lie-Lorentz derivative~(\ref{spinor_lie_thetaN-1_initial_stage_SN_3/2}) we find
\begin{align}\label{LieLorentz_thetaN-1_TYPEII-I_interms_SN-2_3/2}
  \mathbb{L}_{\mathcal{S}}&{\psi}^{(I\!I\text{-}I;+;n \ell m;\rho)}_{\pm  \theta_{N-1}}\nonumber\\
  =&\frac{\tilde{c}^{(I;\tilde{r}=1)}_{N-1}(\ell,m)}{\sqrt{2}}  \begin{pmatrix} i \hat{\tilde{\chi}}_{-m {\rho}}(\bm{\theta}_{N-2}) \,\mathbb{T}^{(I\!I)}_{3}(\theta_{N},\theta_{N-1}) -\hat{\tilde{\chi}}_{+m {\rho}}(\bm{\theta}_{N-2})\, \mathbb{T}^{(I\!I)}_{4}(\theta_{N}, \theta_{N-1})\\ \\
   \pm \hat{\tilde{\chi}}_{-m {\rho}}(\bm{\theta}_{N-2}) \,\mathbb{T}^{(I\!I)}_{1}(\theta_{N},\theta_{N-1}) \pm i \hat{\tilde{\chi}}_{+m {\rho}}(\bm{\theta}_{N-2}) \,\mathbb{T}^{(I\!I)}_{2}(\theta_{N}, \theta_{N-1})
  \end{pmatrix},
\end{align}
where 
\begin{align}
\mathbb{T}_{1}^{(I\!I)} &=~   \left(\mathcal{S}^{\mu}\partial_{\mu}-\cot{\theta_{N}} \cos{\theta_{N-1}}\right)\left[ \phi^{(-1)}_{n \ell} \tilde{\phi}^{(1)}_{\ell m}\right]-\frac{\sin{\theta_{N-1}}}{2\sin{\theta_{N}}}\phi^{(-1)}_{n\ell} \tilde{\psi}^{(1)}_{\ell m}, \label{T1_typeII_relation_define_3/2}\\
\mathbb{T}_{2}^{(I\!I)} &=~   \left(\mathcal{S}^{\mu}\partial_{\mu}-\cot{\theta_{N}} \cos{\theta_{N-1}}\right)\left[ \phi^{(-1)}_{n \ell} \tilde{\psi}^{(1)}_{\ell m}\right]+\frac{\sin{\theta_{N-1}}}{2\sin{\theta_{N}}}\phi^{(-1)}_{n\ell} \tilde{\phi}^{(1)}_{\ell m}, \label{T2_typeII_relation_define_3/2}\\
\mathbb{T}_{3}^{(I\!I)} &=~   \left(\mathcal{S}^{\mu}\partial_{\mu}-\cot{\theta_{N}} \cos{\theta_{N-1}}\right)\left[ \psi^{(-1)}_{n \ell}\tilde{\phi}^{(1)}_{\ell m}\right]+\frac{\sin{\theta_{N-1}}}{2\sin{\theta_{N}}}\psi^{(-1)}_{n\ell} \tilde{\psi}^{(1)}_{\ell m}, \label{T3_typeII_relation_define_3/2}\\
\mathbb{T}_{4}^{(I\!I)} &=~   \left(\mathcal{S}^{\mu}\partial_{\mu}-\cot{\theta_{N}} \cos{\theta_{N-1}}\right)\left[ \psi^{(-1)}_{n \ell} \tilde{\psi}^{(1)}_{\ell m}\right]-\frac{\sin{\theta_{N-1}}}{2\sin{\theta_{N}}}\psi^{(-1)}_{n\ell} \tilde{\phi}^{(1)}_{\ell m} .\label{T4_typeII_relation_define_3/2}
\end{align}
 Then, as in the case of the type-$I$ modes, we prove the following relations: 
\begin{align}
\mathbb{T}^{(I\!I)}_{1} &=~\mathcal{R}^{(I\!I)}k^{(+)}\tilde{k}^{(+)}\phi^{(-1)}_{n\, \ell+1} \tilde{\phi}^{(1)}_{\ell+1\, m} +\mathcal{L}^{(I\!I)}k^{(-)}\tilde{k}^{(-)}\phi^{(-1)}_{n\, \ell-1} \tilde{\phi}^{(1)}_{\ell-1\, m}+\varkappa^{(I\!I)}\psi^{(-1)}_{n \ell} \tilde{\phi}^{(1)}_{\ell m}+\frac{H^{(1)}_{n \ell m}}{2}, \label{T1_typeII_relation_3/2}\\
\mathbb{T}^{(I\!I)}_{2} &=~\mathcal{R}^{(I\!I)}k^{(+)}\tilde{k}^{(+)}\phi^{(-1)}_{n\, \ell+1} \tilde{\psi}^{(1)}_{\ell+1\, m} +\mathcal{L}^{(I\!I)}k^{(-)}\tilde{k}^{(-)}\phi^{(-1)}_{n\, \ell-1} \tilde{\psi}^{(1)}_{\ell-1\, m}-\varkappa^{(I\!I)}\psi^{(-1)}_{n \ell} \tilde{\psi}^{(1)}_{\ell m}+\frac{O^{(1)}_{n \ell m}}{2},\label{T2_typeII_relation_3/2} \\
\mathbb{T}^{(I\!I)}_{3} &=~  \mathcal{R}^{(I\!I)}k^{(+)}\tilde{k}^{(+)}\psi^{(-1)}_{n\, \ell+1} \tilde{\phi}^{(1)}_{\ell+1\, m} +\mathcal{L}^{(I\!I)}k^{(-)}\tilde{k}^{(-)}\psi^{(-1)}_{n\, \ell-1} \tilde{\phi}^{(1)}_{\ell-1\, m}-\varkappa^{(I\!I)}\phi^{(-1)}_{n \ell} \tilde{\phi}^{(1)}_{\ell m}+\frac{E^{(1)}_{n \ell m}}{2}, \label{T3_typeII_relation_3/2}\\
\mathbb{T}_{4}^{(I\!I)} &=~\mathcal{R}^{(I\!I)}k^{(+)}\tilde{k}^{(+)}\psi^{(-1)}_{n\, \ell+1} \tilde{\psi}^{(1)}_{\ell+1\, m} +\mathcal{L}^{(I\!I)}k^{(-)}\tilde{k}^{(-)}\psi^{(-1)}_{n\, \ell-1} \tilde{\psi}^{(1)}_{\ell-1\, m}+\varkappa^{(I\!I)}\phi^{(-1)}_{n \ell} \tilde{\psi}^{(1)}_{\ell m}+\frac{\Sigma^{(1)}_{n \ell m}}{2}\label{T4_typeII_relation_3/2},
\end{align}
where $\varkappa^{(I\!I)}$ is given by eq.~(\ref{varkap_II_coeff_r}) (with $r=1$) and
\begin{align}
    &\mathcal{R}^{(I\!I)}=\frac{\ell +N-2}{2(\ell+\frac{N-1}{2})(\ell+N-1)},\hspace{8mm} \mathcal{L}^{(I\!I)}=\frac{-(1+\ell)}{2\ell(\ell+\frac{N-1}{2})}\label{R,L,II}. 
\end{align}
By substituting eqs.~(\ref{T1_typeII_relation_3/2})-(\ref{T4_typeII_relation_3/2}) into eq.~(\ref{LieLorentz_thetaN-1_TYPEII-I_interms_SN-2_3/2}) we find 
\begin{align}
     \mathbb{L}_{\mathcal{S}} {\psi}^{(I\!I\text{-}I;+;n \ell m;\rho)}_{\pm \theta_{N-1} }  =~&  \mathcal{A}^{(I\!I)} {\psi}^{(I\!I\text{-}I;+;n \,(\ell+1)\,m;\rho)}_{\pm \theta_{N-1} } + \mathcal{B}^{(I\!I)}{\psi}^{(I\!I\text{-}I;+;n \,(\ell-1)\,m;\rho)}_{\pm \theta_{N-1} }\nonumber\\
     &- i \varkappa^{(I\!I)} {\psi}^{(I\!I\text{-}I;-;n \ell m ;\rho)}_{\pm \theta_{N-1} }+ \mathcal{K}^{(I\!I \rightarrow I)} {\psi}^{(I;+;n \ell m ;\rho)}_{\pm \theta_{N-1} },
\end{align}
in precise agreement with the transformation formula~(\ref{transfrmn_type_II-I_unnormalsd_r_SN_even}), where we have defined
\begin{align}
    \mathcal{A}^{(I\!I)}&\equiv \mathcal{R}^{(I\!I)} k^{(+)}\tilde{k}^{(+)}\frac{\tilde{c}^{(I;\tilde{r}=1)}_{N-1}(\ell,m)}{\tilde{c}^{(I; \tilde{r}=1)}_{N-1}(\ell+1,m)}, \\
    \mathcal{B}^{(I\!I)}&\equiv \mathcal{L}^{(I\!I)} k^{(-)}\tilde{k}^{(-)}\frac{\tilde{c}^{(I;\tilde{r}=1)}_{N-1}(\ell,m)}{\tilde{c}^{(I;\tilde{r}=1)}_{N-1}(\ell-1,m)} . 
\end{align}
It easy to verify that these expressions for $\mathcal{A}^{(I\!I)}$ and $\mathcal{B}^{(I\!I)}$ agree with the expressions given by eqs.~(\ref{trnsf_coffcient_raise_unnormtypeII_r}) (with $r=1$) and (\ref{trnsf_coffcient_lower_unnormtypeII_r}) (with $r=1$), respectively. 

\subsection{Derivation of the transformation formulae of type-\texorpdfstring{${I}$}{I} and type-\texorpdfstring{${I\!I}\text{-}{I}$}{II-I} STSSH's of rank 1 and calculation of the normalisation factor \texorpdfstring{${c_{N}^{(I;r=1)}(n,\ell)}$}{cI} for \texorpdfstring{${N}$}{N} odd}\label{Appendix_transfrmn_proeprties_norm_fac_SUB3}

The Lie-Lorentz derivative is given by eqs.~(\ref{spinor_lie_thetaN_initial_stage_SN_3/2}) and (\ref{spinor_lie_thetaN-1_initial_stage_SN_3/2}), where $\gamma^{N}\gamma^{N-1}$ is given by
\begin{equation}\label{N-1_0_generators_odd}
  \gamma^{N} \gamma^{N-1} = \begin{pmatrix} 
  0 & \bm{1} \\
-\bm{1} & 0
    \end{pmatrix} ,
\end{equation}
where $\bm{1}$ is the identity spinorial matrix of dimension $2^{\frac{N-1}{2}}/2$. 

The type-$I$ modes on $S^{N}$ with positive spin projection index on $S^{N-1}$ ($\sigma_{N-1}=+$) are found by substituting eq.~(\ref{form_of_eigenspinors_SN-1even_posspin}) into eqs.~(\ref{TYPE_I_thetaN_Nodd_spin_3/2}) and (\ref{TYPE_I_thetaj_Nodd_spin_3/2}), as
\begin{align}\label{TYPE_I_3/2_thetaN_Nodd_σN-1=+_in_terms_S^(N-2)}
  {\psi}^{(I;n \ell;+;m;\rho)}_{\pm \theta_{N} }&(\theta_{N},\theta_{N-1},\bm{\theta}_{N-2})\nonumber\\
  =& \frac{\tilde{c}_{N-1}( \ell, m)}{\sqrt{2}} \frac{1}{\sqrt{2}}
    \begin{pmatrix}  (1+i)\,i \tilde{\psi}^{(0)}_{\ell m}(\theta_{N-1}) \left[\phi^{(1)}_{n\ell}(\theta_{N})\pm i \psi^{(1)}_{n\ell}(\theta_{N})\right] \tilde{\chi}_{+m \rho}(\bm{\theta}_{N-2})  \\  \\(1-i)\,\tilde{\phi}^{(0)}_{\ell m}(\theta_{N-1})\left[-\phi^{(1)}_{n\ell}(\theta_{N})\pm i \psi^{(1)}_{n\ell}(\theta_{N})\right] \tilde{\chi}_{+m\rho}(\bm{\theta}_{N-2}) \end{pmatrix},
\end{align}
\begin{align}\label{TYPE_I_3/2_thetaN-1_Nodd_σN-1=+_in_terms_S^(N-2)}
  {\psi}&^{(I;n \ell;+;m;\rho)}_{\pm \theta_{N-1} }(\theta_{N},\theta_{N-1},\bm{\theta}_{N-2})\nonumber\\ 
  &= \frac{\tilde{c}_{N-1}( \ell, m)}{\sqrt{2}} \frac{1}{\sqrt{2}}
    \begin{pmatrix}  (1+i)\, \left[i O^{(1)}_{n\ell m}(\theta_{N},\theta_{N-1})\mp  \Sigma^{(1)}_{n\ell m}(\theta_{N},\theta_{N-1})\right] \tilde{\chi}_{+m \rho}(\bm{\theta}_{N-2})  \\  \\(1-i)\,\left[-H^{(1)}_{n\ell m}(\theta_{N},\theta_{N-1})\pm i E^{(1)}_{n\ell m}(\theta_{N}, \theta_{N-1})\right] \tilde{\chi}_{+m\rho}(\bm{\theta}_{N-2}) \end{pmatrix}
\end{align}
(the functions describing the dependence on $\theta_{N}$ and $\theta_{N-1}$ in eq.~(\ref{TYPE_I_3/2_thetaN-1_Nodd_σN-1=+_in_terms_S^(N-2)}) are given by  eqs.~(\ref{function_Onlm})-(\ref{function_Sigmanlm})). The component ${\psi}^{(I;n \ell;-;m;\rho)}_{\pm \theta_{N} }$ is obtained from eq.~(\ref{TYPE_I_3/2_thetaN_Nodd_σN-1=+_in_terms_S^(N-2)}) by making the replacement $\tilde{\chi}_{+m\rho} \rightarrow \tilde{\chi}_{-m\rho}$ and exchanging $i\tilde{\psi}^{(0)}_{\ell m}$ and $\tilde{\phi}^{(0)}_{\ell m}$. The component ${\psi}^{(I;n \ell;-;m;\rho)}_{\pm \theta_{N-1} }$ is obtained from eq.~(\ref{TYPE_I_3/2_thetaN-1_Nodd_σN-1=+_in_terms_S^(N-2)}) by making the replacement $\tilde{\chi}_{+m\rho} \rightarrow \tilde{\chi}_{-m\rho}$ and exchanging $i O^{(1)}_{n\ell m}$ and $H^{(1)}_{n\ell m}$, as well as exchanging $\mp \Sigma^{(1)}_{n\ell m}$ and $\pm i E^{(1)}_{n\ell m}$. The ladder relations for the functions $\phi^{(a)}_{n \ell}(\theta_{N}),\psi^{(a)}_{n \ell}(\theta_{N}), \tilde{\phi}^{(\tilde{a})}_{\ell m}(\theta_{N-1}),\tilde{\psi}^{(\tilde{a})}_{\ell m}(\theta_{N-1})$
are given again by eqs.~(\ref{raising_phi_psi})-(\ref{lowering_tilde}). Equations (\ref{T1_typeI_relation_3/2})-(\ref{T4_typeI_relation_3/2}) hold as in the even-dimensional case.

The type-$I\!I\text{-}I$ modes on $S^{N}$ with positive spin projection index on $S^{N-1}$ ($\sigma_{N-1}=+$) are expressed as
\begin{align}\label{TYPE_II-I_3/2_thetaN_Nodd_σN-1=+_in_terms_S^(N-2)}
  {\psi}&^{(I\!I\text{-}I;n \ell;+;m;\rho)}_{\pm \theta_{N-1} }(\theta_{N},\theta_{N-1},\bm{\theta}_{N-2})\nonumber\\
  &= \frac{\tilde{c}^{(I;\tilde{r}=1)}_{N-1}( \ell, m)}{\sqrt{2}} \frac{1}{\sqrt{2}}
    \begin{pmatrix}  (1+i)\,i \tilde{\psi}^{(1)}_{\ell m}(\theta_{N-1}) \left[\phi^{(-1)}_{n\ell}(\theta_{N})\pm i \psi^{(-1)}_{n\ell}(\theta_{N})\right] \tilde{\chi}_{+m \rho}(\bm{\theta}_{N-2})  \\  \\(1-i)\,\tilde{\phi}^{(1)}_{\ell m}(\theta_{N-1})\left[-\phi^{(-1)}_{n\ell}(\theta_{N})\pm i \psi^{(-1)}_{n\ell}(\theta_{N})\right] \tilde{\chi}_{+m\rho}(\bm{\theta}_{N-2}) \end{pmatrix},
\end{align}
while
${\psi}^{(I\!I\text{-}I;n \ell;-;m;\rho)}_{\pm \theta_{N-1} }$ is obtained from eq.~(\ref{TYPE_II-I_3/2_thetaN_Nodd_σN-1=+_in_terms_S^(N-2)}) by making the replacement $\tilde{\chi}_{+m\rho} \rightarrow \tilde{\chi}_{-m\rho}$ and exchanging $i\tilde{\psi}^{(1)}_{\ell m}$ and $\tilde{\phi}^{(1)}_{\ell m}$. Equations~(\ref{T1_typeII_relation_3/2})-(\ref{T4_typeII_relation_3/2}) hold as in the even-dimensional case.

The rest of the derivation of the transformation formulae is similar to that for the even-dimensional case. We find that the transformation formulae for the type-$I$ and type-$I\!I\text{-}I$ modes are given by eqs.~(\ref{transfrmn_type_I_unnormalsd_r_SN_odd}) and (\ref{transfrmn_type_II-I_unnormalsd_r_SN_odd}), respectively, while the normalisation factor $c_{N}^{(I;r=1)}(n,\ell)$ is given by eq.~(\ref{normalization_fac_TYPE_I_spin3/2}).

\subsection{Transformation properties under spin\texorpdfstring{${(N+1)}$}{N+1} and normalisation factors for STSSH's of rank 2 on \texorpdfstring{${S^{N}}$}{SN}}\label{Appendix_transfrmn_proeprties_norm_fac_SUB4}
As mentioned in the beginning of this Appendix, the calculations needed in order to derive the transformation formulae and determine the normalisation factors for STSSH's of rank 2 on $S^{N}$ have many similarities with the case of rank-1 STSSH's, which was presented above. Therefore, below we just provide a brief description of the basic steps.

Let us begin by determining the normalisation factor for type-$I\!I\!I$ STSSH's of rank 2, $c^{(I\!I\!I;r=2)}_{N}(n ,\ell)$. In the case with $N$ even, we substitute the rank-2 type-$I\!I\!I$ modes~(\ref{TYPE_III_thetaNthetaN_negative_spin_5/2})-(\ref{TYPE_III_thetajthetak_negative_spin_5/2}) into the inner product~(\ref{define_normn_factors_SN_r}), while in the case with $N$ odd we substitute the type-$I\!I\!I$ modes~(\ref{TYPE_III_thetaNthetaN_Nodd_spin_5/2})-(\ref{TYPE_III_thetajthetak_Nodd_spin_5/2}) into the inner product~(\ref{define_normn_factors_SN_r_Nodd}). By working as in the case of rank-1 type-$I\!I$ modes, we readily find (with the use of eq.~(\ref{normalization_SN-1_tensorspinors}))
\begin{align}\label{normalization_fac_TYPE_III_spin5/2}
    \left| \frac{c_{N}^{(I\!I\!I;r=2)}( n,\ell)}{\sqrt{2}}\right|^{2}=&~  \frac{1}{2^{N-5}}\frac{\Gamma(n-\ell+1)  \Gamma(n+\ell+N)}{|\Gamma(n+\frac{N}{2})|^{2}} ,
 \end{align}
for both $N$ even and $N$ odd, which is eq.~(\ref{normln_fac_SN_TYPEB_r}) with $r=\tilde{r}_{(I\!I\!I)}=2$. 
 
Now we will determine the normalisation factor for type-$I\!I$ STSSH's of rank 2, $c^{(I\!I;r=2)}_{N}(n ,\ell)$. For $N$ even we substitute eqs.~(\ref{TYPE_II_thetaNthetaN_negative_spin_5/2})-(\ref{TYPE_II_thetajthetk_negative_spin_5/2}) into the inner product~(\ref{define_normn_factors_SN_r}), while for $N$ odd we substitute eqs.~(\ref{TYPE_II_thetaNthetaN_Nodd_spin_5/2})-(\ref{TYPE_II_thetajthetak_Nodd_spin_5/2}) into the inner product~(\ref{define_normn_factors_SN_r_Nodd}). By performing the integration over $S^{N-1}$ (using eqs.~(\ref{div_(d-psi)-vecspin_N-1}) and (\ref{normalization_SN-1_vectorspinors})), we straightforwardly find 
\begin{align}
   \left|\frac{c^{(I\!I;r=2)}_{N}(n, \ell)}{\sqrt{2}}\right|^{-2}=& ~ 2\int_{0}^{\pi}d\theta_{N}\sin^{N-3}{\theta_{N}}\left[ \left(\phi^{(0)}_{n \ell}(\theta_{N})  \right)^{2}   +\left(\psi^{(0)}_{n \ell}(\theta_{N})  \right)^{2}  \right]  \nonumber\\
    &+2\left[\left(\ell+\frac{N-1}{2} \right)^{2}-\frac{N(N+1)}{4}\right]\nonumber\\
    &\times \int_{0}^{\pi} d\theta_{N}\sin^{N-5}{\theta_{N}}\left[\left(\frac{\Gamma^{(\uparrow)}_{n \ell}(\theta_{N})}{2}  \right)^{2}   +\left(\frac{\Gamma^{(\downarrow)}_{n \ell}(\theta_{N})}{2}  \right)^{2} \right] 
 \nonumber \\
 &+2(N+1)\int_{0}^{\pi} d\theta_{N}\sin^{N-5}{\theta_{N}}\left[\left|\frac{\Delta^{(\uparrow)}_{n \ell}(\theta_{N})}{2}  \right|^{2}   +\left|\frac{\Delta^{(\downarrow)}_{n \ell}(\theta_{N})}{2}  \right|^{2} \right] 
 \nonumber\\
  &+i\left(\ell+\frac{N-1}{2} \right)\nonumber\\
  &\times \int_{0}^{\pi}d\theta_{N} \sin^{N-5}{\theta_{N}}\left[\Gamma^{(\uparrow)}_{n \ell}(\theta_{N})\Delta^{(\uparrow)}_{n \ell}(\theta_{N})   +\Gamma^{(\downarrow)}_{n \ell}(\theta_{N})\Delta^{(\downarrow)}_{n \ell}(\theta_{N})   \right]\label{normln_fac_SN_TYPE_II_5/2_long_int},
\end{align}
where $\Gamma^{(\uparrow)}_{n \ell},\Gamma^{(\downarrow)}_{n \ell},\Delta^{(\uparrow)}_{n \ell}$ and $\Delta^{(\downarrow)}_{n \ell}$ are given by eqs.~(\ref{Gammaup_function}), (\ref{Gammadown_function}), (\ref{Deltaup_function}) and (\ref{Deltadown_function}), respectively. 
The calculations can be significantly simplified by making use of the following relations:
\begin{align}\label{relation_C_and_D_with_Gamma_Delta_functions}
   & \frac{4}{\sin^{2}{\theta}}{\phi^{(0)}_{n \ell}(\theta)}= \phi^{(1)}_{n'\,\ell'}(\theta)\big|_{N\rightarrow N+2}, \nonumber \\
    & \frac{4}{\sin^{2}{\theta}}{\psi^{(0)}_{n \ell}(\theta)}= \psi^{(1)}_{n'\,\ell'}(\theta)\big|_{N\rightarrow N+2}, \nonumber \\
   & \frac{2}{\sin^{2}{\theta}}{\Gamma^{(b)}_{n \ell}(\theta)}= C^{(b)(1)}_{n'\,\ell'}(\theta)\big|_{N\rightarrow N+2}, \nonumber \\
   &\frac{2}{\sin^{2}{\theta}}{\Delta^{(b)}_{n \ell}(\theta)}= D^{(b)(1)}_{n'\,\ell'}(\theta)\big|_{N\rightarrow N+2},\hspace{5mm} (b=\uparrow, \downarrow)
\end{align}
 where $\theta \in [0,\pi]$, $n'=n-1$ and $\ell'=\ell-1$, while on the right-hand sides of the relations in eq.~(\ref{relation_C_and_D_with_Gamma_Delta_functions}) we have denoted the replacement of $N$ by $N+2$ as $N \rightarrow N+2$. The relations in eq.~(\ref{relation_C_and_D_with_Gamma_Delta_functions}) can be readily proved by using eqs.~(\ref{phi_a}), (\ref{psi_a}), (\ref{C1_a_function}), (\ref{C2_a_function}), (\ref{D1_a_function}), (\ref{D2_a_function}), (\ref{Deltaup_function}), (\ref{Deltadown_function}) (\ref{Gammaup_function}) and (\ref{Gammadown_function}).
By comparing eqs.~(\ref{normln_fac_SN_TYPE_II_5/2_long_int}) and (\ref{normln_fac_SN_TYPE_I_3/2_long_int}) and by using eq.~(\ref{relation_C_and_D_with_Gamma_Delta_functions}), we straightforwardly find
\begin{align}\label{normalization_fac_TYPE_II_spin5/2}
    \left|\frac{c^{(I\!I;r=2)}_{N}(n, \ell)}{\sqrt{2}}\right|^{2}&=2^{3} \left|\frac{c^{(I;r=1)}_{N+2}(n-1, \ell-1)}{\sqrt{2}}\right|^{2} \\
    =&~\frac{1}{2^{N}}\frac{\Gamma(n-\ell+1)  \Gamma(n+\ell+N)}{|\Gamma(n+\frac{N}{2})|^{2}} \nonumber\\
   &\times \frac{N(\ell-1) (\ell+N)}{(N+1)\left(\left[ n+N/2\right]^{2}-N^{2}/4  \right)}, 
\end{align}
which is eq.~(\ref{normln_fac_SN_TYPEB_r}) with $r=2$ and $\tilde{r}_{(B)}=\tilde{r}_{(I\!I)}=1$.

As for the normalisation of rank-2 type-$I$ modes, by working as in the case of rank-1 type-$I$ modes, we calculate the normalisation factor for $\ell=n$ using Mathematica 11.2
\begin{align}\label{normln_fac_SN_TYPE_I_5/2_ell=n}
 \left| \frac{c_{N}^{(I;r=2)}(n,n)}{\sqrt{2}}\right|^{2}=~   \frac{(n-1)(N-2)\Gamma(n+\frac{N}{2}+\frac{1}{2})}{4^{2-n}(n+1)(N+1) \sqrt{\pi}\Gamma(n+\frac{N}{2})},
\end{align}
while the normalisation factor $c_{N}^{(I;r=2)}(n, \ell)$ (for $\ell=2,3,...,n-1$) will be determined using the spin$(N+1)$ invariance of the inner product~(\ref{Spin(N+1)-invariance-inner_prod}).

In order to derive the transformation formulae (\ref{transfrmn_type_I_unnormalsd_r_SN_even}), (\ref{transfrmn_type_II-I_unnormalsd_r_SN_even}), (\ref{transfrmn_type_III_unnormalsd_5/2_SN_even}), (\ref{transfrmn_type_I_unnormalsd_r_SN_odd}), (\ref{transfrmn_type_II-I_unnormalsd_r_SN_odd}) and (\ref{transfrmn_type_III-I_unnormalsd_5/2_SN_odd}) for the STSSH's of rank 2 it is sufficient to study the following components of the Lie-Lorentz derivative~(\ref{Lie_Lorentz}):
\begin{align}\label{spinor_lie_thetaN-thetaN}
       \mathbb{L}_{\mathcal{S}} \psi_{\theta_{N} \theta_{N}}  =&~\Big( \mathcal{S}^{\mu}\partial_{\mu}+\frac{\sin{\theta_{N-1}}}{2 \sin{\theta_{N}}}\gamma^{N} \gamma^{N-1} \Big)    \psi_{\theta_{N} \theta_{N}} + \frac{2\,\sin{\theta_{N-1}}}{\sin^{2}{\theta_{N}}}   \psi_{\theta_{N}\theta_{N-1}},
\end{align}
\begin{align}\label{spinor_lie_thetaN-theta(N-1)}
       \mathbb{L}_{\mathcal{S}} \psi_{\theta_{N} \theta_{N-1}}  =&~\Big( \mathcal{S}^{\mu}\partial_{\mu}-\cos{\theta_{N-1}}\, \cot{\theta_{N}}+\frac{\sin{\theta_{N-1}}}{2 \sin{\theta_{N}}}\gamma^{N} \gamma^{N-1} \Big)    \psi_{\theta_{N} \theta_{N-1}} \nonumber\\
       &+ \frac{\sin{\theta_{N-1}}}{\sin^{2}{\theta_{N}}}   \psi_{\theta_{N-1}\theta_{N-1}}-\sin{\theta_{N-1}}\psi_{\theta_{N} \theta_{N}} 
\end{align}
and
\begin{align}\label{spinor_lie_theta(N-1)-theta(N-1)}
       \mathbb{L}_{\mathcal{S}} \psi_{\theta_{N-1} \theta_{N-1}}  =&~\Big( \mathcal{S}^{\mu}\partial_{\mu}-2\cos{\theta_{N-1}}\, \cot{\theta_{N}}+\frac{\sin{\theta_{N-1}}}{2 \sin{\theta_{N}}}\gamma^{N} \gamma^{N-1} \Big)    \psi_{\theta_{N-1} \theta_{N-1}} \nonumber\\
       &-2 \,{\sin{\theta_{N-1}}}  \, \psi_{\theta_{N}\theta_{N-1}} .
\end{align}
By working as in the case of rank-1 STSSH's, we make use of the ladder operators~(\ref{raising_phi_psi})-(\ref{lowering_tilde}) and
(after a long calculation) we find the transformation formulae~(\ref{transfrmn_type_I_unnormalsd_r_SN_even}), (\ref{transfrmn_type_II-I_unnormalsd_r_SN_even}) and (\ref{transfrmn_type_III_unnormalsd_5/2_SN_even}) for $N$ even, and the transformation formulae~(\ref{transfrmn_type_I_unnormalsd_r_SN_odd})-(\ref{transfrmn_type_III-I_unnormalsd_5/2_SN_odd}) for $N$ odd. Then, as in the case of rank-1 type-$I$ modes, the normalisation factor of rank-2 type-$I$ modes is found by combining the spin$(N+1)$ invariance of the inner product between ${\psi}^{(I;\sigma;n \ell m;\rho)}_{\pm  \mu_{1}\mu_{2}}$ and ${\psi}^{\left(I;\sigma;n (\ell+1) m;\rho\right)}_{\pm  \mu_{1} \mu_{2}}$ with eq.~(\ref{normln_fac_SN_TYPE_I_5/2_ell=n}), as
\begin{align}
    \left| \frac{c_{N}^{(I;r=2)}( n,\ell)}{\sqrt{2}}\right|^{2}=&~ \frac{1}{2^{N+3}}\frac{\Gamma(n-\ell+1)  \Gamma(n+\ell+N)}{|\Gamma(n+\frac{N}{2})|^{2}} \nonumber\\
   &\times \frac{N-2}{N+1}\frac{\ell (\ell+N-1)(\ell-1)(\ell+N)}{\left( \left[n+N/2 \right]^{2}-\left[N-2 \right]^{2}/4 \right)\left( \left[n+N/2 \right]^{2}-N^{2}/4 \right)},
 \end{align}
(for both $N$ even and $N$ odd) which is eq.~(\ref{normln_fac_SN_TYPEB_r}) with $r=2$ and $\tilde{r}_{(B)}=\tilde{r}_{(I)}=0$.
\section{Pure gauge modes}\label{appendix_pure_gauge_modes}
In this Appendix, we present details for the derivation of the pure gauge expressions~(\ref{PG_TYPEI_spin3/2}), (\ref{PG_TYPEB_spin5/2}) and (\ref{PG_TYPEI_spin5/2_partially}) for $N$ even. The calculations for $N$ odd are similar and, thus, we do not present them here. 

For later convenience, note that by making the replacements $\theta_{N} \rightarrow x(t)=\pi/2 - i t$, $n \rightarrow \tilde{M}-N/2$ [eq.~(\ref{replacements})] in the formulae~(\ref{psi_to_phi_sphere}) and (\ref{phi_to_psi_sphere}) we find
 \begin{align}
    \left(\frac{d}{dx}+\frac{N+2a-1}{2}\cot{x}+\frac{\ell+(N-1)/2}{\sin{x}}\,\right)\hat{\psi}^{(a)}_{\tilde{M} \ell}(t)&=\tilde{M}\,\hat{\phi}^{(a)}_{\tilde{M} \ell}(t) \label{psi_to_phi_sphere_analcont}\end{align}
    and
    \begin{align}
  \left(\frac{d}{d x}+\frac{N+2a-1}{2}\cot{x}-\frac{\ell+(N-1)/2}{\sin{x}} \,\right)\hat{\phi}^{(a)}_{\tilde{M} \ell}(t)&=-\tilde{M}\hat{\psi}^{(a)}_{\tilde{M} \ell}(t) \label{phi_to_psi_sphere_analcont},
  \end{align}
respectively, where $\cot{x}= i \tanh{t}$ and $\sin{x}=\cosh{t}$.
Also, let us obtain lowering operators for $\tilde{M}$ as follows. By making the replacements $N \rightarrow N+1$, $\theta_{N-1} \rightarrow x(t)=\pi/2 -it$, $\ell \rightarrow \tilde{M}-N/2$, $\tilde{a} \rightarrow a$ and $m \rightarrow \ell$ in the lowering operator~(\ref{lowering_tilde_phi_operator}) we find
\begin{align}\label{lowering_op_mass_phi}
\hat{L}^{(\tilde{M};a)}_{\phi}\hat{\phi}^{(a)}_{\tilde{M} \ell}(t)\equiv&  \left[ \sin{x}\frac{\partial}{\partial x} +\left(-\tilde{M}+\frac{N-1}{2}+a   \right)\cos{x}+\frac{\ell+\frac{N-1}{2}}{2(\tilde{M}-1/2)} \right]\hat{\phi}^{(a)}_{\tilde{M} \ell}(t)\nonumber\\
=&-\frac{\tilde{M}\left(\tilde{M}-\ell-N/2\right)}{\tilde{M}-1/2}\hat{\phi}^{(a)}_{(\tilde{M}-1) \ell}(t),
\end{align}
while by making the same replacements in the lowering operator~(\ref{lowering_tilde_psi_operator}) we find
\begin{align}\label{lowering_op_mass_psi}
 \hat{L}^{(\tilde{M};a)}_{\psi}\hat{\psi}^{(a)}_{\tilde{M} \ell}(t)\equiv&   \left[ \sin{x}\frac{\partial}{\partial x} +\left(-\tilde{M}+\frac{N-1}{2}+a   \right)\cos{x}-\frac{\ell+\frac{N-1}{2}}{2(\tilde{M}-1/2)} \right]\hat{\psi}^{(a)}_{\tilde{M} \ell}(t)\nonumber\\
 =&-\frac{\tilde{M}\left(\tilde{M}-\ell-N/2\right)}{\tilde{M}-1/2}\hat{\psi}^{(a)}_{(\tilde{M}-1) \ell}(t).
\end{align}
\subsection{Pure gauge modes for the strictly massless spin-3/2 field, \texorpdfstring{$N$}{N} even}
The type-$I$ modes~(\ref{PG_TYPEI_spin3/2}) for the strictly massless spin-3/2 field (with $\tilde{M}=\pm (N-2)/2$) are `pure gauge' modes.
In this Subsection, we prove explicitly the $t$-component of eq.~(\ref{PG_TYPEI_spin3/2}) and we describe the calculations needed in order to prove the rest of the components. Let us denote the spinors $\Lambda_{\pm}^{\left(\tilde{\ell}\right)}$ in eq.~(\ref{PG_TYPEI_spin3/2}) as $\Lambda_{\pm}^{(\sigma;\ell;\tilde{\rho})}$, where we have written out explicitly the dependence on the spin projection index $\sigma=\pm$ and the angular momentum quantum number $\ell=1,2,...\,$. Since these spinors satisfy the Dirac equation $(\slashed{\nabla}\pm i N/2)\Lambda_{\pm}^{(\sigma;\ell;\tilde{\rho})}=0$, they are given by~\cite{Letsios}
\begin{equation}\label{PG_TYPEI_spin3/2_negspin_spinor}
    {\Lambda}^{(-;\ell;\tilde{\rho})}_{\pm}(t,\bm{\theta}_{N-1})=\frac{2}{\ell} \begin{pmatrix}\hat{\phi}^{(0)}_{\frac{N}{2}, \ell}(t) \, \chi_{- \ell \tilde{\rho}}(\bm{\theta}_{N-1})  \\\mp i\hat{\psi}^{(0)}_{\frac{N}{2}, \ell}(t) \,\chi_{- \ell \tilde{\rho}}(\bm{\theta}_{N-1})  \end{pmatrix},
    \end{equation}
    \begin{equation}\label{PG_TYPEI_spin3/2_posspin_spinor}
    {\Lambda}^{(+;\ell;\tilde{\rho})}_{\pm}(t,\bm{\theta}_{N-1})=\frac{2}{\ell} \begin{pmatrix}i\hat{\psi}^{(0)}_{\frac{N}{2}, \ell}(t) \, \chi_{+ \ell \tilde{\rho}}(\bm{\theta}_{N-1})  \\\mp \hat{\phi}^{(0)}_{\frac{N}{2}, \ell}(t)\, \chi_{+ \ell \tilde{\rho}}(\bm{\theta}_{N-1})  \end{pmatrix},
    \end{equation}
    where $\hat{\phi}^{(0)}_{\frac{N}{2}, \ell}(t)$ and $\hat{\psi}^{(0)}_{\frac{N}{2}, \ell}(t)$ are found by letting $\tilde{M}=N/2$ in eqs.~(\ref{phi_aM_t}) and (\ref{psi_aM_t}), respectively, while $\chi_{\pm \ell \tilde{\rho}}$ are the eigenspinors~(\ref{eigenspinors on S_(N-1)}) of the Dirac operator on $S^{N-1}$. The factor of $2/\ell$ will be motivated naturally below [it arises from the use of the lowering operators~(\ref{lowering_op_mass_phi}) and (\ref{lowering_op_mass_psi})]. Below we prove the $t$-component of eq.~(\ref{PG_TYPEI_spin3/2}) only for negative spin projection $\sigma=-$. The case with $\sigma=+$ can be proved in the same way.
    
    The type-$I$ modes ${\Psi}^{\left(I;-;(\pm \frac{N-2}{2})\ell;\tilde{\rho}\right)}_{\mu}$ for the strictly massless spin-3/2 field ($\tilde{M}=\pm  (N-2)/2$) are found by combining eq.~(\ref{define_anal_continued_STSSH's}) with eqs.~(\ref{TYPE_I_thetaN_negative_spin_3/2}) and (\ref{TYPE_I_theta_i_negative_spin_3/2}), as
     \begin{equation}\label{TYPE_I_thetaN_negative_spin_3/2_anal_cont}
    {\Psi}^{\left(I;-;(\pm \frac{N-2}{2})\ell;\tilde{\rho}\right)}_{t}(t,\bm{\theta}_{N-1})=-i \begin{pmatrix}\hat{\phi}^{(1)}_{(\frac{N-2}{2}) \ell}(t)  \chi_{- \ell \tilde{\rho}}(\bm{\theta}_{N-1})  \\\mp i\hat{\psi}^{(1)}_{(\frac{N-2}{2}) \ell}(t) \chi_{- \ell \tilde{\rho}}(\bm{\theta}_{N-1})  \end{pmatrix}
\end{equation}
 \begin{align}\label{TYPE_I_theta_i_negative_spin_3/2_anal_cont}
   {\Psi}^{\left(I;-;(\pm \frac{N-2}{2})\ell;\tilde{\rho}\right)}_{  \theta_{j}}&(t,\bm{\theta}_{N-1})  \nonumber\\
   & = \begin{pmatrix} \hat{C}_{(\frac{N-2}{2}) \ell}^{(\uparrow)(1)}(t) \,\tilde{\nabla}_{\theta_{j}}\chi_{- \ell \tilde{\rho}}(\bm{\theta}_{N-1})+ \hat{D}_{(\frac{N-2}{2}) \ell}^{(\uparrow)(1)}(t) \,\tilde{\gamma}_{\theta_{j}}\chi_{- \ell \tilde{\rho}}(\bm{\theta}_{N-1}) \\ \\ \mp i \hat{C}_{(\frac{N-2}{2}) \ell}^{(\downarrow)(1)}(t)\,\tilde{\nabla}_{\theta_{j}}\chi_{- \ell \tilde{\rho}}(\bm{\theta}_{N-1}) \mp i \hat{D}_{(\frac{N-2}{2}) \ell}^{(\downarrow)(1)}(t)\,\tilde{\gamma}_{\theta_{j}}\chi_{- \ell \tilde{\rho}}(\bm{\theta}_{N-1}) \end{pmatrix} , 
\end{align}
where the functions $\hat{C}_{\tilde{M} \ell}^{(b)(1)}(t)$ and $\hat{D}_{\tilde{M} \ell}^{(b)(1)}(t)$ ($b=\uparrow, \downarrow$) are obtained by making the replacements $\theta_{N}\rightarrow \pi/2 - it$, $n \rightarrow \tilde{M}-N/2$, ${\phi}^{(1)}_{n \ell}(\theta_{N}) \rightarrow \hat{\phi}^{(1)}_{\tilde{M} \ell}(t)$, ${\psi}^{(1)}_{n \ell}(\theta_{N}) \rightarrow \hat{\psi}^{(1)}_{\tilde{M} \ell}(t)$ in the functions ${C}_{n \ell}^{(b)(1)}(\theta_{N})$ and ${D}_{n \ell}^{(b)(1)}(\theta_{N})$ ($b=\uparrow, \downarrow$), respectively, in eq.~(\ref{TYPE_I_theta_i_negative_spin_3/2}).

Now, let us prove eq.~(\ref{PG_TYPEI_spin3/2}) for the $t$-component of ${\Psi}^{\left(I;-;(\pm \frac{N-2}{2})\ell;\tilde{\rho}\right)}_{\mu}$. We will show that the two sides of eq.~(\ref{PG_TYPEI_spin3/2}) are equal by making use of the lowering operators~(\ref{lowering_op_mass_phi}) and (\ref{lowering_op_mass_psi}). We want to show
\begin{align}
   {\Psi}^{\left(I;-;(\pm \frac{N-2}{2})\ell;\tilde{\rho}\right)}_{t} = \left(\frac{\partial}{\partial t}  \pm \frac{i}{2} \gamma_{t}  \right)  {\Lambda}^{(-;\ell;\tilde{\rho})}_{\pm}
\end{align}
which is expressed in terms of upper and lower components as
 \begin{equation}
   -i  \begin{pmatrix}\hat{\phi}^{(1)}_{(\frac{N-2}{2}) \ell}(t)  \chi_{- \ell \tilde{\rho}}(\bm{\theta}_{N-1}) \\ \\\mp i \hat{\psi}^{(1)}_{(\frac{N-2}{2}) \ell}(t) \chi_{- \ell \tilde{\rho}}(\bm{\theta}_{N-1})  \end{pmatrix}=\frac{2}{\ell} \begin{pmatrix}\,\left[\frac{\partial}{\partial t}\hat{\phi}^{(0)}_{\frac{N}{2}, \ell}(t)-\frac{i}{2}\hat{\psi}^{(0)}_{\frac{N}{2}, \ell}(t)\right] \, \chi_{- \ell \tilde{\rho}}(\bm{\theta}_{N-1})\\  \\\mp \left[ i\frac{\partial}{\partial t}\hat{\psi}^{(0)}_{\frac{N}{2}, \ell}(t)-\frac{1}{2}\hat{\phi}^{(0)}_{\frac{N}{2}, \ell}(t)\right] \,\chi_{- \ell \tilde{\rho}}(\bm{\theta}_{N-1}) \end{pmatrix},
\end{equation}
[where we have used eq.~(\ref{even_gammas}) and $\gamma^{t}=i \gamma^{N}$] or equivalently
\begin{align}
 &\frac{\ell}{2}\,\hat{\phi}^{(1)}_{(\frac{N-2}{2}) \ell}(t)=\frac{\ell}{\sin{x}}\,\hat{\phi}^{(0)}_{(\frac{N-2}{2}) \ell}(t)=  \frac{\partial}{\partial x}\hat{\phi}^{(0)}_{\frac{N}{2}, \ell}(t)+\frac{1}{2}\hat{\psi}^{(0)}_{\frac{N}{2}, \ell}(t) \label{PG_typeI_tryingtoprove_phi}\\
 &\frac{\ell}{2}\,\hat{\psi}^{(1)}_{(\frac{N-2}{2}) \ell}(t)= \frac{\ell}{\sin{x}}\,\hat{\psi}^{(0)}_{(\frac{N-2}{2}) \ell}(t)=  \frac{\partial}{\partial x}\hat{\psi}^{(0)}_{\frac{N}{2}, \ell}(t)-\frac{1}{2}\hat{\phi}^{(0)}_{\frac{N}{2}, \ell}(t),\label{PG_typeI_tryingtoprove_psi}
\end{align}
where we have used eqs.~(\ref{phi_aM_t}) and (\ref{psi_aM_t}). Then, by using the formulae~(\ref{psi_to_phi_sphere_analcont}) and (\ref{phi_to_psi_sphere_analcont}) we rewrite eqs.~(\ref{PG_typeI_tryingtoprove_phi}) and (\ref{PG_typeI_tryingtoprove_psi}) as
 \begin{align}\label{PG_typeI_tryingtoprove_phi_final}
    \left(\sin{x}\frac{d}{dx}-\frac{1}{2}\cot{x}+\frac{\ell+(N-1)/2}{N-1}\,\right)\hat{\phi}^{(0)}_{\frac{N}{2}, \ell}(t)&=\frac{N\,\ell}{N-1}\hat{\phi}^{(0)}_{(\frac{N-2}{2}) \ell}(t) \end{align}
    and
     \begin{align}\label{PG_typeI_tryingtoprove_psi_final}
    \left(\sin{x}\frac{d}{dx}-\frac{1}{2}\cot{x}-\frac{\ell+(N-1)/2}{N-1}\,\right)\hat{\psi}^{(0)}_{\frac{N}{2}, \ell}(t)&=\frac{N\,\ell}{N-1}\hat{\psi}^{(0)}_{(\frac{N-2}{2}) \ell}(t), \end{align}
    respectively.
  It is easy to verify that eq.~(\ref{PG_typeI_tryingtoprove_phi_final}) is equal to the lowering operator~(\ref{lowering_op_mass_phi}) acting on $\hat{\phi}^{(0)}_{\frac{N}{2}, \ell}(t)$, while eq.~(\ref{PG_typeI_tryingtoprove_psi_final}) is equal to the lowering operator~(\ref{lowering_op_mass_psi}) acting on $\hat{\psi}^{(0)}_{\frac{N}{2}, \ell}(t)$. Hence, the two sides of the time component of eq.~(\ref{PG_TYPEI_spin3/2}) are equal. The rest of the components of eq.~(\ref{PG_TYPEI_spin3/2}), i.e. ${\Psi}^{\left(I;-;(\pm \frac{N-2}{2})\ell;\tilde{\rho}\right)}_{\theta_{j}} = \left(\nabla_{\theta_{j}}  \pm \frac{i}{2} \gamma_{\theta_{j}}  \right)  {\Lambda}^{(-;\ell;\tilde{\rho})}_{\pm}$ ($j=1,...,N-1$), can be proved straightforwardly just by using eqs.~(\ref{PG_typeI_tryingtoprove_phi_final}) and (\ref{PG_typeI_tryingtoprove_psi_final}), as well as formulae~(\ref{psi_to_phi_sphere_analcont}) and (\ref{phi_to_psi_sphere_analcont}).
\subsection{Pure gauge modes for the strictly massless spin-5/2 field, \texorpdfstring{$N$}{N} even}
The type-$I$ and type-$I\!I$ modes for the strictly massless spin-5/2 field (with $\tilde{M}=\pm N/2$ - see eq.~(\ref{PG_TYPEB_spin5/2})) are `pure gauge' modes.
In this Subsection, we briefly describe how to obtain the `pure gauge' expression in eq.~(\ref{PG_TYPEB_spin5/2}). We denote the vector-spinors $\lambda^{\left(B;\tilde{\ell}\right)}_{\pm \nu}(t, \bm{\theta}_{N-1})$ in eq.~(\ref{PG_TYPEB_spin5/2}) as $\lambda^{(B;\sigma;{\ell};\tilde{\rho})}_{\pm \nu}(t, \bm{\theta}_{N-1})$ ($\sigma=\pm$, $B=I , I\!I$ and $\ell=2,3,...$). Since the calculations for $\sigma=-$ and $\sigma=+$ are similar, below we discuss only the case with $\sigma=-$.

\noindent \textbf{Pure gauge modes of type-}$\bm{I}.$ The type-$I$ modes ${\Psi}^{\left(I;-;(\pm \frac{N}{2})\ell;\tilde{\rho}\right)}_{\mu \nu}$ for the strictly massless spin-5/2 field ($\tilde{M}=\pm  N/2$) are found by combining eq.~(\ref{define_anal_continued_STSSH's}) with eqs.~(\ref{TYPE_I_thetaNthetaN_negative_spin_5/2}). The `time-time component' is
     \begin{equation}\label{TYPE_I_thetaNthetaN_negative_spin_5/2_anal_cont}
    {\Psi}^{\left(I;-;(\pm \frac{N}{2})\ell;\tilde{\rho}\right)}_{tt}(t,\bm{\theta}_{N-1})=~(-1)\times \begin{pmatrix}\hat{\phi}^{(2)}_{\frac{N}{2}, \ell}(t)  \chi_{- \ell \tilde{\rho}}(\bm{\theta}_{N-1})  \\\mp i\hat{\psi}^{(2)}_{\frac{N}{2}, \ell}(t) \chi_{- \ell \tilde{\rho}}(\bm{\theta}_{N-1})  \end{pmatrix}.
\end{equation}
Similarly, since the TT vector-spinors $\lambda^{(B;-;{\ell};\tilde{\rho})}_{\pm \mu}(t, \bm{\theta}_{N-1})$ in eq.~(\ref{PG_TYPEB_spin5/2}) satisfy $$\left(\slashed{\nabla}\pm i \frac{N+2}{2} \right)\lambda^{(B;-;{\ell};\tilde{\rho})}_{\pm \mu}=0,$$they are given by the analytic continuation of the type-$I$ STSSH's of rank 1 in eqs.~(\ref{TYPE_I_thetaN_negative_spin_3/2}) and (\ref{TYPE_I_theta_i_negative_spin_3/2}). The `time component' is given by
  \begin{equation}\label{TYPE_I_thetaN_negative_spin_3/2_anal_cont_gauge_fun}
    {\lambda}^{\left(I;-;\ell;\tilde{\rho}\right)}_{\pm t}(t,\bm{\theta}_{N-1})=~-\frac{2i}{\ell-1} \begin{pmatrix}\hat{\phi}^{(1)}_{(\frac{N+2}{2}) \ell}(t)  \chi_{- \ell \tilde{\rho}}(\bm{\theta}_{N-1})  \\\mp i\hat{\psi}^{(1)}_{(\frac{N+2}{2}) \ell}(t) \chi_{- \ell \tilde{\rho}}(\bm{\theta}_{N-1})  \end{pmatrix}.
\end{equation}
(The factor of $2/(\ell-1)$ is inserted for the same reason as the factor of $2/\ell$ in eqs.~(\ref{PG_TYPEI_spin3/2_negspin_spinor}) and (\ref{PG_TYPEI_spin3/2_posspin_spinor}).) Then, by using eqs.~(\ref{TYPE_I_thetaNthetaN_negative_spin_5/2_anal_cont}) and (\ref{TYPE_I_thetaN_negative_spin_3/2_anal_cont_gauge_fun}), we expand the two sides of ${\Psi}^{\left(I;-;(\pm \frac{N}{2})\ell;\tilde{\rho}\right)}_{tt}= \left( \nabla_{t}\pm\frac{i}{2} \gamma_{t}  \right){\lambda}^{\left(I;-;\ell;\tilde{\rho}\right)}_{\pm t}$ [see eq.~(\ref{PG_TYPEB_spin5/2})] and find 
\begin{align}
 &\frac{\ell-1}{\sin{x}}\,\hat{\phi}^{(1)}_{\frac{N}{2}, \ell}(t)=  \frac{\partial}{\partial x}\hat{\phi}^{(1)}_{(\frac{N+2}{2}) \ell}(t)+\frac{1}{2}\hat{\psi}^{(1)}_{(\frac{N+2}{2}) \ell}(t) \label{PG_typeI_tryingtoprove_phi_5/2} \\
 &\frac{\ell-1}{\sin{x}}\,\hat{\psi}^{(1)}_{\frac{N}{2}, \ell}(t)=  \frac{\partial}{\partial x}\hat{\psi}^{(1)}_{(\frac{N+2}{2})\ell}(t)-\frac{1}{2}\hat{\phi}^{(1)}_{(\frac{N+2}{2}) \ell}(t).\label{PG_typeI_tryingtoprove_psi_5/2}
\end{align}
These equations are proved in the same way as eqs.~(\ref{PG_typeI_tryingtoprove_phi}) and (\ref{PG_typeI_tryingtoprove_psi}). Thus, we have verified the `time-time component' of the `pure gauge' expression~(\ref{PG_TYPEB_spin5/2}). The rest of the components of eq.~(\ref{PG_TYPEB_spin5/2}), i.e. ${\Psi}^{\left(I;-;(\pm \frac{N}{2})\ell;\tilde{\rho}\right)}_{t \theta_{j}}= \left( \nabla_{(t}\pm\frac{i}{2} \gamma_{(t}  \right){\lambda}^{\left(I;-;\ell;\tilde{\rho}\right)}_{\pm \theta_{j})}$
and ${\Psi}^{\left(I;-;(\pm \frac{N}{2})\ell;\tilde{\rho}\right)}_{\theta_{k} \theta_{j}}= \left( \nabla_{(\theta_{k}}\pm\frac{i}{2} \gamma_{(\theta_{k}}  \right){\lambda}^{\left(I;-;\ell;\tilde{\rho}\right)}_{\pm \theta_{j})}$, can be proved using eqs.~(\ref{PG_typeI_tryingtoprove_phi_5/2}) and (\ref{PG_typeI_tryingtoprove_psi_5/2}).

\noindent \textbf{Pure gauge modes of type-}$\bm{I\!I}.$ By working as in the case of type-$I$ modes presented above, we find
 \begin{equation}\label{TYPE_II_thetaNthetaN_negative_spin_5/2_anal_cont}
    {\Psi}^{\left(I\!I\text{-}\tilde{A};-;(\pm \frac{N}{2})\ell;\tilde{\rho}\right)}_{t\theta_{j}}(t,\bm{\theta}_{N-1})=~(-i)\times \begin{pmatrix}\hat{\phi}^{(0)}_{\frac{N}{2}, \ell}(t)  \tilde{\psi}^{(\tilde{A}; \ell \tilde{\rho})}_{-\theta_{j}}(\bm{\theta}_{N-1}) \\\mp i\hat{\psi}^{(0)}_{\frac{N}{2}, \ell}(t) \tilde{\psi}^{(\tilde{A}; \ell \tilde{\rho})}_{-\theta_{j}}(\bm{\theta}_{N-1})  \end{pmatrix}
\end{equation}
and
 \begin{equation}
    {\lambda}^{\left(I\!I\text{-}\tilde{A};-;\ell;\tilde{\rho}\right)}_{\pm \theta_{j}}(t,\bm{\theta}_{N-1})=~\frac{4}{\ell-1} \begin{pmatrix}\hat{\phi}^{(-1)}_{(\frac{N+2}{2}) \ell}(t)  \tilde{\psi}^{(\tilde{A}; \ell \tilde{\rho})}_{-\theta_{j}}(\bm{\theta}_{N-1})  \\\mp i\hat{\psi}^{(-1)}_{(\frac{N+2}{2}) \ell}(t) \tilde{\psi}^{(\tilde{A}; \ell \tilde{\rho})}_{-\theta_{j}}(\bm{\theta}_{N-1})  \end{pmatrix}.
\end{equation}
(Recall that for type-$I\!I$ modes we have $   {\Psi}^{\left(I\!I\text{-}\tilde{A};\sigma;(\pm \frac{N}{2})\ell;\tilde{\rho}\right)}_{tt}=0$ and $ {\lambda}^{\left(I\!I\text{-}\tilde{A};\sigma;\ell;\tilde{\rho}\right)}_{\pm t}=0$.) Then, we can verify the `pure gauge' expression~(\ref{PG_TYPEB_spin5/2}) by working as in the case of type-$I$ modes presented above.
\subsection{Pure gauge modes for the partially massless spin-5/2 field, \texorpdfstring{$N$}{N} even}
The type-$I$ modes [eq.~(\ref{PG_TYPEI_spin5/2_partially})] for the partially massless spin-5/2 field (with $\tilde{M}=\pm( N-2)/2$) are `pure gauge' modes.
Below we describe briefly how to obtain the `pure gauge' expression in eq.~(\ref{PG_TYPEI_spin5/2_partially}) for $N$ even. (We present the proof only for the $tt$-component of eq.~(\ref{PG_TYPEI_spin5/2_partially}).) We denote the Dirac spinors $\varphi_{\pm}^{\left(\tilde{\ell} \right)}(t, \bm{\theta}_{N-1})$ in eq.~(\ref{PG_TYPEI_spin5/2_partially}) as $\varphi^{(\sigma;{\ell};\tilde{\rho})}_{\pm }(t, \bm{\theta}_{N-1})$ ($\sigma=\pm$ and $\ell=2,3,...$). Again, the calculations for $\sigma=-$ and $\sigma=+$ are similar and, thus, we discuss only the case with $\sigma=-$.

For later convenience let us write down explicit expressions for lowering operators that lower the parameter $\tilde{M}$ to $\tilde{M}-2$ of the functions $\hat{f}^{(a)}_{\tilde{M} \ell}(t) \in \set{\hat{\phi}^{(a)}_{\tilde{M} \ell}(t), \hat{\psi}^{(a)}_{\tilde{M} \ell}(t)}$. By applying each of the lowering operators~(\ref{lowering_op_mass_phi}), (\ref{lowering_op_mass_psi}) twice, we find
\begin{align}\label{lowering_op_mass_f_M-2}
\hat{L}^{(\tilde{M}-1;a)}_{f}\hat{L}^{(\tilde{M};a)}_{f}\hat{f}^{(a)}_{\tilde{M} \ell}(t)  =&\Bigg[\sin^{2}{x}\,\frac{\partial^{2}}{\partial{x}^{2}} + b_{f}(x)\,\frac{\partial}{\partial{x}}+c_{f}(x)
\Bigg]\hat{f}^{(a)}_{\tilde{M} \ell}(t)\nonumber\\
=&\frac{\tilde{M}(\tilde{M}-1)(\tilde{M}-\ell-\frac{N}{2})(\tilde{M}-1-\ell-\frac{N}{2})}{(\tilde{M}-\frac{1}{2})(\tilde{M}-\frac{3}{2})}\hat{f}^{(a)}_{(\tilde{M}-2) \ell}(t),
\end{align}
(recall $x=\pi/2-it$) where
\begin{align}
    b_{f}(x)=\sin{x}\cos{x}\, \left(-2\tilde{M}+1+2a+N\right) +s_{f}\, \frac{(\ell+\frac{N-1}{2})(\tilde{M}-1)\sin{x}}{(\tilde{M}-1/2)(\tilde{M}-3/2)}
\end{align}
and
\begin{align}
    c_{f}(x)=&\frac{(\ell+\frac{N-1}{2})^{2}}{4(\tilde{M}-1/2)(\tilde{M}-3/2)}\nonumber\\
    &+s_{f}\frac{(\ell+\frac{N-1}{2})\cos{x}}{2}  \left( \frac{-\tilde{M}+a+\frac{N-1}{2}}{\tilde{M}-3/2}+\frac{1-\tilde{M}+a+\frac{N-1}{2}}{\tilde{M}-1/2} \right)\nonumber\\
    &+\left(-\tilde{M}+a+\frac{N-1}{2}\right)\left(1-\tilde{M}+a+\frac{N-1}{2}\right)\nonumber\\
    &-\sin^{2}{x}\,\left(-\tilde{M}+a+\frac{N-1}{2}\right)\left(2-\tilde{M}+a+\frac{N-1}{2}\right),
\end{align}
with $s_{f}=1$ if $\hat{f}^{(a)}_{\tilde{M} \ell}(t) = \hat{\phi}^{(a)}_{\tilde{M} \ell}(t)$ and $s_{f}=-1$ if $\hat{f}^{(a)}_{\tilde{M} \ell}(t) = \hat{\psi}^{(a)}_{\tilde{M} \ell}(t)$. 

Now we will verify the `time-time component' of eq.~(\ref{PG_TYPEI_spin5/2_partially}) with negative spin projection ($\sigma=~-$), i.e.
\begin{align}\label{PG_TYPEI_spin5/2_partially_tt}
    {\Psi}^{\left(I;-;(\pm \frac{N-2}{2}){\ell};\tilde{\rho}\right)}_{ tt}(t, \bm{\theta}_{N-1})= \left( \nabla_{t}\nabla_{t} \pm i  \gamma_{t}\nabla_{t} +\frac{3}{4} g_{t t} \right) \varphi_{\pm}^{(-;{\ell};\tilde{\rho})}(t, \bm{\theta}_{N-1}).
\end{align}
Since the spinors $\varphi^{(\sigma;{\ell};\tilde{\rho})}_{\pm }(t, \bm{\theta}_{N-1})$ satisfy the Dirac equation $\left[\slashed{\nabla}\pm i (N+2)/2\right]\varphi_{\pm}^{(\sigma;\ell;\tilde{\rho})}=0$, they are given by~\cite{Letsios}
\begin{equation}\label{PG_partially_spin5/2_negspin_spinor}
    {\varphi}^{(-;\ell;\tilde{\rho})}_{\pm}(t,\bm{\theta}_{N-1})=\frac{4}{\ell(\ell-1)} \begin{pmatrix}\hat{\phi}^{(0)}_{(\frac{N+2}{2}) \ell}(t) \, \chi_{- \ell \tilde{\rho}}(\bm{\theta}_{N-1})  \\\mp i\hat{\psi}^{(0)}_{(\frac{N+2}{2}) \ell}(t) \,\chi_{- \ell \tilde{\rho}}(\bm{\theta}_{N-1})  \end{pmatrix},
    \end{equation}
    \begin{equation}\label{PG_partially_spin5/2_posspin_spinor}
    {\varphi}^{(+;\ell;\tilde{\rho})}_{\pm}(t,\bm{\theta}_{N-1})=\frac{4}{\ell(\ell-1)} \begin{pmatrix}i\hat{\psi}^{(0)}_{(\frac{N+2}{2}) \ell}(t) \, \chi_{+ \ell \tilde{\rho}}(\bm{\theta}_{N-1})  \\\mp \hat{\phi}^{(0)}_{(\frac{N+2}{2})\ell}(t)\, \chi_{+ \ell \tilde{\rho}}(\bm{\theta}_{N-1})  \end{pmatrix},
    \end{equation}
    where the factor $4/\left(\ell\,[\ell-1]\right)$ is motivated naturally below. On the other hand, the $tt$-component of the type-$I$ mode of the partially massless spin-5/2 field is given by 
     \begin{equation}\label{TYPE_I_thetaNthetaN_negative_spin_5/2_partially}
    {\Psi}^{\left(I;-;(\pm \frac{N-2}{2})\ell;\tilde{\rho}\right)}_{tt}(t,\bm{\theta}_{N-1})=~(-1)\times \begin{pmatrix}\hat{\phi}^{(2)}_{(\frac{N-2}{2}) \ell}(t)  \chi_{- \ell \tilde{\rho}}(\bm{\theta}_{N-1})  \\\mp i\hat{\psi}^{(2)}_{(\frac{N-2}{2}) \ell}(t) \chi_{- \ell \tilde{\rho}}(\bm{\theta}_{N-1})  \end{pmatrix}.
\end{equation}
By substituting eqs.~(\ref{PG_partially_spin5/2_negspin_spinor}) and (\ref{TYPE_I_thetaNthetaN_negative_spin_5/2_partially}) into eq.~(\ref{PG_TYPEI_spin5/2_partially_tt}) we find
\begin{align}
\frac{\ell(\ell-1)}{4}\hat{\phi}^{(2)}_{(\frac{N-2}{2}) \ell}(t)  &=\frac{\ell(\ell-1)}{\sin^{2}{x}}\hat{\phi}^{(0)}_{(\frac{N-2}{2}) \ell}(t)\nonumber\\
&=\left(\frac{\partial^{2}}{\partial x^{2}} +\frac{3}{4}  \right)\hat{\phi}^{(0)}_{(\frac{N+2}{2}) \ell}(t) + \frac{\partial}{\partial x}\hat{\psi}^{(0)}_{(\frac{N+2}{2}) \ell}(t)\label{PG_partially_tryingtoprove_phi_5/2}  \\
\frac{\ell(\ell-1)}{4}\hat{\psi}^{(2)}_{(\frac{N-2}{2}) \ell}(t)  &=\frac{\ell(\ell-1)}{\sin^{2}{x}}\hat{\psi}^{(0)}_{(\frac{N-2}{2}) \ell}(t)\nonumber\\
&=\left(\frac{\partial^{2}}{\partial x^{2}} +\frac{3}{4}  \right)\hat{\psi}^{(0)}_{(\frac{N+2}{2}) \ell}(t) - \frac{\partial}{\partial x}\hat{\phi}^{(0)}_{(\frac{N+2}{2}) \ell}(t)\label{PG_partially_tryingtoprove_psi_5/2}.
\end{align}
Equation~(\ref{PG_partially_tryingtoprove_phi_5/2}) is proved using the lowering operator~(\ref{lowering_op_mass_f_M-2}) as follows. First, we express $\partial \hat{\psi}^{(0)}_{(\frac{N+2}{2}) \ell} \,/ \partial x$ in eq.~(\ref{PG_partially_tryingtoprove_phi_5/2}) in terms of $\partial \hat{\phi}^{(0)}_{(\frac{N+2}{2}) \ell} \,/\partial{x}$ and $\hat{\phi}^{(0)}_{(\frac{N+2}{2}) \ell}$ by making use of the formulae~(\ref{psi_to_phi_sphere_analcont}) and (\ref{phi_to_psi_sphere_analcont}). Then, after a long calculation, we rewrite eq.~(\ref{PG_partially_tryingtoprove_phi_5/2}) as 
\begin{align}
   \frac{\ell(\ell-1)}{\sin^{2}{x}}\hat{\phi}^{(0)}_{(\frac{N-2}{2}) \ell}  =\frac{(N-1)(N+1)}{N(N+2)\sin^{2}{x}}\,\left(\hat{L}^{(\frac{N}{2};a=0)}_{f}\hat{L}^{(\frac{N+2}{2};a=0)}_{f}\hat{\phi}^{(0)}_{(\frac{N+2}{2}) \ell}\right),
\end{align}
which is readily verified using the lowering relation~(\ref{lowering_op_mass_f_M-2}). Equation~(\ref{PG_partially_tryingtoprove_psi_5/2}) is proved in the same way. Thus, we have verified the $tt$-component of the `pure gauge' expression~(\ref{PG_TYPEI_spin5/2_partially}).

Let us now show that our `pure gauge' expression for the type-$I$ modes ${\Psi}^{\left(I;\sigma;(\tilde{M}=+1)\ell;\tilde{\rho}\right)}_{\mu \nu}$ on $dS_{4}$ in eq.~(\ref{PG_TYPEI_spin5/2_partially}) is equal to the gamma-traceless part of the gauge transformation that is proposed in Ref.~\cite{Deser_Waldron_phases} (for a specific choice of the spinor gauge function in the gauge transformation of Ref.~\cite{Deser_Waldron_phases}). In order to compare our results with the results of Ref.~\cite{Deser_Waldron_phases} we let $N=4$ and $\tilde{M}=+(N-2)/2=+1$ in eq.~(\ref{PG_TYPEI_spin5/2_partially}). [Now, the spinors $\varphi_{+}^{(\sigma;{\ell};\tilde{\rho})}$ in eq.~(\ref{PG_TYPEI_spin5/2_partially}) satisfy $\slashed{\nabla}\varphi_{+}^{(\sigma;{\ell};\tilde{\rho})}=-3i\varphi_{+}^{(\sigma;{\ell};\tilde{\rho})}$.] By using units in which the cosmological constant is $\Lambda=3$, the gauge transformation for the partially massless spin-5/2 field $\psi_{\mu \nu}$ in Ref.~\cite{Deser_Waldron_phases} is
\begin{align}
 \delta \psi_{\mu \nu}=& \left(\nabla_{(\mu} \nabla_{\nu)}-\frac{1}{4} \gamma_{(\mu} \nabla_{\nu)}\slashed{\nabla}+\frac{15}{16}g_{\mu \nu} \right) \epsilon \\
 =&\left(\nabla_{(\mu} \nabla_{\nu)}+\frac{3i}{4} \gamma_{(\mu} \nabla_{\nu)}+\frac{15}{16}g_{\mu \nu} \right) \epsilon,
\end{align}
where we have chosen $\epsilon$ to be a solution of the equation $\slashed{\nabla} \epsilon= - 3i \,\epsilon$. (For this choice it is clear that our spinors $\varphi_{+}^{(\sigma;{\ell};\tilde{\rho})}$ are the mode functions corresponding to the field $\epsilon$.) Note that for this choice of $\epsilon$ the gauge transformation of the auxiliary field is zero - see Ref.~\cite{Deser_Waldron_phases}. Also, for this choice of $\epsilon$ it can be readily verified that $g^{\mu \nu} \delta \psi_{\mu \nu}=0$, but $\gamma^{\mu} \delta \psi_{\mu \nu} \neq 0$.   Let $\delta \psi'_{\mu \nu}$ be the gamma-traceless part of $\delta \psi_{\mu \nu}$, i.e.
\begin{align}
    \delta \psi_{\mu \nu}'= \delta \psi_{\mu \nu}-\frac{\gamma_{\mu}}{6} \gamma^{\alpha}\delta \psi_{\alpha \nu}-\frac{\gamma_{\nu}}{6} \gamma^{\alpha}\delta \psi_{\alpha \mu},
\end{align}
where $\gamma^{\alpha} \delta \psi'_{\alpha \nu}=0$ and $g^{\mu \nu}\delta \psi'_{\mu \nu}=0$.
Then, we can straightforwardly show that
\begin{align}
 \delta \psi'_{\mu \nu}=&   \left( \nabla_{(\mu}\nabla_{\nu)} +i  \gamma_{(\mu}\nabla_{\nu)} +\frac{3}{4} g_{\mu \nu} \right) \epsilon,
\end{align}
which is in precise agreement with the expression for our type-$I$ modes in eq.~(\ref{PG_TYPEI_spin5/2_partially}).
\\
\\
\noindent \textbf{DATA AVAILABILITY STATEMENT.} No new data were created or analysed in this study.
\end{widetext}
%

\end{document}